# Regularity Properties and Pathologies of Position-Space Renormalization-Group Transformations

Aernout C. D. van Enter
*Institute for Theoretical Physics*
*Rijksuniversiteit Groningen*
*P.O. Box 800*
*9747 AG Groningen*
*THE NETHERLANDS*
AENTER@RUGTH4.TH.RUG.NL

Roberto Fernández
*Institut de Physique Théorique*
*Ecole Polytechnique Fédérale de Lausanne*
*PHB – Ecublens*
*CH–1015 Lausanne*
*SWITZERLAND*
FERNANDE@ELDP.EPFL.CH

Alan D. Sokal
*Department of Physics*
*New York University*
*4 Washington Place*
*New York, NY 10003*
*USA*
SOKAL@ACF3.NYU.EDU

October 16, 1992

## Abstract

We reconsider the conceptual foundations of the renormalization-group (RG) formalism, and prove some rigorous theorems on the regularity properties and possible pathologies of the RG map. Regarding regularity, we show that the RG map, defined on a suitable space of interactions (= formal Hamiltonians), is always single-valued and Lipschitz continuous on its domain of definition. This rules out a recently proposed scenario for the RG description of first-order phase transitions. On the pathological side, we make rigorous some arguments of Griffiths, Pearce and Israel, and prove in several cases that the renormalized measure is not a Gibbs measure for any reasonable interaction. This means that the RG map is ill-defined, and that the conventional RG description of first-order phase transitions is not universally valid. For decimation or Kadanoff transformations applied to the Ising model in dimension $d \geq 3$, these pathologies occur in a full neighborhood $\{\beta > \beta_0, |h| < \epsilon(\beta)\}$ of the low-temperature part of the first-order phase-transition surface. For block-averaging transformations applied to the Ising model in dimension $d \geq 2$, the pathologies occur at low temperatures for arbitrary magnetic-field strength. Pathologies may also occur in the critical region for Ising models in dimension $d \geq 4$. We discuss in detail the distinction between Gibbsian and non-Gibbsian measures and the possible occurrence of the latter in other situations, and give a rather complete catalogue of the known examples. Finally, we discuss the heuristic and numerical evidence on RG pathologies in the light of our rigorous theorems.

# Contents

# 1 Introduction and Summary of Results

## 1.1 General Introduction

A principal tenet of the renormalization-group (RG) theory of phase transitions [365, 277, 254, 118] is that the RG map, defined on a suitable space of Hamiltonians, is *smooth* (i.e. analytic or at least several-times differentiable), even on phase-transition surfaces. The singularities associated with critical points [365, 277, 254, 118] and first-order phase transitions [280, 217, 119] are then explained in terms of the behavior of the RG map under infinite iteration.

This picture of a smooth RG map has, however, been questioned, particularly as regards the behavior at or near a *first-order* phase transition. On the one hand, the existence of several phases raises the possibility that the RG map may be *discontinuous* or *multi-valued* [173, 197, 305] on the first-order transition surface, as the numerical evidence reported by several groups [33, 233, 72, 163] seems to suggest. On the other hand, Griffiths and Pearce [172, 173, 171] have pointed out some "peculiarities" of the commonly used discrete-spin position-space RG transformations (decimation, majority rule, etc.); in particular they suggested that the RG map for the two-dimensional Ising model must have singularities (or other strange behavior) in a rather large part of $(\beta, h)$-plane (see also [54, 334]).[1] In an important but apparently little-known paper, Israel [207] clarified the nature of the Griffiths-Pearce peculiarities: he showed that in at least one case the renormalized system cannot be described by a Boltzmann-Gibbs prescription for any reasonable Hamiltonian, i.e. the renormalized measure is *non-Gibbsian*.

In this paper[2] we reconsider the conceptual foundations of the RG formalism, and prove some rigorous theorems on the nature of the RG map. On the one hand, we prove two Fundamental Theorems on the single-valuedness and continuity of the RG map; these theorems *rule out* the discontinuous-flow scenario proposed in references [33, 233, 72, 163, 197, 305]. On the other hand, we prove, completing and extending Israel's argument, that in several cases the RG map is ill-defined for a much more basic reason: *the renormalized Hamiltonian may fail to exist altogether*. This implies that the conventional RG description of first-order phase transitions [280, 217, 119] is not valid either (at least in these models and for these RG transformations). Moreover, this pathology can occur in the *vicinity* of — not only *at* — a first-order phase transition: for the Ising model in dimension $d \geq 3$ it occurs in a full neighborhood $\{\beta > \beta_0, |h| < \epsilon(\beta)\}$ of the low-temperature part of the first-order phase-transition surface. Indeed, for certain block-averaging transformations we are able to show that the pathology occurs at low temperature and *all* magnetic fields $h$.

Our point of view is the following: An RG map is defined initially as a rule (which may be either deterministic or stochastic) for generating a configuration $\omega'$ of "block spins" given a configuration $\omega$ of "original spins". Mathematically this is given by a probability kernel $T(\omega \to \omega')$. Using such a map, one can immediately define a probability distribution $\mu'(\omega')$ of block spins from any given probability distribution $\mu(\omega)$ of original spins, namely

$$\mu'(\omega') \;=\; (\mu T)(\omega') \;\equiv\; \sum_{\omega} \mu(\omega)\, T(\omega \to \omega') \;. \tag{1.1}$$

---

[1] Similar peculiarities, and also different ones, have been found by Hasenfratz and Hasenfratz [189]. The phenomena found in Section 4 of their paper are very closely related to those of Griffiths and Pearce, while those in Sections 2 and 3 seem to be quite different.

[2] Brief summaries of our results have appeared previously [352, 353, 354, 355].

In other words, the RG map is easily defined as a map *from measures to measures*. On the other hand, most applications of the renormalization group assume (and in fact need) that the RG map is defined as a map *from Hamiltonians to Hamiltonians*. That is, if $\mu$ is the Gibbs measure for a statistical-mechanical system with Hamiltonian $H$, then one usually assumes that $\mu'$ is the Gibbs measure for a system with some Hamiltonian $H'$; this is taken to define an RG map $\mathcal{R}$ on some suitable space of Hamiltonians, by the diagram

$$\begin{array}{ccc} \mu & \xrightarrow{T} & \mu' \equiv \mu T \\ \uparrow & & \downarrow \\ H & \xrightarrow{\mathcal{R}} & H' \end{array} \tag{1.2}$$

*Formally* the relation between a Hamiltonian and its corresponding Gibbs measure is given by $\mu = \text{const} \times e^{-H}$, and hence the RG map on the space of Hamiltonians is defined *formally* by

$$H'(\omega') \;=\; (\mathcal{R}H)(\omega') \;=\; -\log\left[\sum_{\omega} e^{-H(\omega)}\, T(\omega \to \omega')\right] + \text{const} \;. \tag{1.3}$$

However, *this formula is valid only in finite volume*; in infinite volume, the Hamiltonian $H(\omega)$ is ill-defined (its value is almost surely $\pm\infty$), and the connection between a formal Hamiltonian (more precisely, an *interaction*) and its corresponding Gibbs measure(s) is much more complicated. We emphasize that this is not a mere mathematical nicety, but contains the fundamental *physics* of phase transitions. In finite volume, where the formula $\mu = \text{const} \times e^{-H}$ makes sense, all thermodynamic functions are manifestly analytic functions of the parameters in the Hamiltonian, so a phase transition is impossible. Phase transitions occur *only* for infinite-volume systems. Now, one feature of the infinite-volume limit is the possibility that the Gibbs measure may be non-unique: corresponding to a given formal Hamiltonian (= interaction) there may exist *several distinct* Gibbs measures, each one corresponding to a distinct thermodynamically stable "pure phase" of the system. Indeed, such a multiple-phase coexistence can serve as one definition of a first-order phase transition. Therefore, for Hamiltonians $H$ with a non-unique Gibbs measure (= Hamiltonians lying on a first-order phase-transition surface), the upward vertical arrow in (1.2) may well be a multi-valued map; and one might fear that this could cause the putative RG map $\mathcal{R}$ to become multi-valued as well. (We shall see later, however, that this pathology *cannot* occur.) Even more subtle problems arise from the downward vertical arrow in (1.2): though at most one $H'$ can correspond to a given $\mu'$ [174], it can happen that *no $H'$* corresponds to the given $\mu'$ — that is, *it can occur that the image measure $\mu'$ is not a Gibbs measure for any reasonable Hamiltonian*. In Section 3 we shall show that such non-Gibbsianness is the *only* way that the RG map can become grossly "pathological". In Section 4 we shall show that this pathology does in fact occur in a rather wide variety of examples.

(Of course, we must make precise what we mean by a "reasonable" Hamiltonian, and convince the reader that our class is sufficiently wide to capture fully the intuitive notion of "physical reasonableness". This will be discussed in detail in Sections 2 (especially 2.3.3, 2.4.4 and 2.6.7) and 6.1.2. Suffice it to say now that we allow interactions of arbitrarily long range and involving arbitrarily many spins, subject only to the condition of absolute summability.)

These results leave RG theory in roughly the following situation: The RG map has been proven to be well-defined and analytic at high temperature [207, 212] and, in some cases, at large magnetic field [173] — regions in which phase transitions are absent, and RG theory is unnecessary. The RG map has been proven in some cases to be ill-defined at low temperature (Section 4). Near the critical point — where RG theory is of the most interest — very little is known about the behavior of the RG map, but there are some indications of possible pathologies in dimensions $d \gtrsim 4$ (Sections 4.4 and 5.2). Nevertheless, RG *ideas* have been of great value even in situations in which the strict Wilson prescription (1.2) has not been — and maybe even cannot be — implemented [147, 148, 150, 149, 151, 181, 188, 53, 134, 135, 137, 4, 3, 112, 145, 146, 44, 52, 183, 184, 185, 187, 186]. We discuss these issues further in Section 6.1.

## 1.2 Plan of This Paper (Or, What to Read and What to Skip)

We hope that this paper will be read (and readable) both by theoretical physicists — particularly those doing real-space RG and Monte Carlo RG calculations — and by mathematical physicists interested in the statistical mechanics of lattice systems. For this reason we have given in Section 2 a rather detailed (and, we hope, comprehensible) summary of the general theory of infinite-volume lattice systems, in which we make precise the concepts of "interaction", "Hamiltonian", "Gibbs measure" and "equilibrium measure" and the connections between them. As we have argued, a careful treatment of the infinite-volume problem is essential for a correct *physical* understanding of phase transitions in general, and of the renormalization group in particular. We hope that Section 2 will be useful to physicists who may not be familiar with these ideas. Some abstract mathematics is of necessity involved; we have tried hard to minimize "mathematics for the sake of mathematics", and to introduce only those mathematical objects which correspond to clear *physical* concepts. The reader can judge whether we have succeeded.[3]

---

[3]The "experts" will notice a few innovations and new results in Section 2 and the associated Appendix A: the extensive discussion of physical equivalence (Sections 2.3.5, 2.4.3, 2.4.5, 2.4.6, A.3.4, A.3.5 and A.3.7); some precise estimates on bulk vs. surface effects (Sections 2.4.7, 2.4.8 and A.3.8);

In Section 3 we define our general framework for studying renormalization transformations, and prove the two Fundamental Theorems on single-valuedness and continuity of the RG map. These theorems show that the RG map $\mathcal{R}$ can never become multi-valued or discontinuous; but it can become *non*-valued, which occurs if the image measure $\mu'$ is non-Gibbsian. This focus on non-Gibbsianness — which is the real message of our paper — is a profound insight due to Israel [207]. In Section 4 we complete and extend Israel's argument, and show that in a large class of examples (always at low temperature, but not only on phase-transition surfaces) the image measure $\mu'$ is indeed non-Gibbsian. We also discuss some other operations that can lead to non-Gibbsian measures, including one which is relevant to "large-cell" majority-rule maps; and we give a rather complete catalogue of the known examples of non-Gibbsianness. In Section 5 we discuss the heuristic and numerical evidence on RG pathologies in the light of our rigorous theorems. We also discuss some heuristic arguments for the possible existence of RG pathologies *in the critical region* for Ising-to-Ising RG maps in dimension $d \gtrsim 4$. In Section 6 we summarize our results and discuss their implications. We conclude with a list of open questions.

In Appendix A we supply the proofs of some theorems that are stated without proof in Section 2. In Appendix B we provide a brief summary of Pirogov-Sinai theory, which is needed as a technical tool in Section 4.[4] In Appendix C we solve a Diophantine equation arising in our study of the majority-rule map.

Let us again express our hope that the reader will at least peruse Section 2. (Hey, we spent a long time on it, and we think it is rather good pedagogy.) However, for the reader who is truly allergic to abstract mathematics, we offer the following advice: read the remainder of this Introduction, followed by Sections 3 (skipping the proofs), 4.1.1, 4.4 (skipping the proofs), 5.2 and 6. Finally, for the reader who is allergic both to abstract mathematics and to 250-page papers, we offer "RG lite": read the remainder of this Introduction, and then skip to the Conclusion (Section 6.1).

## 1.3 Summary of First and Second Fundamental Theorems

We would like next to summarize the two Fundamental Theorems and give the physical intuition behind their proofs. Consider, for concreteness, the Ising model in dimension $d \geq 2$ at low temperature ($\beta \gg \beta_c$) and zero magnetic field. At such a point there are precisely two [141] pure phases (extremal translation-invariant Gibbs measures): the positively magnetized (or "+") phase $\mu_+$, and the negatively magnetized (or "−") phase $\mu_-$. These pure phases can be obtained by taking the infinite-volume limit

---

a consistent use of van Hove convergence and *complete* subadditivity (Sections 2.4.1, A.3.3, A.3.4 and A.3.5); and some interesting counterexamples concerning the pressure and entropy (Appendix A.5.2). The first two of these innovations play a crucial role in our proof of the Second Fundamental Theorem (Section 3.3).

[4]The "experts" will notice some small innovations in our presentation of Pirogov-Sinai theory, notably our emphasis on questions of *uniformity*. This plays an important role in our application to the Kadanoff transformation: see Section 4.3.3 and Appendix B.5.4.

using "+" or "−" boundary conditions, respectively. Both of these phases have a large magnetization $\pm M_0$ and a small correlation length $\xi$. Now let us apply some block-spin transformation $T$, such as the majority-rule transformation on blocks of size $2^d$. Then the image measures $\mu'_\pm = \mu_\pm T$ will have a yet larger magnetization $\pm M'_0$ (since minorities tend to get "outvoted") and a yet smaller correlation length $\xi'$ (we expect roughly $\xi' \approx \xi/2$, since distances are being scaled by a factor of 2). We now ask: These image measures $\mu'_\pm$ are typical of what kind of Hamiltonian (if any)?

One possibility — and the one conventionally assumed [280, 217, 119] — is that the RG flow is toward lower temperatures *along the $h = 0$ line*.[5] This picture is certainly consistent with the intuitive idea that magnetization increases and correlation length decreases under the RG map. In this scenario [Figure 1(a)], the two image measures $\mu'_\pm$ would be Gibbsian for the *same* Hamiltonian $H'$, and this Hamiltonian would be invariant under the $\sigma \to -\sigma$ symmetry.

A different possibility was advocated by Decker, Hasenfratz and Hasenfratz [72]. In this scenario [Figure 1(b)], the RG flow is *discontinuous* at the phase-transition line $h = 0$: Hamiltonians $H$ with an infinitesimal positive (resp. negative) magnetic field $h$ get mapped by a single RG step to renormalized Hamiltonians $H'$ having a strictly positive (resp. strictly negative) magnetic field $h'$. Furthermore, *at $h = 0$* the renormalized Hamiltonian $H'$ depends on which pure phase, $\mu_+$ or $\mu_-$, one uses in the top left corner of (1.2): the image measure $\mu'_+$ would be Gibbsian for some Hamiltonian $H'_+$ having (among other couplings) a strictly positive magnetic field, while the image measure $\mu'_-$ would be Gibbsian for some Hamiltonian $H'_-$ having a strictly negative magnetic field. (Obviously $H'_+$ and $H'_-$ would be related by the $\sigma \to -\sigma$ symmetry, i.e. by reversing the signs of all *odd* couplings.) In this scenario, therefore, the RG map $\mathcal{R}$ is *discontinuous* as one approaches the phase-transition line, and *multi-valued* on that line.[6] This picture is also consistent with the intuitive idea that magnetization increases and correlation length decreases under the RG map.

How can we distinguish between these two scenarios? Otherwise put: Suppose we are given a measure $\mu'$ with a large positive magnetization and a small (but nonzero) correlation length. Does this measure come from a Hamiltonian $H'$ with $\beta$ large and $h = 0$, or does it come from a Hamiltonian with $\beta$ not so large (possibly even small) and $h$ large and positive? *Both* of these regions in the $(\beta, h)$-plane correspond to a large positive magnetization and a small correlation length. How can we distinguish between the two?

The answer has to do with the *large-deviation* properties of the measure $\mu'$. Let $\Lambda$ be a large cubical box of side $L$, and let $\mathcal{M}_\Lambda \equiv \sum_{x \in \Lambda} \sigma_x$ be the total spin in $\Lambda$ (a random variable). Clearly there is an overwhelming probability that $\mathcal{M}_\Lambda$ will be positive (and in fact very close to its mean value $L^d M'_0 = L^d \langle \sigma \rangle_{\mu'}$); but how rare is it to have $\mathcal{M}_\Lambda$ *negative*? If $\mu'$ is a Gibbs measure for some Hamiltonian with $h > 0$, then the event $\mathcal{M}_\Lambda < 0$ is suppressed by the *bulk* magnetic field:

$$\text{Prob}_{\mu'}(\mathcal{M}_\Lambda < 0) \sim e^{-O(L^d)} . \tag{1.4}$$

---

[5] More precisely, the flow would take place in an infinite-dimensional space of couplings, but would respect the $\sigma \to -\sigma$ symmetry. That is, second-nearest-neighbor and longer-distance pair couplings, four-spin couplings, six-spin couplings and so forth would certainly be induced; but *no* magnetic fields, three-spin couplings or other *odd* interactions would arise.

[6] This possibility was suggested earlier, in the context of the 3-state Potts model in three dimensions, by Blöte and Swendsen [33] and with especial clarity by Rebbi and Swendsen [305, p. 4099]. It was also suggested, in the context of a mean-field computation, by Hudák [197].

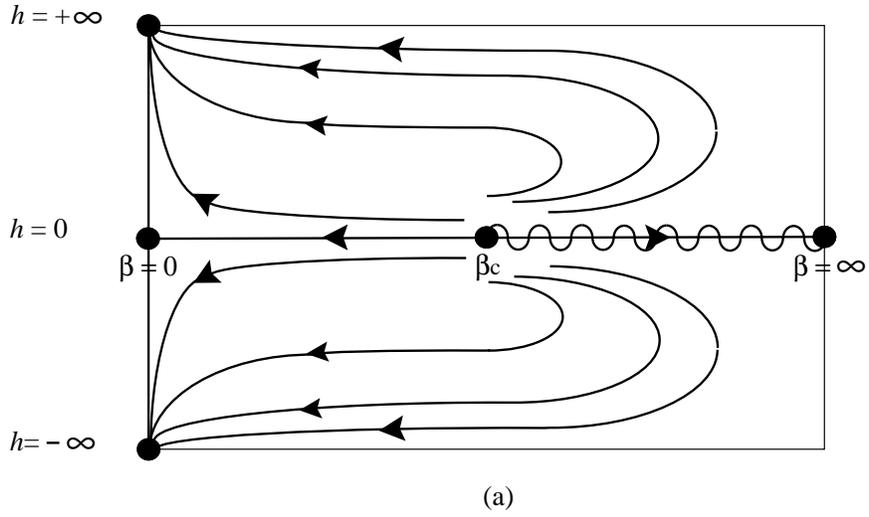

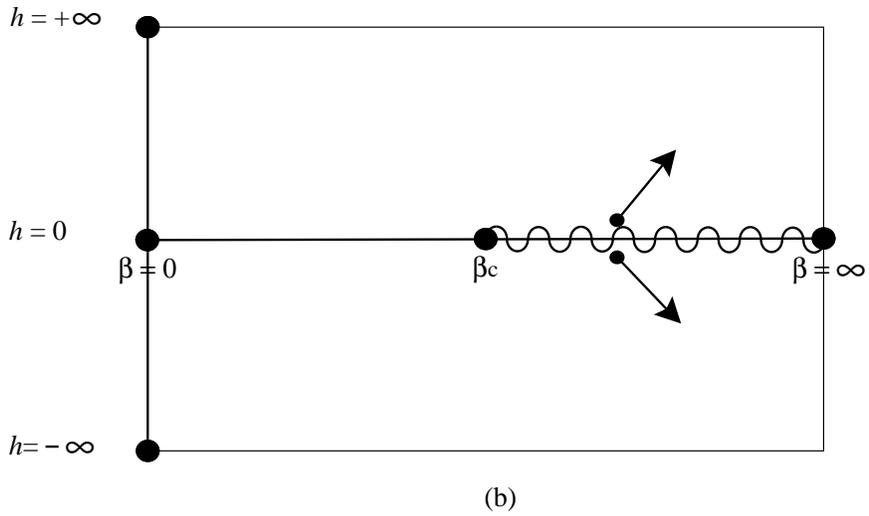

Figure 1: Two possible scenarios for the RG flow in the Ising model at low temperature. (a) RG map is continuous and single-valued on the phase-transition line. (b) RG map is discontinuous and multi-valued on the phase-transition line.

On the other hand, if $\mu'$ is a Gibbs measure for some Hamiltonian with $h = 0$ and $\beta > \beta_c$, then the event $\mathcal{M}_\Lambda < 0$ is suppressed only by a *surface* energy:

$$\text{Prob}_{\mu'}(\mathcal{M}_\Lambda < 0) \sim e^{-O(L^{d-1})} . \tag{1.5}$$

It is now easy to decide between the two scenarios for the RG flow. In the starting measure $\mu_+$, the occurrence of a large region with negative total spin is suppressed only like $e^{-O(L^{d-1})}$; roughly speaking, the measure $\mu_+$ "knows" that it is degenerate with the measure $\mu_-$. But then in the block-spin measure $\mu'_+ = \mu_+ T$, there must also be a probability $\gtrsim e^{-O(L^{d-1})}$ of observing a negative total spin (since a net negative original spin implies, with high probability, a net negative block spin). Since this contradicts (1.4), we conclude that $\mu'_+$ *cannot* be the Gibbs measure of a Hamiltonian with strictly positive magnetic field. Picturesquely, the image measure $\mu'_+$ "remembers" that it arose from an original Hamiltonian $H$ with coexisting phases. Therefore, the discontinuous-flow scenario is impossible; the RG map *cannot* be multi-valued or discontinuous.

It is a relatively short step from these intuitive ideas to a rigorous proof. In Section 3 we prove, in great generality, the following two theorems:

> **First fundamental theorem.** If $\mu$ and $\nu$ are Gibbs measures for the same interaction, then either $\mu T$ and $\nu T$ are both non-Gibbsian, or else there exists an interaction for which both $\mu T$ and $\nu T$ are Gibbs measures. In the latter case, this is the *only* interaction for which either $\mu T$ or $\nu T$ is a Gibbs measure. Therefore, the renormalization-group map $\mathcal{R}$ cannot be multi-valued.

> **Second fundamental theorem.** The renormalization-group map $\mathcal{R}$ is continuous (in fact, Lipschitz continuous) on the domain where it is defined.

Of course, these summaries of the theorems are not quite precise: we need to make clear, for example, in what space of interactions we are working, and in what norm we are defining continuity. The detailed statement of the Fundamental Theorems can be found in Sections 3.2 and 3.3, respectively. The proofs of the Fundamental Theorems are based on the general theory of infinite-volume lattice systems developed in Section 2. These two theorems make clear that the *only* way in which the RG map can become grossly pathological is for it to be *undefined*, i.e. for the image measure $\mu'$ to be non-Gibbsian.

## 1.4 Summary of Griffiths-Pearce-Israel Pathologies

This is not, however, the end of the story: although the discontinuous-flow scenario for the RG map in the low-temperature Ising model is not correct, the traditional scenario is in many cases not correct either! The First Fundamental Theorem leaves open the possibility that the image measure $\mu' = \mu T$ may be *non-Gibbsian*, in which case the RG map $\mathcal{R}$ would be *undefined*. It turns out that *this pathology does in fact occur* in a rather wide variety of examples. The occurrence of non-Gibbsianness for the image measure $\mu'$ was first pointed out by Israel [207] in one of the cases suggested by Griffiths and Pearce [173, 171]. In Section 4 we complete and extend Israel's argument, and show that in a large class of examples (always at low temperature, but not only on phase-transition surfaces) the image measure $\mu'$ is non-Gibbsian.

The non-Gibbsianness arises from the fact — already noted by Griffiths and Pearce — that the "internal spins" (the variables being integrated over in the RG transformation) may undergo a first-order phase transition for some fixed block-spin configuration $\omega'_{\text{special}}$. Moreover, in some cases the different phases (= Gibbs measures) of the internal-spin system can be selected by an appropriate choice of *block-spin* boundary conditions. In this way, information can be transmitted from distant block spins to the block spin at the origin via the internal spins in the intermediate region, even when the block spins in the intermediate region are *fixed*. As a consequence, the renormalized measure $\mu'$ violates a very weak locality condition — *quasilocality*, see Section 2.3.3 — which is obeyed by every Gibbs measure coming from a reasonable interaction. It follows immediately that the renormalized measure $\mu'$ must be non-Gibbsian.

It is at first surprising that the existence of pathologies for a *single* block-spin configuration $\omega'_{\text{special}}$ — which has, of course, probability zero — can nevertheless cause the non-Gibbsianness of the renormalized measure; and indeed, this fact *alone* is *not* sufficient for concluding non-Gibbsianness. Rather, what happens in these examples is that for block-spin configurations which are *near* (in the product topology) to $\omega'_{\text{special}}$ — namely, those which agree with $\omega'_{\text{special}}$ in a large cube and differ from it outside — the internal-spin phase depends sensitively on the block spins *outside* the cube. These configurations have a small but *nonzero* probability, and this turns out to be sufficient for proving non-Gibbsianness. The details of the proof are given in Sections 4.1–4.3.

We prove non-Gibbsianness at low temperature and zero magnetic field in the following examples:

- Decimation with any spacing $b \geq 2$, for the Ising model in any dimension $d \geq 2$.

- The Kadanoff transformation with finite $p$ and arbitrary block size $b \geq 1$, for the Ising model in dimension $d \geq 2$.[7]

- The majority-rule transformation with $7 \times 7$ (or $41 \times 41$, $239 \times 239$, ...) blocks for the two-dimensional Ising model.

---

[7]In earlier versions of this work [352, 353], we claimed this result only for *small p*. Subsequently we found a proof valid for *all* $0 < p < \infty$, which we present here.

- Averaging transformation with any *even* block size $b \geq 2$, for the Ising model in any dimension $d \geq 2$.

Moreover, in several cases we can prove that these pathologies are present also at *nonzero* magnetic field. For the first two examples, we prove non-Gibbsianness in dimension $d \geq 3$ in a full neighborhood $\{\beta > \beta_0, |h| < \epsilon(\beta)\}$ of the low-temperature part of the first-order phase-transition surface. In the last example, the pathologies can be proven in any dimension $d \geq 2$ and for *arbitrary* values of the magnetic field, again at low temperature. These latter results make clear that the Griffiths-Pearce-Israel pathologies are *not* associated with the fact that the original model is *sitting on* a phase-transition surface. Rather, it suffices that a first-order phase transition can be induced in the internal-spin system by choosing an appropriate block-spin configuration. For this we need to work at low temperature but not necessarily close to the phase-transition surface.

Though we have not yet been able to demonstrate non-Gibbsianness for the majority-rule transformation on $2 \times 2$ or $3 \times 3$ blocks, or for any block size in dimension $d \geq 3$, we feel that the obstacles are technical rather than fundamental. Indeed, the results in Section 4 suggest that non-Gibbsianness may be the *normal* situation for RG maps at low temperature and/or near a first-order phase transition.

The reader will probably not be surprised that the decimation transformation is "pathological": this transformation, unlike other RG transformations, does not in any sense integrate out the "high-momentum modes" and leave the "low-momentum modes"; it merely integrates out one sublattice and leaves another. In particular, if the sublattice of internal (integrated-out) spins is *connected*, it is hardly surprising that the internal-spin system can exhibit a phase transition, and that this can give rise to RG pathologies. We therefore want to stress that the same pathology — non-Gibbsianness after one renormalization step — is also present at low temperature for at least some Kadanoff, majority-rule and block-averaging transformations. These latter transformations *do* (at least seemingly) integrate out the "high-momentum modes" and leave the "low-momentum modes", and they have been generally considered to be well-behaved. Indeed, nearly all real-space RG studies of Ising models have used some variant of these transformations. It is thus a highly non-trivial fact that these RG maps can be ill-defined at low temperature.

# 2  Infinite-Volume Lattice Systems: General Formalism

Consider a classical statistical-mechanical system with configuration space $\Omega$, Hamiltonian $H$ and *a priori* measure $\mu_0$. The Boltzmann-Gibbs distribution $\mu_{BG}$ for this system in the canonical ensemble at inverse temperature $\beta$ can be characterized in either of two ways:

(a) *Explicit formula.*
$$d\mu_{BG}(\omega) \;=\; Z^{-1}\, e^{-\beta H(\omega)}\, d\mu_0(\omega)\,, \qquad (2.1)$$

where of course
$$Z = \int e^{-\beta H(\omega)} d\mu_0(\omega) . \qquad (2.2)$$

(b) *Variational principle.* $\mu_{BG}$ is that probability measure which maximizes entropy minus $\beta$ times mean energy:

$$\mu_{BG} \text{ maximizes } S(\mu|\mu_0) - \beta E(H,\mu) , \qquad (2.3)$$

where

$$\begin{aligned} S(\mu|\mu_0) &= -\int \left(\log \frac{d\mu}{d\mu_0}\right) d\mu \\ &= -\int \left(\frac{d\mu}{d\mu_0} \log \frac{d\mu}{d\mu_0}\right) d\mu_0 \end{aligned} \qquad (2.4)$$

and

$$E(H,\mu) = \langle H \rangle_\mu \equiv \int H(\omega) d\mu(\omega) . \qquad (2.5)$$

The equivalence of these two characterizations is a simple computation in the calculus of variations.

Unfortunately, this elementary theory *does not apply* to infinite-volume systems, because the Hamiltonian $H(\omega)$ is ill-defined: for almost any configuration $\omega$ we have $H(\omega) = \pm\infty$. Nevertheless, non-trivial *analogues* of these two characterizations can be developed for infinite-volume systems. The analogue of the explicit formula is the theory of specifications and *Gibbs measures*: an infinite-volume Gibbs measure is one whose conditional probabilities for finite subsystems are given by the Boltzmann-Gibbs formula. The analogue of the variational approach is the theory of *equilibrium measures*: an equilibrium measure is a translation-invariant measure that maximizes entropy *density* minus $\beta$ times mean energy *density*. These approaches are reviewed in Sections 2.3 and 2.5–2.6, respectively. The fundamental feature of infinite-volume systems, which distinguishes them from finite-volume systems, is that the map between "Hamiltonians" (more precisely, interactions) and Gibbs measures (or equilibrium measures) is neither single-valued nor onto: some interactions have multiple Gibbs measures, while some measures are not Gibbsian for any interaction. These facts are at the heart of the theory of phase transitions, and of the renormalization group.

The standard references for the material in this section are the books of Georgii [157], Preston [299] and Israel [206]. Georgii and Preston deal principally with the theory of Gibbs measures, while Israel deals principally with the theory of equilibrium measures.

We assume in this section that the reader has some knowledge of metric spaces and Banach spaces, ideally at the level of Royden [309] or Reed and Simon [306], and of measure theory and probability theory, ideally at the level of Bauer [25] or Krickeberg [224]. However, we realize that for many readers these theories belong to only faintly remembered mathematics courses and are rather distant from their day-to-day work in theoretical physics. Nevertheless, we urge such readers not to be discouraged by the abstract jargon, and to use the examples we provide as a means to grasp the essential *physical* ideas underlying the mathematics.

In this section the emphasis is on concepts and ideas (both physical and mathematical), not on techniques of proof. Therefore, all definitions and theorems are stated precisely, but proofs are omitted. In Appendix A we provide, for each theorem, either a published reference (if the result is known) or a proof (if it is new).

Henceforth we absorb $\beta$ into the Hamiltonian $H$; this simplifies the notation. Let us also remark that although our exposition is couched in the language of the canonical ensemble, the formalism is equally applicable to the grand canonical ensemble: it suffices to interpret our $H$ to mean "$\beta H - \beta \mu N$". In fact, this formalism applies to an arbitrary "generalized (grand) canonical ensemble" with parameters $\beta_1, \ldots, \beta_n$ conjugate to observables $H_1, \ldots, H_n$.

## 2.1 Configurations, Events, Functions, Measures
[8]

Classical statistical mechanics is a branch of probability theory. The basic structures of probability theory are:

- A configuration space $\Omega$ — this is the set of all possible (microscopic) configurations of the system under study.

- A $\sigma$-field $\mathcal{F}$ of subsets of $\Omega$ — this is the set of all events (= yes-no questions) that are measurable by some conceivable (possibly extremely idealized) experiment. Various sub-$\sigma$-fields $\mathcal{A} \subset \mathcal{F}$ may correspond to restricted classes of experiments (e.g. experiments performed within a specified region of space).

- Observables (= random variables = real-valued $\mathcal{F}$-measurable functions on $\Omega$) — these correspond to experiments which give a real number as an answer. Various subclasses of observables (e.g. those measurable with respect to a specified sub-$\sigma$-field $\mathcal{A}$) may correspond to restricted classes of experiments (e.g. experiments performed within a specified region of space).

- A probability measure (= probability distribution) $\mu$ on $(\Omega, \mathcal{F})$ — this describes either our state of partial knowledge of the system (if we take a "subjective" interpretation of probability theory) or an ensemble of "identically prepared" random systems (if we take an "objective" interpretation of probability theory). The mathematics of statistical mechanics does not depend on any particular interpretation of its mathematical objects, so the reader is urged to employ whichever interpretation he/she prefers.

In this section we describe the particular case of this structure that is appropriate for the equilibrium statistical mechanics of an *infinite-volume* classical *lattice* system.

---

[8]A reference for this section is Georgii [157, Introduction and Sections 1.1 and 2.2].

### 2.1.1 Configurations and Events

The configuration space of an infinite-volume lattice system is specified by the following ingredients:

- The *single-spin space* $\Omega_0$. This is the space of possible configurations of the physical variable(s) at a single lattice site. (For brevity we call these variables a "spin".) **Examples:** Ising model, $\Omega_0 = \{-1, 1\}$; $N$-vector model, $\Omega_0 = S^{N-1}$ = unit sphere in $\mathbb{R}^N$; $N$-component Gaussian or $\varphi^4$ model, $\Omega_0 = \mathbb{R}^N$; solid-on-solid (SOS) or discrete Gaussian model, $\Omega_0 = \mathbb{Z}$.

  Since statistical mechanics is based on probability theory, we shall always assume $\Omega_0$ to be equipped with a $\sigma$-field $\mathcal{F}_0$ of "measurable sets". Usually $\Omega_0$ will also come equipped with a physically natural topology; in fact, $\Omega_0$ will almost always be a complete separable metric space, and $\mathcal{F}_0$ will be the $\sigma$-field of Borel sets. If $\Omega_0$ is a *compact* metric space, we say that the system has *bounded spins*; otherwise we say that the system has *unbounded spins*. **Examples:** The Ising and $N$-vector models have bounded spins; the Gaussian, $\varphi^4$, SOS and discrete Gaussian models have unbounded spins.

- The *lattice* $\mathcal{L}$ — a countably infinite set of "sites". For the moment we need not give $\mathcal{L}$ any geometric structure, but for concreteness the reader can imagine $\mathcal{L}$ to be some $d$-dimensional lattice.

The infinite-volume configuration space $\Omega$ is then defined to be the Cartesian product $(\Omega_0)^{\mathcal{L}}$; that is, it is the set of all configurations $\omega = (\omega_x)_{x \in \mathcal{L}}$ with $\omega_x \in \Omega_0$ for each site $x$. The space $\Omega$ is equipped with the product $\sigma$-field $\mathcal{F} = (\mathcal{F}_0)^{\mathcal{L}}$ and with the product topology.[9] The product topology means that a sequence (or net) of configurations $(\omega^n)$ converges to a configuration $\omega$ if and only if $\omega_x^n \to \omega_x$ for all $x \in \mathcal{L}$. If $\Omega_0$ is metrizable (resp. separable, complete metric, compact), then so is $\Omega$.

It is important to understand *physically* what the product topology means. Suppose for simplicity that $\Omega_0$ is a metric space. Then a typical neighborhood of $\omega \in \Omega$ is the set

$$\mathcal{N}_{\omega,\epsilon,\Lambda} = \{\omega': \operatorname{dist}(\omega_x, \omega'_x) < \epsilon \text{ for all } x \in \Lambda\}, \tag{2.6}$$

where $\epsilon > 0$ and $\Lambda$ is a *finite* subset of $\mathcal{L}$.[10] That is, a typical neighborhood of $\omega$ in the product topology is the set of configurations that are close to $\omega$ on some *finite* set of sites $\Lambda$, but are *arbitrary outside* $\Lambda$. In particular, if $\Omega_0$ is discrete (as e.g. in the Ising model), then a neighborhood of $\omega$ is the set of configurations that *agree with* $\omega$ on some finite set of sites $\Lambda$, but are arbitrary outside $\Lambda$. These facts will play an important role in our discussion of non-Gibbsianness for RG image measures (Sections 4.1–4.3).

---

[9]If $\Omega_0$ is a separable metric space, then the product $\sigma$-field of the individual Borel $\sigma$-fields coincides with the Borel $\sigma$-field for the product topology.

[10]More precisely, the sets $\mathcal{N}_{\omega,\epsilon,\Lambda}$ form a *neighborhood basis* of $\omega$, i.e. every neighborhood of $\omega$ contains one of the sets $\mathcal{N}_{\omega,\epsilon,\Lambda}$.

For each subset $\Lambda \subset \mathcal{L}$, we let $\mathcal{F}_\Lambda \subset \mathcal{F}$ be the sub-$\sigma$-field corresponding to events depending only on the spins $\omega_\Lambda = (\omega_x)_{x \in \Lambda}$; that is, $\mathcal{F}_\Lambda$ is the $\sigma$-field of *events measurable within the subset* $\Lambda$. We denote by $\mathcal{S}$ the class of all nonempty *finite* subsets of $\mathcal{L}$. We denote by $\Lambda^c$ the complement of $\Lambda$ in $\mathcal{L}$.

**Remark.** The Cartesian product $(\Omega_0)^\mathcal{L}$ is not the most general configuration space. Often one wishes to study a lattice model with *local constraints* (e.g. hard-core exclusions). One way (not the only one) to treat these constraints is to cut the excluded configurations out of the configuration space: that is, we let the configuration space $\Omega$ be an appropriate *subset* of the product space $(\Omega_0)^\mathcal{L}$. We do not allow this much generality here, but much of the present theory goes through (with some modification) in this situation [313, 269, 15].

### 2.1.2 Functions (= Observables)

An *observable* is simply a real-valued measurable function on $\Omega$. We consider various spaces of such functions:

- The space $B(\Omega) = B(\Omega, \mathcal{F})$ of bounded measurable functions. This is the largest space of functions we shall consider.

- The space $B_{loc}(\Omega) = \bigcup_{\Lambda \in \mathcal{S}} B(\Omega, \mathcal{F}_\Lambda)$ of bounded *local* functions. A function is *local* if it depends on only finitely many spins.

- The space $B_{ql}(\Omega) = \overline{B_{loc}(\Omega)}$ of bounded *quasilocal* functions. A function is *quasilocal* if it is the uniformly convergent limit of some sequence of local functions. Equivalently, a function is quasilocal if it "depends weakly on distant spins" in the sense that[11]

$$\lim_{\Lambda \uparrow \mathcal{L}} \sup_{\substack{\omega, \omega' \in \Omega \\ \omega_\Lambda = \omega'_\Lambda}} |f(\omega) - f(\omega')| = 0 . \tag{2.7}$$

- The space $C(\Omega)$ of bounded *continuous* functions.

- The space $C_{loc}(\Omega) \equiv B_{loc}(\Omega) \cap C(\Omega)$ of bounded *continuous local* functions.

- The space $C_{ql}(\Omega) \equiv B_{ql}(\Omega) \cap C(\Omega)$ of bounded *continuous quasilocal* functions.

**Examples.** 1. For $\Omega_0 = \mathbb{R}$, the function $f(\varphi) = \text{sgn}(\varphi_0)$ is bounded and local (hence quasilocal) but not continuous. Analogous functions can obviously be constructed for $\Omega_0 = S^{N-1}$ or $\mathbb{R}^N$, and indeed on any single-spin space which is not discrete.

---

[11]The statement $\lim_{\Lambda \uparrow \mathcal{L}} F(\Lambda) = \alpha$ (where $\alpha \in \mathbb{R}$ or $\mathbb{C}$) means that for each $\epsilon > 0$, there exists a finite subset $K_\epsilon \subset \mathcal{L}$ such that $|F(\Lambda) - \alpha| < \epsilon$ whenever $\Lambda \supset K_\epsilon$. Statements $\lim_{\Lambda \uparrow \mathcal{L}} F(\Lambda) = +\infty$ or $-\infty$ are to be interpreted analogously. (Mathematicians call this "convergence along the net of finite subsets of $\mathcal{L}$, directed by inclusion".) Please *do not* confuse this notion of convergence with the limit in the sense of van Hove, to be defined in Section 2.4.1.

2. If $\Omega_0 = \mathcal{L} = \mathbb{Z}$, any (bounded) function of $\sigma_{\sigma_0}$ is (bounded and) continuous but not quasilocal. (This example, which was suggested to us by Hans-Otto Georgii, is further discussed in Appendix A.1.)

We equip each of the above spaces with the "supremum norm" (or "uniform norm")

$$\|f\| = \|f\|_\infty \equiv \sup_{\omega \in \Omega} |f(\omega)| . \tag{2.8}$$

So equipped, the spaces $B(\Omega)$, $B_{ql}(\Omega)$, $C(\Omega)$ and $C_{ql}(\Omega)$ are Banach spaces. Let us notice that:

(a) If the single-spin space $\Omega_0$ is a *compact* metric space, then every continuous function is bounded and quasilocal. Hence $C(\Omega) = C_{ql}(\Omega) \subset B_{ql}(\Omega)$.

(b) If the single-spin space $\Omega_0$ is *discrete*, then every quasilocal function is continuous. In particular, $B_{ql}(\Omega) = C_{ql}(\Omega) \subset C(\Omega)$.

(c) If the single-spin space $\Omega_0$ is *finite*, then quasilocality and continuity are equivalent (and imply boundedness). Hence $C(\Omega) = C_{ql}(\Omega) = B_{ql}(\Omega)$.

### 2.1.3 Measures

Next we study measures on $\Omega$. Let $M(\Omega) = M(\Omega, \mathcal{F})$ be the space of finite signed measures on $\Omega$, and $M_{+1}(\Omega) = M_{+1}(\Omega, \mathcal{F}) \subset M(\Omega)$ be the space of probability measures. There is a natural duality between spaces of functions and spaces of measures, namely

$$\langle \mu, f \rangle \equiv \mu(f) \equiv \int f \, d\mu . \tag{2.9}$$

If $\Omega$ is compact, then *every* bounded linear functional on $C(\Omega)$ arises in this way from a finite signed measure (Riesz-Markov theorem); otherwise put, the Banach-space dual of $C(\Omega)$ is exactly $M(\Omega)$.

Let $\mu$ be a probability measure on $\Omega$; then the *support of $\mu$* (denoted supp$\mu$) is a closed subset of $\Omega$ that can be defined in any of three equivalent ways:

(a) The set of all $\omega \in \Omega$ such that every neighborhood of $\omega$ has nonzero measure.

(b) The intersection of all closed sets of measure 1.

(c) The complement of the union of all open sets of measure zero.

The key theorem is: if $\Omega$ is a *separable* metric space, then $\mu(\text{supp}\mu) = 1$, so that supp$\mu$ is the smallest closed set having measure 1.

We need to discuss what it means for a sequence (or net) of measures $\mu_n$ to *converge* to a limiting measure $\mu$; in other words, we need to equip the spaces $M(\Omega)$ and $M_{+1}(\Omega)$ with a *topology*. In fact, there are *several* mathematically natural topologies, each with a distinct *physical* meaning. The simplest topology is the norm topology defined by the *total variation norm*

$$\begin{aligned}
\|\mu - \nu\| &= \sup_{\substack{f \in B(\Omega, \mathcal{F}) \\ \|f\|_\infty \leq 1}} |\mu(f) - \nu(f)| \\
&= \sup_{\substack{f \in C(\Omega) \\ \|f\|_\infty \leq 1}} |\mu(f) - \nu(f)| \\
&= 2 \sup_{A \in \mathcal{F}} |\mu(A) - \nu(A)| \, .
\end{aligned} \quad (2.10a)$$

A sequence (or net) $\mu_n$ converges in variation norm to $\mu$ if $\|\mu_n - \mu\| \to 0$. Physically, norm convergence of $\mu_n$ to $\mu$ means that expectation values in $\mu_n$ converge to those in $\mu$, *uniformly* for all bounded observables $f$. This is an extremely strong notion of convergence, which occurs only rarely in physical applications. Therefore, we introduce also the *weak topologies* induced by the various classes of functions defined in Section 2.1.2:

- The *bounded measurable topology*: $\mu_n \to \mu$ if $\mu_n(f) \to \mu(f)$ for all $f \in B(\Omega, \mathcal{F})$. [If the $\mu_n$ are probability measures, it suffices to check that $\mu_n(A) \to \mu(A)$ for all $A \in \mathcal{F}$.]

- The *bounded quasilocal topology*: $\mu_n \to \mu$ if $\mu_n(f) \to \mu(f)$ for all $f \in B_{ql}(\Omega, \mathcal{F})$. [If the $\mu_n$ are probability measures, it suffices to check convergence for $f \in B_{loc}(\Omega, \mathcal{F})$, or alternatively for all $A \in \bigcup_{\Lambda \in \mathcal{S}} \mathcal{F}_\Lambda$.]

- The (ordinary) *weak topology*: $\mu_n \to \mu$ if $\mu_n(f) \to \mu(f)$ for all $f \in C(\Omega)$.

- The *weak quasilocal topology*: $\mu_n \to \mu$ if $\mu_n(f) \to \mu(f)$ for all $f \in C_{ql}(\Omega, \mathcal{F})$. [If the $\mu_n$ are probability measures, it suffices to check convergence for $f \in C_{loc}(\Omega, \mathcal{F})$.]

We emphasize that the convergence is required to occur for each observable $f$ in the designated class, but the convergence is *not* required to be uniform in $f$. This is important, since $f$ could equally well be the local energy density in New York or the local energy density on Andromeda; and one should not expect, in most situations, the convergence to be uniform on all such observables. This reasoning also suggests that the two quasilocal topologies are likely to be the ones of greatest physical relevance.

**Examples.** 1. Let $\Omega_0 = \mathbb{R}$, and let $\mu_n$ (resp. $\mu$) be the Dirac delta measure concentrated on the configuration in which all the spins take the value $1/n$ (resp. 0). Then $\mu_n \to \mu$ in the weak and weak quasilocal topologies, but not in the bounded measurable or bounded quasilocal topologies.

2. Let $\Omega_0 = \{-1, 1\}$, and let $\mu_n$ be the Dirac delta measure concentrated on the configuration which is $+1$ for all spins at a distance $\leq n$ from the origin and $-1$ for all other spins. Let $\mu$ be the Dirac delta measure concentrated on the configuration which is all +1. Then $\mu_n \to \mu$ in the bounded quasilocal, weak and weak quasilocal topologies, but not in the bounded measurable topology.

Georgii bases his theory on the bounded quasilocal topology (which he calls the "topology of local convergence" or the "$\mathcal{L}$-topology") [157, Chapter 4]; Israel restricts attention to compact metric single-spin spaces, and uses mainly the weak (= weak quasilocal) topology [206, Chapters II and IV].

Finally, let us remark that with respect to the (ordinary) weak topology, $M_{+1}(\Omega)$ is separable and metrizable (resp. complete metrizable, compact metrizable) if and only if $\Omega$ is. Let us also remark that if $\Omega_0$ is separable and metrizable (resp. countable and discrete), then the bounded quasilocal topology is stronger than (resp. equal to) the (ordinary) weak topology; this is true even though these hypotheses do *not* imply that $C(\Omega) \subset B_{ql}(\Omega)$.

## 2.2 Interactions and Hamiltonians

[12]

As discussed in the Introduction to this section, the Hamiltonian $H(\omega)$ for an infinite-volume system is an ill-defined object. Therefore we must proceed more cautiously. We define first the concept of an *interaction*, which corresponds roughly to the idea of a "formal Hamiltonian" or a "set of coupling constants". Then we define the *finite-volume Hamiltonians* corresponding to a given interaction and given boundary conditions.

The (meaningless) Hamiltonian of an infinite-volume system is written formally as a sum of terms corresponding to various finite subsets of the lattice: one-body terms, two-body terms, three-body terms and so forth. Mathematically this idea is made precise as follows:

**Definition 2.1** *An* interaction *(or* interaction potential *or* potential*) is a family* $\Phi = (\Phi_A)_{A \in \mathcal{S}}$ *of functions* $\Phi_A \colon \Omega \to \mathbb{R}$, *such that for each* $A \in \mathcal{S}$, *the function* $\Phi_A$ *is $\mathcal{F}_A$-measurable (i.e. depends only on the spins in the finite subset $A$).*

**Remark.** Note that we do *not* allow the interaction $\Phi_A$ to take the value $+\infty$. Therefore, a "hard-core interaction" is *not* included in our formulation.

**Example.** Consider the Ising model whose formal (i.e. meaningless) Hamiltonian is
$$H(\omega) \text{ "=" } -\sum_{\langle xy \rangle} J_{xy} \omega_x \omega_y - \sum_x h_x \omega_x \, . \tag{2.11}$$
This model is defined (meaningfully!) by the interaction
$$\Phi_A(\omega) = \begin{cases} -h_x \omega_x & \text{if } A = \{x\} \\ -J_{xy} \omega_x \omega_y & \text{if } A = \{x, y\} \\ 0 & \text{otherwise} \end{cases} \tag{2.12}$$

---

[12]References for this section are Georgii [157, Section 2.1] and Israel [206, Sections I.1 and I.2].

The next step is to define the Hamiltonian $H_\Lambda^\Phi$ corresponding to an interaction $\Phi$ acting in a finite volume $\Lambda$. This depends, however, on what boundary conditions one chooses. The simplest case is *free boundary conditions*:

**Definition 2.2** *Let $\Phi$ be an interaction. Then, for each $\Lambda \in \mathcal{S}$, the Hamiltonian $H_{\Lambda,free}^\Phi$ for volume $\Lambda$ with free boundary conditions is the function*

$$H_{\Lambda,free}^\Phi \;=\; \sum_{\substack{A \in \mathcal{S} \\ A \subset \Lambda}} \Phi_A \;. \qquad (2.13)$$

Note that this is always a *finite* sum, so the free-b.c. Hamiltonian is always well-defined. Note also that $H_{\Lambda,free}^\Phi$ depends only on the spins *inside* $\Lambda$.

Free boundary conditions are not, however, sufficient: for many purposes we need Hamiltonians in which the interior of the volume $\Lambda$ is allowed to interact with the exterior. To do this, we must consider the bonds that couple a given volume $\Lambda$ with its exterior; these give a contribution of the form

$$W_{\Lambda,\Lambda^c}^\Phi \;=\; \sum_{\substack{A \in \mathcal{S} \\ A \cap \Lambda \neq \emptyset \\ A \cap \Lambda^c \neq \emptyset}} \Phi_A \;. \qquad (2.14)$$

Note that now we are dealing with an *infinite* sum; therefore we must be careful about its convergence. In any case, the Hamiltonian for volume $\Lambda$ with *general external boundary conditions* corresponds to adding the contributions (2.13) and (2.14):

**Definition 2.3** *Let $\Phi$ be an interaction. Then, for each $\Lambda \in \mathcal{S}$, the Hamiltonian $H_\Lambda^\Phi$ for volume $\Lambda$ with general external boundary conditions is the function*

$$H_\Lambda^\Phi(\omega) \;=\; \sum_{\substack{A \in \mathcal{S} \\ A \cap \Lambda \neq \emptyset}} \Phi_A(\omega) \qquad (2.15a)$$

$$\equiv\; H_{\Lambda,free}^\Phi(\omega) \,+\, W_{\Lambda,\Lambda^c}^\Phi(\omega) \;, \qquad (2.15b)$$

*provided that this sum converges to a finite limit for all $\omega \in \Omega$, in which case we call the interaction $\Phi$ convergent.*[13]

Here the convergence is not required to be absolute, nor is it required to be uniform in $\omega$; we insist only that the finite-volume Hamiltonian $H_\Lambda^\Phi(\omega) \equiv H_\Lambda^\Phi(\omega_\Lambda, \omega_{\Lambda^c})$ be well-defined for all configurations $\omega$ (i.e. all pairs consisting of an internal configuration $\omega_\Lambda$ and an external configuration $\omega_{\Lambda^c}$). This is a very modest requirement. It rules out, however, the use of this formalism for a Coulomb system, in which the interaction decays too slowly to be summable in any reasonable sense.[14]

---

[13]More precisely, what this means is that the net $\left( \sum_{\substack{A \in \mathcal{S} \\ A \cap \Lambda \neq \emptyset \\ A \subset \Delta}} \Phi_A(\omega) \right)_{\Delta \in \mathcal{S}}$ converges to a finite limit (for each $\omega \in \Omega$) as $\Delta \uparrow \mathcal{L}$.

For many purposes it is convenient to think of the configuration outside $\Lambda$ as fixed (the "boundary condition") and the configuration inside $\Lambda$ as variable. Therefore, for any fixed $\tau \in \Omega$, we define the Hamiltonian $H^\Phi_{\Lambda,\tau}$ which uses *boundary condition $\tau$ outside the volume $\Lambda$* to be

$$H^\Phi_{\Lambda,\tau}(\omega) \;=\; H^\Phi_\Lambda(\omega_\Lambda \times \tau_{\Lambda^c}) \,. \tag{2.16}$$

Here $\omega_\Lambda \times \tau_{\Lambda^c}$ is the configuration which agrees with $\omega$ on $\Lambda$ and with $\tau$ on $\Lambda^c$. Note that $H^\Phi_{\Lambda,\tau}(\omega)$ depends only on the behavior of $\omega$ *inside* $\Lambda$.

It is also possible to define the Hamiltonian with other boundary conditions (e.g. periodic), but we shall have no need for these.

The summability properties of the Hamiltonian (2.15a) have important implications for the characteristics of the measures constructed with them. In addition to the notion of *convergence* introduced in Definition 2.3 above, we wish to distinguish two stronger notions of summability:

**Definition 2.4** *We call the interaction $\Phi$*

- **uniformly convergent** *if, for every $\Lambda \in \mathcal{S}$, the sum (2.15a) converges uniformly in $\omega$;*

- **absolutely summable** *if, for every $\Lambda \in \mathcal{S}$, the sum (2.15a) converges in $B(\Omega)$ norm. This is equivalent to the condition that $\sum_{\substack{A \in \mathcal{S} \\ A \ni i}} \|\Phi_A\|_\infty < \infty$ for each $i \in \mathcal{L}$.*

Obviously, absolutely summable implies uniformly convergent, which in turn implies the convergence of (2.15a). Some comments and examples are in order:

1) The physical interpretation of absolute summability is roughly that the *maximum* interaction energy between one spin and the rest of the universe is finite. Alternatively, the flipping of one spin produces always a bounded change in energy.

---

[14]Lattice Coulomb systems admit a partial thermodynamic treatment based on free boundary conditions and a carefully taken infinite-volume limit (ensuring overall neutrality of the plasma) [136, and references therein]. It *may* be possible to cast that theory into a *generalized* version of the Gibbs–DLR framework in which the "bad" external configurations — here the non-neutral ones — are made inaccessible to all Gibbs measures [299, pp. 16–18 and Chapter 6]. If so, many (but not all) of the results discussed in this section would be valid also for such systems. The case of gravitational systems is even worse, because there is no condition analogous to neutrality that can be enforced. These systems are not even thermodynamically stable.

2) An example of a uniformly convergent interaction which is not absolutely summable is the following one-dimensional Ising model [336]:

$$\Phi_A(\omega) = \begin{cases} (-1)^n c_n & \text{if } A \text{ is a non-empty set of } n \text{ adjacent points} \\ & \text{and } \omega_x = +1 \text{ for all } x \in A \\ 0 & \text{otherwise} \end{cases} \quad (2.17)$$

for a suitable sequence of non-negative numbers $(c_n)_{n\geq 1}$. If $nc_n \downarrow 0$, this interaction is uniformly convergent; but it is not absolutely summable unless $\sum_n nc_n < \infty$. Thus, $c_n = n^{-\alpha}$ with $1 < \alpha \leq 2$ provides the desired counterexample. (In fact, if $\sum_n c_n = \infty$ — for example, $c_n = 1/n\log(n+1)$ — the interaction does not even belong to the largest space of interactions considered in the usual thermodynamic formalism, namely the space $\mathcal{B}^0$ introduced in Section 2.4.4 below.)

3) Interactions can also be classified according to the maximum spatial distance over which they extend: the *range* of $\Phi$ is defined to be the supremum of the diameters of the sets $A$ with $\Phi_A \not\equiv 0$. Thus, an interaction $\Phi$ is *of finite range* $R$ ($R < \infty$) if $\Phi_A \equiv 0$ whenever $\text{diam}(A) > R$. For a finite-range interaction, the sum (2.15a) is a finite sum, so $\Phi$ is (trivially) a uniformly convergent interaction. If, in addition, each $\Phi_A$ is a bounded function, then $\Phi$ is absolutely summable.

4) Let us now introduce two natural pieces of terminology: First, we shall call an interaction $\Phi$ *bounded* if each $\Phi_A$ is a bounded function. Note that if $\Phi$ is bounded (resp. absolutely summable), then each Hamiltonian $H^\Phi_{\Lambda,free}$ (resp. $H^\Phi_\Lambda$) is a bounded local (resp. bounded quasilocal) function. A bounded interaction, however, may fail to be absolutely summable if the bounds $\|\Phi_A\|_\infty$ do not decay fast enough.

5) Second: if, as is usual, the space $\Omega_0$ (and hence $\Omega$) comes equipped with a topology, then we call an interaction $\Phi$ *continuous* if each $\Phi_A$ is a continuous function. Note that if $\Phi$ is continuous (resp. continuous and uniformly convergent), then each Hamiltonian $H^\Phi_{\Lambda,free}$ (resp. $H^\Phi_\Lambda$) is a continuous function. All the interactions considered in this work (and an overwhelming majority of those considered elsewhere) are continuous.

6) If $\Omega_0$ (and hence $\Omega$) is *compact*, then every continuous interaction is automatically bounded. This is one reason why systems of bounded spins are easier to work with than systems of unbounded spins.

7) Nevertheless, as we discuss later (Section 2.3.5), all the properties of an interaction must be interpreted modulo physical equivalence. In this regard, the apparent summability properties may turn out to be misleading, as they may change widely from one physically equivalent interaction to another [351].

## 2.3    Specifications and Gibbs Measures

[15]

---

[15] References for this section are Georgii [157, Chapters 1–4] and Preston [299, Chapters 1, 2 and 5].

We now come to the heart of the theory of infinite-volume lattice systems, which is to make precise what we mean by an *infinite-volume Gibbs measure* for a given interaction $\Phi$. We cannot simply use the explicit formula (2.1), because the infinite-volume Hamiltonian $H$ is ill-defined. The traditional solution is to define an infinite-volume Gibbs measure to be any measure which is a limit (in a suitable topology) of finite-volume Gibbs measures with some chosen boundary conditions. The disadvantage of this definition is that it is cumbersome to check: given a measure $\mu$ on the infinite-volume configuration space, how do we determine whether there *exists* some sequence of finite-volume Gibbs measures converging to $\mu$? We would prefer, therefore, to have a more direct condition on the infinite-volume measure $\mu$. Such a condition was first proposed by Dobrushin [83] and Lanford and Ruelle [232]: their idea is to define an infinite-volume Gibbs measure to be one whose *conditional probabilities* for *finite* subsystems $\Lambda$, conditioned on the configuration outside $\Lambda$, are given by the Boltzmann-Gibbs formula based on the Hamiltonian $H_\Lambda^\Phi$. This is the approach we shall take; the traditional approach can then be justified *a posteriori* (Propositions 2.22 and 2.23).

Let us note that, in general, we must condition on the configuration in the *entire* exterior of $\Lambda$ — that is, we must specify a complete "external condition". However, in the special case of a nearest-neighbor interaction (resp. an interaction of finite range $R$), it suffices to specify the spins immediately adjacent to $\Lambda$ (resp. the spins at a distance $\leq R$ from $\Lambda$) — hence the term "boundary condition". We shall usually bow to tradition and call our external configurations "boundary conditions", but we emphasize that in the general case of an infinite-range interaction it is essential to specify the configuration in the entire exterior region.

Let us also remind the reader of the physical role played by boundary conditions: in infinite volume the Gibbs measure (to be defined shortly) may not be unique, and the boundary conditions serve to select a particular Gibbs measure (i.e. a particular "phase"). All this will be described in greater detail in what follows.

### 2.3.1 Specifications

We begin by formalizing the idea of "conditioning on the exterior of a volume $\Lambda$", irrespective of any particular formula for these conditional probabilities. The point is that for a given external configuration $\omega_{\Lambda^c}$, we wish to specify the (conditional) probability distribution of the spins inside the volume $\Lambda$: that is, we want to specify $\text{Prob}_{\omega_{\Lambda^c}}(d\omega_\Lambda)$. Such an object is called a *probability kernel*[16]. In general, a probability kernel $\pi$ from a space $(\Omega, \mathcal{F})$ to another space $(\Omega', \mathcal{F}')$ is an object $\pi(\omega, A)$ with two "slots": an "input" slot that accepts an input configuration $\omega \in \Omega$, and an "output" slot that accepts a set $A \in \mathcal{F}'$ and returns its probability. More formally, a probability kernel from $(\Omega, \mathcal{F})$ to $(\Omega', \mathcal{F}')$ is a map $\pi \colon \Omega \times \mathcal{F}' \to [0,1]$ satisfying:

(a) For each fixed $\omega \in \Omega$, $\pi(\omega, \cdot)$ is a probability measure on $(\Omega', \mathcal{F}')$.

---

[16] For a more extensive introduction to probability kernels and their properties, see [25, Section 56] or [272, Section III–2].

(b) For each fixed $A \in \mathcal{F}'$, $\pi(\,\cdot\,, A)$ is a $\mathcal{F}$-measurable function on $\Omega$.

We shall write such a probability kernel equivalently as $\pi(\omega, A) \equiv \pi(A|\omega) \equiv \pi_\omega(A)$. The first notation emphasizes that $\pi$ is a kind of "transition probability" (as in the theory of Markov processes); the second notation emphasizes that $\pi$ will later be interpreted as a conditional probability; and the third notation emphasizes that $\omega$ is a parameter ("boundary condition") indexing the probability measure on $\Omega'$.

Thus, in our case we need to specify a probability kernel $\pi_\Lambda$ from $(\Omega_{\Lambda^c}, \mathcal{F}_{\Lambda^c})$ to $(\Omega_\Lambda, \mathcal{F}_\Lambda)$. For technical reasons, however, it is convenient to define $\pi_\Lambda$ instead as a probability kernel from the *full* space $(\Omega, \mathcal{F})$ to itself: we then impose explicitly the condition that $\pi(\omega, \,\cdot\,)$ depend on $\omega$ only through its components $\omega_{\Lambda^c}$ (i.e. it is $\mathcal{F}_{\Lambda^c}$-measurable), and that it reproduce the "boundary condition" $\omega_{\Lambda^c}$ when the question fed into its second slot concerns only spins outside $\Lambda$ (i.e. when $A \in \mathcal{F}_{\Lambda^c}$). We are thus led to the following definition[17]:

**Definition 2.5** *A* specification[18] *is a family* $\Pi = (\pi_\Lambda)_{\Lambda \in \mathcal{S}}$ *of probability kernels from* $(\Omega, \mathcal{F})$ *to itself, satisfying the following conditions:*

(a) *For each $A \in \mathcal{F}$, the function $\pi_\Lambda(\,\cdot\,, A)$ is $\mathcal{F}_{\Lambda^c}$-measurable.*

(b) *$\pi_\Lambda$ is $\mathcal{F}_{\Lambda^c}$-proper, i.e. for each $B \in \mathcal{F}_{\Lambda^c}$, $\pi_\Lambda(\omega, B) = \chi_B(\omega)$.*

(c) *If $\Lambda \subset \Lambda'$, then $\pi_{\Lambda'}\pi_\Lambda = \pi_{\Lambda'}$.*[19]

Physically, the idea is that $\pi_\Lambda(\omega, \,\cdot\,)$ is the equilibrium probability distribution for volume $\Lambda$ subject to the boundary condition $\omega$ outside $\Lambda$. Condition (a) states that this measure depends, in fact, only on the behavior of $\omega$ *outside* $\Lambda$. Condition (b) states that for observations *outside* $\Lambda$, this measure equals the delta measure $\delta_\omega$, i.e.

---

[17]See [299, Section 1] for a more leisurely discussion of these points.

[18]In some mathematical-physics literature (e.g. [123]) the term "local specification" is used. We emphasize that this adjective "local" is superfluous; the concepts discussed here and in [123] are identical. In particular, the reader should *not* confuse this (redundant) use of the word "local" with our concept of "quasilocal specification" to be introduced in Section 2.3.3.

[19]The product of two probability kernels is a probability kernel:

$$(\pi_1 \pi_2)(\omega, A) \equiv \int \pi_1(\omega, d\omega')\, \pi_2(\omega', A) \,.$$

For future reference we also define two ways of multiplying a measure by a probability kernel:

$$(\mu\pi)(A) \equiv \int \mu(d\omega)\, \pi(\omega, A)$$

$$(\mu \times \pi)(B) \equiv \int \mu(d\omega)\, \pi(\omega, d\omega') \chi_B(\omega \times \omega')$$

where $A \in \mathcal{F}$ and $B \in \mathcal{F} \times \mathcal{F}'$. Thus, $\mu \times \pi$ is a probability measure on the product space $(\Omega \times \Omega', \mathcal{F} \times \mathcal{F}')$, while $\mu\pi$ is its marginal on the second space $(\Omega', \mathcal{F}')$.

it reproduces the boundary condition. Condition (c) is a compatibility condition for pairs of volumes $\Lambda \subset \Lambda'$: it states that if a volume $\Lambda'$ is in equilibrium with its exterior, then all subsets of $\Lambda'$ are in equilibrium with *their* exteriors.

**Definition 2.6** *A probability measure $\mu$ on $\Omega$ is said to be* consistent *with the specification $\Pi = (\pi_\Lambda)_{\Lambda \in \mathcal{S}}$ if its conditional probabilities for finite subsystems are given by the $(\pi_\Lambda)_{\Lambda \in \mathcal{S}}$: that is,*

$$\text{For each } \Lambda \in \mathcal{S} \text{ and } A \in \mathcal{F}, \quad E_\mu(\chi_A | \mathcal{F}_{\Lambda^c}) = \pi_\Lambda(\,\cdot\,, A) \quad \mu\text{-a.e.} \tag{2.18}$$

*We denote by $\mathcal{G}(\Pi)$ the set of all measures consistent with $\Pi$.*

The following proposition gives two apparently weaker, but in fact equivalent, formulations of the condition (2.18):

**Proposition 2.7** *Let $\Pi = (\pi_\Lambda)_{\Lambda \in \mathcal{S}}$ be a specification, let $\mu$ be a probability measure on $\Omega$, and let $\Lambda \in \mathcal{S}$. Then the following are equivalent:*

(a) *For each $A \in \mathcal{F}$, $E_\mu(\chi_A | \mathcal{F}_{\Lambda^c}) = \pi_\Lambda(\,\cdot\,, A)$ $\mu$-a.e.*

(b) *There exists a measure $\nu_\Lambda$ such that $\mu = \nu_\Lambda \pi_\Lambda$.*

(c) *$\mu = \mu \pi_\Lambda$.*

Physically, (b) states that $\mu$ is the equilibrium probability distribution for volume $\Lambda$ with some (possibly stochastic) boundary condition $\nu_\Lambda$, while (c) states that $\mu$ can itself play the role of $\nu_\Lambda$.

Let us note that $\mathcal{G}(\Pi)$, the set of all measures consistent with $\Pi$, is a *convex* set: if $\mu_1, \ldots, \mu_n$ belong to $\mathcal{G}(\Pi)$, then so does any convex combination of them. The physical interpretation of such convex combinations, and of the extremal points of $\mathcal{G}(\Pi)$, will be discussed in Section 2.3.6.

We also make the (trivial) remark that if the lattice $\mathcal{L}$ were *finite*, then there would be a *unique* measure consistent with $\Pi$, namely the measure $\pi_\mathcal{L}(\omega, \cdot)$ which must be independent of $\omega$. This is one aspect of the fact that phase transitions cannot occur in finite systems.

### 2.3.2 Gibbsian Specifications and Gibbs Measures

An important example of a specification is the *Gibbsian specification* $\Pi^\Phi = (\pi_\Lambda^\Phi)_{\Lambda \in \mathcal{S}}$ corresponding to a given interaction $\Phi$. More precisely, let $\Phi$ be a *convergent* interaction, so that we can define the Hamiltonians $H_\Lambda^\Phi$ with general external boundary conditions. Let $\mu^0 = \prod_{x \in \mathcal{L}} \mu_x^0$ be a probability measure, called the *a priori* measure. We then define the conditional partition function

$$Z_\Lambda^\Phi(\omega_{\Lambda^c}) = \int \exp[-H_\Lambda^\Phi(\omega)] \prod_{x \in \Lambda} d\mu_x^0(\omega_x) \,. \tag{2.19}$$

[Note that the Hamiltonian $H_\Lambda^\Phi(\omega)$ depends on *both* the spins $\omega_\Lambda$ inside $\Lambda$ *and* on the "boundary conditions" $\omega_{\Lambda^c}$. After integrating out the spins $\omega_\Lambda$, we obtain a function of $\omega_{\Lambda^c}$.] Since $H_\Lambda^\Phi$ is everywhere finite, it follows that $Z_\Lambda^\Phi(\omega_{\Lambda^c}) > 0$ for all $\omega$. If moreover $Z_\Lambda^\Phi(\omega_{\Lambda^c}) < +\infty$ for all $\Lambda \in \mathcal{S}$ and all $\omega \in \Omega$, we say that the interaction $\Phi$ is $\mu^0$-*admissible*. Note in particular that if each $H_\Lambda^\Phi$ is bounded below — which certainly occurs if $\Phi$ is absolutely summable, since this makes each $H_\Lambda^\Phi$ bounded — then $\Phi$ is automatically $\mu^0$-admissible. Also, if the single-spin space $\Omega_0$ is finite, then every convergent interaction is automatically $\mu^0$-admissible [because the integral (2.19) is then a finite sum of finite terms].

**Definition 2.8** *Let $\mu^0 = \prod_{x \in \mathcal{L}} \mu_x^0$ be a probability measure, and let $\Phi$ be a convergent, $\mu^0$-admissible interaction. Then the probability measure $\pi_\Lambda^\Phi(\omega, \cdot)$ on $\mathcal{F}$ defined by*

$$\pi_\Lambda^\Phi(\omega, A) \;=\; Z_\Lambda^\Phi(\omega_{\Lambda^c})^{-1} \int \chi_A(\omega) \, \exp[-H_\Lambda^\Phi(\omega)] \prod_{x \in \Lambda} d\mu_x^0(\omega_x) \qquad (2.20)$$

*is called the* Gibbs distribution in volume $\Lambda$ with boundary condition $\omega_{\Lambda^c}$ *corresponding to the interaction $\Phi$ and the a priori measure $\mu^0$.*

It is straightforward to verify that the family $\Pi^\Phi = (\pi_\Lambda^\Phi)_{\Lambda \in \mathcal{S}}$ is indeed a specification; it is called the *Gibbsian specification* for $\Phi$ (and $\mu^0$). A measure consistent with $\Pi^\Phi$ is called a *Gibbs measure* for $\Phi$ (and $\mu^0$). By Proposition 2.7, a measure $\mu$ is a Gibbs measure for $\Pi^\Phi$ if and only if $\mu \pi_\Lambda^\Phi = \mu$ for all $\Lambda$, i.e.

$$\int d\mu(\tau) \, Z_\Lambda^\Phi(\tau_{\Lambda^c})^{-1} \int \chi_A(\omega_\Lambda \times \tau_{\Lambda^c}) \, \exp[-H_\Lambda^\Phi(\omega_\Lambda \times \tau_{\Lambda^c})] \prod_{x \in \Lambda} d\mu_x^0(\omega_x) \;=\; \mu(A) \quad (2.21)$$

for all $A \in \mathcal{F}$ and all $\Lambda \in \mathcal{S}$. The equation (2.21) is called the Dobrushin-Lanford-Ruelle (DLR) equation. A slightly simpler equation is obtained by restricting $A$ to $\mathcal{F}_\Lambda$:

$$\frac{d\mu_\Lambda}{d\mu_\Lambda^0}(\omega_\Lambda) \;=\; \int d\mu(\tau) \, Z_\Lambda^\Phi(\tau_{\Lambda^c})^{-1} \, \exp[-H_\Lambda^\Phi(\omega_\Lambda \times \tau_{\Lambda^c})] \qquad \mu_\Lambda^0\text{-a.e.} \qquad (2.22)$$

In general (2.22) is weaker than (2.21); the former is a necessary but not sufficient condition for $\mu$ to be a Gibbs measure for $\Pi^\Phi$. However, in nearly all practical situations the two conditions are equivalent: see Remark 3 at the end of Section 2.3.3 below.

At this point the reader may be wondering: Why have we bothered to introduce the very general (and abstract) concept of a specification, when virtually all of the concrete models studied in statistical mechanics correspond to *Gibbsian* specifications? We have two reasons: Firstly, non-Gibbsian specifications must be employed in some interesting statistical-mechanical problems, notably those involving hard-core exclusions (which we do not consider in this paper) or zero temperature (Appendix B.2.1). But perhaps more importantly, we want to be consistent with the underlying message of this work, which is that *not everything in the world is Gibbsian*. Therefore, we must introduce concepts which are general enough so that the problems we wish to study will not have been excluded simply *by definition*. Having done so, we will then be able to investigate, without *a priori* preconceptions, which problems give rise to *Gibbsian* specifications and which ones do not.

### 2.3.3 Quasilocality

In all theoretical physics, a fundamental role is played by the concept of an "isolated system". A completely isolated system is of course an idealization, but one can in general render a system *as close to isolated as desired* by moving it a large distance away from all other objects. (Here we neglect cosmological effects, as well as couplings to fields that could carry off radiation.) This asymptotic isolation is possible, of course, because the interaction potentials decay to zero as the spatial separation tends to infinity. One can even argue that this decay of interactions is an essential precondition for the possibility of doing science: without it, the results of experiments on Earth would depend sensitively on conditions on Andromeda, and the repeatability that is fundamental to the scientific method would be absent.

These remarks justify the introduction of a class of specifications that will play a central role in the remainder of this paper:

**Definition 2.9** *A specification* $\Pi = (\pi_\Lambda)_{\Lambda \in \mathcal{S}}$ *is said to be* quasilocal *if, for each* $\Lambda \in \mathcal{S}$, $f \in B_{ql}(\Omega)$ *implies* $\pi_\Lambda f \in B_{ql}(\Omega)$. *[Equivalently:* $f \in B_{loc}(\Omega)$ *implies* $\pi_\Lambda f \in B_{ql}(\Omega)$.*]*

Note that $(\pi_\Lambda f)(\omega) \equiv \int \pi_\Lambda(\omega, d\omega') f(\omega')$ is the mean value of $f$ in the equilibrium probability distribution for volume $\Lambda$ with boundary condition $\omega_{\Lambda^c}$. Therefore, a specification is quasilocal if the mean values of (quasi)local observables depend very weakly on the external spins far from $\Lambda$ (e.g. outside a very large volume $\Lambda'$) *when the external spins in the intermediate region* $\Lambda' \setminus \Lambda$ *are fixed*, i.e.

$$\lim_{\Lambda' \uparrow \mathcal{L}} \sup_{\substack{\omega_1, \omega_2 \in \Omega \\ (\omega_1)_{\Lambda'} = (\omega_2)_{\Lambda'}}} |(\pi_\Lambda f)(\omega_1) - (\pi_\Lambda f)(\omega_2)| = 0 \qquad (2.23)$$

for all $f \in B_{ql}(\Omega)$ [or $B_{loc}(\Omega)$].[20] We emphasize that (2.23) constrains only the *direct* influence of the spins outside $\Lambda'$ (since the spins in the "annulus" $\Lambda' \setminus \Lambda$ are *fixed*). In particular, (2.23) is perfectly compatible with the occurrence of long-range order: it says merely that any influence on $\Lambda$ from the spins outside $\Lambda'$ has to be *transmitted* by the intermediate region. We emphasize also that this condition of "weak dependence" is formulated in the supremum norm, i.e. it is a "worst-case" condition.

---

[20]If the state space $\Omega_0$ is finite, it suffices to check (2.23) for $f \in B(\Omega, \mathcal{F}_\Lambda)$, because any $f \in B_{loc}(\Omega)$ [say, $f \in B(\Omega, \mathcal{F}_{\widetilde{\Lambda}})$ for some $\widetilde{\Lambda} \supset \Lambda$] corresponds to finitely many different functions in $B(\Omega, \mathcal{F}_\Lambda)$ when one fixes the configuration $\omega_{\widetilde{\Lambda} \setminus \Lambda}$. If the state space $\Omega_0$ is infinite, we do not know whether or not this weaker condition is equivalent to (2.23).

**Examples.** 1. If all the Hamiltonians $H_\Lambda^\Phi$ are *local* functions, then $\Pi^\Phi$ is a quasilocal specification. This occurs, in particular, if $\Phi$ is a *finite-range* (and $\mu^0$-admissible) interaction.

2. If all the Hamiltonians $H_\Lambda^\Phi$ are *quasilocal* functions, then $\Pi^\Phi$ is a quasilocal specification. This occurs, in particular, if $\Phi$ is a *uniformly convergent* (and $\mu^0$-admissible) interaction.

3. Although we have not shown explicitly here how to treat models with constraints (e.g. hard-core exclusions), it is easy to see that *local* constraints do not disrupt quasilocality.

Examples 1 and 3 cover all reasonable systems (of either bounded or unbounded spins) with finite-range interactions. Examples 2 and 3 cover all reasonable systems of bounded spins. Therefore, we argue that *all systems of physical interest are quasilocal* with the exception of models of *unbounded spins* with *infinite-range* interactions. These latter systems are, unfortunately, usually *not* quasilocal:

4. Consider a model of real-valued spins $\{\varphi_i\}$ — for example, a Gaussian or $\varphi^4$ model — with formal Hamiltonian

$$H = -\sum_{i,j} J_{ij}\varphi_i\varphi_j \qquad (2.24)$$

where $J$ has infinite range. Then the resulting specification is *not* quasilocal, because an external spin arbitrarily far away from the volume $\Lambda$ can, by taking extremely large values, have large effects inside $\Lambda$. The trouble here is that quasilocality is defined in the supremum norm, which is too strong a condition for systems with unbounded Hamiltonians. (There is in fact a more serious difficulty in this example: for some external conditions the Hamiltonian $H_\Lambda^\Phi$ is divergent. Therefore, to treat these systems it is necessary to enlarge slightly the concept of specification in order to allow some external conditions to be "forbidden" [299, pp. 16–18 and 89] [245, 60], or else to play some minor trickery [157, pp. 261, 264–265 and 424–425].)

We summarize the main conclusion from this discussion:

**Theorem 2.10** *Let $\Phi$ be a uniformly convergent and $\mu^0$-admissible interaction. [In particular this happens if $\Phi$ is absolutely summable, or if $\Phi$ is finite-range and $\mu^0$-admissible.] Then the specification $\Pi^\Phi$ is quasilocal.*

The Gibbsian specification arising from a model with finite (resp. bounded) Hamiltonians has an additional characteristic property:

**Definition 2.11** *A specification $\Pi = (\pi_\Lambda)_{\Lambda \in \mathcal{S}}$ is said to be*

- *nonnull (with respect to $\mu^0$) if, for each $\Lambda \in \mathcal{S}$ and each $A \in \mathcal{F}_\Lambda$,*

$$\mu^0(A) > 0 \implies \pi_\Lambda(\omega, A) > 0 \text{ for all } \omega \in \Omega. \qquad (2.25)$$

- uniformly nonnull (with respect to $\mu^0$) *if, for each $\Lambda \in \mathcal{S}$, there exist constants $0 < \alpha_\Lambda \leq \beta_\Lambda < \infty$ such that*

$$\alpha_\Lambda \, \mu^0(A) \ \leq \ \pi_\Lambda(\omega, A) \ \leq \ \beta_\Lambda \, \mu^0(A) \qquad (2.26)$$

*for all $\omega \in \Omega$ and all $A \in \mathcal{F}_\Lambda$.*

Roughly speaking, "nonnull" means that there are no hard-core exclusions, while "uniformly nonnull" means that moreover the finite-volume Hamiltonians are bounded (as a function of both the interior spins $\omega_\Lambda$ and the exterior spins $\omega_{\Lambda^c}$).

It turns out that the twin properties of being quasilocal and uniformly nonnull exactly characterize the Gibbsian specifications for absolutely summable interactions:

**Theorem 2.12 (Gibbs representation)** *Let $\Pi$ be a specification, and let $\mu^0$ be a product measure. Then the following are equivalent:*

*(a) There exists an absolutely summable interaction $\Phi$ such that $\Pi$ is the Gibbsian specification for $\Phi$ and $\mu^0$.*

*(b) $\Pi$ is quasilocal and is uniformly nonnull with respect to $\mu^0$.*

*Moreover, if the single-spin space $\Omega_0$ is finite, then these are also equivalent to*

*(c) $\Pi$ is quasilocal and is nonnull with respect to $\mu^0$.*

The proof that (a) $\Longrightarrow$ (b) is easy; the nontrivial proof that (b) $\Longrightarrow$ (a) is due to Kozlov [222]. The observation that (c) $\Longrightarrow$ (b) for finite single-spin space was made by both Sullivan [336] and Kozlov [222].

**Definition 2.13** *A measure $\mu$ on $\Omega$ is said to be* quasilocal *if there exists a quasilocal specification with which $\mu$ is consistent. (Sullivan [336] uses the term "almost Markovian" in place of "quasilocal".)*

Theorem 2.12 implies that quasilocality is only slightly more general than Gibbsianness for an absolutely summable interaction: roughly speaking, quasilocality allows for local constraints (e.g. hard-core exclusions) while Gibbsianness does not.

**Remarks.** 1. For further discussion on the Gibbs representation theorem, in connection with translation invariance, see the Remark at the end of Section 2.4.9.

2. Sullivan [336] and Gross [178, pp. 194–195] have introduced a slightly larger class of interactions than those considered here, based on the observation that the only energies which play a role in the definition of the specification $\Pi^\Phi$ are the *relative* energies of pairs of configurations *that differ at only finitely many sites*. Therefore, it is not necessary for the Hamiltonians

$$H_\Lambda^\Phi(\omega) = \sum_{\substack{A \in \mathcal{S} \\ A \cap \Lambda \neq \varnothing}} \Phi_A(\omega) \qquad (2.27)$$

to be well-defined, but only the *relative Hamiltonians*

$$H^{\Phi}_{rel,\Lambda}(\omega,\omega') = \sum_{\substack{A \in \mathcal{S} \\ A \cap \Lambda \neq \varnothing}} [\Phi_A(\omega) - \Phi_A(\omega')] \tag{2.28}$$

for configurations $\omega, \omega'$ that *agree outside* $\Lambda$. It turns out [336, Proposition 3] that for interactions whose *relative* Hamiltonians are uniformly convergent (Sullivan calls these interactions "$\mathcal{L}$-convergent"), the corresponding specification is again quasilocal and nonnull (at least for finite single-spin space). So this generalization does not provide examples of physically interesting non-quasilocal specifications. Indeed, we can combine this result with (c) $\Longrightarrow$ (a) of Theorem 2.12, and conclude that for any "relatively uniformly convergent" interaction $\Phi$ (at least on a finite single-spin space) there is an *absolutely summable* interaction $\Phi'$ such that $\Pi^{\Phi} = \Pi^{\Phi'}$. Roughly speaking this means that $\Phi$ and $\Phi'$ are "physically equivalent" (see Section 2.3.5).

3. If $\Pi$ is a quasilocal specification, then the criterion for $\mu$ to be consistent with $\Pi$ can be weakened slightly: instead of requiring $\mu = \mu\pi_\Lambda$ [Proposition 2.7(c)], it suffices to have $\mu = \mu\pi_\Lambda$ *on the $\sigma$-field $\mathcal{F}_\Lambda$* [157, Remark 4.21]. Thus, if $\Phi$ is a uniformly convergent (and $\mu^0$-admissible) interaction, then the alternate DLR equation (2.22) is *equivalent* to the standard DLR equation (2.21).

4. In rather great generality it can be proven [161, 300, 330] that *every* measure $\mu$ is consistent with *some* specification. However, this specification will in general not be quasilocal. Indeed, in Section 4 we shall give numerous examples of measures that are not consistent with any quasilocal specification.

### 2.3.4 Feller Property

It is useful to single out a class of specifications in which the finite-volume Gibbs measure $\pi_\Lambda(\omega, \cdot)$ depends in a "sufficiently continuous" way on the boundary condition $\omega$:

**Definition 2.14** *A specification $\Pi = (\pi_\Lambda)_{\Lambda \in \mathcal{S}}$ is said to be* Feller *if, for each $\Lambda \in \mathcal{S}$, $f \in C(\Omega)$ implies $\pi_\Lambda f \in C(\Omega)$.*

**Example.** If the interaction $\Phi$ is continuous and uniformly convergent (and $\mu^0$-admissible), then the specification $\Pi^\Phi$ is Feller. Thus, nearly all specifications of physical interest are Feller.

It is worth remarking that the definition of the Feller property formally resembles that of quasilocality: indeed, Definition 2.14 is identical to Definition 2.9, with $B_{ql}(\Omega)$ replaced everywhere by $C(\Omega)$. In particular, if the single-spin space $\Omega_0$ is *finite*, then $B_{ql}(\Omega) = C(\Omega)$, so the concepts of "quasilocal specification" and "Feller specification" coincide.

We can now state a very important uniqueness theorem:

**Theorem 2.15** *Let $\mu$ be a probability measure that gives nonzero measure to every open set $U \subset \Omega$.[21] Then there is at most one Feller specification with which $\mu$ is consistent. In particular, if the single-spin space $\Omega_0$ is finite, then there is at most one quasilocal specification with which $\mu$ is consistent.*

This theorem has an important consequence for the renormalization group: it shows that the downward vertical arrow in (1.2) cannot be a multi-valued map, provided that we interpret $H'$ as standing for a specification.

**Remark.** Such uniqueness does not hold in general for non-Feller specifications. Indeed, if $\mu_1, \mu_2, \ldots$ is *any* finite or countably infinite set of probability measures that are distinguishable at infinity[22], there exists a specification (in general non-Feller and non-quasilocal) with which all these measures are consistent.[23] For example, let $\mu_1, \mu_2, \ldots$ be Gibbs measures of the two-dimensional Ising model at an arbitrary sequence of temperatures $\beta_1, \beta_2, \ldots \in [-\infty, +\infty]$; then there exists a specification with which *all* these measures are consistent! (By Theorem 2.15, such a specification is of necessity non-Feller and non-quasilocal.) This remark shows that non-quasilocal specifications can be extremely pathological and "unphysical"; it is an additional argument for the importance of quasilocality.

### 2.3.5 Physical Equivalence in the DLR Sense

The same physical situation can be described by many different interactions $\Phi$. For example, the interactions

$$\Phi_A(\omega) = \begin{cases} -h\omega_i & \text{if } A = \{i\} \\ -J\omega_i\omega_{i+1} & \text{if } A = \{i, i+1\} \\ 0 & \text{otherwise} \end{cases} \qquad (2.29)$$

and

$$\Phi'_A(\omega) = \begin{cases} -\frac{h}{2}\omega_i - J\omega_i\omega_{i+1} & \text{if } A = \{i, i+1\} \\ 0 & \text{otherwise} \end{cases} \qquad (2.30)$$

both describe the one-dimensional Ising model with nearest-neighbor interaction $J$ and magnetic field $h$; they are obviously "physically equivalent". The reason they are "physically equivalent" is that they *define the same specification* — and it is the specification that determines the physics.

Reflecting a little bit on this and similar examples, one comes to the following definition [157, Section 2.4]:

---

[21]This means, roughly speaking, that every configuration in $\Omega$ is "possible", i.e. there are no "hard-core exclusions".

[22]This means that there exist *disjoint* sets $F_1, F_2, \ldots \in \widehat{\mathcal{F}}_\infty \equiv \bigcap_{\Lambda \in \mathcal{S}} \mathcal{F}_{\Lambda^c}$ such that $\mu_k(F_k) = 1$ for each $k$.

[23]*Proof:* Form the measure $\mu = \sum_k c_k \mu_k$, where $c_1, c_2, \ldots > 0$ is any sequence with sum 1. By [161, 300, 330] there exists a specification $\Pi$ with which $\mu$ is consistent. But then $\mu_k = c_k^{-1} \chi_{F_k} \mu$ is also consistent with $\Pi$ [299, Lemma 2.4].

**Definition 2.16** *Let $\Phi$ and $\Phi'$ be convergent interactions. We say that $\Phi$ and $\Phi'$ are physically equivalent in the DLR sense if, for all $\Lambda \in \mathcal{S}$, the function $H_\Lambda^\Phi - H_\Lambda^{\Phi'}$ is $\mathcal{F}_{\Lambda^c}$-measurable (i.e. depends only on the spins outside $\Lambda$).*

One can then prove the following theorem:

**Theorem 2.17** *Let $\Phi$ and $\Phi'$ be convergent $\mu^0$-admissible interactions. Consider the following statements:*

*(a) $\Phi$ and $\Phi'$ are physically equivalent in the DLR sense.*

*(b) $\Pi^\Phi = \Pi^{\Phi'}$, i.e. the specifications for $\Phi$ and $\Phi'$ coincide.*

*Then (a) $\Longrightarrow$ (b). Moreover, if $\mu^0(U) > 0$ for every open set $U \subset \Omega$,[24] and the interactions $\Phi$ and $\Phi'$ are continuous, then (b) $\Longrightarrow$ (a).*

**Corollary 2.18 (Griffiths–Ruelle)** *Let $\Phi$ and $\Phi'$ be uniformly convergent, continuous, $\mu^0$-admissible interactions; and assume that $\mu^0(U) > 0$ for every open set $U \subset \Omega$. If there exists a measure $\mu$ that is Gibbsian for both $\Phi$ and $\Phi'$, then $\Phi$ and $\Phi'$ are physically equivalent in the DLR sense, and $\Pi^\Phi = \Pi^{\Phi'}$ [hence $\Phi$ and $\Phi'$ have exactly the same Gibbs measures].*

There are several ways to deal with the ambiguity caused by physical equivalence. One way is to select a single "preferred" representative from each class of physically equivalent interactions: in the Ising model this is exemplified by the possibility of using "spin" interactions $\Phi_A = -J_A \sigma^A$ or "lattice-gas" interactions $\Phi_A = -J_A \rho^A \equiv -J_A \left(\frac{1+\sigma}{2}\right)^A$ [206, 351]; and more generally it is exemplified by the concepts of "$\alpha$-normalized" interactions and "gas" interactions [157, Sections 2.3 and 2.4]. However, for interactions which are not finite-range, this approach can give rise to convergence problems [351].

The other approach is to accept the ambiguity as inevitable, and to work with *equivalence classes* of interactions modulo physical equivalence. We shall take this latter approach. The key result here is Corollary 2.18, due originally (albeit in a very slightly weaker form) to Griffiths and Ruelle [174]. This result has an important consequence for the renormalization group: it shows that the downward vertical arrow in (1.2) cannot be a multi-valued map, provided that we interpret $H'$ as standing for an equivalence class of interactions modulo physical equivalence.

To avoid trivialities, *we assume henceforth that $\mu^0(U) > 0$ for every open set $U \subset \Omega$*.

---

[24]This means, roughly speaking, that every configuration in $\Omega$ is "possible". If it were not so, then the *true* configuration space would be a proper closed subset $F = \prod_{x \in \mathcal{L}} \operatorname{supp}\mu_{0x} \subset \Omega$. We could then make the condition hold simply by redefining the configuration space to be $F$ rather than $\Omega$. So the condition means simply that the configuration space does not contain any "useless points". Some such condition is needed for (b) $\Longrightarrow$ (a) to hold, because the interaction $\Phi$ is completely arbitrary at the "useless points" $\omega \in \Omega \setminus F$.

### 2.3.6 Structure of the Space $\mathcal{G}(\Pi)$

Physical systems exhibit in general one or more possible "macrostates"[25], depending on the values of some control parameters. For instance, water can be in a liquid, solid or gaseous "macrostate" depending on temperature and pressure; and there are points on the temperature-pressure phase diagram where two or even all three of these "macrostates" are possible.

The physical relevance of the theory developed in the preceding subsections relies on the assumption that for each physical system there exists a specification $\Pi$ from which all the statistical-mechanical information about the system can be obtained: that is, such that the space $\mathcal{G}(\Pi)$ of measures consistent with $\Pi$ describes all the "macrostates" of the physical system that are possible for the given choice of control parameters. Therefore, we must be able to transcribe all the expected properties of the set of these "macrostates" in terms of properties of the space $\mathcal{G}(\Pi)$. We briefly discuss here this transcription. In consistency with our main message that not everything in the world is Gibbsian, *everything in this subsection holds for general specifications*, which need not be Gibbsian. (This generality will also be useful when discussing statistical mechanics at zero temperature: see Appendix B.2.1.) However, for the sake of brevity and familiarity, we will sometimes refer to the measures consistent with $\Pi$ as the "Gibbs measures" for $\Pi$ — which is a slight abuse of language when $\Pi$ is not Gibbsian.

There are two important properties that characterize the macroscopic systems observed in nature. Firstly, these systems involve a huge number of degrees of freedom, so large that only a statistical description is possible. However, these statistical aspects do not manifest themselves at a macroscopic level: that is, *macroscopic observables do not fluctuate*; the system behaves deterministically with respect to them. The second property refers to the microscopic observables: they do fluctuate, but their fluctuations are only local, not affecting large regions. Equivalently, *local observations made far away one from the other are almost independent*.

To translate these properties into precise mathematical statements, we need first to specify what a macroscopic observable is. As is usual with long-used concepts, there is more than one possible meaning. Some people consider a macroscopic observable to be any translation-invariant measurable function. At this point, however, we would like to remain at a general level, leaving the aspects related to translation-invariance until the next section. So we adopt an alternative definition, which corresponds to what could be called "global" observables, namely observables that do not depend on what happens to finitely many spins. Recall that if $\Lambda$ is a finite subset of the lattice, then $\mathcal{F}_{\Lambda^c}$ is the $\sigma$-field consisting of all events that are measurable by observations made solely *outside* $\Lambda$; that is, they are the events that do not depend on the behavior of the spins inside $\Lambda$. Now consider the events that belong to $\mathcal{F}_{\Lambda^c}$ for *every finite* subset $\Lambda$: these events constitute a $\sigma$-field

$$\widehat{\mathcal{F}}_\infty \equiv \bigcap_{\Lambda \in \mathcal{S}} \mathcal{F}_{\Lambda^c} \;, \tag{2.31}$$

which consists of all those events whose definition is not affected by changes on any finite number of spins. This field is usually called in mathematics the *tail field*, and could be thought as *the field of global events*. The functions measurable with respect to this field are called *observables at infinity* and can be interpreted as *global observables*.

---

[25]These "macrostates" are also referred to as "phases" in the chemical and physical literature. Here, following an established mathematical-physics nomenclature, we reserve the word "phase" for the notion of "pure phase", to be defined in Section 2.4.9 below. For this informal discussion we prefer to use the word "macrostate", but keeping the quotation marks to emphasize the informality of the concept. We do not want to get entangled with the many different senses adopted in the literature for the word "state".

**Examples of global observables.** 1. All "macroscopic averages", for instance observables of the form

$$\overline{f} \equiv \begin{cases} \lim_{n\to\infty} |\Lambda_n|^{-1} \sum_{x \in \Lambda_n} f(\omega_x) & \text{if the limit exists} \\ 0 \text{ (or whatever)} & \text{otherwise} \end{cases} \tag{2.32}$$

where $(\Lambda_n)$ is a suitable increasing sequence of finite subsets of $\mathcal{L}$ which together exhaust $\mathcal{L}$ (we will discuss this further in Section 2.4.1), and $f \colon \Omega_0 \to \mathbb{R}$ is a measurable function. A macroscopic average as in (2.32) is obviously unaffected by altering finitely many spins, so $\overline{f}$ is indeed an observable at infinity.

2. In an Ising model, consider

$$g(\omega) = \begin{cases} 1 & \text{if there exists an infinite connected cluster} \\ & \text{of + spins} \\ 0 & \text{otherwise} \end{cases} \tag{2.33}$$

The existence of an infinite cluster is obviously unaffected by altering finitely many spins, so $g$ is indeed an observable at infinity. (This observable is of particular importance in percolation theory.)

3. In an Ising model, consider the *difference* in magnetization between the even and odd sublattices:

$$\overline{\mathcal{M}}_{stagg} \equiv \begin{cases} \lim_{n\to\infty} |\Lambda_n|^{-1} \sum_{x \in \Lambda_n} (-1)^{|x|} \omega_x & \text{if the limit exists} \\ 0 & \text{otherwise} \end{cases} \tag{2.34}$$

where $(\Lambda_n)$ is as before. This also is obviously an observable at infinity.

Thus, the usual macroscopic measurements performed on real systems correspond to global observables, but the converse is not true: as Example 3 illustrates, our concept of "global observables" includes some quantities that are experimentally not very accessible. For example, in the antiferromagnetic Ising model, the sign of the staggered magnetization is an observable at infinity, which detects which of the two sublattices is positively magnetized and which is negatively magnetized. But it is very unlikely that an experimenter could succeed in reliably labelling the two sublattices, much less in measuring separately their magnetizations.

After the previous discussion, we can now state more precisely which properties a measure $\mu$ representing a "macrostate" of a physical system must have: (i) It must be deterministic on global events, that is $\mu(A)$ can only take the values 0 or 1 for an event $A \in \widehat{\mathcal{F}}_\infty$; and (ii) its expectation for spatially distant events must, in some sense, asymptotically factorize (= short-range correlations = (some type of) cluster property). It turns out that these two properties are equivalent:

**Proposition 2.19** *Let $\mu \in M_{+1}(\Omega)$. Then the following properties are equivalent:*

*(a) $\mu$ has trivial tail field, that is, if $A \in \widehat{\mathcal{F}}_\infty$ then $\mu(A)$ equals either 0 or 1.*

*(b) $\mu$ has short-range correlations, that is, for each $A \in \mathcal{F}$ we have*

$$\lim_{\substack{\Lambda \uparrow \mathcal{L} \\ \Lambda \in \mathcal{S}}} \sup_{B \in \mathcal{F}_{\Lambda^c}} |\mu(A \cap B) - \mu(A)\mu(B)| \;=\; 0 \;. \qquad (2.35)$$

Property (a) states, roughly speaking, that all the observables at infinity (= global observables) take a constant value from the point of view of the measure $\mu$. For instance, the fact that all the sets of the form $\{\omega \in \Omega: \overline{f}(\omega) \in B\}$ have measure either 0 or 1 means that there is a precise value $f_\mu$ such that $\overline{f} = f_\mu$ with $\mu$-probability 1. Property (b) is a strong "cluster property": it states that distant regions of the lattice are asymptotically independent (even if one of the regions involves infinitely many spins), *uniformly* in the observable measured in the second region.

Now fix a specification $\Pi$, and let us consider the structure of the set $\mathcal{G}(\Pi)$. We know that $\mathcal{G}(\Pi)$ is a *convex* set, so it is natural to ask what are its extreme points.[26] The answer is:

**Proposition 2.20** *Let $\mu \in \mathcal{G}(\Pi)$. Then the following properties are equivalent:*

*(a) $\mu$ is an extreme point of $\mathcal{G}(\Pi)$.*

*(b) $\mu$ has trivial tail field.*

*(c) $\mu$ has short-range correlations.*

The upshot of the preceding discussion is that the "macrostates" of a physical system described by a certain specification correspond to the *extremal Gibbs measures* for this specification. What is the interpretation of the non-extremal measures of $\mathcal{G}(\Pi)$? For "nice" convex sets, every point in the set can be represented as the barycenter of a probability measure concentrated on the extreme points (this is a kind of "integral" convex combination). It turns out that $\mathcal{G}(\Pi)$ *is* nice in this sense.[27] Thus, every non-extremal measure in $\mathcal{G}(\Pi)$ is an (integral) convex combination of extremal ones. In fact, a deep result ([299, Theorem 2.2], [157, Theorem 7.26]) states that this decomposition is *unique*, that is, that $\mathcal{G}(\Pi)$ is a *simplex*. These results mean, in experimental terms, that a non-extremal Gibbs measure corresponds simply to the preparation of a randomly chosen extremal Gibbs measure. The probabilities for this choice are given by the "coefficients" of the convex combination. This extra randomness can be interpreted as representing ignorance on the part of the experimenter about the system's "macrostate" (i.e. over and above his/her unavoidable ignorance about its microstate). From this point of view, the physical system itself can always be considered to be in a well-defined "macrostate" described by an *extremal* Gibbs measure. Thus, the extremal Gibbs measures are the "pure" physical objects.

---

[26]We recall that the extreme points of a convex set are those that cannot be written as a non-trivial convex combination of other points in the set.

As a consequence of the preceding discussion, we conclude that the cardinality of the set of extremal measures of $\mathcal{G}(\Pi)$ represents the number of physical "macrostates" available to the system. A change in this number as the control parameters are varied corresponds to a *phase transition* (more precisely, to *one* of the notions of phase transition, see Section 2.6.5); and the variation of this number as a function of these control parameters (temperature, magnetic field, chemical potential, etc.) can be recorded in the form of a *phase diagram*. Therefore, the study of the set of extremal measures of $\mathcal{G}(\Pi)$ is a central problem in statistical mechanics. As a first step, it is essential to determine conditions under which the set $\mathcal{G}(\Pi)$ is *nonempty*, i.e. under which there exists at least one infinite-volume Gibbs measure. Contrary to what one might initially think, this is a non-trivial problem, since there exist physically quite reasonable models for which there are *no* infinite-volume Gibbs measures. The typical examples are the short-range massless Gaussian models (harmonic crystals) in dimension $d \leq 2$, and the solid-on-solid or the discrete Gaussian models in $d = 1$. The essential point here is that the existence of Gibbs measures in these models is equivalent to the breaking of a non-compact symmetry of the single-spin space (the shift of all the spin values by a constant); and, as is well known, it is impossible to break discrete symmetries in $d = 1$ or continuous symmetries in $d \leq 2$. We refer to [157, Chapter 9] for precise statements, references and further examples. In any case, the following theorem suffices for virtually all applications to models of *bounded* spins:

**Proposition 2.21** *Let $\Omega$ be a compact metric space, and let $\Pi = (\pi_\Lambda)_{\Lambda \in \mathcal{S}}$ be a Feller specification. Then $\mathcal{G}(\Pi)$ is nonempty.*

This result is, in fact, an immediate consequence of Proposition 2.22 below: take any sequence whatsoever of boundary conditions $(\nu_n)$; by compactness, the sequence $(\nu_n \pi_{\Lambda_n})$ must have at least one limit point $\mu$, and Proposition 2.22 then guarantees that $\mu \in \mathcal{G}(\Pi)$.

---

[27]For systems of bounded spins this can be proven by appealing to the Choquet theorem [293, 206]. For general systems it can be proven through direct probabilistic arguments [299, pp. 24–32] [157, Section 7.3 and the associated notes] [107].

If there are several "macrostates" available to the system, and an experimenter wants to select a particular one with absolute certainty, how must he/she proceed? There are basically two ways: One approach is to add to the Hamiltonian some additional fields, such that an infinitesimal value of these fields — more precisely, a limit process consisting in turning them on and then slowly off in some appropriate sequence — selects one or the other of the "macrostates". For example, in an Ising model at low temperature, one may add to the Hamiltonian a magnetic field $h$; the limits $h \downarrow 0$ and $h \uparrow 0$ then select the extremal Gibbs measures $\mu_+$ and $\mu_-$ of the *zero-field* Ising model. An alternative approach is to immerse the (finite) sample in a configuration typical of the intended "macrostate" (selection via boundary conditions). For example, in the Ising case we could use boundary conditions in which the spins outside the volume $\Lambda$ are fixed to be all $+$ or all $-$; taking the limit $\Lambda \uparrow \mathcal{L}$ with these boundary conditions again selects $\mu_+$ or $\mu_-$, respectively. In relation with this second point of view we present two propositions, the first of which justifies *a posteriori* the traditional approach to infinite-volume lattice systems based on infinite-volume limits:

**Proposition 2.22** *Let $\Pi = (\pi_\Lambda)_{\Lambda \in \mathcal{S}}$ be a Feller specification. Let $(\Lambda_n)_{n \geq 1}$ be an increasing sequence of finite volumes whose union is $\mathcal{L}$, and let $(\nu_n)_{n \geq 1}$ be an arbitrary sequence of probability measures on $\Omega$ (i.e. arbitrary deterministic or random boundary conditions). Let $\mu$ be any limit point (in the weak topology) of the sequence $(\nu_n \pi_{\Lambda_n})_{n \geq 1}$. Then $\mu$ is consistent with $\Pi$. In particular, $\mathcal{G}(\Pi)$ is a closed subset of $M_{+1}(\Omega)$.*

**Proposition 2.23** *Let $\Omega$ be a compact metric space, let $\Pi = (\pi_\Lambda)_{\Lambda \in \mathcal{S}}$ be a Feller specification, and let $\mu$ be an* extreme *point of $\mathcal{G}(\Pi)$. Then, for $\mu$-a.e. $\omega$,*

$$\lim_{\substack{\Lambda \uparrow \mathcal{L} \\ \Lambda \in \mathcal{S}}} \delta_\omega \pi_\Lambda = \mu \qquad (2.36)$$

*in the weak topology.*

Proposition 2.22 states that any weak limit of finite-volume Gibbs measures, with arbitrary deterministic or random boundary conditions, is an infinite-volume Gibbs measure. This is the link between the DLR approach and the traditional approach via limits of correlations. Proposition 2.23 is a very strong converse statement, for the special case of *extremal* Gibbs measures: it states that if one takes *any* "typical" configuration from the measure $\mu$ and uses it as a boundary condition, in the infinite-volume limit one recovers $\mu$. This is the mathematical transcription of the process of selecting a "macrostate" by preparing the sample with an appropriate boundary condition. In fact, there is a revealing generalization of this, that states that if $\mu$ is *any* Gibbs measure, then if one takes a "typical" configuration from the measure $\mu$ and uses it as a boundary condition, in the infinite-volume limit one recovers one of the extremal Gibbs measures in the decomposition of $\mu$ [156]. This theorem can be interpreted as saying that the result of a measurement on a large (strictly speaking infinite) system will always yield a value characteristic of one of the *extremal* Gibbs measures: for example, a measurement of the magnetization in a low-temperature Ising model at zero magnetic field will always yield $\pm M_0$, not an intermediate value.

Finally, the consistency between the physical picture and the mathematical formalism requires some discussion of the issue of distinguishability of "macrostates". Physically, two "macrostates" should be considered different only if there is some *macroscopic* measurement that can tell the difference. In terms of the formalism discussed so far, this corresponds to the requirement that *global* observables be able to distinguish among the different extremal measures for a given specification. The following theorem shows that even more is true: the global observables uniquely characterize each measure — extremal or not — consistent with a given specification.

**Theorem 2.24** *Let $\Pi$ be an specification. Then:*

*(a) The extremal measures of $\mathcal{G}(\Pi)$ are mutually singular when restricted to the tail field. That is, if $\mu$ and $\nu$ are distinct extremal measures of $\mathcal{G}(\Pi)$, there exists a set $A \in \widehat{\mathcal{F}}_\infty$ such that $\mu(A) = 1$ and $\nu(A) = 0$.*

*(b) Each measure $\mu \in \mathcal{G}(\Pi)$ is uniquely determined [among the measures of $\mathcal{G}(\Pi)$] by the events in the tail field. That is, if $\mu$ and $\nu$ are measures in $\mathcal{G}(\Pi)$ such that $\mu(A) = \nu(A)$ for each $A \in \widehat{\mathcal{F}}_\infty$, then $\mu = \nu$.*

For the proof, see [157, Theorem 7.7].

### 2.3.7 Conditioning on an Arbitrary Subset of Spins

The DLR equations tell us how to condition on the spins in the complement of a *finite* set. However, in Section 4 we shall need to condition on sets of spins which are *not* complements of finite sets. Therefore, we need the following technical construction, which can be skipped on a first reading.

Let $\Pi = (\pi_\Lambda)_{\Lambda \in \mathcal{S}}$ be a specification. Let $\Delta$ be a subset of $\mathcal{L}$ (*not* necessarily co-finite!). Let $\omega \in \Omega$ be a configuration (but only its components $\omega_\Delta$ will play any role). We then define the *system restricted to the volume* $\mathcal{L} \setminus \Delta$, with configuration space $(\Omega_0)^{\mathcal{L}\setminus\Delta}$: the *specification for volume* $\mathcal{L} \setminus \Delta$ *with external spins set to* $\omega_\Delta$ is the family $\Pi^\omega = (\pi_\Lambda^\omega)_{\Lambda \in \mathcal{S}, \Lambda \subset \mathcal{L}\setminus\Delta}$ defined by

$$\pi_\Lambda^\omega(\omega', A) = \pi_\Lambda(\omega_\Delta \times \omega', A) \tag{2.37}$$

where $\omega' \in (\Omega_0)^{\mathcal{L}\setminus\Delta}$ and $A \in \mathcal{F}_{\mathcal{L}\setminus\Delta}$. Clearly the functions $\pi_\Lambda^\omega(\,\cdot\,,A)$ are $\mathcal{F}_{(\mathcal{L}\setminus\Delta)\setminus\Lambda}$-measurable. It is easy to see that the family $\Pi^\omega$ defines a specification on the system with lattice $\mathcal{L} \setminus \Delta$.

Let now $\mu$ be a measure consistent with $\Pi$. Let $\mu^\omega$ be a regular conditional probability for $\mu$ given $\mathcal{F}_\Delta$. (Such regular conditional probabilities always exist if $(\Omega, \mathcal{F})$ is, for example, a standard Borel space. This includes all examples of physical interest.) We then have the following intuitively obvious result:

**Proposition 2.25** *For $\mu$-a.e. $\omega$, the measure $\mu^\omega \!\restriction\! \mathcal{F}_{\mathcal{L}\setminus\Delta}$ is consistent with $\Pi^\omega$.*

## 2.4 Translation Invariance

[28]

Until now the lattice $\mathcal{L}$ has been simply a countably infinite set of sites, devoid of any geometric structure. In most applications, however, $\mathcal{L}$ is a regular $d$-dimensional lattice; this additional structure allows us to define the notion of *translation invariance* for measures, interactions, specifications and so forth. For simplicity we shall take $\mathcal{L}$ to be the simple (hyper)cubic lattice $\mathbb{Z}^d$. This is no real loss of generality, because other regular lattices can be mapped to $\mathbb{Z}^d$ by an appropriate labelling of sites.[29]

### 2.4.1 Van Hove Convergence

An important role in the statistical mechanics of translation-invariant systems is played by sequences of volumes $(\Lambda_n)$ which grow in such a way that the surface-to-volume ratio tends to zero. We therefore make the following definitions:

**Definition 2.26** *Let $r > 0$, and let $\Lambda \subset \mathbb{Z}^d$. We then define*

- *the* inner $r$-boundary $\partial_r^- \Lambda = \{x \in \Lambda: \operatorname{dist}(x, \Lambda^c) \leq r\}$
- *the* outer $r$-boundary $\partial_r^+ \Lambda = \{x \in \Lambda^c: \operatorname{dist}(x, \Lambda) \leq r\}$
- *the* $r$-boundary $\partial_r \Lambda = \partial_r^- \Lambda \cup \partial_r^+ \Lambda$

We can then state the desired condition in a number of equivalent ways:

**Proposition 2.27** *Let $(\Lambda_n)_{n \geq 1}$ be a sequence of nonempty finite subsets of $\mathbb{Z}^d$. Then the following are equivalent:*

(a) $\lim_{n \to \infty} |\partial_1^- \Lambda_n|/|\Lambda_n| = 0$.

(b) $\lim_{n \to \infty} |\partial_1^+ \Lambda_n|/|\Lambda_n| = 0$.

(c) *For each $r > 0$,* $\lim_{n \to \infty} |\partial_r \Lambda_n|/|\Lambda_n| = 0$.

(d) *For each $a \in \mathbb{Z}^d$,* $\lim_{n \to \infty} |\Lambda_n \setminus (\Lambda_n + a)|/|\Lambda_n| = 0$.

(e) *For each $a \in \mathbb{Z}^d$,* $\lim_{n \to \infty} |(\Lambda_n + a) \setminus \Lambda_n|/|\Lambda_n| = 0$.

(f) *For each finite subset $A \subset \mathbb{Z}^d$,* $\lim_{n \to \infty} |\Lambda_n \triangle (\Lambda_n + A)|/|\Lambda_n| = 0$.

---

[28]References for this section are Georgii [157, Chapter 14], Israel [206, Chapter IV], Preston [299, Chapter 4] and Ruelle [313, Chapter 3].

[29]What is really relevant here is not that $\mathcal{L}$ *equals* $\mathbb{Z}^d$, but merely that the additive group $\mathbb{Z}^d$ *acts on* $\mathcal{L}$: that is, there should exist bijections $t_a \colon \mathcal{L} \to \mathcal{L}$ ($a \in \mathbb{Z}^d$) such that $t_a t_b = t_{a+b}$ and $t_0 =$ identity. The formulae below can easily be generalized to this case, by replacing each occurrence of $x - a$ by $t_a(x)$.

*Moreover, all of these conditions imply that:*

*(α)* $\lim_{n\to\infty} |\Lambda_n| = \infty$.

*(β) There exist vectors $a_n \in \mathbb{Z}^d$ such that the translates $\Lambda_n - a_n$ fill out $\mathbb{Z}^d$ in the following sense: for each finite subset $A \subset \mathbb{Z}^d$, there exists $n_0(A) < \infty$ such that $A \subset \Lambda_n - a_n$ for all $n \geq n_0(A)$.*

**Definition 2.28** *A sequence $(\Lambda_n)_{n\geq 1}$ of nonempty finite subsets of $\mathbb{Z}^d$ is said to converge to infinity in the sense of van Hove (denoted $\Lambda_n \nearrow \infty$) if it satisfies any one (hence all) of the equivalent conditions of Proposition 2.27.*

**Definition 2.29** *Let $F$ be a function from $\mathcal{S}$ (the nonempty finite subsets of $\mathbb{Z}^d$) to some metric space $W$, and let $w$ be some element of $W$. We write $\lim_{\Lambda\nearrow\infty} F(\Lambda) = w$ in case $\lim_{n\to\infty} F(\Lambda_n) = w$ for every sequence $(\Lambda_n)$ that tends to infinity in the sense of van Hove.*

### 2.4.2 Translation-Invariant Measures

With these preparations, we now focus attention specifically on translation invariance in lattice spin systems. With $\mathcal{L} = \mathbb{Z}^d$, the translation group $\mathbb{Z}^d$ acts on the infinite-volume configuration space $\Omega = (\Omega_0)^{\mathbb{Z}^d}$ by

$$(T_a\omega)_x = \omega_{x-a} \qquad \text{for all } x \in \mathcal{L} \tag{2.38}$$

where $a \in \mathbb{Z}^d$. (The minus sign here means that $T_a\omega$ is the configuration $\omega$ translated *forward* by $a$.) This action on the configuration space induces in turn an action on functions

$$(T_a f)(\omega) \equiv f(T_a\omega) \qquad \text{for all } \omega \in \Omega \tag{2.39}$$

and on measures

$$(T_a\mu)(A) \equiv \mu(T_a^{-1}[A]) \qquad \text{for all } A \in \mathcal{F}. \tag{2.40}$$

A function $f \in B(\Omega)$ is said to be *translation-invariant* if $T_a f = f$ for all $a \in \mathbb{Z}^d$. A measure $\mu$ is said to be *translation-invariant* if $T_a\mu = \mu$ for all $a \in \mathbb{Z}^d$. We denote by $M_{inv}(\Omega)$ and $M_{+1,inv}(\Omega)$ the spaces of translation-invariant measures. All this is just a precise mathematical statement of the obvious notions that everybody has in mind.

At this point we can repeat the considerations done in Section 2.3.6, this time regarding the relationship between physical "macrostates" and elements of $M_{+1,inv}(\Omega)$. If we take the point of view that the "macroscopic" observables are the *translation-invariant* bounded measurable functions, then the requirements for a measure $\mu$ representing a physical "macrostate" are: (i) Translation-invariant functions in $B(\Omega)$ must not have fluctuations with respect to $\mu$, i.e. they must be constant with $\mu$-probability one; and (ii) the probability of far-away events must factorize in some sense. Once more, the extremal measures are the objects with the right properties. Indeed, $M_{+1,inv}(\Omega)$ is a convex set, and its extreme points are characterized by the following theorem:

**Proposition 2.30** *Let $\mu \in M_{+1,inv}(\Omega)$. Then the following properties are equivalent:*

(a) *$\mu$ is an extreme point of $M_{+1,inv}(\Omega)$.*

(b) *Every translation-invariant function $f \in B(\Omega)$ is $\mu$-a.e. constant.*

(c) *$\lim_{n \to \infty} n^{-d} \sum_{a \in C_n} \mu(f\, T_a g) \;=\; \mu(f)\mu(g)$ for all $f, g \in B(\Omega)$ [or $B_{ql}(\Omega)$ or $B_{loc}(\Omega)$ or $C(\Omega)$], where $C_n$ is a cube of side $n$.*

(d) *$\lim_{\Lambda \nearrow \infty} |\Lambda|^{-1} \sum_{a \in \Lambda} \mu(f\, T_a g) \;=\; \mu(f)\mu(g)$ for all $f, g \in B(\Omega)$ [or $B_{ql}(\Omega)$ or $B_{loc}(\Omega)$ or $C(\Omega)$].*

We notice that the "cluster property" embodied by properties (c) and (d) is much weaker than the one presented in Section 2.3.6 [part (b) of Proposition 2.19]: (c) and (d) state that distant regions of the lattice are asymptotically independent, but only in an averaged sense. A measure $\mu \in M_{+1,inv}(\Omega)$ having the properties listed in Proposition 2.30 is said to be *ergodic*.

Therefore, by considerations analogous to those of Section 2.3.6, if we consider the translation-invariant functions to be the only "macroscopic" observables, then the ergodic measures are associated to physical "macrostates" and their convex combinations to "mixtures" representing ignorance on the part of the experimenter. Note that, as in the first part of Section 2.3.6 (through Proposition 2.19), we have not made any reference to interactions, specifications or Gibbsianness; the present comments have general validity.

We have now introduced two distinct classes of observables that could plausibly be called "macroscopic": the *global* observables (Section 2.3.6) and the *translation-invariant* observables (present section). Which class truly corresponds to the "experimentally accessible" observables? This question does not have a canonical answer: it all depends on the system and the experiments. It is known [157, Proposition 14.9] that for a translation-invariant measure $\mu$, every translation-invariant function is measurable at infinity, modulo a set of $\mu$-measure zero. The converse is *not* true. By limiting ourselves to translation-invariant observables, we eliminate some not-very-accessible global observables, like the staggered magnetization mentioned in Section 2.3.6.

Analogous questions could be posed in relation to whether the extremal measures of $\mathcal{G}(\Pi)$ or the extremal measures of $M_{+1,inv}(\Omega)$ should represent physical "macrostates". We shall comment briefly on this point once we define the notion of translation-invariant specifications (Section 2.4.9). For now, let us comment that the ergodic measures have the additional appeal of being precisely those for which "space averages equal ensemble averages":

**Proposition 2.31 (Ergodic theorem)** *Let $\mu$ be an ergodic translation-invariant probability measure on $\Omega$, and let $f \in L^1(\mu)$. Then:*

(a) $\lim_{\Lambda \nearrow \infty} |\Lambda|^{-1} \sum_{a \in \Lambda} T_a f \;=\; \int f\, d\mu$ *in $L^1(\mu)$ norm.*

(b) $\lim_{n \to \infty} n^{-d} \sum_{a \in C_n} T_a f = \int f \, d\mu$ *pointwise $\mu$-a.e.*

Part (a) is called the $L^1$ (or mean) ergodic theorem; it is easily generalized to $L^p$ for all $p < \infty$. Part (b), which is much deeper, is called the Birkhoff (or individual) ergodic theorem.

The simplex $M_{+1,inv}(\Omega)$ of translation-invariant measures has the property that its extremal elements — namely, the ergodic measures — are *dense* in the whole set, in the bounded quasilocal or weak quasilocal topology. In other words, any translation-invariant measure $\mu$ can be approximated arbitrarily closely, with regard to any *finite* set of *(quasi)local* observables, by ergodic measures. Physically this means that through observations in any *finite* volume, no matter how large, one cannot learn the long-range correlation properties of the measure $\mu$ (ergodicity or the lack thereof). The proof of this fact is really quite simple: Pave $\mathbb{Z}^d$ by cubes of side $n$; let $\mu_n$ be equal to $\mu$ on each cube, but *independent* between cubes (i.e. cut the correlations between distinct cubes); and finally, let $\widetilde{\mu}_n$ be $\mu_n$ averaged over the $n^d$ possible translates (so as to make it translation-invariant). Then it is easy to see that $\widetilde{\mu}_n$ is ergodic, and that $\lim_{n \to \infty} \widetilde{\mu}_n = \mu$ in the bounded quasilocal topology. We have just sketched the proof of:

**Proposition 2.32** *The ergodic measures are a* dense *subset of $M_{+1,inv}(\Omega)$, in the bounded quasilocal topology [and hence also in the weak quasilocal topology].*

The density of the ergodic measures is thus an intrinsic and natural feature of infinite-volume physics. Geometrically, however, a simplex with dense extreme points (a so-called *Poulsen simplex*) is highly unintuitive. Indeed, our usual intuition, derived from finite-dimensional geometry, is that the extreme points should form a *closed* subset (as e.g. the vertices of a triangle, of a tetrahedron, etc.). The unusual behavior of $M_{+1,inv}(\Omega)$ is possible only in infinite dimensions. It will be at the origin of many of the "pathologies" to be discussed in Section 2.6.7.

**Remark.** It is an amazing mathematical fact that a (compact metrizable) simplex with dense extreme points is essentially unique: all Poulsen simplices are affinely homeomorphic to each other [252, 284].

If we think of the ergodic measures as representing all the "macrostates" available to the system, it is natural to inquire whether the *translation-invariant* observables distinguish between different such measures, as is desirable on physical grounds (see the analogous discussion at the end of Section 2.3.6). The answer is yes:

**Theorem 2.33** *(a) The extremal measures of $M_{+1,inv}(\Omega)$ (i.e. the ergodic measures) are mutually singular when restricted to the $\sigma$-field $\mathcal{F}_{inv}$ of translation-invariant events. That is, if $\mu$ and $\nu$ are distinct ergodic measures, there exists a set $A \in \mathcal{F}_{inv}$ such that $\mu(A) = 1$ and $\nu(A) = 0$.*

(b) Each measure $\mu \in M_{+1,inv}(\Omega)$ is uniquely determined [among the measures of $M_{+1,inv}(\Omega)$] by the translation-invariant events. That is, $\mu$ and $\nu$ are measures in $M_{+1,inv}(\Omega)$ such that $\mu(A) = \nu(A)$ for each $A \in \mathcal{F}_{inv}$, then $\mu = \nu$.

For the proof, see [157, Theorem 14.5]. In fact, this theorem is also true with the invariant field $\mathcal{F}_{inv}$ replaced everywhere by the tail field $\widehat{\mathcal{F}}_\infty$; this follows from [157, Proposition 14.9].

### 2.4.3 Dividing Out Translation Invariance

Translation invariance brings along some natural notions of "equivalence". For instance, different observables cannot always be distinguished when looked at in a *translation-invariant* measure. (**Example:** $\sigma_0$ versus $\sigma_{17}$.) In this section we discuss the central object generating all these notions of "equivalence", namely the set of functions that have zero average with respect to all translation-invariant measures.

From now on until the end of Section 2, we shall generally assume that the single-spin space $\Omega_0$ is a compact metric space, i.e. we restrict attention to models of *bounded spins*. The configuration space $\Omega$ is then also compact. This restriction is made primarily to simplify the exposition; in Appendix A we partially remove this restriction.

The functions of interest here are characterized by the following proposition:

**Proposition 2.34** *Let $\Omega_0$ be a compact metric space, and let $f \in C(\Omega)$. Then the following properties are equivalent:*

(a) *$f$ has zero mean with respect to every translation-invariant probability measure, i.e. $\int f \, d\mu = 0$ for all $\mu \in M_{+1,inv}(\Omega)$.*

(b) *$f$ has zero mean with respect to every translation-invariant finite signed measure, i.e. $\int f \, d\mu = 0$ for all $\mu \in M_{inv}(\Omega)$.*

(c) *$f$ lies in the closed linear span of the family of functions $\{g - T_a g\colon g \in C(\Omega),\, a \in \mathbb{Z}^d\}$.*

(d) $\displaystyle \lim_{n\to\infty} n^{-d} \Big\| \sum_{a \in C_n} T_a f \Big\|_\infty = 0.$

(e) $\displaystyle \lim_{\Lambda \nearrow \infty} |\Lambda|^{-1} \Big\| \sum_{a \in \Lambda} T_a f \Big\|_\infty = 0.$

We denote by $\mathcal{I}$ the class of functions having the properties specified in the foregoing proposition; it is a closed linear subspace of $C(\Omega)$, and is exactly the annihilator of $M_{inv}(\Omega)$. The space $\mathcal{I}$ will play a very important role in the theory of translation-invariant equilibrium measures, and in particular in the discussion of "physical equivalence". We define the quotient (semi)norms:

$$\|f\|_{C(\Omega)/const} \equiv \inf_{c \in \mathbb{R}} \|f - c\|_\infty = \tfrac{1}{2}(\sup f - \inf f) \tag{2.41}$$

$$\|f\|_{C(\Omega)/\mathcal{I}} \equiv \inf_{g \in \mathcal{I}} \|f - g\|_\infty \tag{2.42}$$

$$\|f\|_{C(\Omega)/(\mathcal{I}+const)} \equiv \inf_{g \in \mathcal{I}+const} \|f - g\|_\infty \tag{2.43}$$

The quotient (semi)norms in $C(\Omega)/\mathcal{I}$ and $C(\Omega)/(\mathcal{I}+const)$ are given by simple explicit formulae:

**Proposition 2.35** *Let $f \in C(\Omega)$. Then:*

*(a)* $\displaystyle\lim_{\Lambda \nearrow \infty} |\Lambda|^{-1} \Big\| \sum_{a \in \Lambda} T_a f \Big\|_\infty$ *exists and equals* $\|f\|_{C(\Omega)/\mathcal{I}}$.

*(b)* $\displaystyle\lim_{\Lambda \nearrow \infty} |\Lambda|^{-1} \Big\| \sum_{a \in \Lambda} T_a f \Big\|_{C(\Omega)/const}$ *exists and equals* $\|f\|_{C(\Omega)/(\mathcal{I}+const)}$.

### 2.4.4 Spaces of Translation-Invariant Interactions

With $\mathcal{L} = \mathbb{Z}^d$, it also makes sense to discuss translation-invariance of interactions:

**Definition 2.36** *An interaction $\Phi = (\Phi_A)$ is said to be* translation-invariant *if*

$$\Phi_{A+x} = T_x \Phi_A \qquad \text{for all } A \in \mathcal{S},\ x \in \mathbb{Z}^d . \tag{2.44}$$

For example, the Ising interaction (2.12) is translation-invariant iff $J_{xy} = J(x-y)$ and $h_x = h = $ constant.

We now introduce some important Banach spaces of interactions:

**Definition 2.37** *For each $\alpha \geq 0$, we denote by $\mathcal{B}^\alpha$ the space of translation-invariant continuous interactions with norm*

$$\|\Phi\|_{\mathcal{B}^\alpha} \equiv \sum_{X \ni 0} |X|^{\alpha-1} \|\Phi_X\|_\infty < \infty . \tag{2.45}$$

*More generally, for any translation-invariant function $h \colon \mathcal{S} \to [1, \infty)$, we let $\mathcal{B}_h$ be the space of translation-invariant continuous interactions with norm*

$$\|\Phi\|_{\mathcal{B}_h} \equiv \sum_{X \ni 0} \frac{h(X)}{|X|} \|\Phi_X\|_\infty < \infty . \tag{2.46}$$

The most important of these spaces are $\mathcal{B}^0$ ("Israel's big Banach space") and $\mathcal{B}^1$ ("Israel's small Banach space"). Indeed, $\mathcal{B}^0$ is naturally related to $C(\Omega)$ [see Proposition 2.40 below], and so will be the natural space on which to develop the theory of equilibrium measures (Section 2.6); while $\mathcal{B}^1$ is the space of translation-invariant absolutely summable continuous interactions (see Definition 2.3), and so is a natural space for the theory of Gibbs measures. Note that our assumption $h \geq 1$ implies that $\|\Phi\|_{\mathcal{B}_h} \geq \|\Phi\|_{\mathcal{B}^0}$ and hence $\mathcal{B}_h \subset \mathcal{B}^0$; so $\mathcal{B}^0$ is the largest space of interactions that we shall consider.

Let us also introduce the space $\mathcal{B}_{finite}$ consisting of all *finite-range* translation-invariant continuous interactions. $\mathcal{B}_{finite}$ is a dense linear subspace of each of the Banach spaces $\mathcal{B}_h$. It will sometimes be convenient to carry out proofs first for some class of "nice" interactions — e.g. finite-range ones — and then extend to more general interactions by a density argument.

**Remark.** The hypothesis of *continuity* of the interaction plays a role in some but not all of the theorems below (the mathematically inclined reader is invited to figure out which ones). To avoid complicating the notation, we have included continuity as part of the definition of the spaces $\mathcal{B}^\alpha$, $\mathcal{B}_h$ and $\mathcal{B}_{finite}$.

We emphasize that all the spaces $\mathcal{B}^\alpha$ permit two-body (or more generally $n$-body) interactions of *arbitrarily long range*, provided only that they are absolutely summable. Indeed, for a pure $n$-body interaction $\Phi$, the norms $\|\cdot\|_{\mathcal{B}^\alpha}$ are all equivalent: we have $\|\Phi\|_{\mathcal{B}^\alpha} = n^\alpha \|\Phi\|_{\mathcal{B}^0}$. The difference between the spaces $\mathcal{B}^\alpha$ is that lower values of $\alpha$ permit interactions which contain heavier contributions from large $n$, i.e. which are "more strongly many-body". If we want to force $\Phi$ to be "short-range", we must take $h(X)$ to grow to $+\infty$ as the *diameter* of $X$ (and not just its cardinality) tends to infinity [199, 294]:

**Definition 2.38** *We write $h \gtrsim 1$ if, for each $K < \infty$, there exists $R = R(K) < \infty$ such that $h(X) \geq K$ whenever $\mathrm{diam}(X) \geq R$. [Equivalently, for each $K < \infty$, there are only finitely many $X$ (modulo translation) such that $h(X) < K$.] In this case we say that $\mathcal{B}_h$ is a space of* short-range interactions.

The following proposition will be useful in Sections 3.3 and 5.1.2:

**Proposition 2.39** *Fix a translation-invariant weight function $h\colon \mathcal{S} \to [1, \infty)$, and fix $M < \infty$. Then:*

(a) *The ball $\{\Phi\colon \|\Phi\|_{\mathcal{B}_h} \leq M\}$ is a* closed *subset of $\mathcal{B}^0$.*

(b) *If $h \gtrsim 1$ and the single-spin space $\Omega_0$ is finite, then the ball $\{\Phi\colon \|\Phi\|_{\mathcal{B}_h} \leq M\}$ is a* compact *subset of $\mathcal{B}^0$.*

**Remark.** Since we are here using the sup norm $\|\Phi_A\|_\infty$ to measure the strength of an interaction, all of the above spaces consist solely of *bounded* interactions. This is fine for systems of bounded spins, but these spaces are not adequate for treating physically interesting systems of *unbounded* spins (Gaussian model, $\varphi^4$ model, SOS model, etc.). It is an open problem to devise a physically reasonable and adequately comprehensive space of interactions for unbounded-spin systems. We remark that any such space is unlikely to be a linear space, because it is perfectly possible for an interaction $\Phi$ to be reasonable while $-\Phi$ is unstable. Nor can it be a convex cone, because $\Phi$ may be reasonable while $\lambda\Phi$ is unstable for $\lambda$ large and positive. However, such a space could conceivably be a convex subset of an appropriate linear space.

### 2.4.5 The Observable $f_\Phi$ Corresponding to an Interaction $\Phi$

Let $\Phi$ be an interaction in $\mathcal{B}^0$. Then it is useful to define an observable (= function) $f_\Phi$ that corresponds roughly to "the contribution to the energy from the neighborhood of the origin":

$$f_\Phi \;\equiv\; \sum_{X \ni 0} |X|^{-1}\, \Phi_X \;. \tag{2.47}$$

It is obvious from the definition of $\mathcal{B}^0$ that this sum is convergent in $\|\cdot\|_\infty$ norm, and that $\|f_\Phi\|_\infty \leq \|\Phi\|_{\mathcal{B}^0}$. Note also that $f_\Phi$ is a *quasilocal* function, i.e. $f \in C_{ql}(\Omega)$.

This definition of $f_\Phi$ is not unique: one could equally well use instead

$$f'_\Phi \equiv \sum_{X \ni_{min} 0} \Phi_X \qquad (2.48)$$

where $X \ni_{min} 0$ denotes that 0 is the smallest element of $X$ in lexicographic order, or many other definitions [313, Section 3.2]. The important point is that all such definitions give the same value for the mean of $f_\Phi$ with respect to any translation-invariant measure (that is, they give the same "mean energy per site"); in other words, any two such definitions of $f_\Phi$ differ by an element of the space $\mathcal{I}$ defined in Proposition 2.34. Therefore, what is defined naturally is not the map $\Phi \mapsto f_\Phi$ of $\mathcal{B}^0$ into $C(\Omega)$, but rather the map $\Phi \mapsto [f_\Phi]$ of $\mathcal{B}^0$ into the quotient space $C(\Omega)/\mathcal{I}$. We can then define the following subspaces of $\mathcal{B}^0$

$$\begin{align}
Const &= \{\Phi\colon f_\Phi = constant\} & (2.49) \\
\mathcal{J} &= \{\Phi\colon f_\Phi \in \mathcal{I}\} & (2.50) \\
\mathcal{J} + Const &= \{\Phi\colon f_\Phi \in \mathcal{I} + const\} & (2.51)
\end{align}$$

and the corresponding quotient (semi)norms

$$\begin{align}
\|\Phi\|_{\mathcal{B}^0/Const} &= \inf_{\Psi \in Const} \|\Phi - \Psi\|_{\mathcal{B}^0} & (2.52) \\
\|\Phi\|_{\mathcal{B}^0/\mathcal{J}} &= \inf_{\Psi \in \mathcal{J}} \|\Phi - \Psi\|_{\mathcal{B}^0} & (2.53) \\
\|\Phi\|_{\mathcal{B}^0/(\mathcal{J}+Const)} &= \inf_{\Psi \in \mathcal{J}+Const} \|\Phi - \Psi\|_{\mathcal{B}^0} & (2.54)
\end{align}$$

It is then not difficult to verify that:

**Proposition 2.40** *Let $\Omega_0$ be a compact metric space. Then the map $[\Phi] \mapsto [f_\Phi]$ is an isometry of $\mathcal{B}^0/\mathcal{J}$ onto $C(\Omega)/\mathcal{I}$, and of $\mathcal{B}^0/(\mathcal{J} + Const)$ onto $C(\Omega)/(\mathcal{I} + const)$.*

### 2.4.6 Physical Equivalence in the Ruelle Sense

The discussion in the preceding section motivates the following definition:

**Definition 2.41** *Let $\Phi, \Phi' \in \mathcal{B}^0$. We say that $\Phi$ and $\Phi'$ are* physically equivalent in the Ruelle sense *if $\Phi - \Phi' \in \mathcal{J} + Const$, i.e. if $f_\Phi - f_{\Phi'} \in \mathcal{I} + const$.*

Ruelle [313] was the first, to our knowledge, to highlight the central role played by the subspace $\mathcal{I}$ in the variational theory (see also [199, 209]).

We have now defined two distinct notions of "physical equivalence" for interactions:

- The DLR sense (Section 2.3.5), which is defined for arbitrary convergent (but not necessarily translation-invariant) interactions, and which guarantees the equality of the specifications (Theorem 2.17).

- The Ruelle sense, which is defined for arbitrary translation-invariant (but not necessarily absolutely summable or even convergent) interactions in $\mathcal{B}^0$, and which guarantees the equality of the family of equilibrium measures (Proposition 2.65 below).

It is natural to ask, therefore, whether these two notions are equivalent on their common domain of definition. The answer, fortunately, is yes:

**Theorem 2.42** *Let the single-spin space $\Omega_0$ be a complete separable metric space, and let $\Phi, \Phi'$ be interactions in $\mathcal{B}^1$. Then $\Phi$ and $\Phi'$ are physically equivalent in the DLR sense if and only if they are physically equivalent in the Ruelle sense.*

In Sections 3.3 and 5.1.2 we will need a version of Proposition 2.39 "modulo physical equivalence". Unfortunately, we have not been able to prove such a result for $\mathcal{B}^1$ (or any space $\mathcal{B}^\alpha$), and we do not know whether it it true. All we have is a result for spaces $\mathcal{B}_h$ of *short-range* interactions:

**Proposition 2.43** *If $h \gtrapprox 1$ and the single-spin space $\Omega_0$ is finite, then for each $M < \infty$ the sets $\{\Phi: \|\Phi\|_{\mathcal{B}_h/\mathcal{J}} \leq M\}$ and $\{\Phi: \|\Phi\|_{\mathcal{B}_h/(\mathcal{J}+Const)} \leq M\}$ are closed subsets of $\mathcal{B}^0$.*

### 2.4.7 Estimates on Hamiltonians: Bulk versus Surface Effects

We can now prove some estimates on the finite-volume Hamiltonians, which will play a key role both in the variational theory (Section 2.6) and in our applications to the renormalization group (Sections 3.2 and 3.3). The main physical idea in these estimates is to distinguish between "bulk" effects [namely, those which are of order $|\Lambda|$] and "surface" effects [those which are $o(|\Lambda|)$]. The upshot is that, provided one can control the surface contributions, many natural quantities are equivalent "in the bulk": this includes the Hamiltonians $H^\Phi_{\Lambda,free}$ and $H^\Phi_{\Lambda,\tau}$, as well as the "Hamiltonian-like objects" $\sum_{x \in \Lambda} T_x f_\Phi$ and $-\log d\mu_\Lambda/d\mu^0_\Lambda$.

For *free* boundary conditions, it suffices to take $\Phi$ in the "big" Banach space $\mathcal{B}^0$:

**Proposition 2.44** *Let $\Phi \in \mathcal{B}^0$. Then*
*(a)*
$$\|H^\Phi_{\Lambda,free}\|_\infty \leq |\Lambda| \, \|\Phi\|_{\mathcal{B}^0} . \tag{2.55}$$
*(b)*
$$\begin{aligned}\|H^\Phi_{\Lambda,free}\|_\infty &= |\Lambda| \, \|\Phi\|_{\mathcal{B}^0/\mathcal{J}} + o(|\Lambda|) \\ &= |\Lambda| \, \|f_\Phi\|_{C(\Omega)/\mathcal{I}} + o(|\Lambda|)\end{aligned} \tag{2.56}$$

*as $\Lambda \nearrow \infty$ (van Hove).*

(c)
$$\begin{aligned}\|H^\Phi_{\Lambda,free}\|_{C(\Omega)/const} &= |\Lambda|\,\|\Phi\|_{\mathcal{B}^0/(\mathcal{J}+Const)} \;+\; o(|\Lambda|) \\ &= |\Lambda|\,\|f_\Phi\|_{C(\Omega)/(\mathcal{I}+const)} \;+\; o(|\Lambda|)\end{aligned} \qquad (2.57)$$

as $\Lambda \nearrow \infty$ (van Hove).

(d)
$$\left\|H^\Phi_{\Lambda,free} - \sum_{x\in\Lambda} T_x f_\Phi\right\|_\infty \;\le\; o(|\Lambda|) \qquad (2.58)$$

as $\Lambda \nearrow \infty$ (van Hove).

Note, in particular, part (d) of this proposition: since $f_\Phi$ is (roughly) "the contribution to the energy from the neighborhood of the origin", it follows that $\sum_{x\in\Lambda} T_x f_\Phi$ ought to be (roughly) "the contribution to the energy from the volume $\Lambda$". And indeed it is: while this sum does not exactly equal $H^\Phi_{\Lambda,free}$, it differs from it only by a "surface" term. In this sense, $\sum_{x\in\Lambda} T_x f_\Phi$ can be thought of as yet another Hamiltonian for volume $\Lambda$, corresponding to some new type of "boundary condition".

In order to control the Hamiltonians with general external boundary conditions, it is necessary to take $\Phi$ to lie in the "small" Banach space $\mathcal{B}^1$:

**Proposition 2.45** *Let $\Phi \in \mathcal{B}^1$. Then:*

(a) $\Phi$ is absolutely summable, and
$$\|H^\Phi_\Lambda\|_\infty \;\le\; |\Lambda|\,\|\Phi\|_{\mathcal{B}^1} \;. \qquad (2.59)$$

(b)
$$\begin{aligned}\|H^\Phi_\Lambda\|_\infty &= |\Lambda|\,\|\Phi\|_{\mathcal{B}^0/\mathcal{J}} \;+\; o(|\Lambda|) \\ &= |\Lambda|\,\|f_\Phi\|_{C(\Omega)/\mathcal{I}} \;+\; o(|\Lambda|)\end{aligned} \qquad (2.60)$$

as $\Lambda \nearrow \infty$ (van Hove).

(c)
$$\begin{aligned}\|H^\Phi_\Lambda\|_{C(\Omega)/const} &= |\Lambda|\,\|\Phi\|_{\mathcal{B}^0/(\mathcal{J}+Const)} \;+\; o(|\Lambda|) \\ &= |\Lambda|\,\|f_\Phi\|_{C(\Omega)/(\mathcal{I}+const)} \;+\; o(|\Lambda|)\end{aligned} \qquad (2.61)$$

as $\Lambda \nearrow \infty$ (van Hove).

(d)
$$\|W^\Phi_{\Lambda,\Lambda^c}\|_\infty \equiv \|H^\Phi_\Lambda - H^\Phi_{\Lambda,free}\|_\infty = \sup_{\tau\in\Omega}\|H^\Phi_{\Lambda,\tau} - H^\Phi_{\Lambda,free}\|_\infty \;\le\; o(|\Lambda|) \quad (2.62\mathrm{a})$$

$$\left\|H^\Phi_\Lambda - \sum_{x\in\Lambda} T_x f_\Phi\right\|_\infty \;\le\; \sup_{\tau\in\Omega}\left\|H^\Phi_{\Lambda,\tau} - \sum_{x\in\Lambda} T_x f_\Phi\right\|_\infty \;\le\; o(|\Lambda|) \quad (2.62\mathrm{b})$$

as $\Lambda \nearrow \infty$ (van Hove).

In summary, $\Phi \in \mathcal{B}^0$ suffices to control the Hamiltonian with *free* boundary conditions, but $\Phi \in \mathcal{B}^1$ is needed in order to control the Hamiltonian with *external* boundary conditions and hence to apply the theory of specifications and Gibbs measures.

### 2.4.8 How to Obtain an Interaction from a Gibbs Measure

If $\mu$ is a Gibbs measure for an interaction $\Phi \in \mathcal{B}^1$, then the DLR equations permit us to read off the interaction $\Phi$, modulo physical equivalence, from the measure $\mu$:

**Proposition 2.46**

(a) Let $\mu$ be a Gibbs measure (not necessarily translation-invariant) for an interaction $\Phi \in \mathcal{B}^1$. Then
$$\left\| -\log \frac{d\mu_\Lambda}{d\mu_\Lambda^0} - \sum_{x \in \Lambda} T_x f_\Phi \right\|_{C(\Omega)/const} \leq o(|\Lambda|) \tag{2.63}$$
as $\Lambda \nearrow \infty$ (van Hove). In fact, this bound is uniform for $\mu \in \mathcal{G}(\Pi^\Phi)$.

(b) Let $\mu_1, \mu_2$ be Gibbs measures (not necessarily translation-invariant) for interactions $\Phi_1, \Phi_2 \in \mathcal{B}^1$, respectively. Then
$$\left\| \log \frac{d\mu_{1\Lambda}}{d\mu_{2\Lambda}} \right\|_\infty \leq 2|\Lambda| \, \|\Phi_1 - \Phi_2\|_{\mathcal{B}^0/(\mathcal{J}+Const)} + o(|\Lambda|) \tag{2.64}$$
$$\left\| \log \frac{d\mu_{1\Lambda}}{d\mu_{2\Lambda}} \right\|_{C(\Omega)/const} = |\Lambda| \, \|\Phi_1 - \Phi_2\|_{\mathcal{B}^0/(\mathcal{J}+Const)} + o(|\Lambda|) \tag{2.65}$$
as $\Lambda \nearrow \infty$ (van Hove). In fact, this bound is uniform for $\mu_1 \in \mathcal{G}(\Pi^{\Phi_1})$ and $\mu_2 \in \mathcal{G}(\Pi^{\Phi_2})$.

Part (a) of this proposition tells us that the interaction can be reconstructed by taking the logarithm of the finite-volume densities. This corresponds to the fact that Boltzmann factors are exponentials of Hamiltonians. An immediate consequence of this is part (b). One implication of part (b) is that the reconstructed interaction is unique modulo physical equivalence (Griffiths-Ruelle theorem): just take $\mu_1 = \mu_2$ in (2.65) to conclude that $\|\Phi_1 - \Phi_2\|_{\mathcal{B}^0/(\mathcal{J}+Const)} = 0$. In other words, if $\mu$ is a Gibbs measure for interactions $\Phi_1, \Phi_2 \in \mathcal{B}^1$, then $\Phi_1$ and $\Phi_2$ must be physically equivalent in the Ruelle sense. Of course, we already knew this (Corollary 2.18 plus Theorem 2.42).

It is curious that although $\Phi_1, \Phi_2$ are required to belong to the "small" Banach space $\mathcal{B}^1$, the final estimate is in terms of the $\mathcal{B}^0/\mathcal{J}$ norm, hence much stronger. The reason is that $\Phi_1, \Phi_2 \in \mathcal{B}^1$ is needed in order to ensure that the *boundary* energy contributions are indeed $o(|\Lambda|)$; but once this is done, then the *bulk* energy contribution is determined by the $\mathcal{B}^0/\mathcal{J}$ norm, as in Proposition 2.44(b).

### 2.4.9 Translation-Invariant Specifications and Gibbs Measures

We can now examine the theory of specifications and Gibbs measures under the hypothesis of translation invariance.

**Definition 2.47** *A specification* $\Pi = (\pi_\Lambda)_{\Lambda \in \mathcal{S}}$ *is said to be* translation-invariant *if*

$$\pi_\Lambda(\omega, A) = \pi_{\Lambda+a}(T_a\omega, T_a A) \qquad (2.66)$$

*for all* $\Lambda \in \mathcal{S}$, $\omega \in \Omega$, $A \in \mathcal{F}$ *and* $a \in \mathbb{Z}^d$.

In particular, if $\Phi$ is a translation-invariant (and convergent, $\mu^0$-admissible) interaction, then $\Pi^\Phi$ is obviously a translation-invariant specification.

Fix a translation-invariant specification $\Pi$. We denote by $\mathcal{G}_{inv}(\Pi) \equiv \mathcal{G}(\Pi) \cap M_{+1,inv}(\Omega)$ the set of all translation-invariant measures consistent with $\Pi$. $\mathcal{G}_{inv}(\Pi)$ is a convex set, and its extreme points are characterized by the following theorem:

**Proposition 2.48** *Let $\Pi$ be a translation-invariant specification. Then:*

*(a) A measure $\mu \in \mathcal{G}_{inv}(\Pi)$ is extremal in $\mathcal{G}_{inv}(\Pi)$ if and only if it is extremal in $M_{+1,inv}(\Omega)$, i.e. if and only if it is ergodic.*

*(b) $\mathcal{G}_{inv}(\Pi)$ is a face of $M_{+1,inv}(\Omega)$: that is, if $\mu, \nu \in M_{+1,inv}(\Omega)$ and $0 < \lambda < 1$ are such that $\lambda\mu + (1-\lambda)\nu \in \mathcal{G}_{inv}(\Pi)$, then in fact $\mu, \nu \in \mathcal{G}_{inv}(\Pi)$.*

It is now the right moment to make some remarks that may at first seem pedantic, but could actually be helpful to people haunted by an (unfortunately established) terminology that is confusing or at least bothersomely subtle. The situation is as follows. If the specification $\Pi$ is translation-invariant, we have at our disposal two different spaces of measures of physical interest:

- $\mathcal{G}(\Pi)$, the space of *all* measures consistent with $\Pi$, whether or not they are translation-invariant; and

- $\mathcal{G}_{inv}(\Pi)$, the space of all *translation-invariant* measures consistent with $\Pi$.

Physical "macrostates" are interpreted as extremal measures, but the question is: extremal in which space? It is important to observe that we have *three* possibilities:

(i) The extremal points of $\mathcal{G}(\Pi)$. These measures are characterized by very strict properties (Proposition 2.19): they show no fluctuations for the observables measurable at infinity ("global observables") — which, for translation-invariant measures, is larger than the set of translation-invariant observables ("macroscopic observables") — and they exhibit very strong cluster properties (short-range correlations).

(ii) The translation-invariant extremal points of $\mathcal{G}(\Pi)$. This is often a small set, and in many cases it is empty. For example, in the two-dimensional Ising antiferromagnet at low temperature, there are only two extremal Gibbs measures: one has + magnetization on the even sublattice and − magnetization on the odd sublattice (let us call this measure $\mu_\pm$), and the other has the reverse magnetization (call this measure $\mu_\mp$). Neither of these two measures is translation-invariant, so the set in question is empty. More dramatically, there are examples due to van Enter and Miękisz [356] in which there are not even any *periodic* extremal Gibbs measures.

(iii) The extremal points of $\mathcal{G}_{inv}(\Pi)$. This is a much larger set than the one discussed in (ii). In particular it is never empty for compact-spin models (see below). For instance, in the example of the Ising antiferromagnet, there is only one translation-invariant Gibbs measure — $\frac{1}{2}(\mu_\pm + \mu_\mp)$ — which is obviously extremal in $\mathcal{G}_{inv}(\Pi)$ but not in $\mathcal{G}(\Pi)$. These measures satisfy the comparatively weaker properties of Proposition 2.30: they are deterministic for the smaller set of translation-invariant observables, and they exhibit the cluster property only in the weakest (Cesaro-averaged) sense, namely ergodicity. In the mathematical statistical-mechanics literature, these measures — the extremal points of $\mathcal{G}_{inv}(\Pi)$, or equivalently the ergodic elements of $\mathcal{G}_{inv}(\Pi)$ — are called *pure phases* for the specification $\Pi$. (Unfortunately, the term "pure phase" is sometimes used with different but closely related meanings: see e.g. Appendix B.3.1.)

Which set is interpreted as representing the physical "macrostates" is a problem-dependent issue. In problems where non-translation-invariant measures are relevant (interfaces, surface tension, crystal shape, wetting, systems with disorder, quasicrystals), it is mandatory to consider the set $\mathcal{G}(\Pi)$ of *all* Gibbs measures. Then the "macrostates" should correspond to the measures in (i), and the *translation-invariant* "macrostates" should correspond to the measures in (ii). On the other hand, if one limits oneself to measuring *bulk* observables (i.e. macroscopic averages), then it is natural to consider only the *translation-invariant* Gibbs measures $\mathcal{G}_{inv}(\Pi)$ and *their* extreme points: that is, (iii) is the natural choice [(ii) being often too small, e.g. empty]. In this regard, the use of the catchy label "pure phases" for the measures in (iii) is on the one hand natural, given the traditional interest in "macrostates" with symmetry under translations, but on the other hand unfortunate for the current interest in more general phenomena. A nomenclature more consistent with our purposes could be to call *extremal Gibbs measures* those in (i), *translation-invariant extremal Gibbs measures* those in (ii), and just *ergodic Gibbs measures* those of (iii) [or *extremal translation-invariant Gibbs measures*, provided that we pay attention to the subtleties of word-ordering]. In any case, in the remainder of this paper we shall use the term "phase" or "pure phase" to denote the measures in (iii), *with one exception*: in Appendix B (and only there!) we shall succumb to the customary terminology of Pirogov-Sinai theory (as well as brevity) and use the term "pure phase" to denote the measures in (ii) [in fact a slight generalization of them].

Regarding the conditions under which the set $\mathcal{G}_{inv}(\Pi)$ is non-empty, it suffices to mention a result analogous to Proposition 2.21 (Section 2.3.6):

**Proposition 2.49** *Let $\Omega$ be a compact metric space, and let $\Pi = (\pi_\Lambda)_{\Lambda \in \mathcal{S}}$ be a translation-invariant Feller specification. Then $\mathcal{G}_{inv}(\Pi)$ is nonempty.*

Because the translations form an Abelian group, this is an immediate consequence of Proposition 2.21 and the Markov-Kakutani theorem [104, 306]. The idea is that, given a measure $\mu \in \mathcal{G}(\Pi)$, we can construct a measure in $\mathcal{G}_{inv}(\Pi)$ by averaging $\mu$ over translations (and extracting, if necessary, a convergent subsequence).

**Remark.** One would like to have a translation-invariant version of the Gibbs Representation Theorem (Theorem 2.12). That is, if $\Pi$ is a quasilocal, uniformly nonnull and *translation-invariant* specification, one would like to prove that there exists an absolutely summable *translation-invariant* interaction $\Phi$ such that $\Pi = \Pi^\Phi$. However, it seems to be an open question whether this is true or not. Sullivan [336, Corollary to Theorem 2] constructed a translation-invariant $\Phi$ which is "*relatively* absolutely summable" (see Remark 2 at the end of Section 2.3.3), while Kozlov [222, Theorem 3] constructed a translation-invariant absolutely summable $\Phi$ under a condition on $\Pi$ *stronger* than quasilocality.[30]

## 2.5 Entropy, Large Deviations and the Variational Principle: Finite-Volume Case

[31]

We now begin the study of the second approach to classical statistical mechanics, namely the one based on the *variational principle*, which states that the Boltzmann-Gibbs measure is the one that maximizes entropy minus mean energy. The theory developed in this section is applicable to an arbitrary classical-statistical-mechanical system for which the Hamiltonian $H$ makes sense. In practice this usually means a finite-volume system. First we introduce the *free energy*; next we introduce the concept of *relative entropy* and its interpretation in terms of *large deviations*; finally we prove the *variational principle* that connects these two quantities. In Section 2.6 we will develop the analogous theory for translation-invariant infinite-volume lattice systems.

In this section we are working in a completely general classical-statistical-mechanical (= probabilistic) context: $(\Omega, \Sigma)$ is an arbitrary measurable space.

### 2.5.1 Free Energy

**Definition 2.50** *Let $\nu$ be a probability measure on $(\Omega, \Sigma)$, and let $f$ be a bounded measurable function on $\Omega$. We then define*

$$P(f|\nu) \;=\; \log \int e^f \, d\nu \;. \tag{2.67}$$

Physically, $P(f|\nu)$ is minus the free energy for a system with Hamiltonian $H = -f$ and *a priori* measure $\nu$. Our choice of sign convention makes the formulae slightly more elegant.

It is easy to prove the following properties of the free energy:

**Proposition 2.51** *Let $\nu$ be a probability measure on $(\Omega, \Sigma)$. Then $P(\,\cdot\,|\nu)$ has the following properties:*

---

[30]Kozlov's Theorem 3 uses (at least in the English translation) the words "necessary and sufficient", but in fact he proves only the sufficiency.

[31]References for this section are Georgii [157, Section 15.1], Israel [206, Section I.2 and II.2], Preston [299, Chapter 7] and Ellis [110, Chapters I, II, VII and VIII].

(a) $P(\mathbf{0}|\nu) = 0$.

(b) $f \leq g \implies P(f|\nu) \leq P(g|\nu)$.

(c) $P(f + c|\nu) = P(f|\nu) + c$ for any real number $c$.

(d) $\left|P(f|\nu) - P(g|\nu)\right| \leq \|f - g\|_\infty$. That is, $P(\,\cdot\,|\nu)$ is Lipschitz continuous with Lipschitz constant 1.

(e) $P(\,\cdot\,|\nu)$ is convex.

(f) $P(\,\cdot\,|\nu)$ is strictly convex in directions corresponding to functions which are not $\nu$-a.e. constant.

### 2.5.2  Relative Entropy

**Definition 2.52**  Let $\mu$ and $\nu$ be any two probability measures on $(\Omega, \Sigma)$. Then the relative entropy (or information gain or Kullback-Leibler information) of $\mu$ relative to $\nu$ is defined as

$$I(\mu|\nu) = \begin{cases} \int \left(\log \frac{d\mu}{d\nu}\right) d\mu = \int \left(\frac{d\mu}{d\nu} \log \frac{d\mu}{d\nu}\right) d\nu & \text{if } \mu \ll \nu \\ +\infty & \text{otherwise} \end{cases} \tag{2.68}$$

More generally, if $\mathcal{A}$ is any sub-$\sigma$-field of $\Sigma$, then we define

$$I_\mathcal{A}(\mu|\nu) = I\left(\mu\!\restriction\!\mathcal{A} \,\big|\, \nu\!\restriction\!\mathcal{A}\right). \tag{2.69}$$

Actually, our $I(\mu|\nu)$ is the *negative* of the usual relative entropy $S(\mu|\nu)$; but it is more convenient to work with $I$ than with $S$, and it is too cumbersome to keep saying the words "negative of". So we shall just call $I$ the "relative entropy" *tout court*. But this sign difference should be borne in mind when interpreting the variational principle! (See also the Remarks at the end of this subsection for a comparison with the usual physicists' entropy.)

It is not hard to prove the following properties of the relative entropy:

**Proposition 2.53**  Let $\mu, \nu$ be probability measures on $(\Omega, \Sigma)$. Then:

(a) $0 \leq I(\mu|\nu) \leq I_{max} \equiv -\log \nu_{min}$, where $\nu_{min} = \inf\limits_{\emptyset \neq A \in \Sigma} \nu(A)$. [For example, if $\nu$ is normalized counting measure on a finite space $\Omega$, then $I_{max} = \log|\Omega|$. If $\Omega$ is an infinite space, then $I_{max} = +\infty$.]

(b) $I(\mu|\nu) = 0$ if and only if $\mu = \nu$.

(c) $I(\mu|\nu)$ is a convex function of the pair $(\mu, \nu)$.

(d) *For fixed $\nu$, $I(\mu|\nu)$ is "almost" a concave function of $\mu$, in the sense that*

$$I\Big(\sum_{i=1}^{n} \lambda_i \mu_i \,|\, \nu\Big) \;\geq\; \sum_{i=1}^{n} \lambda_i I(\mu_i|\nu) \;+\; \sum_{i=1}^{n} \lambda_i \log \lambda_i \qquad (2.70\text{a})$$

$$\geq\; \sum_{i=1}^{n} \lambda_i I(\mu_i|\nu) \;-\; \log n \qquad (2.70\text{b})$$

*for any probability measures $\mu_1, \ldots, \mu_n$ and numbers $\lambda_1, \ldots, \lambda_n \geq 0$ with $\sum_{i=1}^{n} \lambda_i = 1$.*

(e) *For fixed $\nu$, $I(\mu|\nu)$ is a lower semicontinuous function of $\mu$ in the bounded measurable topology[32], and in the weak topology if $\Omega$ is a complete separable metric space.*

(f) *For fixed $\nu$ and fixed $c < \infty$, the set $\{\mu \colon I(\mu|\nu) \leq c\}$ is compact and sequentially compact in the bounded measurable topology (and hence also in the weak topology).*

(g) *$I_{\mathcal{A}}(\mu|\nu)$ is an increasing function of $\mathcal{A}$.*

(h) *If $\mathcal{A}_1 \subset \mathcal{A}_2 \subset \Sigma$, and $\mu^\omega_{\mathcal{A}_1}$ (resp. $\nu^\omega_{\mathcal{A}_1}$) is a regular conditional probability for $\mu$ (resp. $\nu$) given $\mathcal{A}_1$, then*

$$I_{\mathcal{A}_2}(\mu|\nu) \;=\; I_{\mathcal{A}_1}(\mu|\nu) \;+\; \int d\mu(\omega)\, I_{\mathcal{A}_2}(\mu^\omega_{\mathcal{A}_1} | \nu^\omega_{\mathcal{A}_1})\,. \qquad (2.71)$$

*[This obviously refines (g).]*

(i) *(Strong superadditivity) Let $\mathcal{A}_1, \mathcal{A}_2, \mathcal{A}_3$ be sub-$\sigma$-fields of $\Sigma$ which are* independent *with respect to $\nu$. Then*

$$I_{\mathcal{A}_1 \cup \mathcal{A}_2 \cup \mathcal{A}_3}(\mu|\nu) + I_{\mathcal{A}_2}(\mu|\nu) \;\geq\; I_{\mathcal{A}_1 \cup \mathcal{A}_2}(\mu|\nu) + I_{\mathcal{A}_2 \cup \mathcal{A}_3}(\mu|\nu)\,. \qquad (2.72)$$

**Remarks.** 1. The standard statistical-mechanics textbooks (e.g. [213, Chapters 2, 4 and 5], [307, Section 9.B], [20, Chapter 3]) introduce a quantity which is *apparently* the entropy of a *single* measure $\mu$, *without* reference to a base measure $\nu$:

$$S_{books}(\mu) \text{ "="} \begin{cases} -\sum_{\omega} \mu_\omega \log \mu_\omega & \text{if } \Omega \text{ is discrete} \\ -\int \mu(x) \log \mu(x)\, dx & \text{if } \Omega \text{ is continuous} \end{cases} \qquad (2.73)$$

However, closer examination reveals that a base measure $\nu$ has been introduced *surreptitiously* in these formulae, namely counting measure in the discrete case or Lebesgue measure in the continuous case. This base measure *does* play a physical role in the theory: the physics would be different if counting or Lebesgue measure were replaced by some other measure.[33] Thus, the formulae (2.73), in which the base measure $\nu$ is hidden, are quite misleading. (They are also inelegant, as can be seen from the incompatible treatment given to the discrete and continuous cases.) What is involved here is the common sin of failing to distinguish between a *measure* and a *density* (= Radon-Nikodým derivative): the latter is defined only relative to a specified base measure. In many situations, this sin is harmless, because there is a "natural" and universally-agreed choice of base measure. But not here. We therefore feel strongly that in statistical mechanics the base measure $\nu$ should be introduced *explicitly*.

---

[32]Recall that a net $\{\mu_\alpha\}$ converges to $\mu$ in the bounded measurable topology if $\int f\, d\mu_\alpha \to \int f\, d\mu$ for all $f \in B(\Omega, \Sigma)$.

Note also that the definition (2.73) uses *unnormalized* counting or Lebesgue measure as the base measure, while we always take the base measure $\nu$ to be a *probability* measure. This causes an (irrelevant) additive shift in the entropy: e.g. for $\Omega$ finite,

$$I(\mu|\nu) \;=\; -S_{books}(\mu) \;+\; \log|\Omega| \tag{2.74}$$

when $\nu$ is normalized counting measure [$\nu(\{\omega\}) = 1/|\Omega|$ for each $\omega \in \Omega$]. Thus, both $I(\mu|\nu)$ and $S_{books}(\mu)$ take values in the interval $[0, \log|\Omega|]$, but large values of $I(\mu|\nu)$ correspond to small values of $S_{books}(\mu)$, and vice versa.

The reader is urged to remember the two notational differences — the sign and the additive constant — when interpreting our results.

2. The relative entropy $I(\mu|\nu)$ plays an important role in information theory and in mathematical statistics (large-sample asymptotic theory of hypothesis testing and maximum-likelihood estimation); this follows from the large-deviations theory to be discussed in the next subsection. See e.g. [30, pp. 119–125] and [226, 225, 16]. The relationship with maximum-likelihood estimation is discussed also in Section 5.1.2 below.

### 2.5.3 Large Deviations

The physical interpretation of relative entropy is associated with the problem of *large deviations*, which concerns, roughly speaking, the estimation of the (very small) probabilities of large simultaneous fluctuations in a system consisting of a large number of random variables. In this section we will consider the case of independent, identically distributed (i.i.d.) random variables. So let $X_1, X_2, \ldots$ be a sequence of independent samples from the probability distribution $\nu$; and let $f$ be any bounded real-valued measurable function on $\Omega$. Then $f(X_1), f(X_2), \ldots$ is a sequence of independent, identically distributed real-valued random variables. In such a situation the weak law of large numbers states that the sample mean $S_n^f \equiv n^{-1} \sum_{i=1}^n f(X_i)$ is, with high probability, very close to the theoretical mean value $m \equiv \int f \, d\nu$: more precisely, if $A$ is any closed subset of the real line *not* containing $m$, then $\text{Prob}(S_n^f \in A) \to 0$ as $n \to \infty$. Large-deviation theorems [360, 110, 75] are a strengthening of the weak law of large numbers, in that they give the precise rate of convergence of this probability to zero as $n \to \infty$. It turns out that this probability is *exponentially* small in $n$, that is,

$$\text{Prob}(S_n^f \in A) \sim e^{-n \times \text{const}(f,\nu,A)} \tag{2.75}$$

where $\text{const}(f, \nu, A) > 0$ whenever $A$ is a closed set not containing $m$. More precisely, it can be shown that

$$\lim_{n \to \infty} \frac{1}{n} \log \text{Prob}(S_n^f \in A) \begin{cases} \leq - \inf_{\mu: \int f \, d\mu \in A} I(\mu|\nu) & \text{if } A \text{ is a closed set} \\ \geq - \inf_{\mu: \int f \, d\mu \in A} I(\mu|\nu) & \text{if } A \text{ is an open set} \end{cases} \tag{2.76}$$

where $I(\mu|\nu)$ is the relative entropy.

---

[33]In *some* cases, counting or Lebesgue measure may play a privileged role by virtue of some symmetry: e.g. spin-flip symmetry in the Ising model, or symplectic symmetry in a classical Hamiltonian system. In other cases, however, the privileged measure could be some other measure: e.g. Haar measure on a Lie group is not Lebesgue measure except in some very special parametrizations. This is yet another reason for making the base measure $\nu$ explicit: it clarifies whether or not there is a symmetry argument that privileges one choice of $\nu$ over another.

In the preceding thought-experiment, we looked at only one real-valued observable $f$. More generally, we could look at a vector-valued observable $\mathbf{f} = (f_1, \ldots, f_k)$, and ask for the probability that $S_n^{\mathbf{f}}$ lies in some subset $\mathbf{A} \subset \mathbb{R}^k$. Not surprisingly we have

$$\lim_{n \to \infty} \frac{1}{n} \log \text{Prob}(S_n^{\mathbf{f}} \in \mathbf{A}) \begin{cases} \leq - \inf_{\mu: \int \mathbf{f} \, d\mu \in \mathbf{A}} I(\mu|\nu) & \text{if } \mathbf{A} \text{ is a closed set} \\ \geq - \inf_{\mu: \int \mathbf{f} \, d\mu \in \mathbf{A}} I(\mu|\nu) & \text{if } \mathbf{A} \text{ is an open set} \end{cases} \tag{2.77}$$

These results can be written in a more succinct way by noting the trivial identity

$$\frac{1}{n} \sum_{i=1}^{n} f(X_i) = \left( \frac{1}{n} \sum_{i=1}^{n} \delta_{X_i} \right)(f) \tag{2.78}$$

(here $\delta_x$ is the delta measure at $x$), which can be written as

$$S_n^f = L_n(f) \equiv \int f \, dL_n \tag{2.79}$$

where

$$L_n \equiv n^{-1} \sum_{i=1}^{n} \delta_{X_i} \tag{2.80}$$

is called the *empirical measure*. We emphasize that $L_n$ is a *random* measure: it depends on the random sample $X_1, \ldots, X_n$. In this language, the weak law of large numbers can be reformulated as saying that the empirical measure $L_n$ is, with high probability, very close to the theoretical measure $\nu$, when "closeness" is understood in the bounded measurable topology (that is, the weak topology generated by the bounded measurable functions). More precisely, if $\mathsf{A}$ is any closed subset of $M_{+1}(\Omega)$ *not* containing $\nu$, then $\text{Prob}(L_n \in \mathsf{A}) \to 0$ as $n \to \infty$.[34] The large-deviation theorem [176, 69, 34] then states that this probability is in fact exponentially small in $n$, namely

$$\text{Prob}(L_n \in \mathsf{A}) \sim e^{-n \times \text{const}(\nu, \mathsf{A})} \tag{2.81}$$

---

[34]In this particular topology, a basis for the neighborhoods of $\nu$ is given by the sets

$$\mathsf{B}_{\nu, \mathbf{f}, \epsilon} \equiv \left\{ \mu: \left| \int f_i \, d\mu - \int f_i \, d\nu \right| < \epsilon \text{ for all } i = 1, \ldots, k \right\},$$

where $\text{const}(\nu, \mathsf{A}) > 0$ whenever $\mathsf{A}$ is a closed set of measures not containing $\nu$. More precisely,

$$\lim_{n \to \infty} \frac{1}{n} \log \text{Prob}(L_n \in \mathsf{A}) \begin{cases} \leq - \inf_{\mu:\, \mu \in \mathsf{A}} I(\mu|\nu) & \text{if } \mathsf{A} \text{ is a closed set} \\ \geq - \inf_{\mu:\, \mu \in \mathsf{A}} I(\mu|\nu) & \text{if } \mathsf{A} \text{ is an open set} \end{cases} \quad (2.82)$$

In fact, this result is merely a sophisticated restatement of (2.77), since every closed (resp. open) set of measures $\mathsf{A}$ is contained in (resp. contains) one of the form $\{\mu\colon \int \mathbf{f}\, d\mu \in \mathbf{A}\}$ for some $\mathbf{f} = (f_1, \ldots, f_k)$ and some $\mathbf{A}$ closed (resp. open) $\subset \mathbb{R}^k$.

Formulas (2.81)/(2.82) provide a physical interpretation of the relative entropy. Indeed, we can say (roughly speaking) that the probability that a sample $X_1, \ldots, X_n$, taken from the probability distribution $\nu$, "looks like a typical sample from $\mu$" decays exponentially with rate $I(\mu|\nu)$:

$$\text{Prob}_\nu(X_1, \ldots, X_n \text{ is typical for } \mu) \sim e^{-nI(\mu|\nu)}. \quad (2.83)$$

In the probabilistic literature, (2.76)/(2.77) are called level-1 large-deviation formulae, and (2.82) is called a level-2 large-deviation formula.

### 2.5.4 Variational Principle

The free energy and the relative entropy are related by the following *variational principle*:

**Theorem 2.54 (Variational principle)** *Fix a probability measure $\nu$ on $(\Omega, \Sigma)$. Then $P(\,\cdot\,|\nu)$ and $I(\,\cdot\,|\nu)$ are conjugate convex functions, in the sense that*

$$P(f|\nu) = \sup_{\mu \in M_{+1}(\Omega,\Sigma)} \left[ \int f\, d\mu - I(\mu|\nu) \right] \quad (2.84\text{a})$$

$$I(\mu|\nu) = \sup_{f \in B(\Omega,\Sigma)} \left[ \int f\, d\mu - P(f|\nu) \right] \quad (2.84\text{b})$$

*Moreover, the supremum is achieved if and only if $\mu$ equals the Boltzmann-Gibbs measure for Hamiltonian $H = -f$ (and a priori measure $\nu$), namely*

$$\mu_{BG,f,\nu} \equiv \frac{e^f\, d\nu}{\int e^f\, d\nu}. \quad (2.85)$$

---

where $\mathbf{f} = (f_1, \ldots, f_k)$ runs over all *finite* families of bounded measurable functions, and $\epsilon$ runs over all strictly positive numbers. By the usual weak law of large numbers we have

$$\text{Prob}(L_n \notin \mathsf{B}_{\nu,\mathbf{f},\epsilon}) \leq \sum_{i=1}^{k} \text{Prob}(|S_n^{f_i} - m_i| \geq \epsilon) \to 0$$

as $n \to \infty$, since $k$ is finite. Since any closed set $\mathsf{A} \not\ni \nu$ is contained in the complement of some set $\mathsf{B}_{\nu,\mathbf{f},\epsilon}$, the claim $\text{Prob}(L_n \in \mathsf{A}) \to 0$ is proven.

This complementary pair of variational principles establishes the equivalence of (2.1) and (2.3) for *finite-volume* statistical-mechanical systems. Indeed, $\int f \, d\mu$ is minus the mean energy for a system with Hamiltonian $H = -f$, and $I(\mu|\nu)$ is minus the entropy; therefore, (2.84a) states that the Boltzmann-Gibbs measure is the one that minimizes energy minus entropy, and that the minimum value of energy minus entropy equals the free energy. (In thermodynamic notation, $F = E - TS$; recall that we are taking $\beta = 1$.)

## 2.6 Entropy, Large Deviations and the Variational Principle: Infinite-Volume Case

[35]

The variational approach developed in the preceding section is adequate for finite-volume statistical-mechanical systems, in which the Hamiltonian $H$ is well-defined and finite. But it is (not surprisingly) insufficient for the infinite-volume case, in which all the relevant quantities — Hamiltonian, free energy, mean energy and relative entropy — are almost certainly infinite. Nevertheless, one might hope that for *translation-invariant* infinite-volume systems there would exist an analogous theory in which the concepts of free energy, mean energy and relative entropy are replaced by these same quantities *per unit volume*; one could then define an *equilibrium measure* to be a translation-invariant measure that maximizes the entropy density minus mean energy density. In this section we shall develop such a theory. But this infinite-volume theory is considerably more subtle than its finite-volume counterpart: this subtlety arises from the physical possibility of phase transitions, as well as from additional mathematical pathologies to be explained in Section 2.6.7 below.

The variational approach to infinite-volume lattice systems is less general than the one based on the DLR equations, because of its restriction to translation-invariant measures[36], but within its restricted domain it is equivalent to the DLR theory: the key theorem (Corollary 2.68) states that, for any interaction $\Phi \in \mathcal{B}^1$, the equilibrium measures coincide with the translation-invariant Gibbs measures.

### 2.6.1 Free Energy Density ("Pressure")

We look first at the free energy density, or what is equivalent, the "pressure":

**Definition 2.55** *Let $\nu$ be a translation-invariant probability measure on $\Omega = (\Omega_0)^{\mathbb{Z}^d}$, and let $f$ be a bounded measurable function. Then the* pressure *of $f$ relative to $\nu$ is*

---

[35]References for this section are Georgii [157, Chapters 15 and 16], Israel [206, Chapters I, II and V], Preston [299, Chapters 7 and 8], Ruelle [313, Chapters 3 and 4] and Ellis [110, Chapters IV and V and Appendix C].

[36]Even if the interaction is translation-invariant, there may exist non-translation-invariant Gibbs measures (e.g. for the Ising model in dimension $d \geq 3$ [85, 348]), and these are of interest in describing interfaces.

*defined as*

$$p(f|\nu) \;=\; \lim_{n\to\infty} \frac{1}{n^d} \log \int \exp\left[\sum_{x\in C_n} T_x f\right] d\nu \qquad (2.86)$$

*if this limit exists. Similarly, if $\Phi$ is an interaction in $\mathcal{B}^0$, then the* pressure *of $\Phi$ relative to $\nu$ is defined as*

$$p(\Phi|\nu) \;=\; \lim_{n\to\infty} \frac{1}{n^d} \log \int \exp\left[-H^\Phi_{C_n,free}\right] d\nu \qquad (2.87)$$

*if this limit exists.*

This quantity should really be called "minus the free energy density". The term "pressure" arises from the interpretation of the canonical-ensemble Ising model as equivalent to a grand-canonical-ensemble lattice gas; in the general case the term "pressure" is not really appropriate, but it has become standard among mathematical physicists. It has, at least, the virtue of brevity.

We emphasize that the existence of the limit (2.86) [or (2.87)] is a nontrivial problem; in fact, there exist examples of translation-invariant measures $\nu$ for which the limit does *not* exist, even for simple local functions $f$ (see Appendix A.5.2). Therefore, we shall restrict attention to two cases: when $\nu$ is a product measure, and more generally, when $\nu$ is a Gibbs measure for a translation-invariant interaction.

**Proposition 2.56** *Let $\nu$ be a product measure. Then the pressure $p(f|\nu)$ exists for all bounded quasilocal functions $f$; in fact, the limit exists also in van Hove sense, namely*

$$p(f|\nu) \;=\; \lim_{\Lambda \nearrow \infty} \frac{1}{|\Lambda|} \log \int \exp\left[\sum_{x\in\Lambda} T_x f\right] d\nu \;. \qquad (2.88)$$

*Moreover, $p(\,\cdot\,|\nu)$ has the following properties:*

*(a) $p(\mathbf{0}|\nu) \;=\; 0$.*

*(b) $f \le g \;\Longrightarrow\; p(f|\nu) \le p(g|\nu)$.*

*(c) $p(f + c|\nu) \;=\; p(f|\nu) + c$ for any real number $c$.*

*(d) $\left|p(f|\nu) - p(g|\nu)\right| \;\le\; \|f - g\|_\infty$. That is, $p(\,\cdot\,|\nu)$ is Lipschitz continuous with Lipschitz constant 1.*

*(e) $p(f + h|\nu) \;=\; p(f|\nu)$ for any $h \in \mathcal{I}$.*

*(f) $p(\,\cdot\,|\nu)$ is convex.*

We emphasize, in particular, part (e): the pressure is constant within "subspaces of physical equivalence".

**Proposition 2.57** *Let $\nu$ be a translation-invariant Gibbs measure for an interaction $\Phi \in \mathcal{B}^1$ (and a priori measure $\mu^0$). Then the pressure $p(f|\nu)$ exists for all bounded quasilocal functions $f$; in fact, the limit exists also in van Hove sense, namely*

$$p(f|\nu) \;=\; \lim_{\Lambda \nearrow \infty} \frac{1}{|\Lambda|} \log \int \exp\left[\sum_{x \in \Lambda} T_x f\right] d\nu \;. \tag{2.89}$$

*Moreover, the limit is given by*

$$p(f|\nu) \;=\; p(f - f_\Phi|\mu^0) \;-\; p(-f_\Phi|\mu^0) \;. \tag{2.90}$$

*In particular, $p(\,\cdot\,|\nu)$ has all the properties (a)–(f) of Proposition 2.56.*

The pressure of a *function* $f$ is the simplest object from a mathematical point of view, but the pressure of an *interaction* $\Phi$ is perhaps more familiar to physicists. In fact these two objects are essentially identical:

**Proposition 2.58** *Let $\Phi \in \mathcal{B}^0$, and let $\nu$ be a translation-invariant measure satisfying the conditions of Proposition 2.56 or 2.57. Then:*

(a) *$p(\Phi|\nu)$ exists and equals $p(-f_\Phi|\nu)$. In fact, the limit exists also in van Hove sense, i.e.*

$$p(\Phi|\nu) \;=\; \lim_{\Lambda \nearrow \infty} \frac{1}{|\Lambda|} \log \int \exp\left[-H^\Phi_{\Lambda,free}\right] d\nu \;. \tag{2.91}$$

(b) *If in addition $\Phi \in \mathcal{B}^1$, then for any $\tau \in \Omega$, $\lim\limits_{\Lambda \nearrow \infty} \frac{1}{|\Lambda|} \log \int \exp\left[-H^\Phi_{\Lambda,\tau}\right] d\nu$ also exists and equals $p(-f_\Phi|\nu)$.*

Part (b) states that, for interactions $\Phi \in \mathcal{B}^1$, the pressure is independent of boundary conditions.

The reader will note that we have not asserted the *strict* convexity of $p(\,\cdot\,|\nu)$; this is because, in sharp contrast to the finite-volume case, the infinite-volume pressure is *not* strictly convex (not even modulo physical equivalence). Indeed, this failure of strict convexity is at the origin of some rather surprising pathologies of the infinite-volume variational theory in the "large" space of interactions $\mathcal{B}^0$ (see Section 2.6.7 below). However, in the smaller space $\mathcal{B}^1$ these pathologies do not arise:

**Proposition 2.59 (Griffiths–Ruelle [174])** *Let $\nu$ be a translation-invariant measure satisfying the conditions of Proposition 2.56 or 2.57. Then the pressure $p(\,\cdot\,|\nu)$, restricted to the space of interactions $\mathcal{B}^1$, is strictly convex in directions $\notin \mathcal{J} + Const$.*

Note, in particular, the contrapositive of this proposition: if $p(\,\cdot\,|\nu)$ is *not* strictly convex on $\mathcal{B}^1$ in directions $\notin \mathcal{J} + Const$, then $\nu$ is *not* the Gibbs measure for any interaction in $\mathcal{B}^1$. This gives a method for proving non-Gibbsianness, which will be exploited in Section 4.4.

**Remark.** The failure of strict convexity in $\mathcal{B}^0$ was first pointed out by Fisher [116], who provided a family of exactly soluble one-dimensional Ising models in which the pressure can be explicitly seen to have straight segments. These models are lattice versions of the Fisher-Felderhof [120, 121, 114, 113] cluster models. The failure of strict convexity can here be given a physical interpretation in terms of the formation of a perfectly rigid crystal. This indicates that $\mathcal{B}^0 \setminus \mathcal{B}^1$ does contain some interactions of physical interest, if only for their rather strange thermodynamic properties.

### 2.6.2 Relative Entropy Density

For brevity we henceforth write the relative entropy in volume $\Lambda$ as $I_\Lambda(\mu|\nu)$, instead of the more pedantic $I_{\mathcal{F}_\Lambda}(\mu|\nu)$. We now define the relative entropy density:

**Definition 2.60** *Let $\mu, \nu$ be translation-invariant probability measures on $\Omega = (\Omega_0)^{\mathbb{Z}^d}$. The* relative entropy density *(or* relative entropy per unit volume*) of $\mu$ relative to $\nu$ is defined as*

$$i(\mu|\nu) \;=\; \lim_{n\to\infty} \frac{1}{n^d} I_{C_n}(\mu|\nu) \qquad (2.92)$$

*if this limit exists.*

We emphasize that the existence of the limit (2.92) is a nontrivial problem; in fact, there exist examples of translation-invariant measures $\mu, \nu$ for which the limit does *not* exist (see Appendix A.5.2). Therefore, just as for the pressure, we shall restrict attention to two cases: when $\nu$ is a product measure, and more generally, when $\nu$ is a Gibbs measure for a translation-invariant interaction.

**Proposition 2.61** *Let $\nu$ be a product measure. Then the relative entropy density $i(\mu|\nu)$ exists for all translation-invariant probability measures $\mu$; in fact, the limit exists in van Hove sense and also as a supremum:*

$$i(\mu|\nu) \;=\; \lim_{\Lambda \nearrow \infty} \frac{1}{|\Lambda|} I_\Lambda(\mu|\nu) \qquad (2.93\mathrm{a})$$

$$=\; \sup_{\Lambda \in \mathcal{S}} \frac{1}{|\Lambda|} I_\Lambda(\mu|\nu)\,. \qquad (2.93\mathrm{b})$$

*Moreover, $i(\mu|\nu)$ has the following properties:*

(a) $0 \le i(\mu|\nu) \le i_{max} \equiv -\log \nu_{min,0}$, *where* $\nu_{min,0} = \inf\limits_{\emptyset \ne A \in \mathcal{F}_{\{0\}}} \nu(A)$. *[For example, if $\nu$ is the product of normalized counting measure on a finite single-spin space $\Omega_0$, then $i_{max} = \log |\Omega_0|$. If the single-spin space $\Omega_0$ is infinite, then $i_{max} = +\infty$.]*

(b) $i(\mu|\nu)$ is an affine *function of $\mu$*, i.e.

$$i\Big(\sum_{i=1}^n \lambda_i \mu_i \mid \nu\Big) \;=\; \sum_{i=1}^n \lambda_i\, i(\mu_i|\nu) \tag{2.94}$$

*for any measures* $\mu_1, \ldots, \mu_n \in M_{+1,inv}(\Omega)$ *and numbers* $\lambda_1, \ldots, \lambda_n \geq 0$ *with* $\sum_{i=1}^n \lambda_i = 1$.

(c) *For fixed $\nu$, $i(\mu|\nu)$ is a lower semicontinuous function of $\mu$ in the bounded quasilocal topology*[37], *and in the weak quasilocal topology*[38] *if $\Omega_0$ is a complete separable metric space.*

(d) *For any $\mu$, there exists a sequence $(\mu_n)_{n\geq 1}$ such that $\mu_n \to \mu$ in the bounded quasilocal topology, and $i(\mu_n|\nu) = i_{max}$ for all $n$. It follows that $i(\mu|\nu)$ is a discontinuous function of $\mu$ in the bounded quasilocal topology (and hence also in the weak quasilocal topology) at each $\mu$ satisfying $i(\mu|\nu) < i_{max}$.*

(e) *For any $\mu$, there exists a sequence $(\mu_n)_{n\geq 1}$ of* ergodic *measures such that $\mu_n \to \mu$ in the bounded quasilocal topology, and $i(\mu_n|\nu) \uparrow i(\mu|\nu)$. [This strengthens Proposition 2.32.]*

(f) *For fixed $\nu$ and fixed $c < \infty$, the set $\{\mu \colon i(\mu|\nu) \leq c\}$ is compact and sequentially compact in the bounded quasilocal topology (and hence also in the weak quasilocal topology), at least if $\Omega_0$ is a complete separable metric space.*

It is quite remarkable that the relative entropy density $i(\,\cdot\,|\nu)$ is an *affine* function. This comes from the fact that the relative entropy $I(\,\cdot\,|\nu)$ is not only convex, but also concave within a $\Lambda$-independent additive constant; and this constant disappears when considering the entropy *per unit volume* in the infinite-volume limit. This affineness of $i(\,\cdot\,|\nu)$ makes the infinite-volume variational theory quite different from its finite-volume counterpart.

**Proposition 2.62** *Let $\nu$ be a translation-invariant Gibbs measure for an interaction $\Phi \in \mathcal{B}^1$ (and a priori measure $\mu^0$). Then the relative entropy density $i(\mu|\nu)$ exists for all translation-invariant probability measures $\mu$; in fact, the limit exists also in van Hove sense, namely*

$$i(\mu|\nu) \;=\; \lim_{\Lambda \nearrow \infty} \frac{1}{|\Lambda|} I_\Lambda(\mu|\nu)\;. \tag{2.95}$$

---

[37]Recall that a net $\{\mu_\alpha\}$ converges to $\mu$ in the bounded quasilocal topology if $\int f\, d\mu_\alpha \to \int f\, d\mu$ for all $f \in B_{ql}(\Omega)$. In Georgii [157], this topology is called the "topology of local convergence" or the "$\mathcal{L}$-topology". See [157, Chapter 4] for properties of this topology.

[38]Recall that a net $\{\mu_\alpha\}$ converges to $\mu$ in the weak quasilocal topology if $\int f\, d\mu_\alpha \to \int f\, d\mu$ for all $f \in C_{ql}(\Omega)$. If $\Omega_0$ is a compact metric space, then $C_{ql}(\Omega) = C(\Omega)$, and so the weak quasilocal topology coincides with the usual weak topology.

*Moreover, this limit is given by*

$$i(\mu|\nu) \;=\; i(\mu|\mu^0) \;+\; p(-f_\Phi|\mu^0) \;+\; \int f_\Phi \, d\mu \;. \qquad (2.96)$$

*Moreover, $i(\,\cdot\,|\nu)$ has properties (b) and (c) of Proposition 2.61.*

Note that, by (2.96), the relative entropy density $i(\,\cdot\,|\nu)$ depends on $\nu$ only via the interaction $\Phi$: that is, if $\nu_1$ and $\nu_2$ are translation-invariant Gibbs measures for the same interaction $\Phi \in \mathcal{B}^1$, we have $i(\mu|\nu_1) = i(\mu|\nu_2)$ for all $\mu$.

The reader will note that we have *not* asserted that $i(\mu|\nu) = 0$ if and only if $\mu = \nu$. Indeed, this naive conjecture is *false*: as we have just seen, $i(\mu|\nu) = 0$ also holds whenever $\mu$ and $\nu$ are translation-invariant Gibbs measures for the same interaction. In Section 2.6.6 we shall show that, roughly speaking, $i(\mu|\nu) = 0$ *only* when $\mu$ and $\nu$ are Gibbs measures for the same interaction. This fact will play a crucial role in the proof of the First Fundamental Theorem (see Section 3.2).

**Remark.** We have proven the existence of $i(\mu|\nu)$ when $\nu$ is a Gibbs measure, but this does *not* exhaust the cases for which $i(\mu|\nu)$ exists. Indeed, by combining Theorem 3.4 with our construction in Section 4.1, we provide an explicit example of non-Gibbsian translation-invariant measures $\mu$ and $\nu$ for which $i(\mu|\nu)$ exists (and is in fact zero): namely, $\mu$ (resp. $\nu$) is the image of the $+$ (resp. $-$) phase of the two-dimensional Ising model (at low enough temperature) under the $b = 2$ decimation transformation. It is an interesting (and probably difficult) mathematical problem to characterize the pairs $(\mu, \nu)$ for which $i(\mu|\nu)$ exists.

### 2.6.3 Large Deviations

In Section 2.5.3 we developed the theory of large deviations for independent repetitions of an arbitrary probabilistic experiment. This theory provided a physical (and statistical) interpretation for the concept of relative entropy. It is natural to ask whether there is an analogue, for translation-invariant measures on a classical lattice system, in which "time averages" are replaced by "space averages". That is, instead of considering large deviations for the sample mean in a large number of independent repetitions of the same experiment, one might instead consider large deviations from *spatial* means (physically, large fluctuations of extensive quantities) in a single infinite-volume realization. Such a large-deviation theory would then, it is hoped, provide a physical interpretation of the relative entropy *density*.

In this section we describe (without proof!) the basic features of such a large-deviation theory. We emphasize that this theory is much more subtle than the theory for the independent-repetitions case, because the spins in disjoint regions of space need not be probabilistically independent. Indeed, for general translation-invariant measures on $\Omega$, no satisfactory large-deviation theory is known. Therefore, we shall restrict attention to the case in which $\nu$ is an ergodic translation-invariant Gibbs measure for an interaction $\Phi \in \mathcal{B}^1$. Our exposition is based on the recent work of Föllmer and Orey [124], Olla [282, 283], Comets [65] and Georgii [155] (see also [51]), which in turn is inspired by the pioneering work of Donsker and Varadhan [99, 100, 101, 102, 360]. In the physics literature, the relation between thermodynamics and large deviations was pointed out long ago by Lanford [230].

If $f$ is a bounded measurable function on $\Omega$, then the mean ergodic theorem states that the spatial averages $S_\Lambda^f \equiv |\Lambda|^{-1} \sum_{a \in \Lambda} T_a f$ converge in $L^1(\mu)$ norm to the expected value $m \equiv \int f \, d\nu$, as $\Lambda \nearrow \infty$. In particular, if $A$ is any closed subset of the real line *not* containing $m$, then $\text{Prob}(S_\Lambda^f \in A) \to 0$ as $\Lambda \nearrow \infty$. The mean ergodic theorem is, therefore, a natural generalization of the weak law of large numbers. The large-deviation theorems strengthen the ergodic theorem by giving a precise rate of convergence of $\text{Prob}(S_\Lambda^f \in A)$ to zero as $\Lambda \nearrow \infty$.

If we first restrict attention to *single-site* observables $f$ (i.e. functions of a single spin), then the large-deviation theorems for spatial averages (level 1) and for the single-site empirical measure (level 2) are direct analogues of (2.76) and (2.82):[39]

$$\lim_{\Lambda \nearrow \infty} \frac{1}{|\Lambda|} \log \text{Prob}(S_\Lambda^f \in A) \begin{cases} \leq - \inf_{\mu \in M_{+1,inv}(\Omega): \int f \, d\mu \in A} i(\mu|\nu) & \text{if } A \text{ is a closed set} \\ \geq - \inf_{\mu \in M_{+1,inv}(\Omega): \int f \, d\mu \in A} i(\mu|\nu) & \text{if } A \text{ is an open set} \end{cases} \quad (2.97)$$

and

$$\lim_{\Lambda \nearrow \infty} \frac{1}{|\Lambda|} \log \text{Prob}(L_\Lambda \in \mathsf{A}) \begin{cases} \leq - \inf_{\mu \in M_{+1,inv}(\Omega): \mu \upharpoonright \mathcal{F}_{\{0\}} \in \mathsf{A}} i(\mu|\nu) & \text{if } \mathsf{A} \text{ is a closed set} \\ \geq - \inf_{\mu \in M_{+1,inv}(\Omega): \mu \upharpoonright \mathcal{F}_{\{0\}} \in \mathsf{A}} i(\mu|\nu) & \text{if } \mathsf{A} \text{ is an open set} \end{cases} \quad (2.98)$$

where $i(\mu|\nu)$ is the relative entropy density, and for each configuration $\omega$ the single-site empirical measure in volume $\Lambda$ is defined to be $L_\Lambda \equiv |\Lambda|^{-1} \sum_{i \in \Lambda} \delta_{\omega_i}$.

The empirical measure $L_\Lambda$ is a tool for studying events occurring at a *single site* only. These events would completely characterize the measure if it were a product measure (as in the i.i.d. case studied in Section 2.5.3), but in the general case one clearly needs multi-site observables (i.e. functions of several spins) in order to describe correlations. The study of such observables gives rise to the "level-3" large-deviation theory. It is based on the trivial identity

$$\frac{1}{|\Lambda|} \sum_{a \in \Lambda} (T_a f)(\omega) = \left( \frac{1}{|\Lambda|} \sum_{a \in \Lambda} \delta_{T_a \omega} \right)(f), \quad (2.99)$$

which can be written as

$$S_\Lambda^f = R_\Lambda(f) \equiv \int f \, dR_\Lambda \quad (2.100)$$

where

$$R_\Lambda \equiv |\Lambda|^{-1} \sum_{a \in \Lambda} \delta_{T_a \omega} \quad (2.101)$$

---

[39]In the mathematical literature the large-deviation theorems are usually proven for sequences of cubes, but the same arguments ought to work for general van Hove sequences.

is called the *empirical field*. We emphasize that $R_\Lambda$ is a *random* measure (on the infinite-volume configuration space $\Omega$), since it depends on the random configuration $\omega$. In this language, the ergodic theorem can be reformulated as implying that the empirical field $R_\Lambda$ is, with high probability, very close to the theoretical measure $\nu$, when "closeness" is understood in the bounded quasilocal topology (i.e. the weak topology generated by the bounded quasilocal functions). More precisely, if `A` is any closed subset of $M_{+1}(\Omega)$ *not* containing $\nu$, then $\text{Prob}(R_\Lambda \in \text{A}) \to 0$ as $\Lambda \nearrow \infty$. The large-deviation theorem [155] then states that this probability is in fact exponentially small in $|\Lambda|$, namely

$$\text{Prob}(R_\Lambda \in \text{A}) \;\sim\; e^{-|\Lambda|\text{const}(\nu,\text{A})} \tag{2.102}$$

where $\text{const}(\nu,\text{A}) > 0$ whenever `A` is a closed subset of $M_{+1}(\Omega)$ not containing $\nu$. In detail,

$$\lim_{\Lambda \nearrow \infty} \frac{1}{|\Lambda|} \log \text{Prob}(R_\Lambda \in \text{A}) \begin{cases} \leq - \inf_{\mu:\, \mu \in \text{A} \cap M_{+1,inv}(\Omega)} i(\mu|\nu) & \text{if A is a closed set} \\ \geq - \inf_{\mu:\, \mu \in \text{A} \cap M_{+1,inv}(\Omega)} i(\mu|\nu) & \text{if A is an open set} \end{cases} \tag{2.103}$$

These formulae provide a physical interpretation for the relative entropy *density*. Roughly speaking, the probability that a configuration $\omega$, taken from the probability distribution $\nu$, "looks in $\Lambda$ like a typical configuration from $\mu$" decays exponentially in the volume of $\Lambda$ with rate $i(\mu|\nu)$:

$$\text{Prob}_\nu(\omega_\Lambda \text{ is typical for } \mu) \;\sim\; e^{-|\Lambda|i(\mu|\nu)} \;. \tag{2.104}$$

This interpretation of the relative entropy density will play a key role in motivating the First Fundamental Theorem (Section 3.2).

**Remarks.** 1. Some of the large-deviation theorems use a *periodized* empirical field $R_\Lambda^{(per)}$, which is a *translation-invariant* measure on $\Omega$. One expects $R_\Lambda$ and $R_\Lambda^{(per)}$ to behave in the same way.

2. Our results in Section 4 give examples of some *non-Gibbsian* measures $\nu$ for which a large-deviations theory can be developed, e.g. $\nu = \rho T$ where $\rho$ is a two-dimensional Ising-model Gibbs measure at low temperature, and $T$ is a suitable renormalization map. Of course, one is able to control the large deviations for $\nu$ only by reducing it to the same problem for the better-behaved measure $\rho$.

### 2.6.4 Variational Principle

The pressure and the relative entropy density are related by the following *variational principle*:

**Theorem 2.63 (Variational principle)** *Fix a translation-invariant measure $\nu$ satisfying the conditions of Proposition 2.56 or 2.57. Then $p(\,\cdot\,|\nu)$ and $i(\,\cdot\,|\nu)$ are conjugate* *convex functions, in the sense that*

$$p(f|\nu) = \sup_{\mu \in M_{+1,inv}(\Omega,\mathcal{F})} \left[ \int f \, d\mu - i(\mu|\nu) \right] \tag{2.105a}$$

$$i(\mu|\nu) = \sup_{f \in B_{ql}(\Omega,\mathcal{F})} \left[ \int f \, d\mu - p(f|\nu) \right] \tag{2.105b}$$

*Written in terms of interactions, this reads*

$$p(\Phi|\nu) = \sup_{\mu \in M_{+1,inv}(\Omega,\mathcal{F})} \left[ -\int f_\Phi \, d\mu - i(\mu|\nu) \right] \tag{2.106a}$$

$$i(\mu|\nu) = \sup_{\Phi \in \mathcal{B}^0} \left[ -\int f_\Phi \, d\mu - p(\Phi|\nu) \right] \tag{2.106b}$$

This variational principle gives us another way to associate (infinite-volume) probability measures to a given interaction:

**Definition 2.64** *Let $\Phi \in \mathcal{B}^0$ and $\mu \in M_{+1,inv}(\Omega)$. We say that $\mu$ is an equilibrium measure for $\Phi$ (and a priori measure $\mu^0$) if the pair $(\Phi, \mu)$ saturates the variational principle (2.106) with $\nu = \mu^0$, i.e. if*

$$p(\Phi|\mu^0) + i(\mu|\mu^0) = -\int f_\Phi \, d\mu \ . \tag{2.107}$$

We have now laid out *two distinct* approaches to infinite-volume physics:

1) The *DLR approach*, which says what it means for a (not necessarily translation-invariant) measure $\mu$ to be a *Gibbs measure* for a convergent and $\mu^0$-admissible (but not necessarily translation-invariant) interaction $\Phi$. This is the infinite-volume analogue of the explicit formula (2.1). This approach is constructed purely on the basis of probability theory, and hence it can be called the *statistical-mechanical approach*.

2) The *variational approach*, which says what it means for a translation-invariant measure $\mu$ to be an *equilibrium measure* for a translation-invariant (but not necessarily convergent) interaction $\Phi$. This is the infinite-volume analogue of the variational principle (2.3). This approach is based on optimization of thermodynamic potentials, and hence it can be called the *thermodynamic approach*. However, as remarked by Wightman [363], conventional thermodynamics refers to the optimization of potentials with respect to a rather reduced number of parameters (temperature, chemical potential, etc.). In contrast, the optimization of the previous proposition is with respect to an infinite-dimensional space of possible interactions.

For translation-invariant interactions $\Phi$ and translation-invariant measures $\mu$, this means in practice the following: The DLR approach applies to a more restricted class of interactions, but in return provides much more information on the measures. That is, it requires $\Phi \in \mathcal{B}^1$, but gives strong control on $\mu$ via the DLR equations (2.21)/(2.22). On the other hand, the variational approach needs only $\Phi \in \mathcal{B}^0$, but provides much weaker control over $\mu$. In any case, *the two approaches are equivalent in their common domain of applicability*: if $\Phi$ is a translation-invariant interaction in $\mathcal{B}^1$ and $\mu$ is a translation-invariant measure, then $\mu$ is a Gibbs measure for $\Phi$ if and only if it is an equilibrium measure for $\Phi$. We will prove this in Corollary 2.68 below.

At this point, the reader may be wondering: If the DLR and variational approaches are equivalent (for interactions in $\mathcal{B}^1$), then why bother introducing both of them? Why not stick with one or the other, and shorten this article by at least 30 pages? The answer is that many deep results are based on the *interplay* between DLR and variational ideas. This is the case for Theorem 2.67 below, and it is also the case for many of our RG results (notably those in Sections 3.2, 3.3 and 4.4).

Before leaving the subject of the variational principle, let us note a simple corollary. Let $F(\mu, \Phi)$ be the amount by which the pair $(\mu, \Phi)$ fails to satisfy the variational principle, i.e.

$$F(\mu, \Phi) \equiv p(\Phi|\mu^0) + i(\mu|\mu^0) + \int f_\Phi \, d\mu \geq 0 \, . \tag{2.108}$$

Then it is easy to see that

$$|F(\mu, \Phi) - F(\mu, \Phi')| \leq 2\|\Phi - \Phi'\|_{\mathcal{B}^0/(\mathcal{J}+Const)} \, ; \tag{2.109}$$

indeed, this is an immediate consequence of Propositions 2.56(c)–(e) and 2.58(a). In particular, if $\mu$ is an equilibrium measure for $\Phi$, then the amount by which $\mu$ fails to satisfy the variational principle for $\Phi'$ is at most $2\|\Phi - \Phi'\|_{\mathcal{B}^0/(\mathcal{J}+Const)}$. Note also that if $\Phi \in \mathcal{B}^1$, then $F(\mu, \Phi)$ can be interpreted as a relative entropy:

$$F(\mu, \Phi) = i(\mu|\nu) \quad \text{for any } \nu \in \mathcal{G}_{inv}(\Pi^\Phi) \, . \tag{2.110}$$

This is the content of equation (2.96).

A special case of (2.109) [which is also easy to see directly] is the following:

**Proposition 2.65** *Let $\Phi, \Phi' \in \mathcal{B}^0$ be physically equivalent in the Ruelle sense (i.e. $\Phi - \Phi' \in \mathcal{J} + Const$). Then $\Phi$ and $\Phi'$ have exactly the same equilibrium measures.*

### 2.6.5 What is a Phase Transition?

Informally, the occurrence of a phase transition is associated to one or both of the following phenomena: a singularity of some thermodynamic potential and/or the change in the number of "macrostates" available to the system. Historically, the first point of view was primarily associated with Ehrenfest, while the second point of view was primarily associated with Gibbs. However, the full formalization of the second point of view — in particular, giving a precise meaning to "macrostate" — and the clarification of the relation between these *thermodynamic* concepts and the underlying (microscopic) *statistical-mechanical* concepts had to await the development of the DLR and rigorous variational approaches.

The general interpretation of phase transitions as singularities of the (what turned out to be infinite-volume) free energy (= pressure) gave rise to the Ehrenfest classification: a system is said to exhibit an $n^{th}$-*order phase transition* if some $n^{th}$ derivative of the free energy is discontinuous (and all the derivatives of lower order are continuous). For example, the two-dimensional Ising model at low temperatures undergoes a first-order phase transition as the magnetic field passes through zero, because the magnetization (= first derivative of the free energy with respect to the field) has a discontinuity. On the other hand, if the field is kept equal to zero and the temperature is lowered (starting from a high value), the system undergoes a second-order phase transition at the critical temperature, because the magnetization and energy (= first derivatives of the free energy) remain continuous but the susceptibility and specific heat (= second derivatives of the free energy) blow up. From the point of view of mathematical physics, however, the Ehrenfest classification is both too detailed and too crude for our current level of understanding. It is too detailed because, as we discuss below, only the distinction between first-order and the rest has been put onto a firm basis. Consequently, authors usually group all the transitions of order two or higher into a single class and call all of them *continuous phase transitions* — because the order parameter, e.g. the magnetization (see below), remains continuous. On the other hand, the Ehrenfest classification is too crude, because the possible singularities of the free energy are much too varied to be captured in a single integer $n$. Some examples are:

- The one-dimensional Ising model with $1/r^2$ interaction, in which it is believed [12, 11, 57] that the free energy $f(\beta, h = 0)$ is $C^\infty$ but nonanalytic at the critical point $\beta_c$, at the same time as the spontaneous magnetization $M(\beta, h = 0) = -\partial f/\partial h|_{h=0}$ is *discontinuous* at $\beta_c$ (Thouless effect) [5].

- The two-dimensional $XY$ model (Kosterlitz-Thouless transition), in which it is believed [220] that the free energy $f(\beta, h = 0)$ is $C^\infty$ but nonanalytic at $\beta_c$; here the spontaneous magnetization $M(\beta, h = 0) = -\partial f/\partial h|_{h=0}$ vanishes identically, while the zero-field susceptibility $\chi(\beta, h = 0) = -\partial^2 f/\partial h^2|_{h=0}$ is believed to blow up at $\beta_c$ and remain infinite for all $\beta \geq \beta_c$.

- Systems with disorder, in which it is expected in general (and sometimes proven) that at high temperature the free energy is everywhere $C^\infty$ but nowhere analytic, as a function of temperature and/or magnetic field. This phenomenon is known as a Griffiths singularity [131].

The description of transitions where the number of "macrostates" changes is based on the use of *order parameters*. These are observables acquiring different expectation values for the different "macrostates". Each "macrostate" can be selected either by introducing some extra field that is turned off in the limit, or by using the right boundary conditions. The connection between this point of view and the existence of singularities in the pressure (free energy) was informally known since the beginning of the field: The pressure has to be convex — for the sytem to be stable — hence its only possible discontinuities are the existence of "sharp corners" where the various one-sided derivatives of the pressure take different values. Each of these values defines a different "macrostate". For example, in the case of the Ising model, the right and left derivatives with respect to the magnetic field give the two possible magnetizations. One can select one of the magnetizations by turning off a positive magnetic field (i.e. coming from the right) or a negative one (left limit), or, alternatively, by surrounding the sample by spins polarized in the desired form. It turns out that this intuition can be formalized in the framework of the variational-principle approach. Using the abstract notion of tangent to a convex functional in a Banach space, Gallavotti and Miracle-Solé [139] and Lanford and Robinson [231] showed in the mid-1960's how the existence of more than one pure phase (ergodic equilibrium measure) is equivalent to lack of (Gâteaux) differentiability of the pressure (see e.g. [206] or [157, Chapter 16]). Moreover, in complete agreement with the above example of the Ising model, the direction in which the differentiability fails is precisely the direction of the field conjugate to the relevant order parameter, and the different directional derivatives give the expectations of this observable in the different pure phases.

Therefore, if we restrict ourselves to translation-invariant specifications and measures, we have the important distinction that first-order phase transitions correspond to a change in the number of ergodic equilibrium measures (pure phases), while continuous transitions do not necessarily change this number and correspond to much more subtle phenomena (e.g. slow decay of correlations = fluctuations propagating over macroscopic scales = critical opalescence). The points in parameter space where there is a second- (or higher-) order phase transition are customarily called *critical points*, in analogy to the critical point of liquid-gas systems, which was the earliest-known example of this phenomenon.

For phenomena in which one has to go beyond translation invariance, the connection between free-energy singularities and properties of the set of extremal Gibbs measures is less clear. Nevertheless, transitions involving a change in the number of extremal Gibbs measures are usually called (by analogy rather than logic) "first-order" also in this general case.

Corresponding to the two different notions of "phase transition" mentioned at the beginning of this subsection, there are two different types of result on "absence of phase transitions": On the one hand, there are results proving the uniqueness of the Gibbs measure ($|\mathcal{G}(\Pi)| = 1$) [84, 90, 91, 94] or of the translation-invariant Gibbs measure ($|\mathcal{G}_{inv}(\Pi)| = 1$) [41, 262]. On the other hand, there are results on analyticity of the free energy and correlations [140, 205, 286, 88, 93, 96]. In the last two references, Dobrushin and Shlosman introduced an extremely strong notion of absence of phase transitions, which they call the *complete analyticity* condition. It corresponds roughly to the analyticity of all the finite-volume free energies *uniformly in the volume and in the boundary conditions*.

It is known that in general the different notions of presence and absence of phase transitions are not equivalent. This non-equivalence is probably due to physical reasons in most of the cases, but sometimes it seems an artifact of the mathematical formalism [89, 351].

### 2.6.6 When is the Relative Entropy Density Zero?

We now come to a key question (which will play a crucial role in our RG theory): Under what conditions does $i(\mu|\nu) = 0$? That is, under what conditions is the relative entropy in volume $\Lambda$ a quantity $o(|\Lambda|)$, i.e. a "surface effect"? The answer is simple: if $\nu$ is a Gibbs measure for some interaction, then $i(\mu|\nu) = 0$ when and only when $\mu$ is a Gibbs measure for the *same* interaction. The following two theorems make this precise, in a rather strong form:

**Theorem 2.66** *Let $\mu_1, \mu_2$ be Gibbs measures (not necessarily translation-invariant) for interactions $\Phi_1, \Phi_2 \in \mathcal{B}^1$, respectively. Then*

$$\limsup_{\Lambda \nearrow \infty} \frac{1}{|\Lambda|} I_\Lambda(\mu_1|\mu_2) \;\leq\; 2\|\Phi_1 - \Phi_2\|_{\mathcal{B}^0/(\mathcal{J}+Const)} \;. \tag{2.111}$$

*If $\mu_1$ and $\mu_2$ are translation-invariant, this means that*

$$i(\mu_1|\mu_2) \;\leq\; 2\|\Phi_1 - \Phi_2\|_{\mathcal{B}^0/(\mathcal{J}+Const)} \;. \tag{2.112}$$

*In particular, if $\mu_1$ and $\mu_2$ are translation-invariant Gibbs measures for the* same *interaction $\Phi \in \mathcal{B}^1$, then $i(\mu_1|\mu_2) = 0$.*

**Theorem 2.67** *Let $\Pi$ be a quasilocal specification, let $\nu \in \mathcal{G}_{inv}(\Pi)$, and let $\mu \in M_{+1,inv}(\Omega)$. Suppose that there exists a van Hove sequence $(\Lambda_n)_{n \geq 1}$ such that*

$$\lim_{n \to \infty} \frac{1}{|\Lambda_n|} I_{\Lambda_n}(\mu|\nu) \;=\; 0 \;. \tag{2.113}$$

*Then $\mu \in \mathcal{G}_{inv}(\Pi)$.*

Theorem 2.66 is an immediate consequence of estimate (2.64) in Proposition 2.46(b). Note, again, that although $\Phi_1, \Phi_2$ are required to belong to the "small" Banach space $\mathcal{B}^1$, the final estimate is in terms of the $\mathcal{B}^0/(\mathcal{J} + Const)$ norm, hence much stronger.

Theorem 2.67 is, on the other hand, a deep and surprising (at least to us) result: from a hypothesis on the behavior *per unit volume in the infinite-volume limit* one obtains a conclusion valid *for every volume* (namely $\mu\pi_\Lambda = \mu$).

The combination of Theorems 2.66 and 2.67 will play a key role in the proof of the First Fundamental Theorem (see Section 3.2).

Combining Theorems 2.66 and 2.67, we deduce the key result relating the DLR and variational approaches to classical lattice systems:

**Corollary 2.68** *Let $\Phi \in \mathcal{B}^1$ and let $\mu \in M_{+1,inv}(\Omega)$. Then $\mu$ is a Gibbs measure for $\Phi$ if and only if it is an equilibrium measure for $\Phi$.*

### 2.6.7 Pathologies in Various Interaction Spaces $\mathcal{B}_h$

In Section 2.4.4 we introduced a large class of interaction spaces $\mathcal{B}_h$, of which the most important are $\mathcal{B}^0$ and $\mathcal{B}^1$. Now we would like to discuss the *physical* differences between these spaces. This is an important issue, because we need to justify our view that (roughly speaking) $\mathcal{B}^1$ is the largest "physically reasonable" space of interactions.

Our point of view is that the *fundamental physical principles* of infinite-volume equilibrium statistical mechanics are given by the theory of specifications and Gibbs measures. (We consider the variational theory of translation-invariant equilibrium measures to be only a useful *technical tool*.) Furthermore, we argued in Section 2.3.3 that, at least for systems of *bounded* spins (including, in particular, all models with finite single-spin space), a physically reasonable specification must be *quasilocal*. If then we put aside hard-core interactions, it follows from Theorem 2.12 that a physically reasonable specification must be the Gibbsian specification for some *absolutely summable* interaction. Since $\mathcal{B}^1$ is the space of translation-invariant absolutely summable continuous interactions, this justifies our contention that $\mathcal{B}^1$ is the largest physically reasonable space of interactions.

From a *mathematical* point of view, $\mathcal{B}^0$ is the natural space of interactions on which to develop the variational theory of equilibrium measures. We nevertheless claim that $\mathcal{B}^0$ is, from a *physical* point of view, much too large; even the variational theory on $\mathcal{B}^0$ is "pathological". (This is connected with the fact that interactions in $\mathcal{B}^0 \setminus \mathcal{B}^1$ do not in general define specifications, so there are no DLR equations. For this reason, Corollary 2.18 and Propositions 2.46 and 2.59 do not hold in general in $\mathcal{B}^0$, and the large-deviation theory does not apply to equilibrium measures which are not Gibbs measures.) To emphasize that $\mathcal{B}^0$ is an unphysically large space of interactions, we list here some of the strange phenomena that can be proven for interactions in this space:

1) There is a *dense* set of interactions in $\mathcal{B}^0$ with uncountably many extremal equilibrium measures [206, Theorem V.2.2(c)]. (It is perhaps not surprising that highly frustrated interactions could produce uncountably many pure phases; but in $\mathcal{B}^0$ this happens *arbitrarily close to zero interaction*, i.e. at what ought to correspond to "high temperature".)

2) For any finite family $\mu_1, \ldots, \mu_n$ of ergodic translation-invariant measures of finite entropy density (relative to $\mu^0$), there exists an interaction in $\mathcal{B}^0$ for which *all* of these measures are *simultaneously* equilibrium measures [206, Theorem V.2.2(a)].[40] (We find this result absolutely flabbergasting: it implies, for example, that there exists an interaction in $\mathcal{B}^0$ for which the Gibbs measures of the infinite-temperature and zero-temperature Ising models are coexisting pure phases!) It follows that in $\mathcal{B}^0$ the interaction *cannot* be reconstructed uniquely from the equilibrium measure: for any given measure $\mu$, there are *many* different interactions in $\mathcal{B}^0$ having $\mu$ as an equilibrium measure. This is in sharp contrast to Proposition 2.46, which asserts the uniqueness

---

[40]This result is reminiscent of the corresponding result in the theory of (non-quasilocal) specifications: see the remark at the end of Section 2.3.4.

(modulo physical equivalence) of the interaction (if one exists at all) within $\mathcal{B}^1$.

3) The pressure is nowhere Fréchet-differentiable in $\mathcal{B}^0$ [70]. By contrast, the pressure is Fréchet differentiable of order $n$ in a neighborhood of the origin ("high temperature") in $\mathcal{B}^n$ ($n \geq 2$) [177, 228, 298].[41]

Even the space $\mathcal{B}^1$ is incredibly large, in that it allows interactions which are strongly many-body (though not quite so strongly as in $\mathcal{B}^0$) and of arbitrarily long range (provided only that they are absolutely summable). This means that even in $\mathcal{B}^1$ some rather strange phenomena occur:

4) At low temperature, the Gibbs phase rule is generically violated in all of the spaces $\mathcal{B}^n$. This is because a first-order phase transition can be destroyed by an arbitrarily weak (in $\ell^1$ norm) but very long-range *two-body* interaction [70, 349, 332, 208]. The Gibbs phase rule can hold only in spaces $\mathcal{B}_h$ where the weight $h(X)$ grows sufficiently fast with the *diameter* of $X$ (and not merely its cardinality).

5) The pressure is not analytic in any open set in any of the spaces $\mathcal{B}^n$ [89]; in particular, it is not analytic even at "high temperature" (i.e. a neighborhood of the origin). In fact, for spaces $\mathcal{B}_h$ in which $h(X)$ depends only on the cardinality of $X$, the pressure is analytic in a neighborhood of the origin *if and only if* $h(X) \geq \text{const} \times e^{\epsilon|X|}$ for some $\epsilon > 0$ [205, 89].[42]

**Remark.** In the Ising model, analyticity does hold in $\mathcal{B}^1$ norm for the *subspaces* of $\mathcal{B}^1$ corresponding to interactions written in lattice-gas or spin form ($\Phi_X = J_X \rho^X$ or $\Phi_X = J_X \sigma^X$, respectively) [205]. This is a very surprising result, which we do not completely understand from a physical point of view. It is related to the fact that physically equivalent interactions can have widely differing norms in any given space $\mathcal{B}_h$; in particular, for lattice-gas or spin interactions, one can have $\|\Phi\|_{\mathcal{B}^1} \gg \|\Phi\|_{\mathcal{B}^1/\mathcal{J}}$ [351].

---

[41] It seems to be an open question whether the pressure is once Fréchet differentiable in a neighborhood of the origin in $\mathcal{B}^1$. The proofs of higher-order differentiability in [177, 228, 298] use the Dobrushin uniqueness theorem, which applies only in $\mathcal{B}^2$ or higher. See also [157, Chapter 8 and the corresponding notes].

[42] This statement is a slight lie. What Dobrushin and Martirosyan [89] actually prove is the following: Let the single-spin space $\Omega_0$ be finite; let $h(X)$ depend only on the cardinality of $X$, and *not* satisfy $h(X) \geq \text{const} \times e^{\epsilon|X|}$ for any $\epsilon > 0$; and let $\mathcal{B}_h^{\mathbb{C}}$ be the complexification of $\mathcal{B}_h$. Then, in every open set $U \subset \mathcal{B}_h^{\mathbb{C}}$ containing a real point, there exists a complex interaction $\Phi \in U$ and a sequence of cubes $\Lambda_n \nearrow \infty$ such that the finite-volume partition functions $Z_{\Lambda_n}(\Phi) \equiv \int \exp\left[-H_{\Lambda_n, free}^{\Phi}\right]$ are all *zero*. Thus, the finite-volume free energies have (complex) singularities arbitrarily close to every (real) point in $\mathcal{B}_h$. This result makes it very unlikely that the infinite-volume pressure could be analytic in an open set of $\mathcal{B}_h$; but strictly speaking it does not rule it out, because conceivably the singularities present in finite volume could miraculously disappear in the passage to the infinite-volume limit. [Here is a simple example in one complex variable: Let $Z_n(z) = z - z_0$ for all $n$, where $z_0 \in \mathbb{C} \setminus \mathbb{R}$. Then $\lim_{n\to\infty} n^{-1} \log Z_n(z) = 0$ for all $z \in \mathbb{R}$ (provided that the branch cut is chosen to avoid the real axis). And the function 0 certainly *does* have an analytic continuation from $\mathbb{R}$ to $\mathbb{C}$!] We propose as an *open problem* to mathematical statistical mechanicians: prove that the infinite-volume pressure, which is well-defined on the space $\mathcal{B}_h$ of *real* interactions, has no analytic continuation to any open set $U \subset \mathcal{B}_h^{\mathbb{C}}$ containing a real point. In any case, the result of Dobrushin and Martirosyan does show that the Dobrushin-Shlosman [93, 96] complete analyticity condition does not hold for any open neighborhood in $\mathcal{B}_h$, for the specified class of $h$.

In Section 4 we shall prove that certain renormalized measures are not Gibbsian for any interaction in $\mathcal{B}^1$. The fact that not even in $\mathcal{B}^1$ — a space large enough to support much peculiar behavior — does an interaction exist is an indication of how strong this result is.

# 3 Position-Space Renormalization Transformations: Regularity Properties

In this section we define our general framework for studying renormalization transformations (RTs), and prove the two Fundamental Theorems on single-valuedness and continuity of the RT map.

We consider only a *single* application of the RT map. Therefore, the semigroup property of the "renormalization (semi)group" plays no role for us. In particular, we need not assume that the image system is of the same type as the original system. Nevertheless, we shall occasionally (by abuse of language) use the term "RG map", for reasons of familiarity and brevity.

## 3.1 Basic Set-Up

### 3.1.1 Renormalization Transformation Acting on Measures

We consider a "renormalization map" $T$ from an *original* (or *object*) *system* ($\Omega = \Omega_0^{\mathbb{Z}^d}, \mathcal{F}, \mu^0$) to an *image* (or *renormalized*) *system* ($\Omega' = \Omega_0'^{\mathbb{Z}^{d'}}, \mathcal{F}', \mu^{0'}$). The single-spin spaces $\Omega_0$ and $\Omega_0'$ need not be the same; indeed, we will present an important example in which they are not the same (see Example 5 below, and Section 4.3.5). Although our theory in this section works only when the spatial dimensions $d$ and $d'$ are the same — see the discussion of Example 7 below, as well as Section 4.5.2 — we find it notationally convenient to keep the prime on all image-system quantities, as this makes it easy to see which quantity refers to which system. We assume the following properties for $T$:

T1) $T$ is a probability kernel from $(\Omega, \mathcal{F})$ to $(\Omega', \mathcal{F}')$.

T2) $T$ carries translation-invariant measures on $\Omega$ into translation-invariant measures on $\Omega'$. [That is, if $\mu \in \mathcal{M}_{inv}(\Omega)$, then $\mu T \in \mathcal{M}_{inv}(\Omega')$.]

T3) $T$ is *strictly local* in position space, with asymptotic volume compression factor $K < \infty$. More precisely, there exist van Hove sequences $(\Lambda_n) \subset \mathbb{Z}^d$ and $(\Lambda_n') \subset \mathbb{Z}^{d'}$ such that:

(a) The behavior of the image spins in $\Lambda'_n$ depends only on the original spins in $\Lambda_n$, i.e.

$$\text{For each } A \in \mathcal{F}'_{\Lambda'_n}, \text{ the function } T(\,\cdot\,, A) \text{ is } \mathcal{F}_{\Lambda_n}\text{-measurable.} \quad (3.1)$$

(b) $\limsup\limits_{n\to\infty} \dfrac{|\Lambda_n|}{|\Lambda'_n|} \leq K$.

(T1) allows the renormalization map to be either deterministic or stochastic. In the deterministic case, the configuration $\omega'$ of the image system is a *function* $\omega' = t(\omega)$ of the original configuration. The most conspicuous examples of these type of transformations are decimation, linear block-spin transformations, and majority rule for blocks with an odd number of spins (see Examples 1,2 and 5 below). For the general case of a stochastic transformation, given an original-system configuration $\omega$, we choose an image-system configuration $\omega'$ with a certain probability $T(\omega, d\omega')$. The special case of a deterministic map $t\colon \Omega \to \Omega'$ corresponds to setting $T(\omega, \cdot)$ to be the delta-measure $\delta_{t(\omega)}$ [i.e. the configuration $\omega' = t(\omega)$ is chosen with probability 1]. Examples of stochastic transformations are the majority-rule transformation for blocks with even number of spins, and more generally the Kadanoff transformation (see Examples 2, 3 and 4 below). The main point of (T1) is to exclude transformations with negative weights, which have no sensible probabilistic interpretation.[43]

(T2) is self-explanatory. Typically translations of the image system correspond to some *subgroup* of translations of the original system. That is, there typically exists a homomorphism $R\colon \mathbb{Z}^{d'} \to \mathbb{Z}^d$ such that

$$T(T_{R(x)}\omega,\,\cdot\,) = T_x T(\omega,\,\cdot\,) \quad (3.2)$$

for all $x \in \mathbb{Z}^{d'}$ and all $\omega \in \Omega$.[44] For example, a RT employing $b \times b$ blocks will have $R(x) = bx$. Thus the translation group $\mathbb{Z}^{d'}$ of the image system corresponds to the subgroup $R[\mathbb{Z}^{d'}] \subset \mathbb{Z}^d$ of translations of the original system. Property (T2) trivially follows from this. Some examples are given below.

Properties (T1) and (T2) make rigorous the equation (1.1): the map $\mu \mapsto \mu T$ is a well-defined map from $\mathcal{M}_{+1,inv}(\Omega)$ into $\mathcal{M}_{+1,inv}(\Omega')$. This justifies the claim made in the Introduction, that it is easy to define the RT map *from measures to measures*. The more difficult and subtle problem of defining the RT map *from interactions to interactions* will be discussed in Section 3.1.3.

Property (T3) — the *strict* locality of the renormalization map — is crucial for our proofs of the First and Second Fundamental Theorems. Most often (although we shall not require this) the probability measure $T(\omega, \cdot)$ has a product structure

$$T(\omega, d\omega') = \prod_{x \in \mathbb{Z}^{d'}} \check{T}(\omega_{B_x}, d\omega'_x)\,, \quad (3.3)$$

---

[43]Transformations with negative weights have occasionally been used in the physics literature, not necessarily intentionally: see e.g. [335]. See also the comments in [277, footnote on p. 453 and text on p. 496].

[44]In more detail, $T(T_{R(x)}\omega, A) = T(\omega, T_x^{-1}[A])$ for all $x \in \mathbb{Z}^{d'}$, $\omega \in \Omega$ and $A \in \mathcal{F}'$.

where $B_x$ is the finite set of original spins which together determine the image spin $\omega'_x$. Now let us suppose that $B_x = B_0 + R(x)$ [i.e. $B_0$ translated by $R(x)$], where $R: \mathbb{Z}^{d'} \to \mathbb{Z}^d$ is a homomorphism satisfying $\det R \neq 0$ (obviously this needs $d' = d$). We then claim that (T3) holds with $K = |\det R|$. *Proof:* Let $(\Lambda'_n)$ be any van Hove sequence in $\mathbb{Z}^{d'}$. What sets $(\Lambda_n) \subset \mathbb{Z}^d$ should we take to satisfy (T3)? Clearly the image spins in $\Lambda_n$ depend only on the original spins in the set $\Lambda_n^* \equiv R[\Lambda'_n] + B_0 \subset \mathbb{Z}^d$. So at first one might think to take $\Lambda_n = \Lambda_n^*$. The trouble is that the $(\Lambda_n^*)$ need not form a van Hove sequence, because they may have a nonzero density of "holes". [Consider, for example, decimation with spacing $b > 1$: here $B_0 = \{0\}$ and $R(x) = bx$.] So we take instead

$$\Lambda_n = \mathbb{Z}^d \cap \text{convex hull of } \Lambda_n^* \,. \tag{3.4}$$

Then, using the fact that $\det R \neq 0$, it is not hard to convince oneself that $(\Lambda_n)$ is a van Hove sequence, and that

$$\lim_{n \to \infty} \frac{|\Lambda_n|}{|\Lambda'_n|} = |\det R| \,. \tag{3.5}$$

Two points are relevant here: Firstly, we need $\det R \neq 0$ (and in particular $d' = d$) in order to guarantee that the sets $(\Lambda_n)$ are sufficiently "fat" to form a van Hove sequence (see the discussion of Example 7 below for what can happen if this does not hold).[45] Secondly, the quantity $K \equiv \limsup_{n \to \infty} |\Lambda_n|/|\Lambda'_n|$ is by definition the *asymptotic* volume compression factor: as such, it is determined solely by $R$; it does not depend on the size of $B_0$ as long as $B_0$ is finite.

We *conjecture* that the two Fundamental Theorems hold also for *quasilocal* renormalization maps — i.e. maps in which $\omega'_x$ depends sufficiently *weakly* on distant spins $\omega_y$ — but we are *not* able to prove this with our present methods. Quasilocal renormalization maps are of great practical importance: for example, in "momentum-space" renormalization one often uses a deterministic transformation

$$\omega'_x = \sum_y F(bx - y) \omega_y \tag{3.6}$$

with some length rescaling factor $b > 1$ and some kernel $F$. In particular, if one uses a "soft" cutoff in momentum space [365, 31], then the kernel $F$ is *rapidly decreasing* at infinity in $x$-space (e.g. decreasing faster than any inverse power of its argument). It is an important *open problem* to extend our results to such maps.

### 3.1.2 Examples

1) *Decimation transformation* [210, 364]. Let $\Omega' = \Omega$ and $d' = d$, and let $b$ be an integer $\geq 2$. Define the deterministic RT map

$$\omega'_x = \omega_{bx} \,. \tag{3.7}$$

---

[45]Actually, all we really need is that $R$, considered as a $d \times d'$ matrix, have rank $d$. Thus, we could allow some cases with $d' > d$. But these are of little interest. The interesting cases with $d' \neq d$ have $d' < d$ (Example 7 below), and these do *not* satisfy (T3).

This map is strictly local [in fact, of the product form (3.3)] with asymptotic volume compression factor $K = b^d$. It is of the form (3.2) with $R(x) = bx$.

More generally, let $\Omega' = \Omega$ and $d' = d$ and let $R$ be any homomorphism from $\mathbb{Z}^{d'}$ to $\mathbb{Z}^d$ satisfying $\det R \neq 0$. Define the deterministic RT map

$$\omega'_x = \omega_{R(x)} \,. \tag{3.8}$$

This map is strictly local [in fact, of the product form (3.3)] with asymptotic volume compression factor $K = |\det R|$. Some examples are shown in Figure 2(a)–(b).

2) *Majority-rule transformation for the Ising model* [276, 278, 277]. Let $b$ be an integer $\geq 1$, let $B_0$ be a fixed finite subset of $\mathbb{Z}^d$ (the *block*), and let $B_x = B_0 + bx$ (i.e. $B_0$ translated by $bx$). Define the map

$$\sigma'_x = \begin{cases} +1 & \text{if } \sum_{y \in B_x} \sigma_y > 0 \\ -1 & \text{if } \sum_{y \in B_x} \sigma_y < 0 \\ \pm 1 & \text{if } \sum_{y \in B_x} \sigma_y = 0 \end{cases} \tag{3.9}$$

where "$\pm 1$" denotes a random choice with probabilities of 1/2 each. This transformation is deterministic if $b$ is odd, stochastic if $b$ is even.

3) *Kadanoff transformation for the Ising model* [210]. A large class of nonlinear RT maps for the Ising model $\Omega = \Omega' = \{-1, 1\}^{\mathbb{Z}^d}$ can be represented in the following form: Consider the same blocks $B_x$ as in the previous example, and let $p > 0$. Define the stochastic RT map

$$T(\sigma, \sigma') = \prod_{x \in \mathbb{Z}^{d'}} \frac{\exp\left(p \sigma'_x \sum_{y \in B_x} \sigma_y\right)}{2 \cosh\left(p \sum_{y \in B_x} \sigma_y\right)} \,. \tag{3.10}$$

This map is strictly local [and clearly of the product form (3.3)] with asymptotic volume compression factor $K = b^d$, and is of the form (3.2) with $R(x) = bx$. Many well-known RT maps are special cases of (3.10):

(a) With $B_0 = \{0\}$ and $b = 1$, (3.10) is model I of Griffiths and Pearce [172, 173], a kind of "copying with noise". (This map also arises in applications to image processing [129, 152].) As $p \to \infty$ it tends to the identity transformation.

(b) With $B_0 = \{0\}$ and $b \geq 2$, (3.10) is model II of Griffiths and Pearce [172, 173], a kind of "decimation with noise". As $p \to \infty$ it tends to the ordinary decimation transformation (3.7).

(c) With $B_0 = \{0, 1, \ldots, b-1\}^d$ (a hypercube of side $b$) and $b \geq 2$, (3.10) is the Kadanoff transformation [210]. In the limit $p \to \infty$ it tends to the majority-rule transformation (3.9).

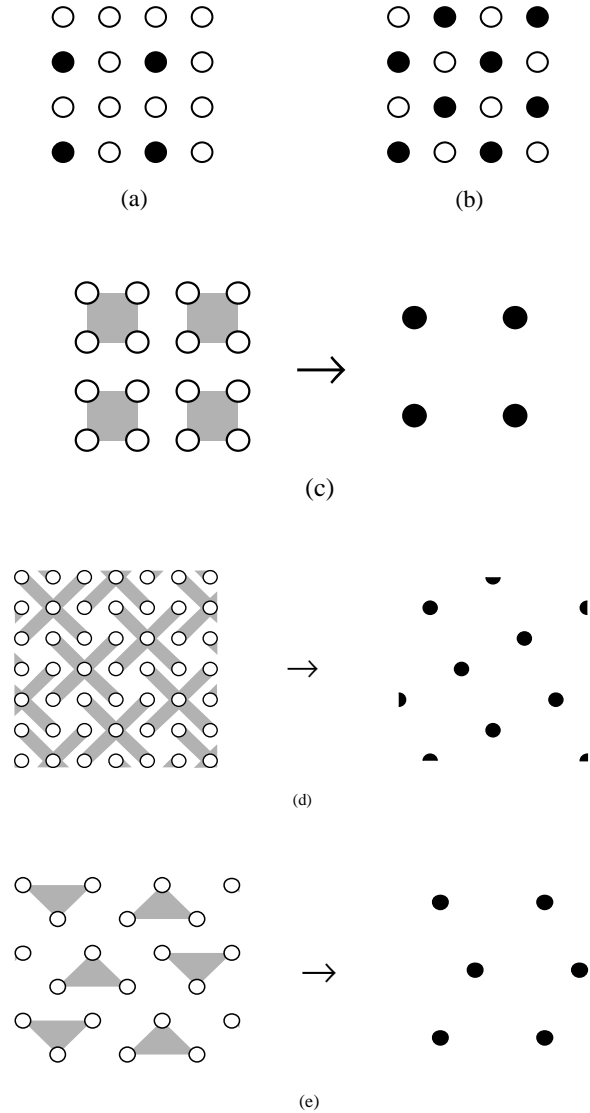

Figure 2: Some examples of RT maps in dimension $d = 2$.

(a) Decimation with $b = 2$ and $K = 4$.

(b) Decimation with $R(x_1, x_2) = (x_1 + x_2, x_1 - x_2)$ and $K = 2$ ("checkerboard decimation").

(c) Block transformation with $b = 2$, $B_0 = \{(0,0), (1,0), (0,1), (1,1)\}$ and $K = 4$.

(d) Block transformation with $R(x_1, x_2) = (2x_1 - x_2, x_1 + 2x_2)$, $B_0 = \{(0,0), (\pm 1, \pm 1)\}$ and $K = 5$ [357].

(e) Block transformation with $R(x_1, x_2) = (2x_1 + x_2, x_1 + 2x_2)$, $B_0 = \{(0,0), (1,0), (1,1)\}$ and $K = 3$ [276, 278].

As in the decimation transformation, we can replace $bx$ by a more general nonsingular homomorphism $R(x)$. Then $K = |\det R|$. Some examples are shown in Figure 2(c)–(e).

4) *Kadanoff transformation for the N-vector model.* For the $N$-vector model, in which the spins are unit vectors in $\mathbb{R}^N$, the natural generalization of the majority-rule transformation is the "rescaled block-spin transformation" [321]

$$\sigma'_x = \frac{\sum_{y \in B_x} \sigma_y}{\left| \sum_{y \in B_x} \sigma_y \right|} , \qquad (3.11)$$

which is deterministic. (In principle one should specify what happens when $\sum_{y \in B_x} \sigma_y = 0$: for example, one could choose some particular value of $\sigma'_x$, or one could let $\sigma'_x$ be uniformly distributed on the unit sphere. But this situation occurs with probability zero, so it is irrelevant what choice one makes.) Similarly, the Kadanoff transformation has a natural generalization [190]:

$$T(\sigma, d\sigma') = \prod_{x \in \mathbb{Z}^{d'}} \frac{\exp\left(p\sigma'_x \cdot \sum_{y \in B_x} \sigma_y\right)}{\mathcal{Z}_N\left(p \sum_{y \in B_x} \sigma_y\right)} \, d\Omega(\sigma'_x) , \qquad (3.12)$$

where

$$\mathcal{Z}_N(h) \equiv \int_{S^{N-1}} e^{h \cdot \sigma} \, d\Omega(\sigma) = \Gamma\left(\frac{N}{2}\right) \left(\frac{2}{|h|}\right)^{\frac{N}{2}-1} I_{\frac{N}{2}-1}(|h|) \qquad (3.13)$$

and $d\Omega$ denotes uniform measure on the unit sphere in $\mathbb{R}^N$. As $p \to \infty$ this tends to the deterministic map (3.11).

Analogous formulae can be used to define a Kadanoff transformation for the $q$-state Potts model, using the representation of Potts spins as unit vectors in $\mathbb{R}^{q-1}$ pointing from the center of a "hypertetrahedron" to its vertices. As $p \to \infty$ this transformation tends to the "plurality-rule" transformation with random tie-breakers. The Potts model with vacancies [279, 305] can also be treated in this framework, by representing the "vacancy" state as the origin in $\mathbb{R}^{q-1}$.

5) *Linear block-spin transformations.* A natural choice of a deterministic linear transformation is the *averaging* transformation

$$\sigma'_x = c \sum_{y \in B_x} \sigma_y , \qquad (3.14)$$

for a suitably chosen *rescaling factor c*. Typically we choose $|B_0|^{-1} \le c \le 1$. We observe that if $c > |B_0|^{-1}$, this transformation does not map any model of *bounded* spins to itself: if $\Omega_0 = [-M, M]$, we must take $\Omega'_0 = [-M', M']$ with $M' \ge |B_0| cM > M$. As

a consequence, the fixed point(s) (if any) for such a transformation must correspond to model(s) of unbounded spins (i.e. $\Omega_0 = \mathbb{R}$). For this reason, it is most natural to consider (3.14) as acting, right from the start, on such a system of real-valued spins. However, in this paper we are not concerned with fixed points; our interest is in whether the *first* application of the RT map is well-defined. For this purpose we may work entirely with models of bounded spins, provided that we are willing to accept $\Omega'_0 \neq \Omega_0$. For example, in the two-dimensional Ising model with $2 \times 2$ blocks (and $c = 1$), we have $\Omega_0 = \{-1, 1\}$ but $\Omega'_0 = \{-4, -2, 0, 2, 4\}$.

For *unbounded* spins with values in $\mathbb{R}$ (or $\mathbb{R}^N$), one can use either a deterministic *linear* block-spin transformation [147]

$$\varphi'_x = c \sum_{y \in B_x} \varphi_y \qquad (3.15)$$

or the stochastic linear block-spin transformation [27, 18]

$$T(\varphi, d\varphi') = \prod_{x \in \mathbb{Z}^{d'}} \text{const} \times \exp\left[-\frac{1}{2\epsilon^2}\left(\varphi'_x - c \sum_{y \in B_x} \varphi_y\right)^2\right] d\varphi'_x, \qquad (3.16)$$

which corresponds to adding Gaussian white noise of variance $\epsilon^2$ to the deterministic block spins (3.15). In both cases, the rescaling factor $c$ must be chosen appropriately if the transformation is to have a fixed point: e.g. for hypercubic blocks of side $b$ one takes

$$c = \begin{cases} b^{-d/2} & \text{to have a high-temperature fixed point} \\ b^{-d} & \text{to have a low-temperature fixed point} \\ b^{-(d+2-\eta)/2} & \text{to have a critical fixed point} \end{cases} \qquad (3.17)$$

This need to fix a parameter is characteristic of *linear* renormalization transformations.

Linear block-spin transformations have attracted the attention of mathematical physicists because of their connections with central-limit theorems: see for example [200, 32, 147, 71, 58].

6) *Linear block-spin transformation with large-field cutoff.* Even when the linear block-spin transformation (3.14) does not map models of bounded spins to themselves, one expects the corresponding fixed-point measure(s) to have rapidly decaying (e.g. Gaussian or faster) densities at large $\varphi$. Therefore, it may be a reasonable approximation to modify (3.14) by cutting off the fields explicitly at $|\sigma| = M$, where $M$ is some fixed large number. That is, on the space $\Omega = \Omega' = [-M, M]^{\mathbb{Z}^d}$ one can consider the deterministic RT map

$$\varphi'_x = \begin{cases} c \sum_{y \in B_x} \varphi_y & \text{if } c\left|\sum_{y \in B_x} \varphi_y\right| \leq M \\ M \, \text{sgn}\left(\sum_{y \in B_x} \varphi_y\right) & \text{if } c\left|\sum_{y \in B_x} \varphi_y\right| > M \end{cases} \qquad (3.18)$$

[This works also for $N$-component spins, if one interprets $\text{sgn}(\varphi) = \varphi/|\varphi|$.] To our knowledge, this transformation has not been considered previously. (But see Cammarota [56] for a related idea.)

7) *Restriction to a hyperplane* [318]. Let $\Omega'_0 = \Omega_0$ but take $d' < d$, and define

$$\omega'_x = \omega_{(x,0)} \tag{3.19}$$

where 0 denotes the origin in $\mathbb{Z}^{d-d'}$. This is an unusual type of decimation transformation in which the original model is restricted to a hyperplane; it has recently elicited some interest (see Section 4.5.2). However, this transformation does *not* satisfy our condition (T3): although the image spins in a volume $\Lambda'_n$ depend only on the original spins in $\Lambda'_n \times \{0\}$ — so that naively one would have a volume compression factor $K = 1$ — the trouble is that the sets $\Lambda'_n \times \{0\}$ do not tend to infinity (in $\mathbb{Z}^d$) in van Hove sense when the $\Lambda'_n$ do so in $\mathbb{Z}^{d'}$. To make them tend to infinity in van Hove sense, one would have to "fatten them out", e.g. by taking $\Lambda_n = \Lambda'_n \times C_{R_n}$ where $C_{R_n}$ is a cube of side $R_n$ in $\mathbb{Z}^{d-d'}$, and $R_n \to \infty$ as $n \to \infty$. But then the volume compression factor $K$ would be infinite. This example makes clear why we need $d' = d$. Indeed, in this example we have $i(\mu_- T | \mu_+ T) > 0$ [see Section 4.5.2], contrary to what would happen if (T3) were to hold [cf. (3.30)].

### 3.1.3 Renormalization Transformation Acting on Interactions

We can now define precisely the renormalization map $\mathcal{R}$ acting on the space of interactions, making rigorous the diagram (1.2). As argued in Section 2.6.7, the largest "physically reasonable" space of interactions is $\mathcal{B}^1$, the space of translation-invariant continuous absolutely summable interactions. Therefore, in defining $\mathcal{R}$, *we shall restrict attention to interactions* $\Phi \in \mathcal{B}^1$ *such that there exists an image interaction* $\Phi' \in \mathcal{B}^1$. Since *a priori* we wish to adopt a completely open-minded definition — allowing for the possibility of multi-valuedness — we must define $\mathcal{R}$ as a relation rather than a function.

**Definition 3.1** *Let $T$ be an RT map satisfying properties (T1) and (T2). We then define the corresponding map $\mathcal{R} = \mathcal{R}_T$ to be the relation*

$$\mathcal{R} = \{(\Phi, \Phi') \in \mathcal{B}^1 \times \mathcal{B}^1 : \text{there exists } \mu \in \mathcal{G}_{inv}(\Pi^\Phi) \text{ such that } \mu T \in \mathcal{G}_{inv}(\Pi^{\Phi'})\} . \tag{3.20}$$

*We can also think of $\mathcal{R}$ as a multi-valued function: we write $\Phi' \in \mathcal{R}(\Phi)$ as a synonym for $(\Phi, \Phi') \in \mathcal{R}$. We define the* domain *of $\mathcal{R}$ to be the set*

$$\begin{aligned}
\operatorname{dom} \mathcal{R} &= \{\Phi : \text{there exists } \Phi' \text{ with } (\Phi, \Phi') \in \mathcal{R}\} \\
&= \{\Phi : \mathcal{R}(\Phi) \neq \varnothing\} .
\end{aligned} \tag{3.21}$$

*A priori* the map $\mathcal{R}$ could be multi-valued. Indeed, the way we have defined it, it surely *is* multi-valued, because of physical equivalence: if $\Phi' \in \mathcal{R}(\Phi)$ and $\Psi' \in \mathcal{B}^1 \cap (\mathcal{J} + Const)$, then also $\Phi' + \Psi' \in \mathcal{R}(\Phi)$. The more interesting question is whether $\mathcal{R}$ can be multi-valued *apart from* the "trivial" multi-valuedness caused by physical equivalence. The scenario proposed in [72] is precisely the claim that this can happen; we shall prove in our First Fundamental Theorem (Theorem 3.4) that in fact it *cannot* happen. That is, we shall prove that the map $\mathcal{R}$ is *single-valued modulo physical equivalence*. We shall moreover prove that the phrase "there exists $\mu$" in (3.20) can be replaced equivalently by "for all $\mu$".

For RT maps satisfying a very mild continuity condition, we can say something about the closure properties of the multi-valued map $\mathcal{R}$. To avoid bothersome topological complexities, we restrict attention to compact metric single-spin spaces $\Omega_0$.

**Theorem 3.2** *Let $\Omega_0$ be a compact metric space, and assume that $T$ satisfies (T1) and (T2) and is Feller (i.e. $Tf$ is continuous if $f$ is). Then $\mathcal{R}$ is a* closed *subset of $\mathcal{B}^1 \times \mathcal{B}^1$ with respect to the $\mathcal{B}^0/(\mathcal{J} + Const) \times \mathcal{B}^0/(\mathcal{J} + Const)$ seminorm.*

PROOF. Assume that $(\Phi_n, \Phi'_n) \in \mathcal{R}$ and $(\Phi_\infty, \Phi'_\infty) \in \mathcal{B}^1 \times \mathcal{B}^1$, with

$$\lim_{n \to \infty} \|\Phi_n - \Phi_\infty\|_{\mathcal{B}^0/(\mathcal{J}+Const)} = \lim_{n \to \infty} \|\Phi'_n - \Phi'_\infty\|_{\mathcal{B}^0/(\mathcal{J}+Const)} = 0 \ . \qquad (3.22)$$

We need to prove that $(\Phi_\infty, \Phi'_\infty) \in \mathcal{R}$.

Choose, for each $n$, a translation-invariant Gibbs measure $\mu_n$ for $\Phi_n$. By passing to a subsequence, we can assume without loss of generality that $\mu_n$ converges weakly to some measure $\mu_\infty$; and since $\Phi_n \to \Phi_\infty$ in $\mathcal{B}^0/(\mathcal{J} + Const)$ seminorm, it is easy to see that $\mu_\infty$ is a translation-invariant Gibbs measure for $\Phi_\infty$. Now the Feller hypothesis on $T$ guarantees that $\mu_n T \to \mu_\infty T$ weakly. Since $\mu_n T$ is a translation-invariant Gibbs measure for $\Phi'_n$, and $\Phi'_n \to \Phi'_\infty$ in $\mathcal{B}^0/(\mathcal{J} + Const)$ seminorm, it follows that $\mu_\infty T$ is a translation-invariant Gibbs measure for $\Phi'_\infty$. But this implies that $(\Phi_\infty, \Phi'_\infty) \in \mathcal{R}$. ■

### 3.1.4 A Remark on Systems of Unbounded Spins

The results to be proven in Sections 3.2 and 3.3 are in principle applicable to systems of either bounded or unbounded spins. But for unbounded spins our results are not of much interest, because we restrict attention to *bounded Hamiltonians* (i.e. absolutely summable interactions). The trouble, as discussed at the end Section 2.4.4, is that we lack at present an adequate general theory of unbounded spin systems: we are unable to specify, for example, a space of interactions that includes all "reasonable" interactions. The development of such a general theory is an important *open problem*; it would be a first step towards putting the standard Wilson-style RG theory [365] on a rigorous footing. In particular, in such a framework one could try to prove analogues of our First and Second Fundamental Theorems.

In this regard it should be remarked that the important work of Gawędzki and Kupiainen [147, 148, 150, 151] on rigorous RG theory does not implement exactly the standard Wilson prescription, at least for bosonic theories: while the small-field part of the Gibbs measure is represented by a Hamiltonian of the usual kind, the large-field part is represented instead by a polymer expansion [148]. (In recent work, Brydges and Yau [53] systematize this idea, and formulate the RG *purely* in terms of a polymer expansion.) For fermionic theories, where there is no "large-field region", Gawędzki and Kupiainen [149] do implement the full Wilson prescription; however, fermionic theories have no direct probabilistic interpretation. Also, for bosons, Koch and Wittwer [218, 219] implement the Wilson prescription, but so far only in the hierarchical model.

## 3.2 First Fundamental Theorem: Single-Valuedness of the RT Map

Among the possible pathologies of the RT applied at the level of Hamiltonians, the following scenario has been proposed [72][46]: Consider a Hamiltonian $H$ lying on a first-order phase-transition surface, that is, one for which there exist at least two distinct pure phases (extremal translation-invariant Gibbs measures), call them $\mu_1$ and $\mu_2$. Now perform a renormalization transformation $T$ as indicated in (1.1). The resulting renormalized measures $\mu'_1 \equiv \mu_1 T$ and $\mu'_2 \equiv \mu_2 T$ may then, it is claimed, be Gibbsian for two *different* renormalized Hamiltonians $H'_1 \neq H'_2$. In other words, the renormalization map $\mathcal{R}$ from Hamiltonians to Hamiltonians, defined by (1.2), may be *multi-valued*.

Here we disprove such a scenario. We show that if two initial Gibbs measures correspond to the same interaction $\Phi$, then the renormalized measures are either both Gibbsian for the *same* renormalized interaction $\Phi'$, or else they are both non-Gibbsian (in which case there is *no* renormalized interaction at all).

This theorem follows from comparing the large-deviation properties of different Gibbs measures according to whether they belong to the same or different interactions. Heuristically, if $\mu$ and $\nu$ are two Gibbs measures corresponding to different interactions, then the probability of finding in $\nu$ a large droplet looking like a typical configuration for the measure $\mu$ is exponentially small in the volume of the droplet:

$$\text{Prob}_\nu(\omega_\Lambda \text{ is typical for } \mu) \sim e^{-O(|\Lambda|)} . \tag{3.23}$$

On the other hand, if $\mu$ and $\nu$ correspond to the same interaction, this probability is sub-exponential:

$$\text{Prob}_\nu(\omega_\Lambda \text{ is typical for } \mu) \sim e^{-o(|\Lambda|)} . \tag{3.24}$$

Mathematically, as seen in Section 2.6, this is expressed in the fact that the relative entropy density satisfies

$$i(\mu|\nu) \begin{cases} > 0 & \text{if } \mu \text{ and } \nu \text{ are Gibbs measures for different interactions} \\ = 0 & \text{if } \mu \text{ and } \nu \text{ are Gibbs measures for the same interaction} \end{cases} \tag{3.25}$$

---

[46]This scenario is stated very clearly in the Monte Carlo paper of Decker, Hasenfratz and Hasenfratz [72, p. 23, lines 2–5]. On the other hand, the analytic arguments in the companion paper of Hasenfratz and Hasenfratz [189] concern "singularities" whose precise nature is unspecified. We are unable to make a connection between the two lines of reasoning.

Now, under renormalization one looks only at the block spins and forgets about the internal spins, hence

$$\text{Prob}_\nu(\text{block spins in } \omega_\Lambda \text{ are typical for } \mu)$$
$$\geq \text{Prob}_\nu(\textit{all} \text{ spins in } \omega_\Lambda \text{ are typical for } \mu). \tag{3.26}$$

Therefore, if initially the probability was subexponential (same interaction), then under renormalization it remains so and we can never obtain the exponential decay (3.23) characteristic of different interactions. Mathematically, this is expressed by the fact that the relative entropy decreases under the application of arbitrary deterministic or stochastic transformations, in particular under the RT:

**Lemma 3.3** *Let $(\Omega, \Sigma)$ and $(\Omega', \Sigma')$ be measurable spaces, and let $T$ be a probability kernel from $(\Omega, \Sigma)$ to $(\Omega', \Sigma')$. Then, if $\mu$ and $\nu$ are probability measures on $\Omega$,*

$$I(\mu T | \nu T) \leq I(\mu | \nu).$$

PROOF. This is a well-known result, although it is rather difficult to find a complete proof in the literature. (Most of the published proofs concern one or another special case: $T$ deterministic, $\nu T = \nu$, discrete state space, etc.) The first complete proof of which we are aware is due to Csiszár (1963) [67]; however, we would not be surprised to learn that this result was known much earlier. See also, for instance, [361] and [61, Theorem 8.1]; and see [64] for some stronger results. For the convenience of the reader, let us give a one-line proof:

$$I(\mu | \nu) \;=\; I(\mu \times T | \nu \times T) \;\geq\; I(\mu T | \nu T). \tag{3.27}$$

Here the first equality is Proposition 2.53(h): the measures $\mu \times T$ and $\nu \times T$ have the same regular conditional probability given $\Sigma$, namely $T$. [The intuitive idea is that the pair $(\mu \times T, \nu \times T)$ contains at least as much information as the pair $(\mu, \nu)$, since the latter is the restriction of the former to the sub-$\sigma$-field $\Sigma \subset \Sigma \times \Sigma'$; but it contains no more information, because the *same* probability kernel has been used to generate both $\mu \times T$ and $\nu \times T$ from $\mu$ and $\nu$.] And the inequality is Proposition 2.53(g), since $\mu T$ (resp. $\nu T$) is the restriction of $\mu \times T$ (resp. $\nu \times T$) to the sub-$\sigma$-field $\Sigma' \subset \Sigma \times \Sigma'$.
■

**Theorem 3.4 (First fundamental theorem)** *Let $\mu$ and $\nu$ be translation-invariant Gibbs measures with respect to the same interaction $\Phi \in \mathcal{B}^1$, and let $T$ be an RT map satisfying properties (T1)–(T3). Then:*

*(a) Either $\mu T$ and $\nu T$ are both non-quasilocal (i.e. not consistent with any quasilocal specification), or else there exists a quasilocal specification $\Pi'$ with which both $\mu T$ and $\nu T$ are consistent. In the latter case, if the single-spin space is finite, then $\Pi'$ is the* unique *quasilocal specification with which either $\mu T$ or $\nu T$ is consistent, and it is translation-invariant.*

(b) *Either $\mu T$ and $\nu T$ are both non-Gibbsian (for absolutely summable interactions), or else there exists an absolutely summable interaction $\Phi'$ for which both $\mu T$ and $\nu T$ are Gibbs measures. In the latter case, if $\Phi'$ is continuous [as it always is e.g. for a discrete single-spin space], $\Phi'$ is the unique continuous absolutely summable interaction (modulo physical equivalence in the DLR sense) for which either $\mu T$ or $\nu T$ is a Gibbs measure.*

PROOF. Let $(\Lambda_n) \subset \mathbb{Z}^d$ and $(\Lambda'_n) \subset \mathbb{Z}^{d'}$ be van Hove sequences having the properties (T3) assumed in Section 3.1. Now, by Theorem 2.66, the fact that $\mu$ and $\nu$ are Gibbs measures for the same interaction implies that

$$\lim_{n \to \infty} \frac{1}{|\Lambda_n|} I_{\Lambda_n}(\mu|\nu) = 0\,. \tag{3.28}$$

On the other hand, the image spins in $\Lambda'_n$ depend only on the original spins in $\Lambda_n$: that is, $(\mu T)\!\restriction\!\mathcal{F}'_{\Lambda'_n}$ is the image under $T$ of $\mu\!\restriction\!\mathcal{F}_{\Lambda_n}$, and likewise for $\nu$. Hence, by Lemma 3.3 we have

$$I_{\Lambda'_n}(\mu T|\nu T) \;\leq\; I_{\Lambda_n}(\mu|\nu)\,. \tag{3.29}$$

It follows that

$$0 \;\leq\; \limsup_{n \to \infty} \frac{1}{|\Lambda'_n|} I_{\Lambda'_n}(\mu T|\nu T) \;\leq\; \lim_{n \to \infty} \frac{K}{|\Lambda_n|} I_{\Lambda_n}(\mu|\nu)$$
$$= \;0\,. \tag{3.30}$$

Therefore, by Theorem 2.67, if $\nu T$ is consistent with a quasilocal specification $\Pi'$, then $\mu T$ must also be consistent with this *same* specification $\Pi'$. The same argument can be made interchanging the roles of $\mu$ and $\nu$. Thus, either $\mu T$ and $\nu T$ are both non-quasilocal, or else there exists a quasilocal specification $\Pi'$ with which *both* $\mu T$ and $\nu T$ are consistent. In the latter case, if the single-spin space is finite, Theorem 2.15 guarantees the uniqueness of $\Pi'$. In particular, since $\mu T$ and $\nu T$ (being translation-invariant) are obviously consistent with any translate of $\Pi'$, we conclude that $\Pi'$ is translation-invariant.

A special case of the foregoing is: if $\nu T$ (resp. $\mu T$) is Gibbsian with respect to an absolutely summable interaction $\Phi'$, then $\mu T$ (resp. $\nu T$) must also be Gibbsian with respect to this *same* interaction $\Phi'$. The uniqueness modulo physical equivalence of $\Phi'$ is then guaranteed by Corollary 2.18. ∎

The First Fundamental Theorem shows that the RT map $\mathcal{R}$ is *single-valued modulo physical equivalence*. It also shows that the phrase "there exists $\mu$" in the definition (3.20) can be replaced equivalently by "for all $\mu$".

**Remarks.** 1. The first step of this proof (using Theorem 2.66) does not require $\mu$ and $\nu$ to be translation-invariant. But the second step (using Theorem 2.67) does seem to require at least $\mu T$ and $\nu T$ to be translation-invariant. So we do not know whether the hypothesis of translation-invariance of $\mu$ and $\nu$ can be omitted in this theorem. (Note: We always assume that the *interaction* $\Phi$ is translation-invariant.)

2. In part (b), the interaction $\Phi'$, if it exists, ought to be physically equivalent in the DLR sense to a *translation-invariant* interaction. Unfortunately, we are not able to prove this. From the uniqueness we know that $\Phi'$ is physically equivalent to all of its translates; but it seems to be an open question whether this guarantees that $\Phi'$ is physically equivalent in the DLR sense to a translation-invariant interaction. An affirmative answer would also allow Kozlov's [222] Gibbs Representation Theorem to be given a satisfactory translation-invariant version (see the Remark at the end of Section 2.4.9).

## 3.3 Second Fundamental Theorem: Continuity Properties of the RT Map

A second aspect of the scenario proposed by Decker, Hasenfratz and Hasenfratz [72] is that the RT map may be *discontinuous* at an original Hamiltonian $H_0$ lying on a first-order phase-transition surface: namely, for original Hamiltonians $H$ arbitrarily close to $H_0$ on opposite sides of the phase-transition surface, it is claimed that the corresponding renormalized Hamiltonians $H'$ may be a finite distance apart.

Here we disprove this scenario too. We show that *the RT map is always continuous* (in a suitable norm) *on the set of Hamiltonians where it is well-defined*, that is, on the set of Hamiltonians for which the image measures are Gibbsian.

The key idea underlying our proof is the fact that, if $\mu$ is a Gibbs measure for an interaction $\Phi \in \mathcal{B}^1$, then the DLR equations allow the reconstruction of the interaction $\Phi$ (modulo physical equivalence) from the measure $\mu$:

$$\log \frac{d\mu_\Lambda}{d\mu_\Lambda^0} = -\sum_{x \in \Lambda} T_x f_\Phi + \mathrm{const}(\Phi) + o(|\Lambda|) \tag{3.31}$$

(see Section 2.4.8). Therefore, if $\mu_1$ and $\mu_2$ are Gibbs measures for interactions $\Phi_1, \Phi_2 \in \mathcal{B}^1$, we have

$$\log \frac{d\mu_{1\Lambda}}{d\mu_{2\Lambda}} = -\sum_{x \in \Lambda} T_x f_{\Phi_1 - \Phi_2} + \mathrm{const}(\Phi_1) - \mathrm{const}(\Phi_2) + o(|\Lambda|) \tag{3.32}$$

and in particular

$$\left\| \log \frac{d\mu_{1\Lambda}}{d\mu_{2\Lambda}} \right\|_{B(\Omega)/\mathrm{const}} = |\Lambda| \, \|\Phi_1 - \Phi_2\|_{\mathcal{B}^0/(\mathcal{J}+\mathit{Const})} + o(|\Lambda|) \ . \tag{3.33}$$

Now the probability densities of renormalized measures are (particular) weighted averages of the original densities, so the supremum of the renormalized density cannot exceed that of the original density. That is, $\| \log(d\mu_1/d\mu_2) \|_{B(\Omega)/\mathrm{const}}$ can only decrease under the RT:

**Lemma 3.5** *Let $(\Omega, \Sigma)$ and $(\Omega', \Sigma')$ be measurable spaces, and let $T$ be a probability kernel from $(\Omega, \Sigma)$ to $(\Omega', \Sigma')$. Let $\mu$ and $\nu$ be probability measures on $\Omega$, with $\mu \ll \nu$. Then $\mu T \ll \nu T$ and in fact*

$$\left\|\log \frac{d(\mu T)}{d(\nu T)}\right\|_{L^\infty(\nu T)} \leq \left\|\log \frac{d\mu}{d\nu}\right\|_{L^\infty(\nu)} \tag{3.34}$$

$$\left\|\log \frac{d(\mu T)}{d(\nu T)}\right\|_{L^\infty(\nu T)/const} \leq \left\|\log \frac{d\mu}{d\nu}\right\|_{L^\infty(\nu)/const} \tag{3.35}$$

PROOF. Suppose that the Radon-Nikodỳm derivative (= density) $d\mu/d\nu$ satisfies

$$0 \leq a \leq \frac{d\mu}{d\nu} \leq b \leq +\infty \qquad \nu\text{-a.e.} \tag{3.36}$$

Then $a\nu \leq \mu \leq b\nu$ (in the sense of the usual ordering on positive measures), so obviously $a(\nu T) \leq \mu T \leq b(\nu T)$. It follows that

$$a \leq \frac{d(\mu T)}{d(\nu T)} \leq b \qquad (\nu T)\text{-a.e.} \tag{3.37}$$

Since

$$\left\|\log \frac{d\mu}{d\nu}\right\|_{L^\infty(\nu)} = \max(\log b, -\log a) \tag{3.38}$$

$$\left\|\log \frac{d\mu}{d\nu}\right\|_{L^\infty(\nu)/const} = \tfrac{1}{2}(\log b - \log a) \tag{3.39}$$

[where $a$ and $b$ are the sharpest values making (3.36) true], with an analogous formula for $d(\mu T)/d(\nu T)$, the lemma is proven. ∎

**Theorem 3.6 (Second fundamental theorem)** *Let $T$ be an RT map satisfying properties (T1)–(T3), and let $\Phi_1, \Phi_2 \in \mathrm{dom}\,\mathcal{R}$. Then, for all $\Phi_1' \in \mathcal{R}(\Phi_1)$ and $\Phi_2' \in \mathcal{R}(\Phi_2)$,*

$$\|\Phi_1' - \Phi_2'\|_{\mathcal{B}^0/(\mathcal{J}+Const)} \leq K\|\Phi_1 - \Phi_2\|_{\mathcal{B}^0/(\mathcal{J}+Const)}. \tag{3.40}$$

*That is, on its domain the map $\mathcal{R}$ is Lipschitz continuous (with Lipschitz constant $\leq K$) in the $\mathcal{B}^0/(\mathcal{J} + Const)$ norm.*

PROOF. Let $(\Lambda_n) \subset \mathbb{Z}^d$ and $(\Lambda_n') \subset \mathbb{Z}^{d'}$ be van Hove sequences having the properties (T3) assumed in Section 3.1. Let $\mu_1 \in \mathcal{G}_{inv}(\Pi^{\Phi_1})$ and $\mu_2 \in \mathcal{G}_{inv}(\Pi^{\Phi_2})$. By the First Fundamental Theorem (Theorem 3.4) we have $\mu_1 T \in \mathcal{G}_{inv}(\Pi^{\Phi_1'})$ and $\mu_2 T \in \mathcal{G}_{inv}(\Pi^{\Phi_2'})$. Now the image spins in $\Lambda_n'$ depend only on the original spins in $\Lambda_n$: that is, $(\mu_1 T)\restriction\mathcal{F}_{\Lambda_n'}'$ is the image under $T$ of $\mu_1\restriction\mathcal{F}_{\Lambda_n}$, and likewise for $\mu_2$. Therefore, by Lemma 3.5 we have

$$\left\|\log \frac{d(\mu_1 T)_{\Lambda_n'}}{d(\mu_2 T)_{\Lambda_n'}}\right\|_{B(\Omega')/const} \leq \left\|\log \frac{d(\mu_1)_{\Lambda_n}}{d(\mu_2)_{\Lambda_n}}\right\|_{B(\Omega)/const}. \tag{3.41}$$

(Since the measures $\mu_2$ and $\mu_2 T$ are both Gibbsian, they give nonzero measure to every open set; and moreover the interactions $\Phi_1$, $\Phi_2$, $\Phi'_1$ and $\Phi'_2$ are all continuous. Therefore we can replace the essential sup norms by the true sup norms.) Then

$$\begin{aligned} \|\Phi'_1 - \Phi'_2\|_{\mathcal{B}^0/(\mathcal{J}+Const)} &= \lim_{n\to\infty} \frac{1}{|\Lambda'_n|} \left\|\log \frac{d(\mu_1 T)_{\Lambda'_n}}{d(\mu_2 T)_{\Lambda'_n}}\right\|_{B(\Omega')/const} \\ &\leq \lim_{n\to\infty} \frac{K}{|\Lambda_n|} \left\|\log \frac{d(\mu_1)_{\Lambda_n}}{d(\mu_2)_{\Lambda_n}}\right\|_{B(\Omega)/const} \\ &= K \|\Phi_1 - \Phi_2\|_{\mathcal{B}^0/(\mathcal{J}+Const)} \,, \end{aligned} \qquad (3.42)$$

where we have twice used Proposition 2.46(b). ∎

It is curious that although all the interactions $\Phi_1$, $\Phi_2$, $\Phi'_1$ and $\Phi'_2$ are here required to belong to $\mathcal{B}^1$, the Lipschitz estimate (3.40) is stated in $\mathcal{B}^0$ norm. This is because $\mathcal{B}^0$ (or more precisely its quotient by $\mathcal{J}$ or $\mathcal{J} + Const$) is the natural norm for measuring *bulk* energy contributions, as discussed in Section 2.4.8. The restriction to $\mathcal{B}^1$ is needed solely to ensure that the *boundary* energy contributions are $o(|\Lambda|)$, so as to avoid the pathologies discussed in Section 2.6.7. In any case, we would like to emphasize that all the $\mathcal{B}^\alpha$ norms are equivalent (up to a bounded factor) for interactions involving boundedly many spins at a time (e.g. two-spin interactions), even when they are of arbitrarily long range. The difference between the $\mathcal{B}^\alpha$ norms concerns how they treat interactions that are *very strongly multi-body*.

Theorem 3.6 constrains very strongly the ways in which the RT map can blow up as $\Phi$ approaches the boundary of its domain. Indeed, suppose that $(\Phi_n)_{n\geq 1}$ is a sequence in $\operatorname{dom}\mathcal{R} \subset \mathcal{B}^1$ that converges *in $\mathcal{B}^0$ norm* [or more generally, in $\mathcal{B}^0/(\mathcal{J}+Const)$ seminorm] to $\Phi_\infty \in \mathcal{B}^0$. (We need not require convergence in $\mathcal{B}^1$ norm, nor need we require that $\Phi_\infty$ belong to $\mathcal{B}^1$.) Next let $(\Phi'_n)_{n\geq 1}$ be any choice of renormalized interactions [i.e. $\Phi'_n \in \mathcal{R}(\Phi_n) \subset \mathcal{B}^1$]; here the choice concerns the selection of representatives modulo physical equivalence. Then (3.40) guarantees that $(\Phi'_n)$ is a Cauchy sequence in the $\mathcal{B}^0/(\mathcal{J}+Const)$ seminorm, hence converges in $\mathcal{B}^0/(\mathcal{J}+Const)$ seminorm to some $\Phi'_\infty \in \mathcal{B}^0$; moreover, this limit is unique modulo physical equivalence (i.e. modulo $\mathcal{J} + Const$). Now, if $\Phi_\infty$ and $\Phi'_\infty$ (or any interactions in their physical-equivalence classes) are *both* in $\mathcal{B}^1$, then it follows from Theorem 3.2 that $(\Phi_\infty, \Phi'_\infty) \in \mathcal{R}$, hence $\Phi_\infty \in \operatorname{dom}\mathcal{R}$. Therefore, if $\Phi_\infty \in \mathcal{B}^1 \setminus \operatorname{dom}\mathcal{R}$, it must be that $\Phi'_\infty$ is not physically equivalent to any interaction in $\mathcal{B}^1$.

One would like to conclude from this that the $\mathcal{B}^1$ (semi)norms $\|\Phi'_n\|_{\mathcal{B}^1/(\mathcal{J}+Const)}$ must diverge as $n \to \infty$. Unfortunately, we are not quite able to prove this, because we have not been able to prove a version of Proposition 2.39(a) modulo physical equivalence (cf. Proposition 2.43). The best we have been able to prove is the following:

**Corollary 3.7** *Let $\Omega_0$ be a compact metric space, and assume that $T$ satisfies (T1)–(T3) and is Feller. Let $(\Phi_n)_{n\geq 1}$ be a sequence in $\operatorname{dom}\mathcal{R} \subset \mathcal{B}^1$ that converges in $\mathcal{B}^0/(\mathcal{J}+Const)$ seminorm to $\Phi_\infty \in \mathcal{B}^1 \setminus \operatorname{dom}\mathcal{R}$. For each $n$, let $\Phi'_n$ be any interaction in $\mathcal{R}(\Phi_n) \subset \mathcal{B}^1$. Then:*

(a) Either $(\Phi'_n)_{n \geq 1}$ fails to converge in $\mathcal{B}^0$, or else $\|\Phi'_n\|_{\mathcal{B}^1} \to \infty$.

If the single-spin space $\Omega'_0$ is finite, then we also have:

(b) For any $h \gtrapprox 1$, $\|\Phi'_n\|_{\mathcal{B}_h/(\mathcal{J}+Const)} \to \infty$.

PROOF. (a) Suppose that $\Phi'_n \to \Phi'_\infty$ in $\mathcal{B}^0$, but $\|\Phi'_n\|_{\mathcal{B}^1} \not\to \infty$. So there is a subsequence of $(\Phi'_n)$ on which the $\mathcal{B}^1$ norm is bounded, say by $M$; and Proposition 2.39(a) then implies that $\Phi'_\infty \in \mathcal{B}^1$ (with $\|\Phi'_\infty\|_{\mathcal{B}^1} \leq M$). But by Theorem 3.2 this means that $(\Phi_\infty, \Phi'_\infty) \in \mathcal{R}$, contrary to the hypothesis that $\Phi_\infty \notin \operatorname{dom} \mathcal{R}$.

(b) As argued above, the equivalence classes $[\Phi'_n] \equiv \Phi'_n + \mathcal{J} + Const$ converge in $\mathcal{B}^0/(\mathcal{J}+Const)$ to some equivalence class $[\Phi'_\infty]$. It follows that one can choose new representatives $\hat{\Phi}'_n \in [\Phi'_n]$ and $\hat{\Phi}'_\infty \in [\Phi'_\infty]$ such that $\hat{\Phi}'_n \to \hat{\Phi}'_\infty$ in $\mathcal{B}^0$. Now suppose that $\|\hat{\Phi}'_n\|_{\mathcal{B}_h/(\mathcal{J}+Const)} \equiv \|\Phi'_n\|_{\mathcal{B}_h/(\mathcal{J}+Const)} \not\to \infty$. Then there is a subsequence of $(\hat{\Phi}'_n)$ on which the $\mathcal{B}_h/(\mathcal{J}+Const)$ seminorm is bounded, say by $M$; and Proposition 2.43 then implies that $\hat{\Phi}'_\infty \in \mathcal{B}_h + \mathcal{J} + Const$ (with $\|\hat{\Phi}'_\infty\|_{\mathcal{B}_h/(\mathcal{J}+Const)} \leq M$). But this means that there exists $\hat{\hat{\Phi}}'_\infty \in \mathcal{B}_h \subset \mathcal{B}^1$ (with $\|\hat{\hat{\Phi}}'_\infty\|_{\mathcal{B}_h} \leq M$) such that $\hat{\hat{\Phi}}'_\infty \in [\hat{\Phi}'_\infty] = [\Phi'_\infty]$. And by Theorem 3.2 this means that $(\Phi_\infty, \hat{\hat{\Phi}}'_\infty) \in \mathcal{R}$, contrary to the hypothesis that $\Phi_\infty \notin \operatorname{dom} \mathcal{R}$. ∎

Our inability to prove the divergence of the $\mathcal{B}^1$ seminorm is not as serious as it may seem: as will be discussed in Section 6.1.2, one probably wants anyway to formulate RG theory in a space $\mathcal{B}_h$ of *short-range* interactions, and for such a space our result (b) is sufficient (when $\Omega'_0$ is finite).

# 4 Provably Pathological Renormalization Transformations

## 4.1 Griffiths-Pearce-Israel Pathologies I: Israel's Example

### 4.1.1 Introduction

Griffiths and Pearce [172, 173, 171] were the first to point out the possible existence of what they called "peculiarities" of the RT. These peculiarities were exhibited in models in which the internal spins undergo a phase transition for some fixed block-spin configuration. They observed that in such a situation the correlation functions of the internal-spin system could become discontinuous functions of the block spins, which implies that each of the terms of the (formal) expansion yielding the renormalized Hamiltonian (1.3) could be discontinuous. This casts doubts on the convergence of such an expansion, and hence on either the existence or the continuity properties of the renormalized Hamiltonian.

This situation was further clarified by Israel [207] in the particular case of the $b = 2$ decimation transformation. He argued that when such peculiarities exist, a very weak locality condition is violated by the renormalized measure: the conditional expectation for a single site is a discontinuous function (in the product topology) of the boundary conditions. That is, it is possible to fix the block-spin configuration in an arbitrarily large volume around the origin in such a way that what happens at the origin depends strongly on the block spins which are *outside* of the volume. The set of configurations for which this pathology occurs is improbable, but not of zero measure. In our terminology, the renormalized measure is *non-quasilocal*: that is, it is not consistent with any quasilocal specification. In particular, the renormalized measure is not the Gibbs measure for any uniformly convergent interaction.

In this section we fill in the technical details of Israel's argument, thereby converting it into a rigorous proof. In the following sections we shall generalize Israel's argument to other models and other renormalization transformations. In all cases, the underlying physical mechanism causing the non-Gibbsianness of the renormalized measure is the same: the influence from the block spins outside the specified volume is transmitted to the origin via the internal spins in the intermediate region, by-passing the block spins in the finite environment of the origin. This occurs because the internal spins have a phase transition, and the block-spin boundary conditions can pick different phases of these internal spins.

### 4.1.2 Israel's Example: Decimation in $d = 2$

Let us present now Israel's example — the two-dimensional Ising model at low temperature and zero magnetic field, using the $b = 2$ decimation transformation — together with the proof that *after one renormalization step* the renormalized measure is no longer Gibbsian.[47] The strategy of the proof is to show that the renormalized measure exhibits grossly non-local correlations, in the sense that the conditional probability distribution of the spin at the origin, as a function of all the other spins, depends strongly on the spins arbitrarily far away from the origin. More precisely, we shall show that if we take an arbitrarily large cube and fix all the block spins inside, except the origin, in a fully alternating configuration, then the renormalized magnetization at the origin depends strongly on the block-spin configuration outside of the cube.

The ferromagnetic nearest-neighbor Ising model in $\mathbb{Z}^2$ is defined by the formal Hamiltonian

$$H = -J \sum_{\langle ij \rangle} \sigma_i \sigma_j \,, \tag{4.1}$$

where $\langle ij \rangle$ denotes nearest-neighbor pairs and $J$ plays the role of an inverse temperature. We shall use the decimation transformation with scale factor $b = 2$. The *image* (or *block*) spins are those spins with both coordinates even, while the remaining spins are the *internal* spins. We denote by $(\mathbb{Z}^2)^{image}$ (resp. $(\mathbb{Z}^2)^{int}$) the set of image (resp. internal) spins. More generally, if $\Lambda$ is a subset of $\mathbb{Z}^2$, we denote by $\Lambda^{image} \equiv \Lambda \cap (\mathbb{Z}^2)^{image}$ (resp. $\Lambda^{int} \equiv \Lambda \cap (\mathbb{Z}^2)^{int}$) the set of image (resp. internal) spins in $\Lambda$.

---

[47]It is well known that the decimation transformation is badly behaved *in the limit of infinitely many decimations* [210, 364]: for example, any fixed point must have a two-point correlation function $\langle \sigma_0; \sigma_x \rangle$ which is independent of $x$ (so in particular doesn't decay as $|x| \to \infty$). But the present example is much more drastic, as the problems appear after a single step.

The proof of non-quasilocality of the renormalized measure goes in four steps:

*Step 0. Computation of the conditional probabilities for the image system.* These conditional probabilities turn out to be related to expectation values in a system of internal spins, with *fixed* image spins $\omega'$.

*Step 1. Selection of an image-spin configuration $\omega'_{\text{special}}$.* We find an image-spin configuration $\omega'_{\text{special}}$ such that the corresponding system of internal spins has a non-unique Gibbs measure (i.e. a first-order phase transition).

*Step 2. Study of a neighborhood of $\omega'_{\text{special}}$.* We study the internal-spin system for image-spin configurations $\omega'$ in a neighborhood of $\omega'_{\text{special}}$, and show that the internal-spin order parameter is a discontinuous function of $\omega'$. In physical terms, this means that the internal-spin order parameter depends sensitively on the image-spin configuration arbitrarily far from the origin, if the image-spin configuration in the intermediate region is set to $\omega'_{\text{special}}$.

*Step 3. "Unfixing" of the spin at the origin.* This is a technical step relating the image spin at the origin to the internal spins nearby. (After all, we want the conditional probabilities for image spins, not internal spins.)

Let us now discuss these steps in detail:

*Step 0. Computation of the conditional probabilities for the image system.* Let $\mu$ be *any* Gibbs measure for the ferromagnetic Ising model in $\mathbb{Z}^2$ with nearest-neighbor coupling $J$. Our goal is to show that, for $J$ sufficiently large, the image (decimated) measure $\mu T$ has non-quasilocal conditional probabilities. Therefore, our first order of business must be to compute these conditional probabilities. To do this, we use the only fact we know about the measure $\mu$, namely that it satisfies the DLR equations for the ferromagnetic nearest-neighbor Ising model.

The present case is relatively simple, because the image spins are simply a subset of the original spins. Let, therefore, $\Lambda'$ be a finite subset of $\mathbb{Z}^2$; we wish to compute the conditional probabilities $E_{\mu T}(f | \{\sigma'_j\}_{j \in \Lambda'^c})$ for functions $f$ of the spins $\{\sigma'_i\}_{i \in \Lambda'}$. But these are just the conditional probabilities $E_\mu(f | \{\sigma_l\}_{l \in 2(\Lambda'^c)})$ for functions $f$ of the spins $\{\sigma_k\}_{k \in 2\Lambda'}$. There is a slight complication now, because the set $2(\Lambda'^c) = (\mathbb{Z}^2)^{image} \setminus 2\Lambda'$ is not the complement of a finite set; its complement consists of the image spins in $2\Lambda'$ *plus all the internal spins*. Therefore, the DLR equations for the original model do not immediately tell us how to condition on $\{\sigma_l\}_{l \in 2(\Lambda'^c)}$. However, we have studied this problem in Section 2.3.7; the conclusion (Proposition 2.25) is that the conditional probability measure $\mu(\,\cdot\,|\{\sigma_l\}_{l \in 2(\Lambda'^c)})$ is, for $\mu$-almost-every $\{\sigma_l\}_{l \in 2(\Lambda'^c)}$, a Gibbs measure for the Ising model restricted to volume $(2\Lambda') \cup (\mathbb{Z}^2)^{int}$ with external spins set to $\{\sigma_l\}_{l \in 2(\Lambda'^c)}$. This latter system is specified by the same formal Hamiltonian (and hence the same interaction) as the original Ising model, except that now only the spins in $(2\Lambda') \cup (\mathbb{Z}^2)^{int}$ are considered to be random variables, and the spins $\{\sigma_l\}_{l \in 2(\Lambda'^c)}$ are considered to be fixed.

Note that we know only that $\mu(\,\cdot\,|\{\sigma_l\}_{l \in 2(\Lambda'^c)})$ is *some* Gibbs measure for the re-stricted interaction: if the restricted interaction happens to have more than one Gibbs measure, then we have no way of knowing which one is $\mu(\,\cdot\,|\{\sigma_l\}_{l\in 2(\Lambda'^c)})$. Therefore, we shall have to prove bounds which are valid uniformly for *all* Gibbs measures of the restricted interaction. This is what we shall do in Steps 2 and 3 below.

Note also that this computation of the conditional probabilities is asserted to be valid only for $\mu$-*almost-every* $\{\sigma_l\}_{l\in 2(\Lambda'^c)}$; indeed, conditional probabilities are *only* defined up to modifications on a set of measure zero. Therefore, in order to prove non-quasilocality we must prove not only that this *particular* version of the conditional probabilities is a discontinuous function, but that *no* function obtained from this one by modification on a set of $\mu$-measure zero can be continuous. That is, we must prove that the conditional probabilities are *essentially discontinuous*. We shall do this in Steps 2 and 3 below.

It is convenient to study first the system of internal spins alone, i.e. the system in $(\mathbb{Z}^2)^{int}$ with *all* image spins $\{\sigma'_j\}_{j\in\mathbb{Z}^2}$ set to fixed values. We call this system the *modified object system for image-spin configuration* $\{\sigma'_j\}_{j\in\mathbb{Z}^2}$. In Step 3 below we will "unfix" the image spins in the volume $\Lambda'$. In fact, it suffices to consider just one particular volume $\Lambda'$, which we shall take to be $\{0\}$.

*Step 1. Selection of an image-spin configuration* $\omega'_{\text{special}}$. Our goal is to show that the conditional probabilities $\mu(\,\cdot\,|\{\sigma_l\}_{l\in 2(\Lambda'^c)})$ are essentially discontinuous functions of $\{\sigma_l\}_{l\in 2(\Lambda'^c)}$. Therefore, we must find a point $\omega' = \{\sigma'_j\}_{j\in\mathbb{Z}^2}$ of essential discontinuity. A good candidate would be an image-spin configuration $\omega'_{\text{special}}$ such that the corresponding system of internal spins has a non-unique Gibbs measure. Indeed, non-uniqueness of the Gibbs measure means that the internal spins in volume $2\Lambda'$ depend sensitively on the *internal spins* arbitrarily far from the volume $2\Lambda'$ (albeit with the intermediate internal spins free to fluctuate); so it is a reasonable guess that the Gibbs measure might depend sensitively also on the *image spins* arbitrarily far away (but with the intermediate image spins held fixed at $\omega'_{\text{special}}$), and this is precisely the statement of essential discontinuity (see Step 2 below).

For the $b=2$ decimation transformation, such a configuration $\omega'_{\text{special}}$ was found by Griffiths and Pearce [173, 171]: it is the fully alternating configuration $\omega'_{alt}$ defined by

$$\sigma'_{i_1,i_2} \equiv \sigma_{2i_1,2i_2} = (-1)^{i_1+i_2} \qquad (4.2)$$

[see Figure 3(a)]. Notice that each internal spin is adjacent either to two image spins *of opposite sign* — in which case the effective magnetic fields cancel — or else to no image spin. Therefore, the modified object system is simply a ferromagnetic Ising model in zero field on a *decorated lattice* [342], as shown in Figure 3(b). Now we can explicitly integrate out the spins in the decorated lattice that have exactly two neighbors, yielding an effective coupling $J' = \frac{1}{2}\log\cosh 2J$ between those two neighbors. The result is an ordinary ferromagnetic Ising model on $\mathbb{Z}^2$, with nearest-neighbor coupling $J'$ and zero magnetic field [Figure 3(c)]. If $J' > J_c = \frac{1}{2}\log(1+\sqrt{2}) = 0.440686\ldots$, that is, $J > \frac{1}{2}\cosh^{-1}(1+\sqrt{2}) = 0.764285\ldots \approx 1.73 J_c$, then the modified object system for image-spin configuration $\omega'_{alt}$ has two distinct Gibbs measures, a "+" phase and a $-$ phase (obtainable by using "+" or $-$ boundary conditions, respectively).

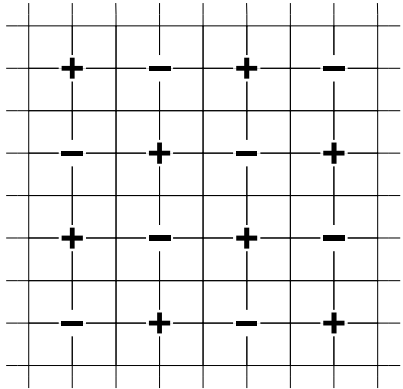

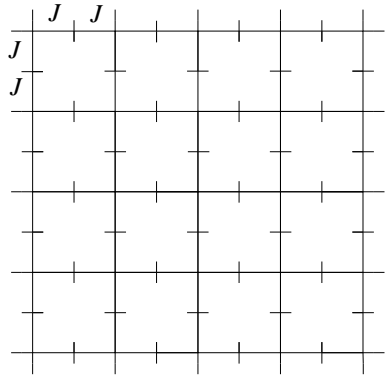

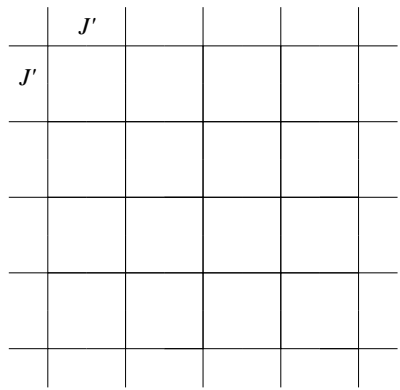

Figure 3: (a) The fully alternating image-spin configuration $\omega'_{alt}$. (b) The decorated system of internal spins. (c) The equivalent nearest-neighbor interaction on $(2\mathbb{Z})^2$.

*Step 2. Study of a neighborhood of $\omega'_{\text{special}} = \omega'_{alt}$.* The next step is to study image-spin configurations in a neighborhood (in the product topology) of $\omega'_{alt}$. To show that the order parameter $\langle \sigma_i \rangle_{\omega'}$ is an essentially discontinuous function of the image-spin configuration $\omega'$, it suffices to show that there exists a constant $\delta > 0$ such that in each neighborhood of $\omega'_{alt}$ the essential oscillation of $\langle \sigma_i \rangle_{\omega'}$ is at least $\delta$. More precisely, it suffices to show that there exists $\delta > 0$ such that in each neighborhood $\mathcal{N} \ni \omega'_{alt}$ there exist nonempty open sets $\mathcal{N}_+, \mathcal{N}_- \subset \mathcal{N}$ and constants $c_+ > c_-$ with $c_+ - c_- \geq \delta$ such that

$$\langle \sigma_i \rangle_{\omega'} \geq c_+ \quad \text{whenever } \omega' \in \mathcal{N}_+ \tag{4.3a}$$
$$\langle \sigma_i \rangle_{\omega'} \leq c_- \quad \text{whenever } \omega' \in \mathcal{N}_- \tag{4.3b}$$

Now a basis for the neighborhoods $\mathcal{N} \ni \omega'_{alt}$ is given by sets of the form

$$\mathcal{N}_R = \{\omega' \colon \omega' = \omega'_{alt} \text{ on } \Lambda_R, \ \omega' = \text{ arbitrary outside } \Lambda_R\}, \tag{4.4}$$

where $\Lambda_R$ is a square of side $2R + 1$ centered at the origin (here $R$ is an *unprimed* distance). We shall take $\mathcal{N}_+, \mathcal{N}_-$ to be sets of the form

$$\mathcal{N}_{R,R',+} = \{\omega' \colon \omega' = \omega'_{alt} \text{ on } \Lambda_R, \ \omega' = +1 \text{ on } \Lambda_{R'} \setminus \Lambda_R, \ \omega' = \text{ arbitrary outside } \Lambda_{R'}\} \tag{4.5a}$$
$$\mathcal{N}_{R,R',-} = \{\omega' \colon \omega' = \omega'_{alt} \text{ on } \Lambda_R, \ \omega' = -1 \text{ on } \Lambda_{R'} \setminus \Lambda_R, \ \omega' = \text{ arbitrary outside } \Lambda_{R'}\} \tag{4.5b}$$

with $R'$ chosen appropriately as a function of $R$ ($R < R' < \infty$). The motivation behind this choice is that setting the image spins in $\Lambda_{R'} \setminus \Lambda_R$ to be all $+$ (resp. all $-$) is expected to push the system into its $+$ (resp. $-$) phase. The remainder of Step 2 is devoted to proving that this is in fact the case.

Since we know only that the conditional distribution $\langle \cdot \rangle_{\omega'}$ is *some* Gibbs measure for the modified object system, we need to prove the bounds (4.3) *uniformly* for *all* Gibbs measures for this system. It suffices to show that there exists $R'' < \infty$ such that the Gibbs measure for the modified object system in the *finite* volume $\Lambda_{R''}^{int}$, with image spins $\omega' \in \mathcal{N}_{R,R',+}$ (or $\mathcal{N}_{R,R',-}$) *and arbitrary internal-spin boundary condition* $\{\overline{\sigma}_l\}_{l \in (\mathbb{Z}^2)^{int} \setminus \Lambda_{R''}^{int}}$, satisfies the bounds (4.3). For simplicity we shall take $R'' = R'$. In fact, in this two-dimensional example (but *not* in higher dimensions) we can take $R' = R + 2$. Therefore, we are led to the following situation:

Let $\Lambda_R$ be the square of side $2R + 1$ centered at the origin, and let $\Gamma_R \equiv \Lambda_R \setminus \Lambda_{R-1}$ be the $R^{th}$ layer. Now choose an even number $R$, and consider all the configurations with the image spins in $\Lambda_R$ fixed in the alternating configuration $\omega'_{alt}$, and those in the second layer outside $\Lambda_R$ (that is, in $\Gamma_{R+2}^{image}$) fixed to be "+". The spins outside $\Lambda_{R+2}$ (*both image and internal*) are fixed in some arbitrary configuration. The situation is depicted in Figure 4(a), where a circle represents an internal spin which fluctuates over all possible values. Now consider all the resulting systems of internal spins in $\Lambda_{R+2}^{int}$ given the above configuration of image spins in $\Lambda_{R+2}^{image}$ and an arbitrary fixed configuration (of both image and internal spins) outside $\Lambda_{R+2}$. We want to convince the reader that all the measures so obtained have local magnetizations $\langle \sigma_{i_1,i_2} \rangle$ for $(i_1, i_2) \in \Lambda_{R+1}^{int}$ which are bounded below by a strictly positive constant, uniformly in $R$ (sufficiently large) and uniformly in the boundary condition outside $\Lambda_{R+2}$. The sequence of bounds used in our proof is summarized in Figure 4.

Each internal spin in $\Lambda_{R+2}^{int}$ feels an "effective magnetic field" $\pm J$ from each image spin adjacent to it; but because the image-spin configuration in $\Lambda_R$ is alternating, these "effective magnetic fields" are *all zero* except at some sites in layers $\Gamma_{R+1}$ and $\Gamma_{R+2}$:

(i) An internal spin in layer $\Gamma_{R+1}$ feels an effective field $+2J$ if it is adjacent to two "+" image spins.

(ii) An internal spin in layer $\Gamma_{R+2}$ feels an effective field $+3J$ or $+J$ depending on whether the adjacent spin in layer $\Gamma_{R+3}$ (which is always an internal spin) happens to be "+" or "−".

We therefore consider the system of internal spins in $\Lambda_{R+2}^{int}$ with the magnetic fields described in (i) and (ii) above [Figure 4(b)].

Next we notice that by the FKG inequality [126] (or alternatively the Griffiths II inequality [341]), the local magnetizations $\langle \sigma_{i_1,i_2} \rangle$ for $(i_1, i_2) \in \Lambda_{R+2}^{int}$ are bounded below by the values that they would take if the magnetic fields $+2J$ in (i) were changed to zero, and the fields $+3J$ in (ii) changed to $+J$. We now have a system consisting of the spins in $\Lambda_{R+2}^{int}$, with a magnetic field $+J$ on each spin in layer $\Gamma_{R+2}^{int}$ [Figure 4(c)].

This latter system lives on a finite subset of the decorated lattice. We can explicitly integrate out the spins in $\Lambda_{R+1}^{int}$ that have exactly two neighbors (namely, the spins that have one coordinate even and one coordinate odd), yielding an effective coupling $J' = \frac{1}{2} \log \cosh 2J$ between those two neighbors. Similarly, we can integrate out the spins in $\Gamma_{R+2}^{int}$, yielding an effective magnetic field $h' = \frac{1}{2} \log \cosh 2J > 0$ on each remaining spin in $\Gamma_{R+1}^{int}$, except that the field is $2h'$ at the corners [Figure 4(d)]. But this last system is equivalent to a square lattice of size $(R+2) \times (R+2)$ with nearest-neighbor coupling $J'$ and + boundary conditions [Figure 4(e)]. As $R \to \infty$ this system tends to the "+" phase for an Ising model with coupling $J'$. In particular, the magnetization $\langle \sigma_{i_1,i_2} \rangle$ of any spin remaining in this system (i.e. any spin with $i_1$ and $i_2$ both odd) tends to the spontaneous magnetization $M_0(J')$, which is $> 0$ if $J' > J_c$. We can now return to the decorated lattice, to compute the magnetization $\langle \sigma_{i_1,i_2} \rangle$ on the internal spins that got integrated out (i.e. the ones with $i_1$ even and $i_2$ odd or vice versa):

$$\begin{aligned} \langle \sigma_{i_1,i_2} \rangle_{\Lambda_{R+2}^{int}} &= \langle \tanh(J(\sigma' + \sigma'')) \rangle_{R+2} \\ &= \langle (\tfrac{1}{2} \tanh 2J)(\sigma' + \sigma'') \rangle_{R+2} \\ &\longrightarrow (\tanh 2J) M_0(J') > 0 \end{aligned} \quad (4.6)$$

where $\sigma'$ and $\sigma''$ are the two internal spins adjacent to $\sigma_{i_1,i_2}$. We have therefore proven our claim that the magnetizations $\langle \sigma_{i_1,i_2} \rangle$ for $(i_1, i_2) \in \Lambda_{R+1}^{int}$ are bounded below by a strictly positive constant [namely $(1-\epsilon)(\tanh 2J) M_0(J')$ for any $\epsilon > 0$], uniformly in $R \geq R_0(i_1, i_2)$ and uniformly in the boundary condition outside $\Lambda_{R+2}$. Repeating the argument but with the image spins in $\Gamma_{R+2}$ chosen as "$-$",[48] we obtain the "$-$" phase for the internal spins and thus a strictly negative upper bound on $\langle \sigma_{i_1, i_2} \rangle$. This proves that the local magnetization, say for the four internal spins neighboring the origin, is determined by the image spins at faraway distances.

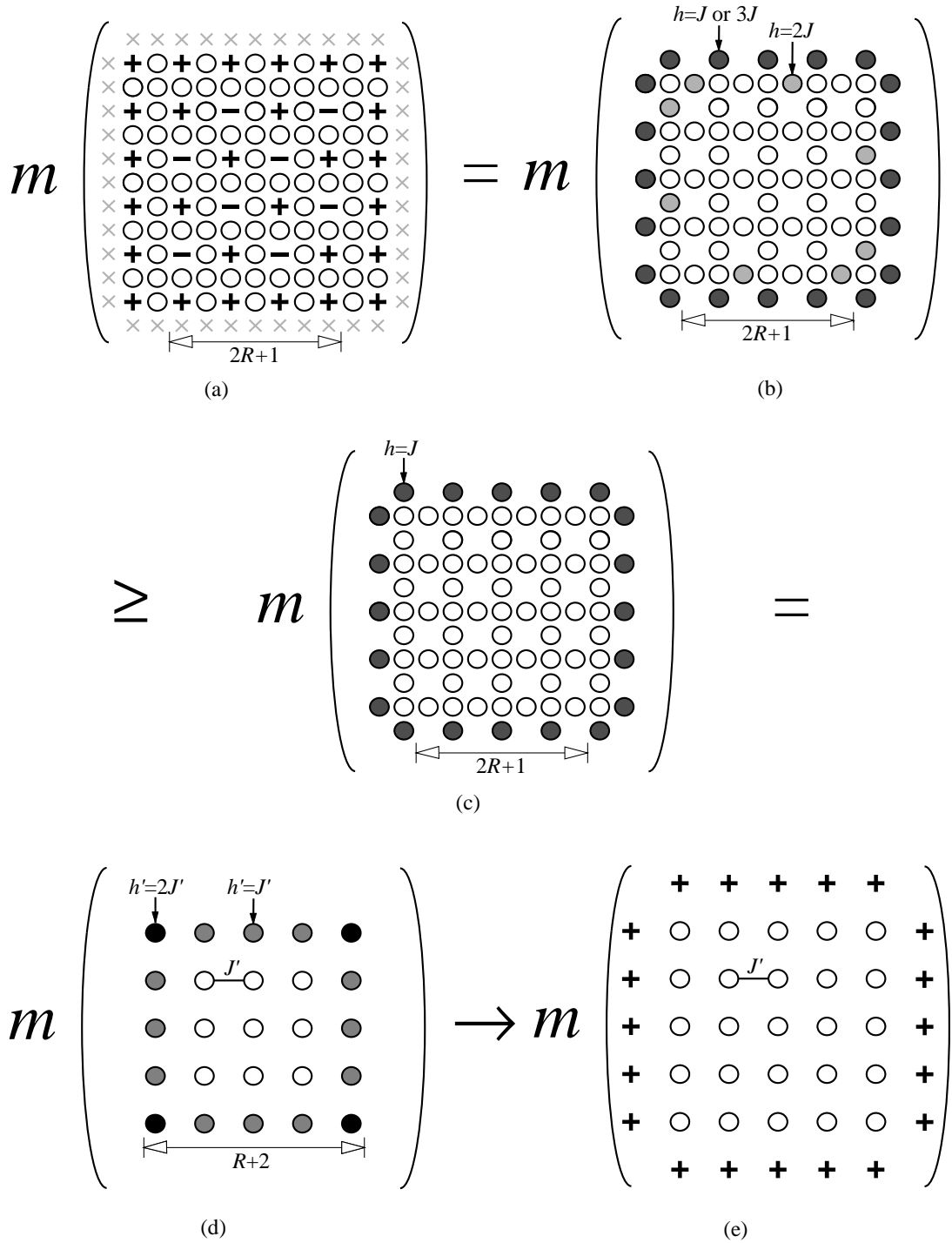

Figure 4: Sequence of bounds proving a lower bound on the magnetization for the internal spins.

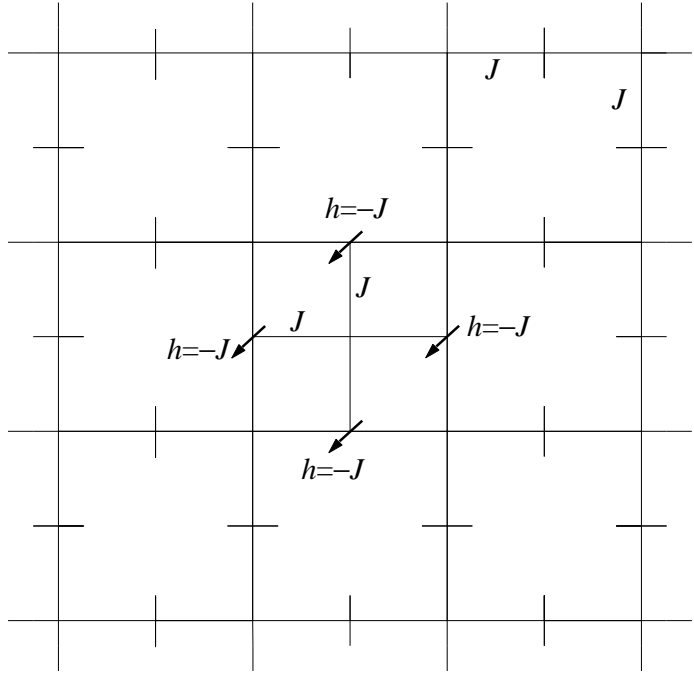

Figure 5: Decorated lattice when the spin at the origin is free to fluctuate.

*Step 3. Unfixing of the spin at the origin.* We now have to make a slight modification in the preceding argument, as the system we really want to study is the system consisting of the internal spins in $\Lambda_{R+2}$ *and the spin at the origin*, with the image spins in $\Lambda_R^{image}$ *other than the one at the origin* fixed in the alternating configuration $\omega'_{alt}$, the image spins in layer $\Gamma_{R+2}^{image}$ set to be "+", and the spins outside $\Lambda_{R+2}$ (both image and internal) fixed in some arbitrary configuration. By the same reasoning as before, we obtain a system on the decorated lattice *plus the origin*, with a coupling $J$ between the origin and its four neighbors *and an additional magnetic field $-J$ on the neighbors of the origin* [Figure 5]. Denoting by $\langle \cdot \rangle_+$ (resp. $\langle \cdot \rangle_+^\sim$) the expectation in the old (resp.

---

[48]Because we have fixed the image spin at the origin to be +, the two situations are not quite symmetric. But the only change is a shift in the *location* of the internal spins in $\Gamma_{R+1}$ which feel a

new) decorated system, it is easy to see that

$$\langle \sigma_{0,0} \rangle_+^\sim = \frac{\sum_{\sigma_{0,0}=\pm 1} \langle \sigma_{0,0} \exp[J(\sigma_{0,0}-1)(\sigma_{0,1}+\sigma_{0,-1}+\sigma_{1,0}+\sigma_{-1,0})] \rangle_+}{\sum_{\sigma_{0,0}=\pm 1} \langle \exp[J(\sigma_{0,0}-1)(\sigma_{0,1}+\sigma_{0,-1}+\sigma_{1,0}+\sigma_{-1,0})] \rangle_+}$$
$$= \frac{1 - \langle \exp[-2J(\sigma_{0,1}+\sigma_{0,-1}+\sigma_{1,0}+\sigma_{-1,0})] \rangle_+}{1 + \langle \exp[-2J(\sigma_{0,1}+\sigma_{0,-1}+\sigma_{1,0}+\sigma_{-1,0})] \rangle_+} . \quad (4.7)$$

Similarly, for the analogous system with the image spins in $\Gamma_{R+2}^{image}$ set to "$-$", we have

$$\langle \sigma_{0,0} \rangle_-^\sim = \frac{1 - \langle \exp[-2J(\sigma_{0,1}+\sigma_{0,-1}+\sigma_{1,0}+\sigma_{-1,0})] \rangle_-}{1 + \langle \exp[-2J(\sigma_{0,1}+\sigma_{0,-1}+\sigma_{1,0}+\sigma_{-1,0})] \rangle_-}$$
$$= \frac{1 - \langle \exp[+2J(\sigma_{0,1}+\sigma_{0,-1}+\sigma_{1,0}+\sigma_{-1,0})] \rangle_+}{1 + \langle \exp[+2J(\sigma_{0,1}+\sigma_{0,-1}+\sigma_{1,0}+\sigma_{-1,0})] \rangle_+} .$$
$$(4.8)$$

Therefore,

$$\langle \sigma_{0,0} \rangle_+^\sim - \langle \sigma_{0,0} \rangle_-^\sim = \frac{2(y-x)}{(1+x)(1+y)} , \quad (4.9)$$

where

$$x = \langle \exp[-2J(\sigma_{0,1}+\sigma_{0,-1}+\sigma_{1,0}+\sigma_{-1,0})] \rangle_+ \quad (4.10)$$
$$y = \langle \exp[+2J(\sigma_{0,1}+\sigma_{0,-1}+\sigma_{1,0}+\sigma_{-1,0})] \rangle_+ \quad (4.11)$$

Now

$$y - x = 2 \langle \sinh 2J(\sigma_{0,1}+\sigma_{0,-1}+\sigma_{1,0}+\sigma_{-1,0}) \rangle_+$$
$$= 2 \sum_{\substack{k=1 \\ k \text{ odd}}}^{\infty} \frac{(2J)^k}{k!} \langle (\sigma_{0,1}+\sigma_{0,-1}+\sigma_{1,0}+\sigma_{-1,0})^k \rangle_+$$
$$\geq 4J \langle (\sigma_{0,1}+\sigma_{0,-1}+\sigma_{1,0}+\sigma_{-1,0}) \rangle_+$$
$$= 16J \langle \sigma_{0,1} \rangle_+ , \quad (4.12)$$

since the contributions from $k = 3, 5, \ldots$ are all nonnegative by Griffiths' first inequality. On the other hand, the denominator in (4.9) is bounded between 1 and $(1+e^{8J})^2$. Since we proved previously that for $J' > J_c$, the local magnetization $\langle \sigma_{0,1} \rangle_+$ is bounded below by a strictly positive constant, uniformly in $R$ (sufficiently large) and in the configuration outside $\Lambda_{R+2}$, we can conclude that

$$\langle \sigma_{0,0} \rangle_+^\sim - \langle \sigma_{0,0} \rangle_-^\sim \geq \delta > 0 \quad (4.13)$$

uniformly in $R$ (sufficiently large) and in the configuration outside $\Lambda_{R+2}$.

---
nonzero effective field; and this is irrelevant, since we replace these fields by zero anyway.

*Conclusion of the argument.* In summary, we have shown that at zero magnetic field and any sufficiently low temperature, given any Gibbs measure $\mu$, the renormalized measure $\mu T$ has the following property: Let $\omega'_{alt}$ be the fully alternating configuration $\sigma'_{i_1,i_2} = (-1)^{i_1+i_2}$; let $A_{R,+} \equiv \mathcal{N}_{R/2,(R/2)+1,+}$ be the set of all configurations $\{\sigma'\}$ that are alternating in $\Lambda_{R/2}$, "+" in $\Gamma_{(R/2)+1}$ and arbitrary outside; and let $A_{R,-}$ be analogously defined but with "−" in $\Gamma_{(R/2)+1}$. In the product topology $A_{R,+}$ and $A_{R,-}$ are open sets — in particular, they have strictly positive $(\mu T)$-measure — and given any neighborhood of $\mathcal{N} \ni \omega'_{alt}$ we always choose $R$ large enough so that $A_{R,+} \cup A_{R,-} \subset \mathcal{N}$. Moreover, we have proven that for all $\omega'_1 \in A_{R,+}$ and $\omega'_2 \in A_{R,-}$, we have

$$E_{\mu T}\left(\sigma'_{0,0} \mid \{\sigma'_{i_1,i_2}\}_{(i_1,i_2)\neq(0,0)}\right)(\omega'_1) - E_{\mu T}\left(\sigma'_{0,0} \mid \{\sigma'_{i_1,i_2}\}_{(i_1,i_2)\neq(0,0)}\right)(\omega'_2) \geq \delta > 0 \;. \tag{4.14}$$

This means — as first pointed out by Israel [207] — that the conditional expectations of $\sigma'_{0,0}$ are discontinuous as a function of the boundary conditions. More precisely, they are *essentially discontinuous*: no modification on a set of $(\mu T)$-measure zero can make them continuous at $\omega'_{alt}$. Now, for systems with a finite single-spin space (such as the Ising model), continuity is equivalent to quasilocality. Therefore, what we have really proven is that *the renormalized measure $\mu T$ is not consistent with any quasilocal specification*. (In our language, $\mu T$ is *non-quasilocal*; in the terminology of Sullivan [336], it is *non-almost Markovian*.) In particular, $\mu T$ is not Gibbsian for any uniformly convergent interaction.

**Theorem 4.1** *Let $\mu$ be any Gibbs measure for the two-dimensional Ising model with nearest-neighbor coupling $J > \frac{1}{2}\cosh^{-1}(1+\sqrt{2}) = 0.764285\ldots \approx 1.73 J_c$ and zero magnetic field. Let $T$ be the decimation transformation with spacing $b=2$. Then the measure $\mu T$ is not consistent with any quasilocal specification. In particular, it is not the Gibbs measure for any uniformly convergent interaction.*

In physical terms, we have shown that the value of the renormalized spin at the origin, $\sigma'_{0,0}$, depends strongly on the values of the renormalized spins arbitrarily far from the origin, if the renormalized spins in the intermediate region are fixed to be alternating. Such a long-range dependence is incompatible with the measure $\mu T$ being Gibbsian for any reasonable interaction.

## 4.2 Griffiths-Pearce-Israel Pathologies II: General Method

In this section we abstract the essential features of the Griffiths-Pearce-Israel argument, in order to prepare the way for generalizations to more complicated examples.

*Step 0. Computation of the conditional probabilities.*

This step is technical and messy, but the final result is the obvious one [cf. (4.22)/(4.23) below]. The reader is therefore invited to skip this step on a first reading.

For decimation, the computation of the conditional probabilities of $\mu T$ was an immediate application of Proposition 2.25. For more general RT maps, it will be a more complicated application of this same proposition: the idea is to consider first a *joint* system of interacting spins $\omega$ and $\omega'$, and then decimate this system to the space $\Omega'$.

If $\mu$ is any measure on the system of original spins (i.e. on $\Omega$), and $T$ is any probability kernel from $\Omega$ to $\Omega'$, then the joint measure $\mu \times T$ on $\Omega \times \Omega'$ is well-defined by

$$(\mu \times T)(A) \;=\; \int d\mu(\omega) \int T(\omega, d\omega') \chi_A(\omega, \omega') \tag{4.15}$$

for measurable sets $A \subset \Omega \times \Omega'$. Now, if $\Pi = \{\pi_\Lambda\}$ is a specification for the system of original spins, we wish to define a specification $\Pi \otimes T = \{(\Pi \otimes T)_{\Lambda,\Lambda'}\}$ for the joint system, with the property that

$$\mu \text{ consistent with } \Pi \;\Longrightarrow\; \mu \times T \text{ consistent with } \Pi \otimes T \;. \tag{4.16}$$

(In fact, the converse should also hold, i.e. a measure $\nu$ on $\Omega \times \Omega'$ should be consistent with $\Pi \otimes T$ if and only if $\mu \equiv \nu \!\restriction\! \mathcal{F}$ is consistent with $\Pi$ and $\nu = \mu \times T$.)

For simplicity let us assume that the probability kernel $T$ has the following form:

$$T(\omega, d\omega') \;=\; \prod_{x \in \mathcal{L}'} \widetilde{T}_x(\omega_{B_x}, \omega'_x) \, d\nu_x(\omega'_x) \tag{4.17}$$

where the $\nu_x$ are probability measures, and the $B_x$ are finite sets of original spins which together determine the image spin $\omega'_x$. We also assume that the family of sets $\{B_x\}_{x \in \mathcal{L}'}$ is locally finite[49], i.e. only finitely many image spins $x$ depend on any given original spin $y$. Now, to motivate the construction, suppose that $\mu$ is a Gibbs measure for an interaction $\Phi$ (and *a priori* measure $\mu^0$). Then, *formally* the measure $\mu \times T$ is given by

$$(\mu \times T)(d\omega, d\omega') \text{ "="} \text{ const} \times$$
$$\prod_{X \subset \mathcal{L}} e^{-\Phi_X(\omega)} \prod_{x \in \mathcal{L}'} \widetilde{T}_x(\omega_{B_x}, \omega'_x) \prod_{x \in \mathcal{L}} d\mu^0_x(\omega_x) \prod_{x \in \mathcal{L}'} d\nu_x(\omega'_x) \;. \tag{4.18}$$

Of course, the first two infinite products (the ones over functions $e^{-\Phi_X(\omega)}$ and $\widetilde{T}_x$) are meaningless, but we know what to do: to describe the *conditional* probability distribution $\mu \times T$, with $\omega$ fixed outside a finite set $\Lambda$ and $\omega'$ fixed outside a finite set $\Lambda'$, we retain in the products only those terms that intersect $\Lambda$ and/or $\Lambda'$, i.e.

$$(\mu \times T)(d\omega_\Lambda, d\omega'_{\Lambda'} | \omega_{\Lambda^c}, \omega'_{\Lambda'^c}) \;=\; \text{const}(\omega_{\Lambda^c}, \omega'_{\Lambda'^c}) \times$$
$$\prod_{X \cap \Lambda \neq \emptyset} e^{-\Phi_X(\omega)} \prod_{x:\, x \in \Lambda' \text{or } B_x \cap \Lambda \neq \emptyset \text{or both}} \widetilde{T}_x(\omega_{B_x}, \omega'_x) \prod_{x \in \Lambda} d\mu^0_x(\omega_x) \prod_{x \in \Lambda'} d\nu_x(\omega'_x) \;. \tag{4.19}$$

---

[49]That is, the set $\{x \colon B_x \ni y\}$ is finite for each $y \in \mathcal{L}$.

Now, the first product is just $e^{-H_\Lambda^\Phi(\omega_\Lambda, \omega_{\Lambda^c})}$, and the first and third products together yield (when properly normalized) the kernel $\pi_\Lambda(\omega_{\Lambda^c}, d\omega_\Lambda)$. Therefore, the specification $\Pi \otimes T$ should be defined as

$$(\Pi \otimes T)(d\widetilde{\omega}_\Lambda, d\widetilde{\omega}'_{\Lambda'} | \omega_{\Lambda^c}, \omega'_{\Lambda'^c}) = \bar{Z}_{\Lambda,\Lambda'}(\omega_{\Lambda^c}, \omega'_{\Lambda'^c})^{-1} \times$$

$$\pi_\Lambda(\omega_{\Lambda^c}, d\widetilde{\omega}_\Lambda) \prod_{x:\, x \in \Lambda' \text{ or } B_x \cap \Lambda \neq \emptyset \text{ or both}} \widetilde{T}_x\left((\omega_{\Lambda^c} \times \widetilde{\omega}_\Lambda)_{B_x}, (\omega'_{\Lambda'^c} \times \widetilde{\omega}'_{\Lambda'})_x\right) \prod_{x \in \Lambda'} d\nu_x(\omega'_x) \,, \tag{4.20}$$

where

$$\bar{Z}_{\Lambda,\Lambda'}(\omega_{\Lambda^c}, \omega'_{\Lambda'^c})^{-1} =$$

$$\int_{\Omega_\Lambda \times \Omega'_{\Lambda'}} \pi_\Lambda(\omega_{\Lambda^c}, d\widetilde{\omega}_\Lambda) \prod_{x:\, x \in \Lambda' \text{ or } B_x \cap \Lambda \neq \emptyset \text{ or both}} \widetilde{T}_x\left((\omega_{\Lambda^c} \times \widetilde{\omega}_\Lambda)_{B_x}, (\omega'_{\Lambda'^c} \times \widetilde{\omega}'_{\Lambda'})_x\right) \prod_{x \in \Lambda'} d\nu_x(\omega'_x) \tag{4.21}$$

and we have assumed, of course, that $\bar{Z}_{\Lambda,\Lambda'}(\omega_{\Lambda^c}, \omega'_{\Lambda'^c}) > 0$. [If $\bar{Z}_{\Lambda,\Lambda'}(\omega_{\Lambda^c}, \omega'_{\Lambda'^c}) = 0$, then $(\omega_{\Lambda^c}, \omega'_{\Lambda'^c})$ is a "forbidden boundary condition", which has to be dealt with as in the theory of lattice systems with hard-core constraints [299, 313].] We must now check that:

(a) $\Pi \otimes T$, thus defined, is indeed a specification.

(b) If $\mu$ is any measure consistent with $\Pi$, then $\mu \times T$ is consistent with $\Pi \otimes T$.

These two verifications are messy calculations, which the authors are convinced will work out (although mental exhaustion prevented them from writing out the full details).

Things become much simpler when $\Pi$ is the Gibbsian specification for an interaction $\Phi$ and *a priori* measure $\mu^0$, and the $\widetilde{T}_x$ are all nonvanishing. Then it is easy to see that $\Pi \otimes T$ is the Gibbsian specification for the interaction $\widetilde{\Phi}$ (on the lattice $\mathcal{L} \cup \mathcal{L}'$) defined by

$$\widetilde{\Phi}_{X,X'}(\omega, \omega') = \begin{cases} \Phi_X(\omega) & \text{if } X' = \emptyset \\ -\log \widetilde{T}_x(\omega_{B_x}, \omega'_x) & \text{if } X = B_x \text{ and } X' = \{x\} \\ 0 & \text{otherwise} \end{cases} \tag{4.22}$$

and *a priori* measure $\mu^0 \times \nu$. (In particular, it follows immediately from the general theory in Section 2.3.2 that $\Pi \otimes T$ is indeed a specification.) This is the interaction corresponding to the formal Hamiltonian

$$\begin{aligned} H_{joint}(\omega, \omega') &= \sum_{X \subset \mathcal{L}} \Phi_X(\omega) - \sum_{x \in \mathcal{L}'} \log \widetilde{T}_x(\omega_{B_x}, \omega'_x) \\ &= H_{original}(\omega) - \sum_{x \in \mathcal{L}'} \log \widetilde{T}_x(\omega_{B_x}, \omega'_x) \,. \end{aligned} \tag{4.23}$$

If the $\widetilde{T}_x$ can vanish, then $\widetilde{\Phi}$ may take the value $+\infty$, which is not (strictly speaking) permitted in our formulation; but the same algebra shows that $\Pi \otimes T$ is indeed a specification, at least when $\bar{Z}_{\Lambda,\Lambda'}(\omega_{\Lambda^c}, \omega'_{\Lambda'^c}) > 0$.[50]

Having constructed the specification $\Pi \otimes T$ on the lattice $\mathcal{L} \cup \mathcal{L}'$, we can now apply the same argument as in the decimation case, based on Proposition 2.25 (see Section 4.1.2, Step 0). Indeed, the renormalized measure $\mu T$ is obtained by decimating the joint measure $\mu \times T$, i.e. restricting it to the lattice $\mathcal{L}'$.

We hope that someone will come along and simplify our "abstract nonsense" concerning Step 0. But we have no doubt that our concrete arguments in this paper are correct.

*Step 1. Selection of an image-spin configuration $\omega'_{\text{special}}$ for which the corresponding internal-spin system has a non-unique Gibbs measure.*

We need to find an image-spin configuration $\omega'_{\text{special}}$ such that the resulting system of internal spins (the "modified object system") has at least two distinct Gibbs measures, call them $\mu_+$ and $\mu_-$. How we do this depends on the details of the model and the renormalization transformation. For the $b = 2$ decimation transformation on the nearest-neighbor Ising model, the fully alternating configuration $\omega'_{alt}$ does the trick. For the majority-rule transformation we shall need a more complicated configuration (Section 4.3.4).

Now let $f$ be a local observable such that $\mu_+(f) > \mu_-(f)$; we shall call $f$ the "internal-spin order parameter". (For the decimation example, $f$ is the spin at a neighbor of the origin.)

*Step 2. Discontinuity of the internal-spin order parameter as a function of the image-spin configuration in a neighborhood of $\omega'_{\text{special}}$.*

The next step is to study the behavior of the internal-spin system for image-spin configurations in a *neighborhood* (in the product topology) of $\omega'_{\text{special}}$. Our goal is to show that the order parameter for the internal-spin system is *essentially discontinuous* as a function of the image-spin configuration $\omega'$.

To do this, we first choose image-spin configurations $\omega'_+$ and $\omega'_-$ which we hope will "select the phases $\mu_+$ and $\mu_-$". We then study image-spin configurations $\omega'$ which are equal to $\omega'_{\text{special}}$ on some large box $\Lambda_R$, which are equal to $\omega'_+$ [or $\omega'_-$] on some annulus $\Lambda_{R'} \setminus \Lambda_R$ ($R < R' < \infty$), and which are arbitrary outside $\Lambda_{R'}$. Our goal is to show that, no matter how large $R$ is, the internal-spin phase is selected by the behavior of the image spins in $\Lambda_{R'} \setminus \Lambda_R$ — for a suitable choice of $R'$ depending on $R$ — no matter what happens outside $\Lambda_{R'}$.

---

[50]Many of our concrete examples *do* have configurations for which $\bar{Z}_{\Lambda,\Lambda'}(\omega_{\Lambda^c}, \omega'_{\Lambda'^c}) = 0$: for example, in the case of decimation, one obviously cannot insist that a certain image spin be $+1$ and simultaneously insist that the corresponding original spin be $-1$! But each of these concrete cases has a simple resolution: for example, in the case of decimation, we called *internal spin* only those original spins which are *not* (locked to) image spins; of course, the original spins which are locked to image spins don't even need to be considered.

In mathematical terms, our goal is to show that there exists a number $\delta > 0$ such that in each neighborhood $\mathcal{N} \ni \omega'_{\text{special}}$ (in the product topology) there exist nonempty open sets $\mathcal{N}_+, \mathcal{N}_- \subset \mathcal{N}$ and numbers $c_+, c_-$ with $c_+ - c_- \geq \delta$ such that for every $\omega' \in \mathcal{N}_+$ [resp. $\omega' \in \mathcal{N}_-$] and every Gibbs measure $\mu$ for the internal-spin system with image spins set to $\omega'$, we have $\mu(f) \geq c_+$ [resp. $\mu(f) \leq c_-$]. Now a basis for the neighborhoods $\mathcal{N} \ni \omega'_{alt}$ is given by sets of the form

$$\mathcal{N}_R = \{\omega'\colon \omega' = \omega'_{alt} \text{ on } \Lambda_R,\ \omega' = \text{ arbitrary outside } \Lambda_R\}, \qquad (4.24)$$

We shall take $\mathcal{N}_+, \mathcal{N}_-$ to be sets of the form

$$\mathcal{N}_{R,R',+} = \{\omega'\colon \omega' = \omega'_{alt} \text{ on } \Lambda_R,\ \omega' = \omega'_+ \text{ on } \Lambda_{R'} \setminus \Lambda_R,\ \omega' = \text{ arbitrary outside } \Lambda_{R'}\} \quad (4.25a)$$

$$\mathcal{N}_{R,R',-} = \{\omega'\colon \omega' = \omega'_{alt} \text{ on } \Lambda_R,\ \omega' = \omega'_- \text{ on } \Lambda_{R'} \setminus \Lambda_R,\ \omega' = \text{ arbitrary outside } \Lambda_{R'}\} \quad (4.25b)$$

We then have to prove that $R'$ can be chosen as a function of $R$ ($R < R' < \infty$) so that $\mu(f)$ satisfies the claimed bounds.

In practice, the only way we shall be able to prove the existence of such an $R' < \infty$ is to prove that the internal-spin system with $R' = \infty$ and $\omega' \in \mathcal{N}_{R,\infty,+}$ or $\mathcal{N}_{R,\infty,-}$ has a *unique* Gibbs measure, and that this measure satisfies the required bounds. It will then follow fairly easily that the (possibly non-unique) Gibbs measures for $R' < \infty$ tend to this unique limit as $R' \to \infty$, and satisfy the bounds (with a slightly reduced $\delta$) for some sufficiently large $R'$.

We emphasize that since we know only that the conditional distribution $\langle \cdot \rangle_{\omega'}$ is *some* Gibbs measure for the internal-spin system, we need to prove the claimed bounds on $\mu(f)$ *uniformly* for *all* Gibbs measures for this system. To do this, it suffices to show that the bounds are satisfied for a *finite-volume* internal-spin system, for some sufficiently large volume, uniformly in the (internal-spin) boundary conditions; that is, it suffices to show that there exists $R'' < \infty$ such that the Gibbs measure for the internal-spin system in the volume $\Lambda_{R''}^{int}$, with image spins $\omega' \in \mathcal{N}_{R,R',+}$ (or $\mathcal{N}_{R,R',-}$) *and arbitrary internal-spin boundary condition* $\{\overline{\sigma}_l\}_{l \in (\mathbb{Z}^2)^{int} \setminus \Lambda_{R''}^{int}}$, satisfies the claimed bounds. For simplicity we shall take $R'' = R'$.

Let us emphasize once again that both the image spins and the internal spins are arbitrary outside $\Lambda_{R'}$, but *for different reasons*. The image spins are arbitrary outside $\Lambda_{R'}$ because our computation of the conditional probabilities $\mu(\cdot|\omega')$ is valid only for $(\mu T)$-almost-every $\omega'$; therefore, to prove that these conditional probabilities are *essentially discontinuous* (i.e. cannot be made continuous by modification on a set of $(\mu T)$-measure zero), we must prove our bounds for a nonempty open set of configurations $\omega'$. The internal spins are arbitrary outside $\Lambda_{R''}$ ($= \Lambda_{R'}$) because we know only that the conditional measure $\mu(\cdot|\omega')$ is *some* Gibbs measure for the modified object system, but we have no idea which one (in case it is non-unique); therefore, we must prove bounds valid for *all* infinite-volume Gibbs measures of the modified object system.

*Step 3. Unfixing of the spin at the origin.*

The final step is to show that if the system of internal spins is slightly modified by changing the interaction with a few (in the our examples just one) image spins close to the origin, the order parameter at these extra spins differs little from the value at internal spins close to the origin. This is the step of "unfixing" some image spins discussed above.

*Conclusion of the argument.*

Combining the conclusions of Steps 2 and 3, we have that for all possible image-spin configurations outside $\Lambda_{R'}$, the order parameter at image spins close to the origin is determined by the image spins in the arbitrarily faraway annulus $\Lambda_{R'} \setminus \Lambda_R$. In mathematical terms, the conditional probability distribution of the image spin at the origin is an *essentially discontinuous* function of the other image spins, in a neighborhood of $\omega'_{specific}$. Thus, the renormalized measure has non-quasilocal conditional probabilities: it is not consistent with any quasilocal specification, and in particular is not the Gibbs measure of any uniformly convergent interaction.

## 4.3 Griffiths-Pearce-Israel Pathologies III: Some Further Examples

In this section we apply the Griffiths-Pearce-Israel method to prove non-Gibbsianness of the renormalized measure in the following additional examples:

- $b = 2$ decimation for the Ising model in dimension $d \geq 3$.

- Decimation with spacing $b \geq 3$, for the Ising model in any dimension $d \geq 2$.

- The Kadanoff transformation with finite $p$ and arbitrary block size $b \geq 1$, for the Ising model in any dimension $d \geq 2$.

- Some cases of the majority-rule transformation for the Ising model in dimension $d = 2$.

- Block-averaging, with *even* block size $b$, for the Ising model in any dimension $d \geq 2$.

Finally, and most strikingly, we can show that in all of these examples except (and this probably only for technical reasons) the majority-rule case, there is in fact an *open region in the $(J, h)$-plane* for which the renormalized measures are non-Gibbsian. Therefore, the Griffiths-Pearce-Israel pathologies are *not* associated with the fact that the original model is *sitting on* a phase-transition surface. Rather, it suffices that a first-order phase transition can be induced in the internal-spin system by choosing an appropriate block-spin configuration. For this we need to work at low temperature but not necessarily at zero magnetic field.

### 4.3.1 Israel's Example in Dimension $d \geq 3$

In this section we study the $b = 2$ decimation transformation for the Ising model in dimension $d \geq 3$.

*Step 0. Computation of the conditional probabilities.* This has already been done.

*Step 1. Choice of $\omega'_{\text{special}}$.* As in the two-dimensional case, we choose $\omega'_{\text{special}}$ to be the fully alternating configuration $\omega'_{alt}$. The system of internal spins for a fully alternating image-spin configuration again corresponds to a periodically diluted ferromagnet: an internal spin with all but one of its coordinates even — that is, one which is adjacent to two image spins — has two less neighbors coupled to itself, while all other internal spins are unaffected. The only difference from the two-dimensional case is that the resulting lattice is not merely a decorated version of an exactly soluble Ising model, so we cannot write an explicit formula for its critical temperature. Nevertheless, it is easy to show that the internal-spin system does have a phase transition, and that at low enough temperature there exist distinct Gibbs measures $\mu_+$ and $\mu_-$ with strictly positive and strictly negative magnetization, respectively; these phases can be selected by using, for example, "+" or "−" boundary conditions. These claims follow easily from a Peierls argument (for a description of such arguments, see e.g. [169]). They can alternatively be proven by observing that the diluted system is a collection of $(d-1)$-dimensional diluted and undiluted Ising models, ferromagnetically coupled. In particular, the $d$-dimensional diluted system is more ferromagnetic than the $(d-1)$-dimensional undiluted Ising model, and hence [169] exhibits spontaneous magnetization for all temperatures below the critical temperature $J_{c,d-1}$ of the $(d-1)$-dimensional undiluted Ising model.

*Step 2. Study of a neighborhood of $\omega'_{\text{special}} = \omega'_{alt}$.* Next we must find image-spin configurations $\omega'_+$ and $\omega'_-$ that will "select" the phases $\mu_+$ and $\mu_-$ of the internal-spin system. The choice is obvious: as in the two-dimensional case, we take $\omega'_+$ (resp. $\omega'_-$) to be the configuration with all spins + (resp. all spins −). We need then to show that if the image spins in $\Lambda_R^{image}$ are fixed in a fully alternating configuration, and those in an annulus $\Lambda_{R'}^{image} \setminus \Lambda_R^{image}$ are set to all + (or all −), then for $R'$ large enough (depending on $R$) the image spins in the annulus are capable of determining the internal-spin phase.

In two dimensions we were able to take $R' = R + 2$. That is, we were able to shield off a volume by fixing around it a *single* layer of image spins: namely, by setting the *image spins only* in layer $\Gamma_{R+2}$ to be +, we were able to guarantee that the effective magnetic fields felt by the internal spins in $\Lambda_{R+2}^{int}$ are all nonnegative, even if all the spins (both image and internal) outside $\Lambda_{R+2}$ are set to be − [see Figures 4(a)–(b)]. This situation does not, however, persist in higher dimensions: a layer $\Gamma_{R+2}^{image}$ of + image spins does *not* protect all of the internal spins in $\Gamma_{R+2}^{int}$ from the possible − spins in layer $\Gamma_{R+3}$ [see Figure 6]. Therefore, we have to resort to a more general argument to show that there exists a shielding layer, though thicker. Consider, therefore, the system of internal spins in volume $\Lambda_{R'}^{int}$, with the image spins in $\Lambda_R^{image}$ fixed in the alternating ($\pm$) configuration, the image spins in the annulus $\Lambda_{R'}^{image} \setminus \Lambda_R^{image}$ fixed to be all $+$, and the spins outside $\Lambda_{R'}$ (both image and internal) fixed in an arbitrary configuration $\overline{\sigma}$. We denote expectations in this system by $\langle \cdot \rangle_{\pm,+,\overline{\sigma}}^{R,R'}$. We want to show that, for $J$ sufficiently large, there exists $c > 0$ such that for all $R > 0$ there exists $R' > R$ (depending on $R$) such that

$$\langle \sigma_i \rangle_{\pm,+,\overline{\sigma}}^{R,R'} \geq \langle \sigma_i \rangle_{\pm,+,-}^{R,R'} \geq c > 0 \qquad (4.26)$$

and by symmetry

$$\langle \sigma_i \rangle_{\pm,-,\overline{\sigma}}^{R,R'} \leq \langle \sigma_i \rangle_{\pm,-,+}^{R,R'} \leq -c < 0 \;, \qquad (4.27)$$

for every configuration $\overline{\sigma}$ outside $\Lambda_{R'}$ and every $i \in \Lambda_R^{int}$. This will be proven using correlation inequalities together with the uniqueness of the Gibbs measure for the internal-spin system with image spins set to all $+$ or all $-$.

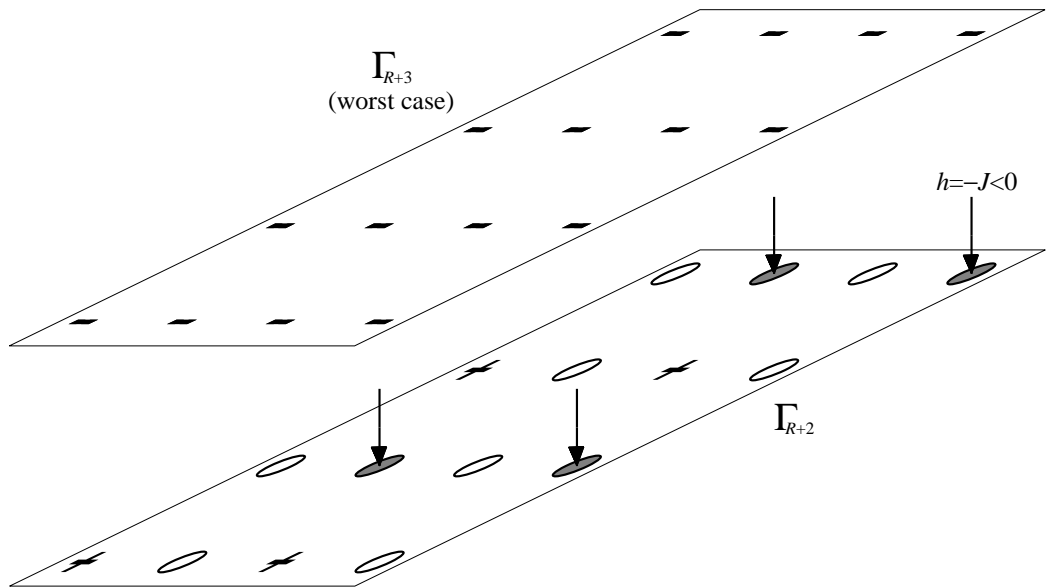

Figure 6: Why a single layer does not work in $d \geq 3$. For the "worst" configuration of the next external layer (image and internal spins all "$-$"), some of the internal spins in layer $\Gamma_{R+2}^{int}$ pick up a negative magnetic field.

More precisely, the proof of (4.26) will involve a sequence of inequalities comparing the following systems of internal spins:

- The system of internal spins in volume $\Lambda_{R'}^{int}$ described above, which we denote by
$$\begin{pmatrix} R, & R' \\ \pm, & +, & \overline{\sigma} \end{pmatrix}.$$

- The *infinite-volume* system of internal spins $\Lambda_\infty^{int} \equiv (\mathbb{Z}^d)^{int}$, with the image spins in $\Lambda_R^{image}$ fixed in the alternating ($\pm$) configuration and the image spins outside $\Lambda_R$ fixed to be all $+$. We denote this system by $\begin{pmatrix} R, & \infty \\ \pm, & + \end{pmatrix}$.

- The *infinite-volume* system of internal spins $\Lambda_\infty^{int} \equiv (\mathbb{Z}^d)^{int}$, with the image spins everywhere fixed in the alternating ($\pm$) configuration. We denote this system by
$$\begin{pmatrix} \infty \\ \pm \end{pmatrix} = \begin{pmatrix} R, & \infty \\ \pm, & \pm \end{pmatrix}.$$

We shall prove the following:

Step 2.1) $\mu_{\pm,+,\overline{\sigma}}^{R,R'}$ converges as $R' \to \infty$ to a Gibbs measure for the system $\begin{pmatrix} R, & \infty \\ \pm, & + \end{pmatrix}$.

Step 2.2) The system $\begin{pmatrix} R, & \infty \\ \pm, & + \end{pmatrix}$ has a *unique* Gibbs measure, call it $\mu_{\pm,+}^{R,\infty}$.

Step 2.3) The measure $\mu_{\pm,+}^{R,\infty}$ is larger (in FKG sense) than *all* Gibbs measures for the system $\begin{pmatrix} \infty \\ \pm \end{pmatrix}$.

Step 2.4) Let $\mu_\pm^{\infty(+)}$ be the $+$ phase (i.e. the maximal Gibbs measure in FKG sense) for the system $\begin{pmatrix} \infty \\ \pm \end{pmatrix}$. Then $\mu_\pm^{\infty(+)}(\sigma_i) \geq c > 0$.

From these results we will then deduce (4.26).

*Step 2.1. The limit $R' \to \infty$.* We wish to consider the limit as $R' \to \infty$ of the measures $\mu_{\pm,+,\overline{\sigma}}^{R,R'}$ (for fixed $R$). By compactness, this sequence of measures has at least one limit point in the weak topology (in fact, any subsequence has a limit point). We claim that any limit point of the measures $\mu_{\pm,+,\overline{\sigma}}^{R,R'}$ (with arbitrary $\overline{\sigma}$) is necessarily a Gibbs measure for the system $\begin{pmatrix} R, & \infty \\ \pm, & + \end{pmatrix}$. The proof is trivial: for any volume $\Lambda \subset \Lambda_{R'-1}$, the DLR equations for the systems $\begin{pmatrix} R, & R' \\ \pm, & +, & \overline{\sigma} \end{pmatrix}$ and $\begin{pmatrix} R, & \infty \\ \pm, & + \end{pmatrix}$ are identical (i.e. the $\pi_\Lambda$'s are the same); so for large enough $R'$, the measure $\mu_{\pm,+,\overline{\sigma}}^{R,R'}$ satisfies the DLR equation in volume $\Lambda$ also for the system $\begin{pmatrix} R, & \infty \\ \pm, & + \end{pmatrix}$. Since the latter system's specification is Feller, the DLR equations are preserved under a weak limit.

Note that we have not yet proven that the limit as $R' \to \infty$ exists; different convergent subsequences might *a priori* have different limits. But in the next step we will prove that the Gibbs measure for the system $\begin{pmatrix} R, & \infty \\ \pm, & + \end{pmatrix}$ is *unique*, so in fact the limit *does* exist.

*Step 2.2. Unique Gibbs measure for the system $\begin{pmatrix} R, & \infty \\ \pm, & + \end{pmatrix}$.* Consider the *infinite-volume* system of internal spins $(\mathbb{Z}^d)^{int}$, with the image spins in $\Lambda_R^{image}$ fixed in the alternating ($\pm$) configuration, and the image spins outside $\Lambda_R$ fixed to be all $+$. We claim that this system has a *unique* Gibbs measure. (We only need uniqueness at low enough temperature, but in fact the Gibbs measure is unique at all temperatures.) This uniqueness is intuitively obvious: the effective magnetic fields induced by the $+$ image spins outside $\Lambda_R$ are sufficient to push the system into the $+$ phase. Unfortunately, the proof we have to offer is a bit too complicated for our taste. It goes as follows. First, we notice that it is enough to prove uniqueness of the Gibbs measure when *all* the image spins (including those inside $\Lambda_R$) are set in the "+" position. Indeed, changing the image spins inside $\Lambda_R$ amounts to a finite-volume perturbation of the system and hence it does not alter the number of Gibbs measures [157, section 7.4]. [In fact, every Gibbs measure $\mu'$ for the perturbed interaction comes from a uniquely defined Gibbs measure $\mu$ of the unperturbed interaction: if $W$ is the perturbation, then $\mu'(\,\cdot\,) = \mu(\,\cdot\, e^{-W})/\mu(e^{-W})$.]

To prove the uniqueness of the Gibbs measure for the system with all image spins "+", we provide two arguments. First argument, proving uniqueness only at low temperature: Pirogov-Sinai theory [328, 260, 322] implies that the phase diagram at low enough temperature is a small deformation of that at zero temperature, but in this case there is only one ground state (namely, all spins "+"). Second argument, proving uniqueness at all temperatures: The internal-spin system is an Ising model on a periodic lattice, with nearest-neighbor coupling $J > 0$ and a periodic magnetic field $h_x = h\delta_x^{\text{n.i.}}$ (here $\delta_x^{\text{n.i.}} = 1$ if $x$ neighbors an image spin, and 0 otherwise), specialized to $h = +J$. By the Lee-Yang theorem ([160, Section 4.5] or [250] and references cited therein) and a result of Lebowitz and Penrose [242] (see also [169, Theorem 4.4]), it can be shown that the pressure of such an Ising model is a jointly analytic function of $J$ and $h$ on the domain $J, h > 0$. It follows that all periodic Gibbs measures give the same mean value to the observables conjugate to $J$ and $h$: these observables are, respectively, $\sum \sigma_i \sigma_j$ where the sum runs over all nearest-neighbor pairs $\langle ij \rangle$ in a unit cell of the periodic lattice, and $\sum \sigma_k$ where the sum runs over all sites $k$ in this unit cell that are nearest neighbor to an image spin. By Griffiths' comparison inequality, it follows that

$$\mu_-(\sigma_i \sigma_j) = \mu_+(\sigma_i \sigma_j) \qquad (4.28a)$$
$$\mu_-(\sigma_k) = \mu_+(\sigma_k) \qquad (4.28b)$$

for every pair $\langle ij \rangle$ of nearest neighbors and for every site $k$ neighboring an image spin; here $\mu_+$ and $\mu_-$ are the measures corresponding to "+" and "−" boundary conditions, respectively. We then resort to the inequality [236]

$$\mu_+(\sigma^A) - \mu_-(\sigma^A) \geq \left| \mu_+(\sigma^B) \mu_-(\sigma^A \sigma^B) - \mu_-(\sigma^B) \mu_+(\sigma^A \sigma^B) \right| \qquad (4.29)$$

valid for any sets $A, B \subset \mathbb{Z}^d$ (we denote $\sigma^A = \prod_{i \in A} \sigma_i$). From (4.28) and (4.29) we conclude that

$$\mu_-(\sigma^A) = \mu_+(\sigma^A) \qquad (4.30)$$

whenever $\sigma^A$ is a product of functions of the form $\sigma_i \sigma_j$ with $i, j$ nearest neighbors and $\sigma_k$ with $k$ being a neighbor to a image-spin site. (In other words, $A$ must be the symmetric difference of a family of such sets $\{i, j\}$ and/or $\{k\}$.) But it is not hard to see that *all* sets $A \subset (\mathbb{Z}^d)^{int}$ are of this form, hence

$$\mu_- = \mu_+ . \qquad (4.31)$$

Now by the FKG inequality $\mu_- \leq \rho \leq \mu_+$ in FKG sense[51] for *every* Gibbs measure $\rho$, hence there is a unique Gibbs measure at all temperatures. (This argument is essentially due to Lebowitz [236, 237], with minor alterations to accommodate periodic systems.)

*Step 2.3. Comparison to the* $\begin{pmatrix} \infty \\ \pm \end{pmatrix}$ *system.* We claim that the measure $\mu_{\pm,+}^{R,\infty}$ is larger (in FKG sense) than *all* Gibbs measures for the system $\begin{pmatrix} \infty \\ \pm \end{pmatrix}$. This is an immediate consequence of the FKG inequality combined with the uniqueness proved in Step 2.2. Indeed, by the FKG inequality, the finite-volume Gibbs measure for the $\begin{pmatrix} R, & \infty \\ \pm, & + \end{pmatrix}$ system with any (internal-spin) boundary condition is larger in FKG sense than the finite-volume Gibbs measure for the $\begin{pmatrix} \infty \\ \pm \end{pmatrix}$ system with the same boundary conditions. This inequality passes directly to the infinite-volume limit.

---

[51] We write $\sigma \leq \sigma'$ in case $\sigma_i \leq \sigma'_i$ for all sites $i$. An observable $f$ is said to be *increasing* if $f(\sigma) \leq f(\sigma')$ whenever $\sigma \leq \sigma'$. We say that $\mu \leq \nu$ in FKG sense in case $\mu(f) \leq \nu(f)$ for all increasing local observables $f$.

*Step 2.4. Spontaneous magnetization for the + phase of the* $\begin{pmatrix} \infty \\ \pm \end{pmatrix}$ *system.* The $\begin{pmatrix} \infty \\ \pm \end{pmatrix}$ system is precisely the Ising model on a periodically diluted lattice. As discussed in Step 1, this model has spontaneous magnetization for $J$ sufficiently large.

*Step 3. Unfixing of the spin at the origin.* Finally, we can "unfix" the spin at the origin in the same way as in the 2-dimensional example.

*Conclusion of the argument.* We conclude that in every neighborhood of $\omega'_{alt}$ there are open sets $\mathcal{N}_+, \mathcal{N}_-$ such that

$$E_{\mu T}\left(\sigma'_0 \mid \{\sigma'_i\}_{i \neq 0}\right)(\omega_1) - E_{\mu T}\left(\sigma'_0 \mid \{\sigma'_i\}_{i \neq 0}\right)(\omega_2) \geq \delta > 0 \qquad (4.32)$$

for $\omega_1 \in \mathcal{N}_+$ and $\omega_2 \in \mathcal{N}_-$. As in the 2-dimensional case, this implies the non-quasilocality of the renormalized measure $\mu T$, for any original Gibbs measure $\mu$. This works for any temperature below the critical temperature of the undiluted $(d-1)$-dimensional Ising model. We have therefore proven:

**Theorem 4.2** *Let $d \geq 2$. Then for all $J > J_{c,d-1}$, the following holds: Let $\mu$ be any Gibbs measure for the $d$-dimensional Ising model with nearest-neighbor coupling $J$ and zero magnetic field. Let $T$ be the decimation transformation with spacing $b = 2$. Then the measure $\mu T$ is not consistent with any quasilocal specification. In particular, it is not the Gibbs measure for any uniformly convergent interaction.*

### 4.3.2 Decimation with Spacing $b \geq 3$

The conclusions of Theorem 4.2 for decimation with spacing $b = 2$ hold also for larger spacings. The main difference from the $b = 2$ case is that for $b \geq 3$ the system of internal spins obtained with $\omega' = \omega'_{alt}$ is no longer simply a periodically diluted Ising model in zero magnetic field; rather, it contains a periodic alternating magnetic field which is nonzero at the sites neighboring an image spin. As a consequence, we need a more sophisticated technique to conclude that there is indeed a phase transition (Step 1). The appropriate tool for this purpose is Pirogov-Sinai theory [328, 329], which is summarized in Appendix B. The upshot of P-S theory is that the phase diagram of a lattice system at *low* temperature can in some cases be deduced from the phase diagram at *zero* temperature. More precisely, if there are a finite number of periodic ground states, and these ground states satisfy a suitable "Peierls condition", then the phase diagram of periodic Gibbs measures at low temperature is a small perturbation of the phase diagram of ground states. In the case at hand, one can show that for the fully alternating block-spin configuration, the system of internal spins has only two periodic ground states — namely, the one with all internal spins +, and the one with all internal spins − − and that these ground states satisfy the Peierls condition. It follows from P-S theory that at low temperature there are precisely two periodic Gibbs measures, $\mu_+$ and $\mu_-$, characterized respectively by a strictly positive or strictly negative magnetization. The details of this part of the argument are presented in the Appendix B (Section B.5.3). Steps 2 and 3 are then proven in a manner exactly identical to the $b = 2$ case. The analysis of Section B.5.3 yields a (very weak) estimate of the range of temperatures for which the pathologies are present [formula (B.79)].

We have thus proven the following:

**Theorem 4.3** *Let $d \geq 2$ and $b \geq 2$. Then for all $J$ sufficiently large (depending on $d$ and $b$), the following holds: Let $\mu$ be any Gibbs measure for the $d$-dimensional Ising model with nearest-neighbor coupling $J$ and zero magnetic field. Let $T$ be the decimation transformation with spacing $b$. Then the measure $\mu T$ is not consistent with any quasilocal specification. In particular, it is not the Gibbs measure for any uniformly convergent interaction.*

**Remark.** Checkerboard decimation, as shown in Figure 2(b), is a very different situation: the internal spins are not connected, and hence they cannot cooperate to have a phase transition. In fact, in this case the first iteration of the transformation is well-defined [364] [346, p. 193]. However, the *second* iteration of this transformation corresponds to a single iteration of the $b = 2$ decimation transformation, and so is ill-defined at low enough temperature.

### 4.3.3 Kadanoff Transformation with $p$ Finite

In some sense the results thus far should not be surprising: the decimation transformation, unlike other RG transformations, does not in any sense integrate out the "high-momentum modes" and leave the "low-momentum modes"; it merely integrates out one sublattice and leaves another. In particular, if the sublattice of internal (integrated-out) spins is *connected*, it is hardly surprising that the internal-spin system can exhibit a phase transition, and that this can give rise to RG pathologies.

In this section we show something considerably more surprising: that the same pathology — non-Gibbsianness after one renormalization step — is present at low temperature for the Kadanoff transformation with any finite (but nonzero) $p$.[52] This result is in clear conflict with the RG ideology, which states that integration over high-momentum modes cannot produce singularities. (Indeed, our proof makes no distinction between block sizes $b \geq 2$ and $b = 1$ — and for $b = 1$ one is not integrating over *any* "modes", high-momentum or otherwise!) In the next subsection we shall prove a similar result for some majority-rule transformations (i.e. Kadanoff with $p = \infty$).

---

[52]In earlier versions of this work [352, 353], we claimed this result only for *small* $p$. Subsequently we found a proof valid for *all $0 < p < \infty$*, which we present here.

Consider the Kadanoff transformation (3.10) with block size $b$ and parameter $p$. From (3.10) one readily concludes [55] that for each choice of block spins $\sigma'$ the conditional probabilities of the internal spins $\sigma$ correspond to a Hamiltonian

$$H_{\text{eff}}(\sigma) \;=\; -J \sum_{\langle ij \rangle} \sigma_i \sigma_j - p \sum_x \sigma'_x \sum_{i \in B_x} \sigma_i + \sum_x \log 2 \cosh\left( p \sum_{i \in B_x} \sigma_i \right). \qquad (4.33)$$

This is the original Ising-model Hamiltonian perturbed by a block-dependent magnetic field and an antiferromagnetic multi-spin coupling. To obtain non-trivial results we consider blocks at least of size 2 in each coordinate direction. It is natural to expect that, for any fixed $p < \infty$, for sufficiently large $J$ (i.e. low enough temperature) the perturbation become effectively small, and the phase diagram a small deformation to that of the original Ising model.

We notice, however, that there is a small difference with the original perturbative setting in that *the last two terms in (4.33) do not include a temperature factor*. In the study of deformations of phase diagrams, one considers a fixed value of $\beta$ multiplying *all* the terms of the Hamiltonian, and analyzes the consequences of changing (perturbing) some of the remaining parameters. The proof that the deformations are smooth usually requires that the size of this perturbation not exceed a certain $\beta$-dependent bound. In our case, after pulling out a common factor $\beta$, the parameters of the perturbation acquire a $\beta$-dependence and one is confronted with the problem of verifying that this $\beta$-dependent size is smaller than the $\beta$-dependent bound. This problem is especially serious in the case of the last term in (4.33), which does not have any small parameter preceding it, so that its size decreases only as $1/\beta$. We conclude that to successfully complete Step 1 we need a slight strengthening of the usual PS theory, involving *families* of interactions, and showing that the deformations of the phase diagram are small *uniformly* in members of this family. Such a strengthening is discussed in Appendix B (Corollaries B.25 and B.29).

For Step 1, then, we choose a configuration $\omega'_{\text{special}}$ for the block spins so that the middle term in the RHS of (4.33) does not favor any overall internal spin orientation — for example, a fully alternating configuration. The "uniform" version of PS theory implies (Appendix B.5.4) that at low enough temperature there are two coexisting phases $\mu_+$ and $\mu_-$. This is the end of Step 1 of the Griffiths-Pearce-Israel argument. Steps 2 and 3 are then completed almost identically to the previous examples.

In this way we conclude:

**Theorem 4.4** *Let $d \geq 2$, $b \geq 1$ and $0 < p < \infty$. Then there exists a $J_0$ (depending on $d$, $b$ and $p$) such that for all $J > J_0$ the following holds: Let $\mu$ be any Gibbs measure for the $d$-dimensional Ising model with nearest-neighbor coupling $J$ and zero magnetic field. Let $T$ be the Kadanoff transformation with parameter $p$ and block size $b$. Then the renormalized measure $\mu T$ is not consistent with any quasilocal specification. In particular, it is not the Gibbs measure for any uniformly convergent interaction.*

A (poor) estimate of the smallness of the temperature is given in formula (B.87). We emphasize that our estimate $J_0$ is *nonuniform* in $p$. As a result, we are *not* able to take $p \to \infty$ at any fixed $J$, and thereby treat the majority-rule map. (Our partial results on the majority-rule map, obtained by a *different* method, will be described in the next subsection.)

**Remarks.** 1. The occurrence of "peculiarities" in the Kadanoff transformation at small $p$ and low temperature was suggested already by Griffiths and Pearce [172, 173].

2. Interesting applications of the Kadanoff transformation (with block size $b = 1$!) arise in image processing [152, 154, 63, 158, 129], speech recognition [302] and other fields of applied probability theory. The basic theoretical construct in these fields is a class of models termed *hidden Markov models* [302, 153, 248]; in our language these are simply the images of Markovian (i.e. nearest-neighbor) spin models under local renormalization transformations. It has been long recognized that such measures can be very far from Markovian; here we have shown that they can even be non-Gibbsian.

Consider, for example, an Ising-model Gibbs measure corrupted by white noise: with probability $\epsilon$ a spin is observed incorrectly, independently at each site. This is model I of Griffiths and Pearce [172, 173], and is equivalent to the Kadanoff transformation with $p = \tanh^{-1}(1 - 2\epsilon)$ on blocks of size $b = 1$. For any $\epsilon > 0$, we have proven that the image measure is non-Gibbsian for $(J, h)$ in an ($\epsilon$-dependent) open neighborhood of the low-temperature zero-field region. This system is of interest in applications to image processing [152, 129].

### 4.3.4 Majority-Rule Transformation

Next we wish to show that Griffiths-Pearce-Israel pathologies occur also for the majority-rule transformation (i.e. the Kadanoff transformation with $p = \infty$). For simplicity let us consider the case of an *odd* block size $b$, so as to avoid the complications caused by ties. In view of the foregoing examples, it is natural to try a fully alternating block-spin configuration $\omega'_{alt}$. Using Pirogov-Sinai theory, one might hope to prove that at low temperature the internal-spin system has precisely two extremal periodic Gibbs measures: a "+" phase in which the internal spins show an overwhelming majority of + spins in blocks where the block spin is + but only a weak majority of − spins where the block spin is −, and a "−" phase with the reverse behavior. This result would in fact follow if one could show that there are precisely two periodic *ground* states: a "+"-like state in which the internal spins are unanimously + in blocks where the block spin is + and show a bare − majority where the block spin is −, and a "−"-like state with the reverse behavior [Figure 7(a)]. Unfortunately, neither the shape nor the position within a block of these "minimal islands" of minority spins is in general uniquely determined [Figure 7(b)]; therefore, this family of states is infinitely degenerate, and we cannot apply P-S theory (at least in its usual form). Moreover, it turns out that these configurations are not even ground states: there are "strip-like" states of lower energy density [Figure 7(c)]. We believe that these strip-like states are truly ground states (though we have not proven it); and since they too are infinitely degenerate, P-S theory cannot be applied.

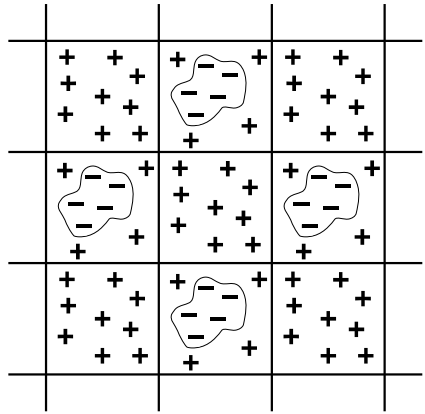

(a)

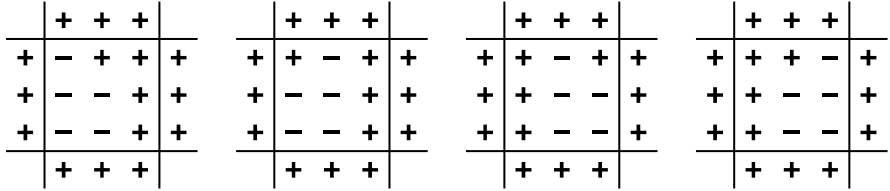

(and rotated)

(b)

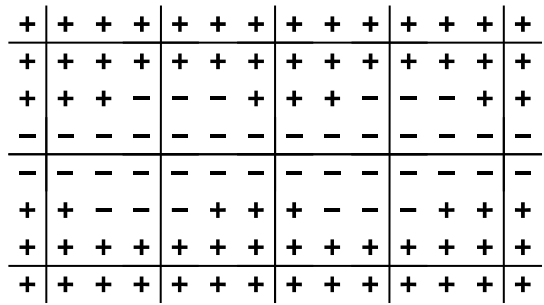

(c)

Figure 7: (a) The hoped-for structure of the "+"-like ground state. (b) Indeterminacy of the shape and position of the minimal islands of − spins, for the case of a $3 \times 3$ block. The energy per island is $20J$, irrespective of its shape. The energy density is $10J$ per block. (c) Strip-like states with an energy density of $8J$ per block. These states also have an indeterminacy in each block.

We suspect that for each odd $b \geq 3$ there do exist block-spin configurations (more complicated than $\omega'_{alt}$) for which the Griffiths-Pearce-Israel argument can be carried through, but for $b = 3, 5$ we have been unable to find any. The simplest case in which we managed to avoid these problems is $b = 7$. Here a bare majority in a $7 \times 7$ block consists of 25 spins, and the unique minimal-energy configuration for an island of 25 or more spins is a $5 \times 5$ square. By taking a *doubly-alternating* block-spin configuration [Figure 8(a)], we can force these $5 \times 5$ squares to be positioned in a unique minimal-energy way [Figure 8(b)]. The energy of this arrangement is $80J$ per group of eight blocks, or $10J$ per block. On the other hand, strip-like states would cost at least $14J$ per block. Therefore, with this block-spin configuration the internal-spin system has precisely two ground states: the "+"-like state depicted in Figure 8(b), and the reverse "−"-like state. It then follows from P-S theory that at low enough temperature the internal-spin system has two extremal periodic Gibbs measures, $\mu_+$ and $\mu_-$, characterized by a nonzero position-dependent magnetization of opposite signs. The ingredients of the rigorous argument showing that indeed the "+"- and "−"-like configurations of the type of Figure 8(b) are the only ground states, and that PS theory is applicable, are summarized in Section B.5.5. This completes Step 1, which is the hard part of the proof.

The proof of Step 2 relies again on P-S theory and the FKG inequality. Step 2.1 is proven in the usual way. The fact that the system with + block magnetization outside a square $\Lambda_R$ has a unique Gibbs measure is a consequence of P-S theory: at zero temperature this system has a unique ground state, namely the state with all spins +, and P-S implies (see Section B.5.2) that this trivial phase diagram persists at low enough temperature. This proves Step 2.2. Finally, we claim that the system with + block spins outside a square $\Lambda_R$ (and doubly alternating block spins inside) has a larger magnetization than the system with doubly alternating block spins everywhere. Indeed, the constraint that the majority of internal spins in a block $B$ be − (resp. +) can be imposed by including in the Hamiltonian a term $-h_B \operatorname{sgn}\left(\sum_{i \in B} \sigma_i\right)$ with $h_B \to -\infty$ (resp. $h_B \to +\infty$). Since $\operatorname{sgn}\left(\sum_{i \in B} \sigma_i\right)$ is an increasing function of the spins, the FKG inequality implies that the magnetization at any site is an increasing function of $h_B$. This proves Step 2.3.

Step 3 is proven in the usual way. We therefore conclude:

**Theorem 4.5** *For all $J$ sufficiently large, the following holds: Let $\mu$ be any Gibbs measure for the two-dimensional Ising model with nearest-neighbor coupling $J$ and zero magnetic field. Let $T$ be the majority-rule transformation on $7 \times 7$ square blocks. Then the measure $\mu T$ is not consistent with any quasilocal specification. In particular, it is not the Gibbs measure for any uniformly convergent interaction.*

It is of course unnatural and unpleasant for this result to be restricted to the special case of $7 \times 7$ blocks. This restriction was necessary only in Step 1 (the proof of a phase transition for some fixed block-spin configuration); it arose from the necessity to obtain a finite number of periodic ground states in order to apply P-S theory. All the other steps in the proof remain valid for blocks of arbitrary size and for Ising models in arbitrary dimension. Digging a little deeper we see that the "rigidity" in the shape and position of the islands of minority spins, and hence the boundedness of the number of periodic ground states, is a consequence of the following numerological "miracle": the block size $b = 7$ and the island size $c = 5$ satisfy the Diophantine equation

$$1 + b^2 = 2c^2 \:. \tag{4.34}$$

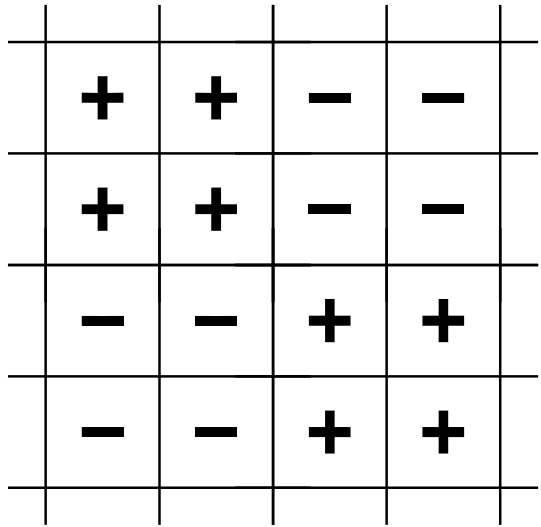

(a)

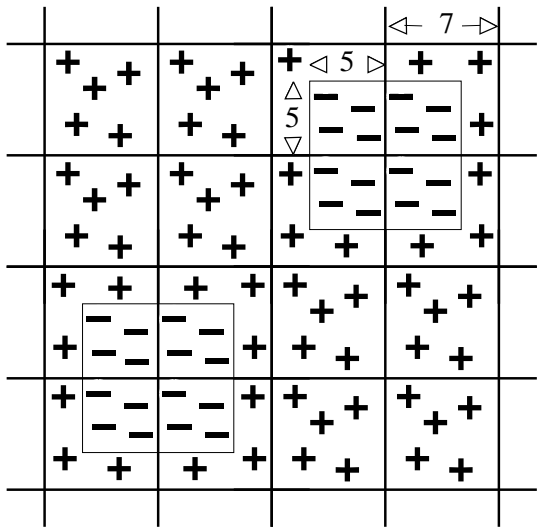

(b)

Figure 8: Majority rule for $7 \times 7$ blocks. (a) The doubly-alternating block-spin configuration. (b) The unique minimal-energy arrangement of islands of $-$ spins in a $+$ sea. The energy is $80J$ per 8 blocks, or $10J$ per block.

The proof extends automatically to any block size $b$ for which $c$ defined by (4.34) is an integer. In Appendix C we find the general solution to this Diophantine equation: the admissible block sizes turn out to be

$$\widetilde{b}_k = \frac{1}{2}\left[(1 + \sqrt{2})^{2k+1} + (1 - \sqrt{2})^{2k+1}\right] \tag{4.35}$$

for $k = 1, 2, 3, \ldots$. The first few $\widetilde{b}_k$ are 7, 41, 239, 1393, 8119, .... For other block sizes, a proof of non-Gibbsianness using our methods would require either a more clever choice of block-spin configuration $\omega'_{\text{special}}$, or else a more sophisticated version of P-S theory capable of dealing with infinitely many periodic ground states [46, 179, 22, 23, 48]. Irrespective of these technical details, it seems plausible to expect that the *conclusion* of Theorem 4.5 remains valid for *all* block sizes $b$.

**Remark.** Griffiths and Pearce [172, 173] and later Hasenfratz and Hasenfratz [189, Section 4] have presented a rather different class of cases in which the majority-rule transformation is expected to have "peculiarities": in these examples the block-spin configuration $\omega'_{\text{special}}$ is taken to be purely +, and the magnetic field is taken to be negative (with an order-1 strength chosen to exactly compensate the effect of the block spins). Our scheme of proof does not apply in these examples, for two reasons: Firstly, there are infinitely many periodic ground states, so P-S theory in its usual form does not apply. Secondly (and perhaps more seriously), in the configuration $\omega'_{\text{special}}$ all of the block spins are *already* +, and by construction the corresponding internal-spin system does *not* have a unique Gibbs measure; so it is clearly impossible to "select" the + phase (i.e. make it unique) by setting the block spins in an annulus to be +. This latter fact was already noted by Israel [207, p. 597].

### 4.3.5 Block-Averaging Transformations

In contrast to our previous example, in this case our proof works for *even* block sizes (and only these) precisely *because* of the possibility of ties. We discuss here the simplest case, namely the $2 \times 2$ block-averaging transformation for the two-dimensional nearest-neighbor Ising model at low temperatures. We divide $\mathbb{Z}^2$ into $2 \times 2$ blocks $B_j$, and define

$$\sigma'_j = \sum_{i \in B_j} \sigma_i \:. \tag{4.36}$$

We notice that although the original variables $\sigma$ take two values ($\pm 1$), the renormalized spins $\sigma'$ take five values $(0, \pm 2, \pm 4)$. Usually the average spins are rescaled, but such a rescaling is irrelevant for our discussion because we only consider a *single* application of the transformation and do not iterate.

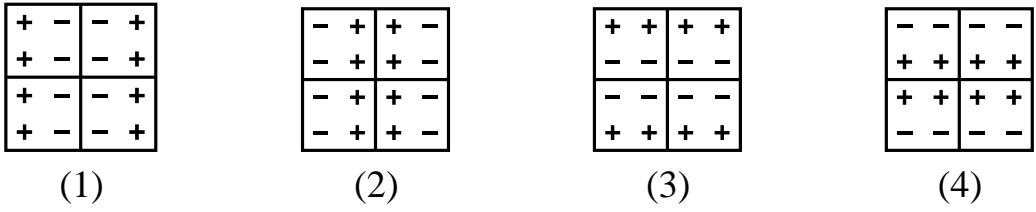

Figure 9: The four periodic ground states of the internal-spin system obtained by constraining the block spins to be zero.

*Step 1.* We choose the configuration $\omega'_{\text{special}}$ defined by $\sigma'_j = 0$ for all $j \in \mathbb{Z}^2$. The resulting system of internal spins has, at low temperatures, four periodic Gibbs measures corresponding to four ground states formed by infinite alternating strips of thickness 2 (see Fig. 9). This follows immediately from Pirogov-Sinai theory (see Appendix B.5.7).

*Step 2.* Let $\Lambda$ be a $4N \times 4N$ square. Take block-spin boundary conditions as follows: $+4$ for the rows of block spins immediately above and below $\Lambda$, $+2$ for the columns immediately to the right and left of $\Lambda$, and $+4$ for the columns just to the right and left of these [see Fig. 10(a)]. A slight modification of the usual Peierls argument proves that these boundary conditions induce at low temperature the Gibbs measure associated to ground-state #3 in Fig. 9 [see Fig. 10(b)].

*Step 3.* We unfix *two* nearest-neighbor block spins: the one at the origin and the one immediately above it. Then, at sufficiently low temperature, one has with high probability the boundary condition of Fig. 11 for the two-block system $(0',0')$–$(0',1')$ (for a suitable positioning of the volume $\Lambda$). Notice that this boundary condition has eight $+$ spins and only four $-$ spins; therefore, it is clear that the spins inside the two blocks are biased towards $+$, so that

$$\mu'_{\Lambda,\omega'_{\text{special}};\partial\Lambda_{4;4,2}}(\sigma'_{(0',0')}) = \mu'_{\Lambda,\omega'_{\text{special}};\partial\Lambda_{4;4,2}}(\sigma'_{(0',1')}) \geq c_+(J) > 0 \qquad (4.37)$$

at zero magnetic field. [Indeed, at low temperature there is a probability $\approx \frac{1}{2}$ of having a strip configuration with $\sigma'_{(0',0')} = \sigma'_{(0',1')} = 0$ and a probability $\approx \frac{1}{2}$ of having an all-$+$ configuration with $\sigma'_{(0',0')} = \sigma'_{(0',1')} = +4$, so that $\lim_{J\to\infty} c_+(J) = +2$.] Similarly, by reversing the sign of the block spins on the boundary, we obtain

$$\mu'_{\Lambda,\omega'_{\text{special}};\partial\Lambda_{4;4,2}}(\sigma'_{(0',0')}) = \mu'_{\Lambda,\omega'_{\text{special}};\partial\Lambda_{4;4,2}}(\sigma'_{(0',1')}) \leq c_-(J) < 0 \qquad (4.38)$$

where of course $c_-(J) = -c_+(J)$ by symmetry. This completes the argument.

**Remark.** Notice that it does not suffice to unfix a single block spin to distinguish among the four Gibbs measures, because in all four measures the boundary condition on the unfixed block would be symmetric between + and − (i.e. four + spins and four − spins), and the expectation of the block spin would be zero.

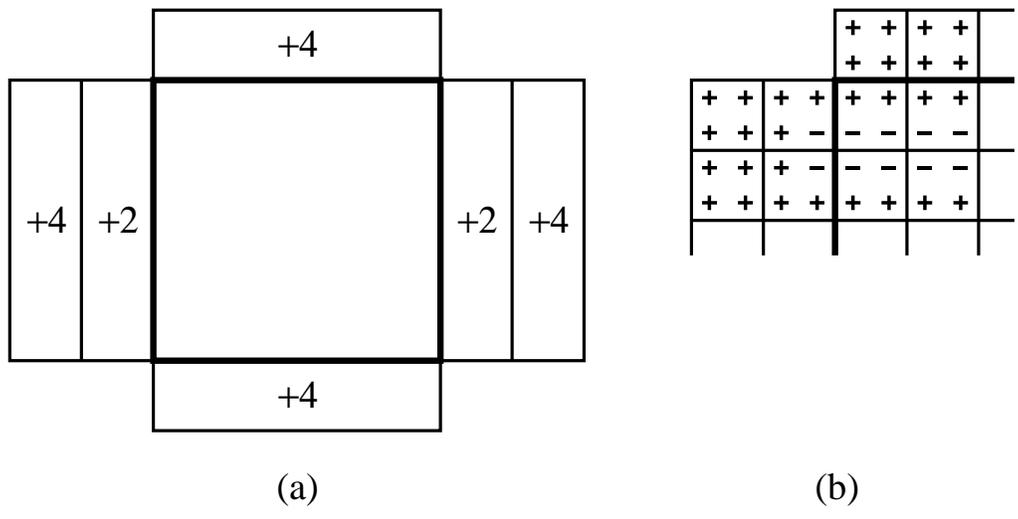

Figure 10: Block-spin boundary conditions chosen for Step 2.

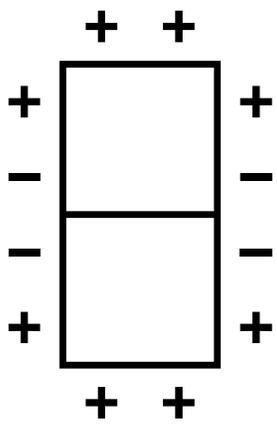

Figure 11: Boundary conditions for the block-spin observable of Step 3.

The argument given here clearly works for any *even* block size $b \geq 2$, in any lattice dimension $d \geq 2$. In this way we conclude:

**Theorem 4.6** *Let $d \geq 2$, and let $b \geq 2$ be even. Then for all $J$ sufficiently large (depending on $d$ and $b$), the following holds: Let $\mu$ be any Gibbs measure for the $d$-dimensional Ising model with nearest-neighbor coupling $J$ and zero magnetic field. Let $T$ be the block-averaging transformation with block size $b$. Then the measure $\mu T$ is not consistent with any quasilocal specification. In particular, it is not the Gibbs measure for any uniformly convergent interaction.*

### 4.3.6 Generalization to Nonzero Magnetic Field

It might appear from our results thus far that the RG pathology is somehow associated with the fact that the original Hamiltonian $H$ lies *on* the phase-coexistence curve (which in the Ising model means zero magnetic field). This is in fact *not* the case. In this section we include a magnetic field $h$, and show that in dimension $d \geq 3$ there is an *open region in the $(J,h)$-plane* — namely, low enough temperature and small enough field — where the decimation transformation produces a non-Gibbsian measure after one iteration. (We suspect that the result is true also for $d = 2$, but it will require a different proof.) Our argument works for decimation with arbitrary scale factor $b$, and for the Kadanoff transformation with any $p < \infty$. Moreover, for block-averaging we have an even stronger result: the renormalized measures are non-Gibbsian *for arbitrary strength of the field* in dimensions $d \geq 2$, at low temperatures. We conclude that the Griffiths-Pearce-Israel pathologies are *not* associated with the fact that the original model is *sitting on* a phase-transition surface. Rather, it is the system of internal spins constrained by the configuration $\omega'_{\text{special}}$ which must have a phase transition. If this transition is of a similar type as that of the original system, then it is natural to expect that the original system must at least be *close* to a phase transition, in some sense. But even this need not be the case, as the example of block-averaging transformations will show, if these two transitions are of different nature.[53]

Let us first treat the case of decimation transformations. Consider, therefore, an interaction
$$H \;=\; -J \sum_{\langle ij \rangle} \sigma_i \sigma_j - h \sum_i \sigma_i \;. \qquad (4.39)$$

(Note that in our normalization, the magnetic field is *not* explicitly multiplied by any factor of $J$ or $\beta$.) The idea is that for suitable (small) values of $\widetilde{h} \equiv h/J$, we can find an image-spin configuration $\omega'_{\text{special}}$ for which the corresponding internal-spin system has a non-unique Gibbs measure. Roughly speaking, $\omega'_{\text{special}}$ must be such that it "compensates" the effect of the magnetic field, so that the system of internal spins subjected both to the homogeneous field $h$ and to the inhomogeneous field due to the image spins has two or more extremal Gibbs measures.

---

[53]For this reason, the title of our earlier report [353] is in retrospect somewhat misleading.

As in all the previous cases, we formalize this idea in two steps: we first show that it works at zero temperature, namely that there are choices of $\widetilde{h}$ and $\omega'_{\text{special}}$ for which the *ground state* is not unique; and second we show that Pirogov-Sinai theory is applicable so that it implies nonuniqueness of Gibbs measures at low but nonzero temperatures. The simplest case is to let $\omega'_{\text{special}}$ be *periodic*.[54] In this situation one can find the "compensating" field $\widetilde{h}_0$ needed to obtain more than one ground state by studying configurations inside a period. We do not want to enter into the details, as we later offer a better and more general procedure, but we make the rather obvious remark that the field must be taken in a direction opposite to that of the majority of internal spins within a period. The value of the field must be such that if the internal spins follow it, the energy gain is exactly compensated by the penalty paid by the internal spins neighboring the image spins of opposite sign. (In other words, the sum of all the fields — external or due to image spins — felt by the internal spins in a period must be zero.) This field strength $\widetilde{h}_0$ is too weak to favor the flipping of small regions of internal spins, and only a collective flip is energetically acceptable. Of course, this delicate balance is broken if the field is changed, no matter how little. Therefore, we conclude that this value $\widetilde{h}_0 \equiv \widetilde{h}_0(\omega'_{\text{special}})$ is such that for $\widetilde{h} > \widetilde{h}_0$ (resp. $\widetilde{h} < \widetilde{h}_0$) the internal-spin system has only one periodic ground state, namely all spins $+$ (resp. all spins $-$), while for $\widetilde{h} = \widetilde{h}_0$ there are precisely two periodic ground states, namely all spins $+$ and all spins $-$.

For the second step — relying on Pirogov-Sinai theory — we already have two of the required conditions (see Appendix B): a periodic (internal-spin) interaction and a finite number of ground states. We need in addition to verify the Peierls condition, but this can be done basically following the same energy-cost arguments outlined above for the determination of ground states. The conclusion is that there exists $J_0 \equiv J_0(\omega'_{\text{special}})$ such that for $J > J_0$ there is a continuous curve $\widetilde{h} = \widetilde{h}^*(J)$, with $\lim_{J \to +\infty} \widetilde{h}^*(J) = \widetilde{h}_0$, on which the internal-spin system has precisely two periodic extremal Gibbs measures, namely a "+" phase and a "−" phase. As long as $\omega'_{\text{special}}$ is not all + or all −, Step 2 can be proven using the FKG inequality, as in Section 4.3.1. We therefore conclude that for $\widetilde{h} = \widetilde{h}^*(J)$ the renormalized measure $\mu T$ is non-Gibbsian.

The chief limitation of this procedure is that it produces only *rational* values of $\widetilde{h}_0$ and that there is no uniformity in $\widetilde{h}_0$ for the range of temperatures for which the nonuniqueness persists. Hence, by letting $\omega'_{\text{special}}$ range over *all* periodic configurations, we prove non-Gibbsianness only for a region of the phase diagram formed by countably many curves $\widetilde{h} = \widetilde{h}^*(J)$. We can prove that the set of $\widetilde{h}_0$ values is dense in some interval $|\widetilde{h}| < \epsilon$ but, unfortunately, we *cannot* conclude non-Gibbsianness for any dense subset of an open set in the $(J, h)$-plane: the trouble is that the curves $\widetilde{h}^*(J)$ corresponding to configurations $\omega'_{\text{special}}$ of very high period may survive only to very low temperatures (i.e. we have no uniform control on $J_0$).

---

[54]This construction was already suggested by Israel [207].

If we want to extend this argument to more general choices of $\omega'_{\text{special}}$, we are confronted with the limitation imposed by the present versions of Pirogov-Sinai theory. One possible generalization of this construction is to let $\omega'_{\text{special}}$ be *quasiperiodic*. Then one can use an extension of Pirogov-Sinai theory due to Koukiou, Petritis and Zahradník [221]. (Actually, these authors require the quasiperiodic part of the interaction to be small; so we cannot handle decimation, but can handle the Kadanoff transformation with $p$ small.) In this way we obtain uncountably many curves $\tilde{h} = \tilde{h}^*(J)$ on which the renormalized measure is non-Gibbsian. (If the results of Koukiou *et al.* can be extended to frequencies which are Diophantine of arbitrary type $l < \infty$ — at present they treat only $l = 2$ — then the corresponding set of $\tilde{h}_0$ values would contain some interval $|\tilde{h}| < \epsilon$ except for a subset of Lebesgue measure zero.) However, we still cannot conclude non-Gibbsianness for any dense subset of an open set in the $(J, h)$-plane: the trouble is again that we lack uniform control on $J_0$.

At any rate, we are able to overcome these technicalities, and we present here an argument proving the existence of Griffiths-Pearce-Israel pathologies for an *open region* $\{J > J_0, |h| < \delta_0 J\}$ in the $(J, h)$-plane, as originally conjectured by Griffiths and Pearce [172, 173] and Israel [207]. The key ingredient is a mechanism to generate a continuum of image-spin configurations $\omega'_{\text{special}}$ such that P-S theory is applicable to the resulting internal-spin system. At present this is only possible if we resort to randomness: Zahradník [369, 370, 371], and with less generality Bricmont and Kupiainen [43, 44], extended P-S theory to systems with superimposed (small) random interactions for dimensions $d \geq 3$. Our construction will, therefore, be based on a (slightly) random choice of the configuration $\omega'_{\text{special}}$ and will be limited to $d \geq 3$.

Consider, for starters, decimation with some spacing $b \geq 2$, applied to an Ising model with ferromagnetic nearest-neighbor interaction $J$ and magnetic field $h = J\tilde{h} > 0$. We consider a block-spin configuration which is equal to the fully alternating configuration except that the spins that would correspond to a "+" have a probability $\epsilon/2J$ of becoming a "−", independently for each such spin. We wish to show that for each sufficiently small positive $\tilde{h}$, there exists an $\epsilon$ such that the random magnetic field induced by the block spins (whose net effect is negative) exactly compensates the positive uniform field, in the sense that for almost all such image-spin configurations there are two distinct Gibbs measures $\mu_+$ and $\mu_-$. To do this, we apply an as-yet-unpublished theorem of Zahradník [370, 371], which generalizes Pirogov-Sinai theory to small random interactions, if the lattice dimension is $\geq 3$. (In the preprint [370], the random interactions are assumed to be small uniformly in all realizations of the randomness. This condition is not satisfied in our case, as one has large terms (of strength $\sim J$), albeit occurring with small probability ($\epsilon/2J$). In a private communication [371], Zahradník has informed us that minor modifications of his proofs suffice to cover also this case.)

We apply Zahradník's theory with the original Hamiltonian $H_0$ taken to be the system of internal spins with fully alternating image spins (i.e. a ferromagnetic nearest-neighbor Ising model in a periodically diluted lattice and with a periodic magnetic field of mean zero, see Section 4.3.2); the symmetry-breaking "fields" are taken to be the uniform magnetic field, and the random negative magnetic fields coming from those block spins that were flipped from "+" to "−" according to the procedure explained above. The analysis of the ground-state structure of $H_0$, and the proof of the Peierls condition for it, were already carried out in Sections 4.3.2 and B.5.3. Zahradník's theory (Theorem B.31) then assures us (Section B.5.7) that for each $J$ sufficiently large and each $\epsilon$ sufficiently small, the phase diagram is, with probability 1, a small deformation of that of the Hamiltonian $H_0$; that is, for each such pair $(J, \epsilon)$ there exists a unique $\widetilde{h}^*(J, \epsilon) > 0$ such that the system has two distinct Gibbs measures $\mu_+$ and $\mu_-$ (which can be obtained, for example, by taking $\widetilde{h} \downarrow \widetilde{h}^*$ or $\widetilde{h} \uparrow \widetilde{h}^*$, respectively). Moreover, the value $\widetilde{h}_0(\epsilon)$ at which the "+" and "−" configurations are simultaneously ground states is a strictly increasing linear function of $\epsilon$. (This follows from an argument similar to, albeit more elaborate than, the one presented at the beginning of the section for periodic choices of $\omega'_{\text{special}}$. The slope depends on the block-size $b$ and the dimensionality $d$.) As Zahradník's theory tells us that the low-temperature phase diagram is a smooth deformation of the zero-temperature one, we conclude that that $\widetilde{h}^*$ is a continuous and strictly increasing function of $\epsilon$. Obviously the case $h < 0$ can be handled by the same argument with "+" and "−" reversed.

The bottom line is, therefore, that there exists — for each $J$ sufficiently large — a continuous and monotonic curve $\epsilon^*(\widetilde{h})$ through the origin, defined for $|\widetilde{h}|$ small, such that for almost all choices of the random block-spin configuration the system presents multiple Gibbs measures on the curve and a unique Gibbs measure to each side of the curve (Figure 12). Thus, for the Ising model with $J$ sufficiently large and $|\widetilde{h}|$ sufficiently small, we can prove Step 1 by chosing as $\omega'_{\text{special}}$ any one of the configurations from the probability-1 set corresponding to $\epsilon = \epsilon^*(\widetilde{h})$. The proof of the validity of Steps 2 and 3 is essentially identical to that of the case $h = 0$ (Sections 4.3.1 and 4.3.2). We notice that due to the smoothness of the phase diagram deformations, the bound $|\widetilde{h}| < \delta(J)$ for which these steps, and hence the existence of RG pathologies, can be proven is given by a *continuous* function $\delta(J)$. Moreover, we have $\liminf_{J \to \infty} \delta(J) \geq \delta_0 > 0$.

The final result is the following:

**Theorem 4.7** *For each $d \geq 3$ and $b \geq 2$, there exists a $J_0 < \infty$ and a $\delta_0 > 0$ (depending on $d$ and $b$) such that for all $J > J_0$ and $|h| < \delta_0 J$ the following is true: Let $\mu$ be any Gibbs measure for the $d$-dimensional Ising model with nearest-neighbor coupling $J$ and magnetic field $h$. Then the renormalized measure $\mu T$ arising from the decimation transformation with spacing $b$ is not consistent with any quasilocal specification. In particular, it is not the Gibbs measure for any uniformly convergent interaction.*

Similar results are valid, by a similar argument, for the Kadanoff transformation with any fixed $0 < p < \infty$. However, we are not able to apply such an argument to the majority-rule example because we need dimension $d \geq 3$. This seems to be only a technical reason.

The proof for block-averaging transformations in a field is much simpler: we do not need randomness in the choice of $\omega'_{\text{special}}$. In fact, the same steps detailed in Section 4.3.5 above can be applied *regardless of whether or not a field is present*. Steps 1 and 2 are exactly the same for all values of the field, because the constraint of zero block spins removes all field-dependence inside such blocks. (The point is that the ground-state configurations, as well as the block-spin boundary conditions needed to select them, are the *same* for all values of the magnetic field.) In Step 3, the presence of a field causes an asymmetry between "+" and "−" boundary conditions — i.e. we no longer have $c_-(J) = -c_+(J)$ — as well as a smaller value for the difference $c_+(J) - c_-(J)$ between the two magnetizations. But this difference is still bounded away from zero uniformly in $\Lambda$ (the extra factor involved depends only on the fields at the eight sites of the two-block observable), so the result is still valid. Alternatively, one could "unfix" a strip of $N \times 2$ blocks.

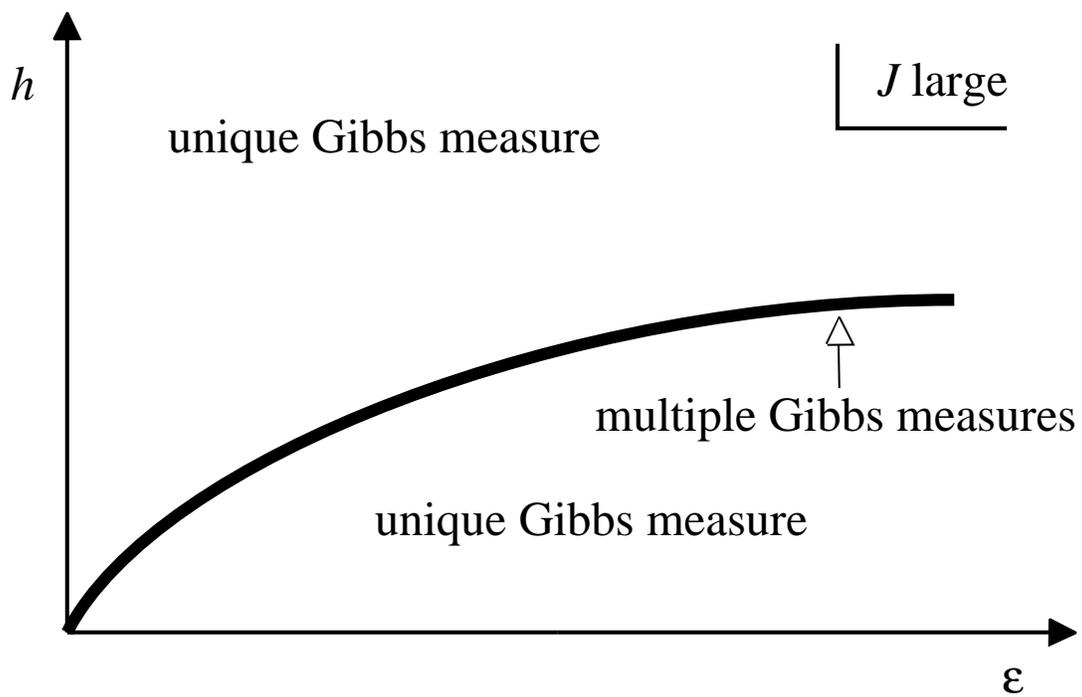

Figure 12: Phase diagram for a random-field Ising model at low temperatures ($d \geq 3$).

Therefore, we have:

**Theorem 4.8** *For each $d \geq 2$ and each even $b \geq 2$, there exists a $J_0$ (depending on d and b) such that the following is true: For any Gibbs measure $\mu$ of the d-dimensional Ising model with nearest-neighbor coupling $J > J_0$ and arbitrary magnetic field h, the renormalized measure $\mu T$ arising from the block-averaging transformation is not consistent with any quasilocal specification. In particular, it is not the Gibbs measure for any uniformly convergent interaction.*

This result is in contrast with the results obtained above for decimation and Kadanoff transformations, where we were able to prove non-Gibbsianness for $h \neq 0$ only for $d \geq 3$ and only for $|h|/J$ small. The restriction to weak fields is, for these examples, essential, because it is known that in a strong field the renormalized measure is Gibbsian [172, 173, 207]. Moreover, Martinelli and Olivieri [258] have proven that for *any* $(J, h)$ with $h \neq 0$, the decimation transformation results in a Gibbsian measure when the spacing $b$ is *large enough* (how large depends, of course, on $J$ and $h$).

Finally, we note an interesting consequence of our Theorem 4.7: for the Ising model in dimension $d \geq 3$, in the region $\{J > J_0, |h| < \delta_0 J\}$, the Dobrushin-Shlosman [94, 96] complete analyticity condition is violated.

## 4.4 Large-Cell Renormalization Maps in Dimension $d \gtrsim 4$

~~Four~~ Five years ago, Lebowitz and Maes [239] constructed a very different example of a non-Gibbsian measure, arising in the study of entropic repulsion of a surface by a wall. Subsequently, Dorlas and van Enter [103] generalized this example, and pointed out its relevance for the renormalization-group theory of Ising-like models in dimension $d \gtrsim 4$. In this section we present a slightly generalized version of the Lebowitz-Maes-Dorlas-van Enter theorem on non-Gibbsianness, and then discuss its relevance for RG theory. The reader interested primarily in the results (resp. in the application to RG theory) should read up through the statement of Theorem 4.9, and then skip directly to Section 4.4.3 (resp. to Section 4.4.4).

### 4.4.1 Non-Gibbsianness of the Sign Field of an (An)harmonic Crystal

Consider a system of *real-valued* spins $\{\varphi_x\}_{x\in\mathbb{Z}^d}$, and define $\sigma_x = \mathrm{sgn}(\varphi_x)$.[55] Clearly $\{\sigma_x\}_{x\in\mathbb{Z}^d}$ is a field of Ising spins. We shall show that for certain massless Gibbs measures on the system of $\{\varphi\}$ spins, the projection of such a measure on the $\{\sigma\}$ spins is non-Gibbsian.

The measures we have in mind are those possessing a spontaneously broken global shift symmetry $\varphi \to \varphi + c$. More precisely, consider a system defined formally by the Hamiltonian

$$H(\varphi) = \frac{1}{2} \sum_{x \neq y} V_{xy}(\varphi_x - \varphi_y) , \qquad (4.40)$$

where the functions $V_{xy}$ are even, and $V_{xy} = V_{x+a,y+a}$ for all $x, y, a \in \mathbb{Z}^d$. Such a system is termed an *anharmonic crystal* (or if the functions $V_{xy}$ are all quadratic, a *harmonic crystal*). [More rigorously, such a system is defined by the interaction

$$\Phi_A(\varphi) = \begin{cases} V_{xy}(\varphi_x - \varphi_y) & \text{if } A = \{x,y\} \\ 0 & \text{otherwise} \end{cases} \qquad (4.41)$$

where the *a priori* measure $d\mu_x^0(\varphi_x)$ is taken to be Lebesgue measure. Lebesgue measure is not normalizable, but if the potentials $V_{xy}$ are chosen suitably, then one has $Z_\Lambda(\varphi_{\Lambda^c}) < \infty$, and the specification is then well-defined. In the infinite-range case there are some subtleties associated with rapidly growing boundary conditions, as discussed in Example 4 of Section 2.3.3.]

For an (an)harmonic crystal, an infinite-volume Gibbs measure need not exist; and indeed, it will not exist in low enough dimension, e.g. $d \leq 2$ for short-range interactions [39, 92, 133]. However, if a Gibbs measure $\mu$ does exist, then it possesses a spontaneously broken global shift symmetry in the sense that $\tau_c\mu$ is also a Gibbs measure for the same interaction (here $\tau_c$ is the map that shifts all spins by a constant $c$), but $\tau_c\mu \neq \mu$ for $c \neq 0$. That $\tau_c\mu$ is a Gibbs measure is an immediate consequence of the DLR equations, while $\tau_c\mu \neq \mu$ follows from the impossibility of the probability distribution of $\varphi_0$ being invariant under a non-trivial shift. Further information on the properties of (an)harmonic crystals can be found in references [39, 42].

Every harmonic crystal is a massless Gaussian model, and the converse is very nearly true. To see this, consider a translation-invariant Gaussian measure $\mu$ on $\mathbb{R}^{\mathbb{Z}^d}$ with mean $m$ and covariance

$$\langle \varphi_x ; \varphi_y \rangle = C_{xy} = (2\pi)^{-d} \int\limits_{[-\pi,\pi]^d} \widehat{c}(p) e^{ip\cdot(x-y)} dp , \qquad (4.42)$$

---

[55]Strictly speaking we must define $\sigma_x$ also in the ambiguous case $\varphi_x = 0$. The simplest choice is to set $\sigma_x = +1$ by fiat; the most elegant choice is to set $\sigma_x = \pm 1$ with probabilities $\frac{1}{2}$. However, this choice will in fact play no role, as every measure that we will consider has the property $\mathrm{Prob}(\varphi_x = 0$ for at least one $x) = 0$.

where $\widehat{c}(p)$ is a nonnegative, even, integrable function of $p \in [-\pi, \pi]^d$. Now let us define

$$B_{xy} = (2\pi)^{-d} \int_{[-\pi,\pi]^d} \widehat{c}(p)^{-1} e^{ip \cdot (x-y)} dp, \qquad (4.43)$$

assuming that $\widehat{c}(p)^{-1}$ is an integrable function of $p$. Then $B$ is the inverse matrix of the covariance matrix $C$. Now, suppose that

$$\sum_y |B_{xy}| < \infty, \qquad (4.44)$$

as will occur if $\widehat{c}(p)^{-1}$ is at least modestly smooth.[56] In that case, there is a well-defined specification corresponding to the formal Hamiltonian

$$H(\varphi) = \tfrac{1}{2} \sum_{x,y} B_{xy} \varphi_x \varphi_y - h \sum_x \varphi_x \qquad (4.45)$$

with $h = m/\widehat{c}(0)$ and *a priori* measure taken to be Lebesgue measure; and $\mu$ is a Gibbs measure for this specification. In particular, if $\widehat{c}(0) = \infty$ — this is the "massless" case — then the couplings $B_{xy}$ satisfy

$$\sum_y B_{xy} = 0. \qquad (4.46)$$

This means that the Hamiltonian can be rewritten as

$$H(\varphi) = \tfrac{1}{4} \sum_{x \neq y} B_{xy} (\varphi_x - \varphi_y)^2 \qquad (4.47)$$

(note that here $h = 0$). Thus, every massless Gaussian measure (satisfying mild regularity conditions) is the Gibbs measure for some harmonic crystal, and conversely. Further details on the Gibbs representation of Gaussian measures can be found in references [86, 227] and [157, Chapter 13].

Now let $\mu$ be any translation-invariant Gibbs measure of the (an)harmonic crystal, and let $\widetilde{\mu}$ be its Ising projection. Under mild technical conditions on the potentials $V_{xy}$, we will prove that $\widetilde{\mu}$ is a *non-Gibbsian* measure. The basic idea of the proof is to use the spontaneously broken shift symmetry to show that

$$\text{Prob}_\mu(\varphi_x > 0 \text{ for all } x \in A) \geq e^{-o(|A|)} \qquad (4.48)$$

as $A \nearrow \infty$ (van Hove). That is, the probability that *all* the spins in a region $A$ are simultaneously positive is exponentially suppressed at a rate *slower* than the volume of $A$ (roughly speaking, it is suppressed by a "surface term"). This means that

$$i(\delta^+ | \widetilde{\mu}) = 0, \qquad (4.49)$$

---

[56]A necessary condition for $\sum_y |B_{xy}| < \infty$ to hold is that $\widehat{c}(p)^{-1}$ be a continuous function of $p \in [-\pi, \pi]^d$. However, this condition is not sufficient; for some sufficient conditions in the case $d = 1$, see [108, Section 10.6] and [211].

where $\delta^+$ is the delta measure concentrated on the configuration with all spins $+$. If $\widetilde{\mu}$ were Gibbsian, then by Proposition 2.67, $\delta^+$ would have to be Gibbsian for the same interaction. But $\delta^+$ is obviously non-Gibbsian (no absolutely summable interaction can force a spin to be $+$), so $\widetilde{\mu}$ must also be non-Gibbsian.[57]

The proof of (4.48) proceeds in three steps:

*Step 1.* Using the DLR equations, one proves the identity

$$\int F(\varphi)\, d\mu(\varphi) \;=\; \int F(\varphi + k\chi_A)\, e^{-H_{rel}(\varphi + k\chi_A;\varphi)}\, d\mu(\varphi) \qquad (4.50)$$

for any bounded function $F$ and any Gibbs measure $\mu$. Here $A$ is an arbitrary finite set of sites, $\chi_A$ is its indicator function, $k$ is an arbitrary real number, and $H_{rel}$ denotes the energy difference between the two configurations. (Since the two configurations differ on a finite set of sites, this energy difference is finite $\mu$-a.e.) In essence, this identity says that a configuration $\varphi + k\chi_A$ has a probability $e^{-H_{rel}(\varphi+k\chi_A;\varphi)}$ times as large as that of the configuration $\varphi$.

*Step 2.* One estimates the energy difference $H_{rel}$, and attempts to remove the factor $e^{-H_{rel}}$ from the right-hand side of (4.50) at the price of a prefactor $e^{-o(|A|)}$.

*Step 3.* Specializing to the case $F(\varphi) = \chi(\varphi > 0 \text{ on } A)$, one attempts to prove a lower bound on $\int F(\varphi + k\chi_A)\, d\mu(\varphi)$ that is of the form $e^{-|A|f(k)}$, where $f(k) \to 0$ as $k \to +\infty$. Since the left-hand side of (4.50) is independent of $k$, we can take $k \to +\infty$ and thus complete the proof.

Unfortunately, Steps 2 and 3 are slightly tricky (though not terribly complicated), and the details of the proof depend on the exact form of the potentials $V_{xy}$. In fact, we have *three* distinct proofs, each one valid for a distinct class of $V_{xy}$:

(a) Each $V_{xy}$ is convex, and $\sum_{z \neq 0} \|V'_{0z}\|_\infty < \infty$.

(b) Each $V_{xy}$ is quadratic (of either sign), and the measure $\mu$ is a massless Gaussian satisfying (4.44) and (4.46).

(c) Each $V_{xy}$ is convex, the model is finite-range (i.e. only finitely many of the $V_{0z}$ are nonzero), and the model is dominated by a stable Gaussian in the sense that the vectors $\{z \colon \inf_\varphi V''_{0z}(\varphi) > 0\}$ span a subspace of $\mathbb{R}^d$ of dimension $> 2$.

(We conjecture that these technical conditions can be removed or at least weakened.) Case (a) is the easiest case, as the energy shift is uniformly bounded; unfortunately, the sup norm condition on $V'$ does not allow potentials growing faster than linearly at infinity (such as Gaussians!). All the technical details in cases (b) and (c) are attempts to control an energy shift that is bounded only in some *average* sense. We urge the reader to study first the proof for case (a), before proceeding to cases (b) and (c). Case (b) is the one treated by Dorlas and van Enter [103]; we follow their proof almost verbatim. Case (c) is a minor generalization of the one treated by Lebowitz and Maes [239]; the proof we give is slightly different from theirs, but the underlying ideas and tricks are the same.

---

[57]*Alternative argument:* If $\widetilde{\mu}$ were Gibbsian, then by Proposition 2.67, $\delta^+$ would have to be Gibbsian for the same interaction, and moreover $i(\widetilde{\mu}|\delta^+)$ would have to be zero. But in fact $i(\widetilde{\mu}|\delta^+) = \infty$. *Second alternative argument:* If $\widetilde{\mu}$ were Gibbsian, then by Proposition 2.59, the pressure $p(g|\widetilde{\mu})$ would have to be *strictly* convex in directions $g = f_\Phi \notin \mathcal{I} + const$ arising from interactions $\Phi \in \mathcal{B}^1$. But for $g(\sigma) = \sigma_0$ (i.e. a magnetic field), it is easy to see from (4.48) that $p(\lambda g|\widetilde{\mu}) = \lambda$ for all $\lambda \geq 0$, contradicting the strict convexity. (For a more general version of this latter argument, see Section 4.4.2.)

**Theorem 4.9** *Let $\mu$ be a translation-invariant Gibbs measure for an (an)harmonic crystal satisfying one of the conditions (a)–(c) listed above. In case (c), assume in addition that $\mu$ is symmetric around its mean. Then for each $M < \infty$, we have*

$$\mathrm{Prob}_\mu(\varphi_x > M \text{ for all } x \in A) \geq e^{-o(|A|)} \tag{4.51}$$

*as $A \nearrow \infty$ (van Hove). It follows that $\widetilde{\mu}$, the projection of $\mu$ on the Ising spins $\sigma_x \equiv \mathrm{sgn}(\varphi_x)$, is not the Gibbs measure for any interaction in $\mathcal{B}^1$.*

**Remarks.** 1. One consequence of this theorem is that an arbitrarily weak perturbation of the form $H \to H - \sum_x f(\varphi_x)$, where $f$ is nondecreasing and nonconstant, will drive the spins $\varphi_x$ to $+\infty$. As a result, thus the perturbed model will have no infinite-volume translation-invariant Gibbs measures. This is the phenomenon of entropic repulsion of a surface by a soft wall, studied by Lebowitz and Maes [239].

2. It is natural to ask whether $\widetilde{\mu}$ is non-quasilocal (and not merely non-Gibbsian). We discuss this question, in somewhat greater generality, in Section 4.4.3.

PROOF OF THEOREM 4.9. Since the hypotheses of the theorem are invariant under a uniform shift $\varphi \to \varphi + c$, it suffices to consider the case $M = 0$; this lightens the notation. (For our application to the sign function, we need only $M = 0$ anyway. But we will exploit the formulation with general $M$ in Section 4.4.2.)

*Step 1.* Let $\mu$ be any Gibbs measure for *any* model of real-valued spins (not necessarily an anharmonic crystal). Then the DLR equations for volume $\Lambda$ say that

$$d\mu_\Lambda(\varphi_\Lambda|\varphi_{\Lambda^c}) = Z_\Lambda(\varphi_{\Lambda^c})^{-1} e^{-H_\Lambda(\varphi_\Lambda,\varphi_{\Lambda^c})} d\varphi_\Lambda . \tag{4.52}$$

Now let $F$ be any bounded (for simplicity) measurable function, and let $\psi$ be any field which vanishes outside $\Lambda$. Then

$$\begin{aligned}
\int F(\varphi) \, d\mu_\Lambda(\varphi_\Lambda|\varphi_{\Lambda^c}) &= Z_\Lambda(\varphi_{\Lambda^c})^{-1} \int F(\varphi) \, e^{-H_\Lambda(\varphi_\Lambda,\varphi_{\Lambda^c})} \, d\varphi_\Lambda \\
&= Z_\Lambda(\varphi_{\Lambda^c})^{-1} \int F(\varphi + \psi) \, e^{-H_\Lambda(\varphi_\Lambda + \psi_\Lambda,\varphi_{\Lambda^c})} \, d\varphi_\Lambda \\
&= \int F(\varphi + \psi) \, e^{-[H_\Lambda(\varphi_\Lambda + \psi_\Lambda,\varphi_{\Lambda^c}) - H_\Lambda(\varphi_\Lambda,\varphi_{\Lambda^c})]} \, d\mu_\Lambda(\varphi_\Lambda|\varphi_{\Lambda^c}) ,
\end{aligned} \tag{4.53}$$

where in the middle line we used the shift invariance of Lebesgue measure. Now integrate over $d\mu_{\Lambda^c}(\varphi_{\Lambda^c})$: we obtain

$$\int F(\varphi) \, d\mu(\varphi) = \int F(\varphi + \psi) \, e^{-H_{rel}(\varphi + \psi; \varphi)} \, d\mu(\varphi) , \tag{4.54}$$

where
$$H_{rel}(\varphi + \psi; \varphi) \equiv H_\Lambda(\varphi_\Lambda + \psi_\Lambda, \varphi_{\Lambda^c}) - H_\Lambda(\varphi_\Lambda, \varphi_{\Lambda^c}) \,. \tag{4.55}$$

Note that $H_{rel}$ is independent of $\Lambda$ as soon as $\Lambda \supset \text{supp}\,\psi$. The identity (4.54) is thus valid for any $\psi$ of bounded support. In particular, if we take $\psi = k\chi_A$, we obtain (4.50). In the case of the anharmonic crystal (4.40) we have the following expression for $H_{rel}$:

$$
\begin{align}
H_{rel}(\varphi + k\chi_A; \varphi) &= \sum_{\substack{x \in A \\ y \in A^c}} [V_{xy}(\varphi_x + k - \varphi_y) - V_{xy}(\varphi_x - \varphi_y)] \tag{4.56a} \\
&= \sum_{\substack{x \in A \\ y \in A^c}} \int_0^k V'_{xy}(\varphi_x - \varphi_y + \psi)\,d\psi \,. \tag{4.56b}
\end{align}
$$

*Step 2.* The goal of this step is to prove that

$$\int F(\varphi)\,d\mu(\varphi) \geq e^{-o_k(|A|)} \int F(\varphi + k\chi_A)\,d\mu(\varphi) \tag{4.57}$$

(or some similar formula) for some suitable class of nonnegative functions $F$. Here $o_k(|A|)$ denotes a term that may depend in an arbitrary way on $k$, but for each real $k$ it should be $o(|A|)$ as $A \nearrow \infty$.

*Case (a):* This is the easy case, as the energy shift (4.56) can be bounded in sup norm:

$$
\begin{align}
\|H_{rel}\|_\infty &\leq \sum_{\substack{x \in A \\ y \in A^c}} |k|\,\|V'_{xy}\|_\infty \\
&\leq |k|\,o(|A|) \quad \text{as } A \nearrow \infty \tag{4.58}
\end{align}
$$

by the usual argument based on $\sum_{z \neq 0} \|V'_{0z}\|_\infty < \infty$ (see e.g. the proof of Proposition 2.45 in Appendix A.3.8). Substituting (4.58) into the identity (4.50), we conclude that

$$\int F(\varphi)\,d\mu(\varphi) \geq e^{-|k|o(|A|)} \int F(\varphi + k\chi_A)\,d\mu(\varphi) \tag{4.59}$$

uniformly for all nonnegative bounded functions $F$.

*Case (b):* Here we apply the Schwarz inequality to the right-hand side of the identity (4.50):

$$\int F(\varphi + k\chi_A)\,e^{-H_{rel}(\varphi + k\chi_A; \varphi)}\,d\mu(\varphi) \geq \frac{\left[\int F(\varphi + k\chi_A)^{1/2}\,d\mu(\varphi)\right]^2}{\int e^{+H_{rel}(\varphi + k\chi_A; \varphi)}\,d\mu(\varphi)} \tag{4.60}$$

for any $F \geq 0$. In particular, if $F$ is the indicator function of some set, then $F^{1/2} = F$. Now in case (b) we have

$$H_{rel}(\varphi + k\chi_A; \varphi) = k(\varphi, B\chi_A) + \frac{k^2}{2}(\chi_A, B\chi_A) \,, \tag{4.61}$$

and $\mu$ is a Gaussian measure with mean $m$ and covariance matrix $C = B^{-1}$. We can therefore calculate exactly

$$\int e^{+H_{rel}(\varphi+k\chi_A;\varphi)}\,d\mu(\varphi) = \exp[k^2(\chi_A, B\chi_A) + km(\mathbf{1}, B\chi_A)]$$
$$= \exp[k^2(\chi_A, B\chi_A)] \qquad (4.62)$$

since $B\mathbf{1} = 0$ by (4.46). Now

$$k^2(\chi_A, B\chi_A) = 2k^2 \sum_{\substack{x \in A \\ y \in A^c}} B_{xy}$$
$$\leq k^2 o(|A|) \quad \text{as } A \nearrow \infty \qquad (4.63)$$

by the usual argument based on $\sum_{z \neq 0} |B_{0z}| < \infty$. Hence

$$\int F(\varphi)\,d\mu(\varphi) \geq e^{-k^2 o(|A|)} \left[\int F(\varphi + k\chi_A)\,d\mu(\varphi)\right]^2 \qquad (4.64)$$

uniformly for all indicator functions $F$. This is a slight variant of (4.57).

*Case (c):* This case is a little bit trickier. Let $F$ be any nonnegative function supported on the set $\{\varphi: a \leq \varphi \leq b \text{ on } A \text{ and } a' \leq \varphi \leq b' \text{ on } \partial_r^+ A\}$, where $r$ is the range of the interaction. Then on the right-hand side of (4.50) the integrand is nonvanishing only when $a - k \leq \varphi_x \leq b - k$ for $x \in A$, and $a' \leq \varphi_y \leq b'$ for $y \in \partial_r^+ A$. Now, since $V_{xy}$ is convex, $V'_{xy}$ is increasing, so $V_{xy}(\varphi_x - \varphi_y + k) - V_{xy}(\varphi_x - \varphi_y)$ is an increasing (resp. decreasing) function of $\varphi_x - \varphi_y$ for $k \geq 0$ (resp. $k \leq 0$), as seen from (4.56b). Therefore, for $k \geq 0$ (which is the case that will interest us) we have

$$H_{rel}(\varphi + k\chi_A; \varphi) \leq \sum_{\substack{x \in \partial_r^- A \\ y \in \partial_r^+ A}} [V_{xy}(b - a') - V_{xy}(b - a' - k)]$$
$$\leq C(a', b, k)\,|\partial_r^- A|\,, \qquad (4.65)$$

where $C(a', b, k) \equiv \sum_z [V_{0z}(b - a') - V_{0z}(b - a' - k)]$ is finite for all $a', b, k$ (since only finitely many terms in this sum are nonzero). Hence

$$\int F(\varphi)\,d\mu(\varphi) \geq e^{-C(a',b,k)\,|\partial_r^- A|} \int F(\varphi + k\chi_A)\,d\mu(\varphi) \qquad (4.66)$$

uniformly for all nonnegative $F$ satisfying the support condition. This, too, is a variant of (4.57).

*Step 3, Case (a):* We apply (4.59) to $F(\varphi) = \chi(\varphi > 0 \text{ on } A)$, so that $F(\varphi + k\chi_A) = \chi(\varphi > -k \text{ on } A)$. Since the $V_{xy}$ are convex, it follows immediately from the DLR equation that the FKG inequality [24] holds for every Gibbs measure $\mu$. Therefore we have

$$\langle \chi(\varphi > -k \text{ on } A) \rangle_\mu \geq \prod_{x \in A} \langle \chi(\varphi_x > -k) \rangle_\mu = \text{Prob}_\mu(\varphi_0 > -k)^{|A|} \qquad (4.67)$$

Combining this with (4.59), we get

$$\liminf_{A \nearrow \infty} \frac{1}{|A|} \log \text{Prob}_\mu(\varphi > 0 \text{ on } A) \geq \log \text{Prob}_\mu(\varphi_0 > -k). \qquad (4.68)$$

But taking $k \to +\infty$, the right-hand side goes to zero.

*Step 3, Case (b):* We apply (4.64) to $F(\varphi) = \chi(\varphi > 0 \text{ on } A)$. We control $\text{Prob}_\mu(\varphi > -k \text{ on } A)$ using the Brascamp-Lieb inequality [37, 38], which is valid for arbitrary Gaussian measures, combined with the Chebyshev inequality:

$$\begin{aligned}
\text{Prob}_\mu(\varphi > -k \text{ on } A) &\geq \text{Prob}_\mu(|\varphi - m| < m + k \text{ on } A) \\
&= \prod_{i=1}^{n} \text{Prob}_\mu\Big(|\varphi_{x_i} - m| < m + k \Big| |\varphi_{x_j} - m| < m + k \text{ for } 1 \leq j < i\Big) \\
&\quad [\text{where } A = \{x_1, \ldots, x_n\}] \\
&\geq \prod_{i=1}^{n} \left(1 - \frac{E_\mu\Big((\varphi_{x_i} - m)^2 \Big| |\varphi_{x_j} - m| < m + k \text{ for } 1 \leq j < i\Big)}{(m+k)^2}\right) \\
&\quad [\text{by Chebyshev}] \\
&= \prod_{i=1}^{n} \left(1 - \frac{\text{var}_\mu\Big(\varphi_{x_i} \Big| |\varphi_{x_j} - m| < m + k \text{ for } 1 \leq j < i\Big)}{(m+k)^2}\right) \\
&\quad [\text{since conditioning a Gaussian on a set} \\
&\quad\text{ symmetric about the mean preserves the mean}] \\
&\geq \prod_{i=1}^{n} \left(1 - \frac{\text{var}_\mu(\varphi_{x_i})}{(m+k)^2}\right) \\
&\quad [\text{by Brascamp-Lieb}] \\
&= \left(1 - \frac{C_{00}}{(m+k)^2}\right)^{|A|}. \qquad (4.69)
\end{aligned}$$

Combining this with (4.64), we get

$$\liminf_{A \nearrow \infty} \frac{1}{|A|} \log \text{Prob}_\mu(\varphi > 0 \text{ on } A) \geq 2 \log \left(1 - \frac{C_{00}}{(m+k)^2}\right). \qquad (4.70)$$

Now take $k \to +\infty$.

*Step 3, Case (c):* We apply (4.66) to $F(\varphi) = \chi(a \leq \varphi \leq b \text{ on } A \text{ and } a' \leq \varphi \leq b' \text{ on } \partial_r^+ A)$, with the choices $a = 0$, $b = 2m + 2k$, $a' = -k$, $b' = 2m + k$, $k \geq 0$. We therefore need to control

$$\begin{aligned}
&\text{Prob}_\mu(a - k \leq \varphi \leq b - k \text{ on } A \text{ and } a' \leq \varphi \leq b' \text{ on } \partial_r^+ A) \\
&= \text{Prob}_\mu(|\varphi - m| \leq m + k \text{ on } A \cup \partial_r^+ A). \qquad (4.71)
\end{aligned}$$

To do this, we employ the Brascamp-Lieb and Chebyshev inequalities as in case (b) [the Brascamp-Lieb inequality is valid because all the $V_{xy}$ are convex]. Here it is important that $\mu$ be even about its mean, because Brascamp-Lieb refers to variances rather than to expectations of squares; we need to know that conditioning $\mu$ on a set symmetric around the mean does not displace the mean. The only other change from case (b) is that $\text{var}_\mu(\varphi_{x_i}|\cdots)$ is bounded above not by $\text{var}_\mu(\varphi_{x_i})$, but rather by the variance of $\varphi_{x_i}$ in the dominating Gaussian, which by hypothesis is finite (call it $C_{00}$). Thus, we have

$$\text{Prob}_\mu(|\varphi - m| \leq m + k \text{ on } A \cup \partial_r^+ A) \geq \left(1 - \frac{C_{00}}{(m+k)^2}\right)^{|A \cup \partial_r^+ A|}. \qquad (4.72)$$

Combining this with (4.66), we get

$$\liminf_{A \nearrow \infty} \frac{1}{|A|} \log \text{Prob}_\mu(\varphi > 0 \text{ on } A)$$
$$\geq \liminf_{A \nearrow \infty} \frac{1}{|A|} \log \text{Prob}_\mu(0 < \varphi < 2m + 2k \text{ on } A \text{ and } -k < \varphi < 2m + k \text{ on } \partial_r^+ A)$$
$$\geq \log\left(1 - \frac{C_{00}}{(m+k)^2}\right). \qquad (4.73)$$

Now take $k \to +\infty$. ∎

**Remark.** We do not know whether there can exist translation-invariant Gibbs measures for the anharmonic crystal that fail to be symmetric around their mean. (In the Gaussian case such measures cannot exist.) That is, we do not know whether the reflection symmetry can be spontaneously broken. If the answer is no, then our additional hypothesis in case (c) is superfluous.

The technical condition in case (c) — that the model be dominated by a stable Gaussian — unfortunately excludes some interesting models, such as the $(\nabla \varphi)^4$ model. The need for this technical condition arises from the use of Brascamp-Lieb inequalities to bound the conditional probability $\text{Prob}_\mu(|\varphi_{x_i} - m| < m+k \, | \, |\varphi_{x_j} - m| < m+k$ for $1 \leq j < i)$. An alternate approach would be to use the FKG inequalities as in case (a), but then we would be forced to work with increasing functions, i.e. to take $b = b' = +\infty$. Unfortunately, the cutoff $b < \infty$ was necessary in case (c) in order to control the energy shift $H_{rel}$, which otherwise could be unbounded above.

How can we escape from this dilemma? Let us first note that the large energy shift arises from applying the shift $\varphi_x \to \varphi_x + k$ to fields $\varphi_x$ that are *already* large and positive, hence have *no need* to be shifted farther upwards in order to bring them above the level $\varphi = 0$. This suggests that instead of applying a uniform shift $\varphi_x \to \varphi_x + k$ in the region $A$, we should apply a *nonlinear* map $\varphi_x \to f(\varphi_x)$ that would produce a large upward shift when $\varphi_x$ is negative, but a smaller shift when $\varphi_x$ is large and positive. In this way we may hope to have an energy shift that is uniformly bounded above. Of course, in the case of a nonlinear map $f$ we must also deal with a Jacobian, but this turns out to be manageable. The idea may be crazy, but it seems to work, at least for some rather large class of potentials $V_{xy}$. However, this paper is already much too long, and we have not had time to work out all the details, so we leave further development of this circle of ideas to the interested reader.

### 4.4.2 Non-Gibbsianness of Local Nonlinear Functions of an (An)harmonic Crystal

The method of the preceding section applies, in fact, to local nonlinear functions much more general than the sign. Indeed, let $\Omega_0'$ be a compact metric space, and let $f\colon \mathbb{R} \to \Omega_0'$ be any function (not necessarily continuous) such that $\lim_{\varphi \to +\infty} f(\varphi) = \omega^*$ exists. We shall show that for the class of massless Gibbs measures on the system of $\{\varphi\}$ spins considered in the preceding section, the projection of such a measure on the $\{\omega\}$ spins is non-Gibbsian.

**Theorem 4.10** *Let $\mu$ be any translation-invariant measure on $\mathbb{R}^{\mathbb{Z}^d}$ satisfying the estimate (4.51) for all $M < \infty$. Let $\Omega_0'$ be a compact metric space, and let $f\colon \mathbb{R} \to \Omega_0'$ be a function (not necessarily continuous) such that $\lim_{\varphi \to +\infty} f(\varphi) = \omega^*$ exists. Let $\widetilde{\mu}$ be the image measure of $\mu$ under the map $f$ applied to each spin. Then $\widetilde{\mu}$ is not the Gibbs for any interaction in $\mathcal{B}^1$, with respect to any a priori measure supported on more than one point.*

PROOF. Let $U, V$ be open sets in $\Omega_0'$ satisfying $\omega^* \in U \subset \bar{U} \subset V$. Then let $g_0 \colon \Omega_0' \to [0,1]$ be a continuous function satisfying $g_0 \restriction \bar{U} \equiv 1$ and $g_0 \restriction V^c \equiv 0$; the existence of such a function is guaranteed by Urysohn's lemma. Now define $g \colon {\Omega_0'}^{\mathbb{Z}^d} \to [0,1]$ by $g(\{\omega_x\}_{x \in \mathbb{Z}^d}) = g_0(\omega_0)$. That is, $g$ is the function $g_0$ applied to the spin at the origin.

Now let us compute the pressure $p(\lambda g | \widetilde{\mu})$ for $\lambda \geq 0$:

$$\begin{aligned}
p(\lambda g | \widetilde{\mu}) &\equiv \lim_{n \to \infty} n^{-d} \log \int \exp\left[\lambda \sum_{x \in C_n} g_0(\omega_x)\right] d\widetilde{\mu}(\omega) \\
&= \lim_{n \to \infty} n^{-d} \log \int \exp\left[\lambda \sum_{x \in C_n} g_0(f(\varphi_x))\right] d\mu(\varphi) \quad (4.74)
\end{aligned}$$

if this limit exists. Since $g_0 \leq 1$, clearly the lim sup is $\leq \lambda$. On the other hand, the lim inf is

$$\begin{aligned}
&\geq \liminf_{n \to \infty} n^{-d} \log \left[e^{\lambda n^d} \mathrm{Prob}_\mu(f(\varphi_x) \in U \text{ for all } x \in C_n)\right] \\
&\geq \liminf_{n \to \infty} n^{-d} \log \left[e^{\lambda n^d} \mathrm{Prob}_\mu(\varphi_x > M \text{ for all } x \in C_n)\right] \\
&= \lambda \,, \quad (4.75)
\end{aligned}$$

where $M$ is chosen so that $\varphi > M$ implies $f(\varphi) \in U$; here the final equality uses the fundamental estimate (4.51). So we have

$$p(\lambda g | \widetilde{\mu}) \;=\; \lambda \qquad \text{for all } \lambda \geq 0 \;. \tag{4.76}$$

But this violates the strict convexity of the pressure which must hold if $\widetilde{\mu}$ is a Gibbs measure for an interaction in $\mathcal{B}^1$ (Griffiths-Ruelle theorem, Proposition 2.59). Hence $\widetilde{\mu}$ is non-Gibbsian. ∎

### 4.4.3 Physical Interpretation

We have proven that $\widetilde{\mu}$ is not the Gibbs measure for any interaction in $\mathcal{B}^1$, but is this enough? We know that non-Gibbsianness can sometimes occur for "trivial" reasons, e.g. if there are hard-core exclusions, or for "semi-trivial" reasons, e.g. if the Hamiltonian $H_\Lambda^\Phi$ is quasilocal but unbounded. (This latter can happen only when the single-spin space is infinite.) If we contend that $\widetilde{\mu}$ is "pathological", then we really ought to prove not merely that $\widetilde{\mu}$ is non-Gibbsian, but also that it is *non-quasilocal*.

We are not able at present to prove non-quasilocality, but we can argue heuristically that in at least some cases the non-Gibbsianness does involve some strongly non-local effect. Consider the sign of the (an)harmonic crystal. Recalling Theorem 4.9 together with Remark 1 following it, it is natural to conjecture that

$$\lim_{R' \to \infty} E_\mu \left( \text{sgn}(\varphi_0) \,\Big|\, \varphi_x > 0 \text{ for all } x \text{ having } R \leq |x| \leq R' \right) \;=\; 1 \tag{4.77}$$

for all $R$, no matter how large. (At least in case (c) of Section 4.4.1, we are able to prove this using the FKG inequality, via a slight extension of the arguments of Lebowitz and Maes [239].) That is, if we condition on the spins in an annulus $R \leq |x| \leq R'$ being all $> 0$, as $R' \to \infty$ this drives all the spins to $+\infty$, and in particular *forces* the sign of the spin at the origin to be $+$ (with probability 1!). For the Ising measure $\widetilde{\mu}$, this means heuristically that the spin at the origin is feeling an infinite energy. However, since the effect occurs for all $R$, no matter how large, this infinite energy must arise from the interaction between the spin at the origin and *arbitrarily distant spins*. (Crudely speaking, the interaction, if it exists, is non-summable.) Thus, we do not have here merely the "semi-trivial" situation of a Hamiltonian which is quasilocal but unbounded (which anyway is impossible for a model with finite single-spin space); some strongly non-local effect is taking place. It may even be that (4.77) implies non-quasilocality; or it may be that non-quasilocality can be proven by a different argument. These are open questions.

A similar situation probably holds in the setup of Section 4.4.2, whenever the image single-spin space $\Omega_0'$ is finite.

A very different situation arises if $f$ is a *bijective* map of $\mathbb{R}$ onto $\Omega_0'$ (of course $\Omega_0'$ must then be uncountably infinite!). In this case $f$ is merely a one-to-one relabelling of spin values; the physics of the image measure $\widetilde{\mu}$ is obviously *identical* to that of the original measure $\mu$. In particular, if the original (an)harmonic crystal has *finite-range* interactions, then $\widetilde{\mu}$ is consistent with a *Gibbsian* specification for a particular finite-range but *unbounded* interaction, namely the one gotten by mapping the (an)harmonic-crystal specification via the function $f$. Such a specification is always quasilocal; the interaction is uniformly convergent but not absolutely summable.

Finally, let us remark that the local nonlinear maps considered here are a special case of the renormalization transformations considered in Sections 3 and 4.1–4.3: namely, one in which the blocks are single sites, the transformation is deterministic, and the image space is in general different from the original space. Such transformations trivially obey properties (T1)–(T3) of Section 3.1. Of course, if $f$ is one-to-one, then the transformation is trivial (just a relabelling of spin configurations). However, if $f$ is many-to-one, then the transformation is not so different in nature from the usual (block-spin) renormalization transformations: both "discard details" from the original spin configuration. These details may be in the fine structure of a single spin, or in the local fine structure of a small block of spins, but qualitatively there does not seem to be any great intrinsic difference. Our theorems both in Sections 4.1–4.3 and in the current subsection are of the general type: an RT map which discards (important) information makes the image measure (sometimes) non-Gibbsian (and possibly even non-quasilocal).

### 4.4.4 Application to the Renormalization Group

In this section we apply Theorem 4.9 to the RG, following closely Dorlas and van Enter [103]. Let us consider an Ising model in dimension $d > 4$ at the critical point, and apply block-averaging transformations on various block sizes $b$. Then DeConinck and Newman [71] and Shlosman [323, and private communication] have shown that there exists a $b$-dependent choice of normalization such that the block-spin measures converge as $b \to \infty$ to a massless Gaussian measure[58]; this is a slight variant of the Aizenman-Fröhlich triviality theorem.

Now the key observation is that a block-averaging transformation followed by a projection on Ising configurations is identical to a majority-rule transformation. So consider applying the majority-rule transformation using larger and larger block sizes $b$. Since the block-averaged spins (with a suitable $b$-dependent normalization) converge as $b \to \infty$ to a massless Gaussian, it is not difficult to show that the majority-rule image spins converge as $b \to \infty$ to the sign of this same massless Gaussian. But by Theorem 4.9, this latter measure is non-Gibbsian! (For details, see [103].)

This non-Gibbsian scaling limit is not a fixed point in the strict sense, as the sequence of majority-rule transformations lacks the semigroup property: the majority rule on block size $b^2$ is not equal to the second iteration of majority rule on block size $b$ (as politicians well know!). Therefore, the existence of pathologies for the fixed point arising from the $b \to \infty$ limit does not *guarantee* that the corresponding pathologies will occur for the fixed point arising from iteration of a majority-rule map with a fixed block size $b$. But it does make it plausible: there does not seem to be *so much* difference between majority rule on a block of size $b^n$ and $n$ iterations of majority rule on a block of size $b$. And, in any case, the "large-cell majority-rule" approach is clearly part of the RG enterprise [128, 249], so it is interesting to see that it can fail. Finally, as we discuss in Section 5.2, there are other reasons to expect that this behavior is in some sense typical. Indeed, we conjecture that the fixed-point measures of nonlinear RG transformations for $d \gtrsim d_u$ ($\equiv$ upper critical dimension of the model) will be non-Gibbsian in considerable generality.

---

[58]Conventional wisdom holds that the normalization can be chosen to be $b^{-p}$ for a suitable power $p$ [in fact one predicts $p = (d + 2 - \eta)/2 = (d + 2)/2$]. If this is the case, then the limiting measure can also be obtained by repeated application of the block-averaging transformation with a *fixed* block size $b$, and hence is a *self-similar* Gaussian measure [327, 17, 32]. However, this conventional wisdom has not yet (as far as we know) been proven rigorously.

Finally, we remark that the results discussed here for $d > 4$ are expected to hold also for $d = 4$, provided that the "triviality conjecture" [365, 115] is true.

## 4.5 Other Results on Non-Gibbsianness and Non-Quasilocality

In Sections 4.1–4.4, we have given a number of examples of non-Gibbsian (or what is slightly stronger, non-quasilocal) measures, with particular attention to those arising in RG theory. It is natural to ask whether the phenomenon of non-Gibbsianness (or non-quasilocality) is more widespread. Unfortunately, very little is known at present about the properties of non-quasilocal measures, and very few examples of non-quasilocal measures are known. In this section we try to make a complete survey of all known physically interesting examples of non-quasilocality. (The list is short enough that such a comprehensive survey is feasible.)

### 4.5.1 Trivial Example: Convex Combination of Gibbs Measures for Different Interactions

These are perhaps rather silly examples: if one makes a convex combination of Gibbs measures for the Ising model at two different temperatures, then it is hardly surprising that the resulting measure will not be Gibbsian at all. The proof says roughly that if the resulting measure *were* Gibbsian for some interaction $\Phi$, then the two original measures would also have to be Gibbsian for $\Phi$. But this is impossible, because the Griffiths-Ruelle theorem tells us that a measure can be Gibbsian for at most one interaction (modulo physical equivalence).

We need a preliminary result, concerning the conditions under which a "reweighting" of a Gibbs measure remains a Gibbs measure:

**Lemma 4.11** *Let $\Pi$ be a specification and $\mu$ a measure in $\mathcal{G}(\Pi)$. A measure $\nu$ of the form $\nu = f\mu$ belongs also to $\mathcal{G}(\Pi)$ if and only if $f$ is $\widehat{\mathcal{F}}_\infty$-measurable (modulo $\mu$-null sets).*

(We recall that $\widehat{\mathcal{F}}_\infty \equiv \bigcap_{\Lambda \in \mathcal{S}} \mathcal{F}_{\Lambda^c}$ is the $\sigma$-field of observables at infinity: see Section 2.3.6.) The proof of this lemma is given, for instance, in [299, Lemma 2.4] and in [157, Theorem 7.7].

We can now prove the main result:

**Proposition 4.12** *Let $\mu_1, \mu_2, \ldots$ be a finite or countably infinite family of measures (not necessarily translation-invariant) which are are distinguishable at infinity, i.e. there exist disjoint sets $F_1, F_2, \ldots \in \widehat{\mathcal{F}}_\infty$ such that $\mu_k(F_k) = 1$ for each $k$. Assume further that each of the measures $\mu_1, \mu_2, \ldots$ gives nonzero measure to every open set in $\Omega$. Now form a convex combination $\mu = \sum_k c_k \mu_k$ with all $c_k > 0$. If $\mu$ is consistent with a specification $\Pi$, then so are $\mu_1, \mu_2, \ldots$; and if $\Pi$ is Feller, then this is the* only *Feller specification with which any of these measures is consistent.*

Thus, if some two of the $\{\mu_k\}$ — say, $\mu_i$ and $\mu_j$ — happen to be consistent with *different* Feller specifications ($\Pi_i \neq \Pi_j$), then it follows that $\mu$ is not consistent with *any* Feller specification. In particular, $\mu$ is not a Gibbs measure for any continuous, uniformly convergent interaction. If the single-spin space $\Omega_0$ is finite, this means that $\mu$ is not consistent with any quasilocal specification, and in particular that $\mu$ is not a Gibbs measure for any uniformly convergent interaction.

**Remark.** It is not difficult to show that if the measures $\mu_1, \mu_2, \ldots$ are *pairwise* distinguishable at infinity, then they are jointly distinguishable at infinity in the sense of Proposition 4.12. Here it is crucial that we are dealing with a *countable* family.

PROOF OF PROPOSITION 4.12. Suppose that $\mu$ is consistent with a specification $\Pi$. Then, by Lemma 4.11, the measures $\mu_k = c_k^{-1} \chi_{F_k} \mu$ are also consistent with $\Pi$. The uniqueness follows from Theorem 2.15. ∎

In order to apply Proposition 4.12, we need to verify that the measures $\mu_1, \mu_2, \ldots$ are distinguishable at infinity (the support hypothesis is usually trivial to check). One easy way to obtain such measures is to recall that distinct *ergodic* translation-invariant measures are distinguishable at infinity (Theorem 2.33 and the remark following it). We therefore have:

**Corollary 4.13** *Let $\mu_1, \mu_2, \ldots$ be a finite or countably infinite family of ergodic translation-invariant Gibbs measures for interactions $\Phi_1, \Phi_2, \ldots \in \mathcal{B}^1$, respectively. Now form a convex combination $\mu = \sum_k c_k \mu_k$ with all $c_k > 0$. If $\mu$ is consistent with a* Feller *specification $\Pi$, then all the interactions $\Phi_k$ must be physically equivalent in the DLR sense (and hence also in the Ruelle sense).*

PROOF. If $\mu_i = \mu_j$, then $\Phi_i$ and $\Phi_j$ must be physically equivalent in the DLR sense (Corollary 2.18). So we can assume without loss of generality that the measures $\mu_1, \mu_2, \ldots$ are all distinct. Since distinct ergodic measures are distinguishable at infinity, and Gibbs measures for an absolutely summable interaction always give nonzero measure to every open set, we can apply the preceding proposition to conclude that $\Pi = \Pi^{\Phi_1} = \Pi^{\Phi_2} = \ldots$. The rest follows from Theorems 2.17 and 2.42. ∎

Therefore, *(non-trivial) finite or countably infinite convex combinations of ergodic translation-invariant Gibbs measures for non-physically-equivalent interactions cannot be Gibbsian*; and for finite single-spin space they cannot even be quasilocal.

### 4.5.2 Restriction of the Two-Dimensional Ising Model to an Axis

Schonmann [318] gave another example of a non-Gibbsian measure that can be obtained by applying a simple transformation to a well-known Gibbsian measure. He proved that if $\mu_+$ is the "+" phase of the two-dimensional Ising model at zero field and at any temperature below critical, then its restriction $\mu_+ P$ to the axis $\{(i,0): i \in \mathbb{Z}\}$ is a *non-Gibbsian* one-dimensional Ising model. His argument is based on two results:

R1) For all temperatures below the critical temperature for the $d=2$ Ising model, $i(\mu_- P | \mu_+ P) \neq 0$.

R2) Let $\sigma'_{n,N}$ denote the spin configuration on the "annulus" $\{(i,0): n \leq |i| \leq N\}$. Then for each $n$ there exists an $N(n)$ such that

$$\mu(\,\cdot\,|\sigma'_{n,N(n)} = -1) \;\to\; \mu_- \qquad \text{as } n \to \infty \qquad (4.78)$$

for all Gibbs measures $\mu$ of the original model.[59] As a consequence

$$(\mu_+ P)(\,\cdot\,|\sigma'_{n,N(n)} = -1) \;\to\; \mu_- P \qquad \text{as } n \to \infty \,. \qquad (4.79)$$

Result (4.79) implies that if $\mu_+ P$ is consistent with some quasilocal (= Feller) specification, then $\mu_- P$ must be consistent with that *same* specification. Heuristically this is due to the fact that a measure obtained just by a change in the boundary conditions must be a different phase for the same interaction. To see it mathematically, let $\Pi = (\pi_\Lambda)$ be a quasilocal specification with which $\mu_+ P$ is consistent. Then for each set $\Lambda$ contained in the interval $(-n, n)$ we have that

$$(\mu_+ P)(\,\cdot\,|\sigma'_{n,N(n)} = -1)\,\pi_\Lambda \;=\; (\mu_+ P)(\,\cdot\,|\sigma'_{n,N(n)} = -1) \qquad (4.80)$$

by property (b) of Definition 2.5; and passing to the limit $n \to \infty$ (since $\Pi$ is Feller) we obtain

$$(\mu_- P)\pi_\Lambda \;=\; \mu_- P \,. \qquad (4.81)$$

Therefore, if $\mu_+ P$ were a Gibbs measure for some (uniformly convergent) interaction, then so would be $\mu_- P$. But this contradicts the result (R1), because Gibbs measures for the same (absolutely summable translation-invariant) interaction have zero relative entropy density.

---

[59]This statement easily follows from Schonmann's Lemma 1.

Schonmann's restriction $P$ does not fit into the framework considered in Section 3, because the volume compression factor $K$ is not finite (see Example 7 in Section 3.1). On the other hand, Schonmann's proof of non-Gibbsianness seems to be rather different from our proofs in Sections 4.1–4.3. We show here that, nevertheless, his result can be obtained by following basically the steps discussed in Sections 4.1–4.3 (although at present we are able to do it only for temperatures low enough). This will prove that $\mu_+ P$ is not merely non-Gibbsian, but in fact non-quasilocal. The proof will use (R2) but not (R1).

In our language, the image spins for this transformation are the spins on the horizontal line, and the internal spins are all the spins of the plane except those of the line. We first notice that Schonmann's result (R2) corresponds exactly to our Step 2: that is, (4.78) shows that the annulus $[-N, -n] \cup [n, N]$ of image spins selects the phase of the internal spins. Physically, this is a kind of wetting phenomenon: imposing $-$ spins on a large segment of the axis (of size $\sim N$) give rise to a droplet of the $-$ phase in a neighborhood of the axis, *even when the bulk boundary conditions are $+$*; as $N \to \infty$ the width of the droplet grows to infinity, and moreover the left and right droplets join, thereby enforcing the $-$ phase throughout the infinite system.

We sketch now how our Step 1 can be proven via a contour argument, so that we obtain the non-quasilocality of the image system without making use of the large-deviation estimate (R1). We consider the origin unfixed from the start (so Step 3 is superfluous), and consider $\omega'_{\text{special}}$ to be an alternating configuration such that the neighbors of the origin are of opposite sign:

$$(\omega'_{\text{special}})_{(i,0)} = \begin{cases} (-1)^i & \text{if } i > 0 \\ (-1)^{i+1} & \text{if } i < 0 \end{cases}. \tag{4.82}$$

We shall prove the following: there exists $\epsilon > 0$ such that for all $k$ there exist $n(k)$ and $N(k)$ such that

$$\mu_+(\sigma_0 | \sigma'_{1,k} = \omega'_{\text{special}}, \sigma'_{n(k),N(k)} = +1) \geq \epsilon > 0 \tag{4.83a}$$

$$\mu_+(\sigma_0 | \sigma'_{1,k} = \omega'_{\text{special}}, \sigma'_{n(k),N(k)} = -1) \leq -\frac{\epsilon}{2} < 0 \tag{4.83b}$$

It is clear that (4.83a) and (4.83b) together imply the non-quasilocality of $\mu_+ P$, for they show that in an arbitrarily small neighborhood of $\omega'_{\text{special}} \in \{-1,1\}^{\mathbb{Z}}$ (namely, $\mathcal{N}_k \equiv \{\sigma': \sigma'_{1,k} = \omega'_{\text{special}}\}$), there exist open subsets

$$\mathcal{N}_{k,+} = \{\sigma': \sigma'_{1,k} = \omega'_{\text{special}} \text{ and } \sigma'_{n(k),N(k)} = +1\} \tag{4.84a}$$

$$\mathcal{N}_{k,-} = \{\sigma': \sigma'_{1,k} = \omega'_{\text{special}} \text{ and } \sigma'_{n(k),N(k)} = -1\} \tag{4.84b}$$

such that the $(\mu_+ P)$-average value of $E_{\mu_+ P}(\sigma'_0 | \{\sigma'_x\}_{x \neq 0})$ over $\mathcal{N}_{k,+}$ (resp. $\mathcal{N}_{k,-}$) is $\geq \epsilon$ (resp. $\leq -\epsilon/2$). This is incompatible with $E_{\mu_+ P}(\sigma'_0 | \{\sigma'_x\}_{x \neq 0})$ having any continuous ($\equiv$ quasilocal) version.

In order to prove (4.83a) and (4.83b), we shall prove the following intermediate result: there exists $\epsilon > 0$ such that

$$\mu_+(\sigma_0 | \sigma'_{1,k} = \omega'_{\text{special}}) \geq \epsilon \qquad (4.85)$$

for all $k$. This trivially implies (4.83a), by the FKG inequality, for *any* choice of $n$ and $N$. To see that it also implies (4.83b), we use (4.78) with $\mu = \mu_+$ and applied to the functions

$$f_k = \chi(\sigma'_{1,k} = \omega'_{\text{special}}) \qquad (4.86a)$$
$$g_k = \sigma_0 \chi(\sigma'_{1,k} = \omega'_{\text{special}}) \qquad (4.86b)$$

We obtain

$$\lim_{n\to\infty} \mu_+(f_k | \sigma'_{n,N(n)} = -1) = \mu_-(f_k) \qquad (4.87a)$$

$$\lim_{n\to\infty} \mu_+(g_k | \sigma'_{n,N(n)} = -1) = \mu_-(g_k) \qquad (4.87b)$$

Dividing (4.87b) by (4.87a) we get

$$\lim_{n\to\infty} \mu_+(\sigma_0 | \sigma'_{1,k} = \omega'_{\text{special}}, \sigma'_{n,N(n)} = -1) = \mu_-(\sigma_0 | \sigma'_{1,k} = \omega'_{\text{special}}) . \qquad (4.88)$$

Now, by (4.85) and spin-flip symmetry, the RHS is $\leq -\epsilon$.[60] Therefore, for each $k$ there exists an $n(k)$ such that

$$\mu_+(\sigma_0 | \sigma'_{1,k} = \omega'_{\text{special}}, \sigma'_{n(k),N(n(k))} = -1) \leq -\frac{\epsilon}{2} , \qquad (4.89)$$

which is (4.83b).

So now let us prove (4.85) — at low enough temperatures — by a more-or-less standard Peierls argument [169]. Here the contours are defined as the boundaries (in the dual lattice) of regions where the spins differ from the ground-state configuration (that is, all "+" except for the required alternating "−"). In counting the energy of such contours one must subtract the energy of the contours already existing in the ground state (squares surrounding the alternating "−"). After some thought, one concludes that the energy of the contours is at least proportional to $N_v + \frac{1}{2} N_h - 2$, where $N_v$ (resp. $N_h$) is the the number of vertical (resp. horizontal) bonds in the contour. On the other hand, the number of possible contours is even less than that for the unconditioned Ising model. As in the standard Peierls argument, these facts imply that the probability of finding a contour surrounding the origin — that is, of having a "−" at the origin — goes to zero as $\beta$ goes to infinity.

---

[60]If we apply spin-flip symmetry to (4.85), we not only change $\mu_+$ to $\mu_-$ and $\epsilon$ to $-\epsilon$, but must also change $\omega'_{\text{special}}$ to $-\omega'_{\text{special}}$. But this latter is just $\omega'_{\text{special}}$ reflected in the $x_2$-axis (i.e. $x_1 \to -x_1$), and the measures $\mu_+$ and $\mu_-$ are invariant under this reflection.

**Remarks.** 1. In contrast to the RG examples given in Sections 4.1–4.3, here the non-Gibbsianness occurs only for interactions *on* the first-order phase-transition curve, i.e. zero magnetic field. Indeed, Maes and van de Velde [256] have proven that if either $h \neq 0$ or $\beta$ is sufficiently small, the restriction of the two-dimensional Ising model to an axis *is* Gibbsian.

2. It is natural to generalize this example: consider a $d$-dimensional Ising model and a $d'$-dimensional coordinate plane ($1 \leq d' < d$). It seems to be an open question, for all cases other than $(d, d') = (2, 1)$, whether the restricted measure is non-Gibbsian at low temperatures.

### 4.5.3 Fortuin-Kasteleyn Random-Cluster Model

In 1972 Fortuin and Kasteleyn [127] introduced a correlated bond-percolation model which has since become known as the *Fortuin-Kasteleyn random-cluster model*. For a *finite* graph $G = (V, B)$ having vertex set $V$ and edge (or "bond") set $B$, the model is defined as follows: On each bond $b$ there is a variable $n_b$ taking the value 0 ("bond vacant") or 1 ("bond occupied"). The probability of a configuration $\mathbf{n} = \{n_b\}$ is defined to be

$$\text{Prob}(\mathbf{n}) = \text{const} \times p^{\mathcal{N}_1(\mathbf{n})}(1-p)^{\mathcal{N}_0(\mathbf{n})} q^{\mathcal{C}(\mathbf{n})}, \qquad (4.90)$$

where $0 < p < 1$ and $q > 0$ are parameters; here $\mathcal{N}_0(\mathbf{n})$ [resp. $\mathcal{N}_1(\mathbf{n})$] is the number of bonds $b$ with $n_b = 0$ [resp. $n_b = 1$], and $\mathcal{C}(\mathbf{n})$ is the number of "clusters" (i.e. connected components of vertices) in the graph $G_\mathbf{n}$ whose vertex set is $V$ and whose edges are the *occupied* ($n_b = 1$) bonds. For $q = 1$ this model reduces to ordinary (independent) bond percolation, while for integer $q \geq 2$ there are identities relating the random-cluster model to the $q$-state Potts model [127, 125, 109].

Let us now try to formulate the random-cluster model on a *countably infinite* graph $G = (V, B)$ [for example, $V = \mathbb{Z}^d$ and $B = $ nearest-neighbor bonds in $\mathbb{Z}^d$], following the DLR approach. The "lattice" is here $B$, and the configuration space is $\{0,1\}^B$. Let $\Lambda$ be a *finite* subset of $B$, and let $\Lambda^* \subset V$ be the set of all vertices touching at least one bond $b \in \Lambda$. We need to specify the conditional probabilities of $\{n_b\}_{b \in \Lambda}$ given $\{n_{b'}\}_{b' \in B \setminus \Lambda}$. But this is easy, by the same method as for spin systems: we write down the *formal* (meaningless) Boltzmann factor for the infinite lattice, and then drop all terms that don't involve $\{n_b\}_{b \in \Lambda}$. The result is simple: it is

$$\text{Prob}(\{n_b\}_{b \in \Lambda} | \{n_{b'}\}_{b' \in B \setminus \Lambda}) = \text{const}(\{n_{b'}\}_{b' \in B \setminus \Lambda}) \times p^{\mathcal{N}_1(\mathbf{n}_\Lambda)}(1-p)^{\mathcal{N}_0(\mathbf{n}_\Lambda)} q^{\mathcal{C}_{\Lambda^*}(\mathbf{n})}, \quad (4.91)$$

where $\mathcal{N}_0(\mathbf{n}_\Lambda)$ [resp. $\mathcal{N}_1(\mathbf{n}_\Lambda)$] is the number of bonds $b \in \Lambda$ with $n_b = 0$ [resp. $n_b = 1$], while $\mathcal{C}_{\Lambda^*}(\mathbf{n})$ is the number of clusters *containing at least one element of* $\Lambda^*$, in the graph whose edges are the *occupied* ($n_b = 1$) bonds (both those inside and outside $\Lambda$).

It is easy to see that (4.91) defines a specification (i.e. it is consistent for different $\Lambda$). It is also easy to see that the dependence on $\{n_{b'}\}_{b' \in B \setminus \Lambda}$ is only via the set of answers to the following questions: for each pair $x, y \in \Lambda^*$, one wants to know whether $x$ and $y$ can be connected by a path of occupied bonds lying in $B \setminus \Lambda$. Note, however, that the answer to this question could depend on bonds $n_{b'}$ arbitrarily far away from $\Lambda$ (provided that the graph $G$ contains arbitrarily large closed loops). Therefore, for $q \neq 1$, *the specification defined by (4.91) is not quasilocal* (as was previously noted in [5]).

Aizenman, Chayes, Chayes and Newman [5] have proven the existence of the infinite-volume limit for the "Gibbs" measures of the random-cluster model taken with either free ($n_{b'} \equiv 0$) or wired ($n_{b'} \equiv 1$) boundary conditions. However, since the specification (4.91) is not quasilocal (hence not Feller), it is not immediate that these limiting measures $\mu_f$ and $\mu_w$ are indeed consistent with the specification (4.91) [since Proposition 2.22 does not apply], although it seems very plausible. Indeed, it is not clear that there exist *any* measures consistent with the specification (4.91). We therefore pose the following *open question*: Prove that the infinite-volume limit measures taken with free or wired boundary conditions are consistent with the specification (4.91).

Assuming that there do exist measures consistent with the specification (4.91), we can now prove that all these measures are non-quasilocal (hence non-Gibbsian).

**Definition 4.14** *Let $\Omega$ be a metric space. We call a function $f: \Omega \to \mathbb{R}$ strongly discontinuous if every continuous function differs from $f$ on a set having nonempty interior. [In detail: for every $g \in C(\Omega)$, the set $\{\omega: f(\omega) \neq g(\omega)\}$ has nonempty interior.]*

*We call a specification strongly non-Feller if there exists $\Lambda \in \mathcal{S}$ and $f \in C(\Omega)$ such that $\pi_\Lambda f$ is strongly discontinuous.*

A sufficient condition for strong discontinuity of a function $f$ is the following: there exists an $\omega^* \in \Omega$ and an $\epsilon > 0$ such that for every neighborhood $\mathcal{N} \ni \omega^*$ there exist open sets $\mathcal{N}_+, \mathcal{N}_- \subset \mathcal{N}$ such that $\inf_{\omega \in \mathcal{N}_+} f(\omega) - \sup_{\omega \in \mathcal{N}_-} f(\omega) \geq \epsilon$.

It is now easy to prove that the specification (4.91) is strongly non-Feller. To avoid uninteresting graph-theoretic complexities, we prove the theorem for the special case $V = \mathbb{Z}^d$ and $B =$ nearest-neighbor bonds in $\mathbb{Z}^d$. The reader can easily generalize this to a suitable class of countably infinite graphs $G$.

**Proposition 4.15** *Let $q \neq 1$. Then the specification (4.91) for the random-cluster model is strongly non-Feller, when $V = \mathbb{Z}^d$ and $B =$ nearest-neighbor bonds in $\mathbb{Z}^d$.*

PROOF. Let $\Lambda$ be a set containing a single bond $b_0 = \{x_0, x_1\}$, and let $f(\mathbf{n}) = n_{b_0}$. Now let $\omega^*$ be the configuration which sets $n_b = 1$ on parallel rays running from $x_0$ and $x_1$ to infinity, perpendicular to the bond $b_0$, and which sets $n_b = 0$ on all other bonds. Now any neighborhood $\mathcal{N} \ni \omega^*$ (in the product topology) contains the particular neighborhood
$$\mathcal{N}_R = \{\mathbf{n}: \mathbf{n} = \omega^* \text{ on } \Lambda_R\}, \qquad (4.92)$$
where $\Lambda_R$ is the set of all bonds in a square of side $2R+1$ centered at the origin. We then choose $\mathcal{N}_{R,+}$ to be the subset of $\mathcal{N}_R$ in which an occupied bond in $\Lambda_{R+1} \setminus \Lambda_R$ connects the two parallel rays; and we choose $\mathcal{N}_{R,-}$ to be the subset of $\mathcal{N}_R$ in which *all* the bonds in $\Lambda_{R+1} \setminus \Lambda_R$ are vacant (so that the two parallel rays cannot be connected, no matter what happens outside $\Lambda_{R+1}$). It is easy to see that

$$(\pi_{\{b_0\}}f)(\omega) \begin{cases} p & \text{for all } \omega \in \mathcal{N}_{R,+} \\ \frac{p}{p+(1-p)q} & \text{for all } \omega \in \mathcal{N}_{R,-} \end{cases} \quad (4.93)$$

for all $R$. Since $0 < p < 1$ and $q \neq 1$, it follows that $\pi_{\{b_0\}}f$ is strongly discontinuous. ∎

**Proposition 4.16** *(a) Let $\Pi$ be a strongly non-Feller specification, and let $\mu$ be any measure consistent with $\Pi$ that gives nonzero measure to every open set. Then $\mu$ is not consistent with any Feller specification.*

*(b) Let $\Pi$ be a strongly non-Feller specification, and assume further that $\Pi$ is nonnull with respect to an a priori measure $\mu^0$ that gives nonzero measure to every open set. Let $\mu$ be any measure consistent with $\Pi$. Then $\mu$ is not consistent with any Feller specification.*

PROOF. (a) Let $\Lambda \in \mathcal{S}$ and $f \in C(\Omega)$ be such that $\pi_\Lambda f$ is strongly discontinuous. If now $\Pi'$ is a Feller specification, by definition $\pi'_\Lambda f$ is continuous, and therefore differs from $\pi_\Lambda f$ on a set having nonempty interior. But since $\mu$ gives nonzero measure to every open set, $\pi_\Lambda f$ and $\pi'_\Lambda f$ cannot be equal $\mu$-a.e.; $\mu$ cannot be consistent with both $\Pi$ and $\Pi'$.

(b) is an immediate consequence of (a), once we realize that any measure consistent with a nonnull specification (Definition 2.11) must give nonzero measure to every open set. ∎

Since the FK specification (4.91) is clearly nonnull (for $0 < p < 1$ and $q > 0$), we conclude:

**Corollary 4.17** *Let $q \neq 1$, and let $\mu$ be any measure consistent with the FK specification (4.91) [for $V = \mathbb{Z}^d$ and $B = $ nearest-neighbor bonds in $\mathbb{Z}^d$]. Then $\mu$ is not consistent with any Feller ($\equiv$ quasilocal) specification.*

We note that the method used here to prove non-quasilocality is essentially the same as that used in Sections 4.1–4.3 on the RG examples. The only difference is that here we are working with an explicit specification, so that we can prove bounds over the *whole* sets $\mathcal{N}_+$ and $\mathcal{N}_-$; whereas in Sections 4.1–4.3 we were working with the conditional probabilities of a given measure $\mu'$, which are defined only up to modification on $\mu'$-null sets, and therefore we could only prove the bounds over $\mathcal{N}_+$ and $\mathcal{N}_-$ in the $\mu'$-a.e. sense.

Finally, we remark that for integer $q \geq 2$, there exists a *joint* model of interacting Potts spins and bond occupation variables — that is, a model whose state space is $\{1,\ldots,q\}^V \times \{0,1\}^B$ — whose marginals on the spin and bond variables are the Potts and random-cluster models, respectively [109]. This joint model has *local* interactions, so its specification obviously quasilocal. (The only reason it isn't Gibbsian is that there are some exclusions.) The identities relating the joint, Potts and random-cluster models are easily proven in finite volume, but they can presumably be made rigorous in infinite volume by methods like those sketched in Section 4.2, Step 0. If so, then any Gibbs measure of the joint model would produce, upon "decimation" to the bond variables, a non-quasilocal measure (namely, a measure consistent with the random-cluster-model specification).[61] This would then be another example in which "decimation" of a quasilocal measure yields a non-quasilocal measure.

### 4.5.4 Stationary Measures in Nonequilibrium Statistical Mechanics

Consider an infinite-volume lattice system evolving stochastically, in either continuous time or discrete time, according to (quasi)local rules which do *not* satisfy detailed balance. Thus, in continuous time we have in mind an *interacting particle system* [251]: for example, a system of the spin-flip (resp. spin-exchange) type, in which each spin flips (resp. each nearest-neighbor pair of spins exchanges values) independently, at Poisson random times, with rates depending in a (quasi)local way on the other spins. Examples of such dynamics include:

(a) The voter model [251]: independently at each site $x$, at Poisson random times the spin ("voter") at $x$ changes its value to that of a randomly chosen neighbor.

(b) An Ising model with competing dynamics: for example, a mixture of Glauber dynamics for two different temperatures [144], or a mixture of Glauber dynamics for one temperature and Kawasaki dynamics for a different temperature [362]. (The latter model has been considered by Lebowitz and his collaborators in connection with the hydrodynamic limit [238].)

In discrete time we have in mind a *probabilistic cellular automaton* (PCA) [162, 240]: simultaneously at each clock tick, each spin attempts independently to flip, again with rates depending in a (quasi)local way on the other spins. An example is:

(c) The Toom model [347, 240]: each spin changes its value, with probability $p$, to the majority of its northern neighbor, its eastern neighbor, and itself, and with probabilities $(1-p)/2$ to $\pm 1$.

Thus, the PCAs are the discrete-time analogue of the *spin-flip* interacting particle systems.

Lebowitz and Schonmann [246, p. 50] have argued that in both the continuous-time and discrete-time cases, the stationary measure(s) should generally be expected to be non-Gibbsian and indeed non-quasilocal: for "systems maintained in a nonequilibrium state by contacts with outside sources ... [the measures describing] stationary non-equilibrium states cannot be expected to behave in a quasi-Markovian [in our language, quasilocal] way — isolating a part [of the system from the rest] will generally change its behavior drastically."

---

[61]This would probably also give a method for proving that $\mu_f$ and $\mu_w$ are consistent with the random-cluster-model specification, at least for *integer* $q$.

This conjecture has been proven by Lebowitz and Schonmann [246] in the case of the voter model. More precisely, they have proven [246, equation (3.8)] that $i(\delta_+|\nu_\rho) = 0$ where $\nu_\rho$ ($0 < \rho < 1$) is an extremal translation-invariant stationary measure of the voter model in $\mathbb{Z}^d$ ($d \geq 3$). This shows that $\nu_\rho$ is non-Gibbsian (as remarked also in [240]). It is interesting to note that this is the same large-deviations argument employed in the Lebowitz-Maes-Dorlas-van Enter examples (Section 4.4).

Martinelli and Scoppola [259] have given another example of a dynamics in which the stationary measure is non-Gibbsian: again the probability of a region in which *all* the spins are + decays more slowly than exponentially in the volume of the region, so the measure cannot be Gibbsian. However, the Martinelli-Scoppola dynamics is highly non-local — it involves collective flips of arbitrarily large clusters — so perhaps the non-Gibbsianness is not so surprising. (The Martinelli-Scoppola dynamics superficially resembles the Swendsen-Wang [340] dynamics; but in truth the resemblance is not so close, since the stationary measure of the former is non-Gibbsian, while the stationary measure of the latter is the nearest-neighbor Ising model!)

Finally, Maes and Redig [255] have described an (anisotropic) local spin-*exchange* dynamics in which the stationary measure is expected to have non-summable long-range correlations in the "high-noise" regime (i.e. at what ought to correspond to "high temperature"). Such unusual behavior would suggest, though it would not prove, that the stationary measure is non-Gibbsian. The long-range spatial correlations are indicated in this model by a perturbation calculation, but a more general physical intuition seems to be the following: Transport properties for spin-exchange processes are diffusive, and the correlation functions are expected to exhibit slow (power-law) decay in time ("long-time tails"). Now, one expects spatial and temporal correlations to have roughly similar decay — i.e. both exponential or both power-law — except in very special cases such as models satisfying detailed balance. This suggests that spin-exchange processes not satisfying detailed balance should have, quite generally, stationary measures with long-range correlations, and very likely, stationary measures that are non-Gibbsian.

In the PCA models, the probability measure on the space-time histories is the Gibbs measure for a $(d+1)$-dimensional lattice model with interactions which can be expressed in terms of the transition rules of the PCA model [162, 240]. The stationary measure of the PCA model thus corresponds to the restriction of this space-time measure to a $d$-dimensional (equal-time) hyperplane. When the PCA is in the "high-noise" regime — so that the associated $(d+1)$-dimensional equilibrium model is in the Dobrushin-Shlosman high-temperature regime — the stationary measure is known to be unique and Gibbsian [240]. (A similar theorem has recently been proven also for continuous-time *spin-flip* systems [257].) However, by analogy with the Schonmann example (Section 4.5.2), one may suspect that in the "non-ergodic" (phase-transition) regime of the PCA model — where the stationary measure is not unique — each stationary measure would typically be non-Gibbsian. In particular, one may conjecture that this is so for the Toom model. We thus suspect that Liggett's conjecture [251, p. 224], to the effect that every translation-invariant finite-range dynamics with *strictly positive* rates has a Gibbsian stationary measure, is most likely false.

**Remark.** The foregoing considerations are for rates that do *not* satisfy detailed balance. If the rates satisfy detailed balance, then one expects all the stationary measures to be Gibbsian (for an explicit Gibbsian specification that is easy to write down given the rates); however, this has not yet been proven rigorously even in the Glauber dynamics for the nearest-neighbor Ising model in dimension $d \geq 3$ [251, Problem IV.7.1].

Finally, let us quote a result of Künsch [229] for continuous-time local spin-flip processes with strictly positive rates: if there exists a translation-invariant stationary measure which is Gibbsian for some (absolutely summable) interaction, then every other translation-invariant stationary measure must be Gibbsian for the same interaction.

### 4.5.5 Comparison of Methods for Proving Non-Gibbsianness

Any theorem of the form "every Gibbs measure has the property $\mathcal{P}$" provides a method for proving non-Gibbsianness via the contrapositive: a measure not having the property $\mathcal{P}$ must be non-Gibbsian. We have seen four properties of this sort:

(i) A Gibbs measure (for an absolutely summable interaction) must be uniformly non-null. This is a consequence of the "easy half" of the Gibbs representation theorem [Theorem 2.12 (a) $\Longrightarrow$ (b)].

(ii) A Gibbs measure (for a uniformly convergent interaction) must be quasilocal [Theorem 2.10].

(iii) A measure can be Gibbsian for at most one (uniformly convergent, continuous) interaction, up to "physical equivalence" [Corollary 2.18].

(iv) Translation-invariant Gibbs measures (for translation-invariant absolutely summable interactions) have "good" large-deviation properties: the probability that spins in a certain region fluctuate into a configuration characteristic of another translation-invariant measure decreases exponentially in the volume of the region, *except* if this other measure is also Gibbsian for the same (absolutely summable) interaction. In precise mathematical terms: a translation-invariant measure $\mu$ has zero relative entropy density respect to another translation-invariant measure $\nu$ which is Gibbsian for an interaction $\Phi$, if and only if $\mu$ is also Gibbsian for the same interaction $\Phi$. This is one of the consequences of the discussion of Section 2.6.6. It is also closely related to the strict convexity of the pressure [Proposition 2.59].

For each of these conditions, we have seen examples in which the non-Gibbsianness is proven by its violation:

(i) *Lack of uniform nonnullness.* This has two manifestations: A measure can be nonnull but not uniformly so (see Definition 2.11). This typically means that the Hamiltonians are unbounded functions and one cannot use the formalism developed for absolutely summable interactions. This is the generic situation for unbounded-spin models, and it gets delicate for infinite-range interactions. In these cases, often the notion of Gibbsianness can be preserved if one excludes "by hand" problematic configurations [245, 60]. On the other hand, the measure may fail to be nonnull, which means that some cylinder sets have zero measure. This is a rather simple case of non-Gibbsianness in which the Gibbsianness can be restored by allowing hard-core interactions or working on a more restricted configuration space (see for example [313]). We mention that, in the setting of *complex* interactions, there are examples of Gibbsian measures that after one renormalization step remain quasilocal but lose nonnullness [14].

(ii) *Violation of quasilocality.* Most of the cases of pathological renormalization transformations analyzed above (Sections 4.1–4.3 and 4.5.2) fall into this category. This phenomenon appears when there are some "hidden spins" that transmit information from arbitrarily far away even if the "non-hidden" spins are fixed. In the renormalization transformations the "hidden variables" are the fluctuations of the original or internal spins that remain once the block spins are fixed. In Schonmann's example (Section 4.5.2), the "hidden variables" are all the spins outside the $x$-axis, which are "hidden" by the process of restriction.

(iii) *Threatened violation of uniqueness.* We used this method to study the "trivial" examples of non-Gibbsianness discussed in Section 4.5.1. Consider a finite or countable family of different (non-physically-equivalent) interactions and pick for each one an ergodic translation-invariant Gibbs measure. Then a nontrivial convex combination of these measures cannot be Gibbsian for any (uniformly convergent, continuous) interaction, because if it were, then each of the original measures would be a Gibbs measure also for this new interaction, violating uniqueness. In the case of a finite single-spin space, this method also proves non-quasilocality.

(iv) *Wrong large deviation properties.* There seem to be two rather different types of "bad" large-deviation properties:

($\alpha$) Sub-exponential decay for events whose probability "should" decay exponentially in the volume. This applies to the sign field of the (an)harmonic crystal (Section 4.4), and the stationary measures for the voter and Martinelli-Scoppola models mentioned in Section 4.5.4. Here one shows that the probability of *all* the spins in a large region becoming $+$ decays sub-exponentially in the volume of the region; this is incompatible with being Gibbsian for any absolutely summable interaction. In other words, one shows that the measure $\mu$ satisfies $i(\delta_+|\mu) = 0$, where $\delta_+$ is the delta-measure concentrated on the all-$+$ configuration. As this measure is obviously non-Gibbsian (it is *not* nonnull!), neither is $\mu$.

($\beta$) Exponential decay for events whose probability "should" decay sub-exponentially. The original proof of Schonmann's example [318] is based on an argument of this kind. Here one shows that, in the + phase, the probability of having a net negative magnetization in a large region decays exponentially in the volume of the region. In other words, the measures obtained via "+" and "−" boundary conditions have a strictly positive relative entropy. If either of these measures were Gibbsian (for an absolutely summable interaction), the other would have to be Gibbsian for the same interaction (because they differ only by boundary conditions); but then the relative entropy would have to be zero (Theorem 2.66). Therefore, they cannot be Gibbsian.

Often we would like to prove not only that a measure is non-Gibbsian, but also that it is non-quasilocal (which is stronger). In nearly all cases we have done this "by hand": that is, by proving bounds on the conditional probabilities which are incompatible with their having any quasilocal version (see Sections 4.1–4.3 and 4.5.2). In only one case were we able to prove non-quasilocality by an abstract "trick": this was the "trivial" convex-combination example (Section 4.5.1), where we used method (iii) above. It would be interesting to have available other methods for proving non-quasilocality.

### 4.5.6 Are "Most" Measures Non-Gibbsian?

The traditional belief among physicists (including ourselves until recently) is that all (or nearly all) physically interesting measures are Gibbsian. Indeed, this belief is so much taken for granted[62] that it is rarely stated explicitly.[63] The profound message of Israel's pioneering work [207], and of the examples given here, is that this traditional belief is false: *many physically interesting measures are non-Gibbsian*. In fact, we now suspect that Gibbsianness should be considered to be the exception rather than the rule — that, in some sense, *most* measures are non-Gibbsian.

It is therefore of at least mathematical interest to study the set $\mathcal{G} \equiv \bigcup_{\Phi \in \mathcal{B}^1} \mathcal{G}_{inv}(\Pi^\Phi)$ of all translation-invariant measures which are Gibbsian for some translation-invariant absolutely summable continuous interaction. Is $\mathcal{G}$ a "big" or a "small" subset of the space $M_{+1,inv}(\Omega)$ of all translation-invariant measures?

---

[62]There are many examples of this in the physics literature: see, for example, [54, 49].

[63]One exception is the recent statement by a noted mathematical physicist that "every good random field is Gibbsian" [324]. In a similar vein, a mathematician says: "the Gibbsian form of local conditional distributions is a rather weak condition, but it is difficult to check it." [229, p. 410] A related though somewhat weaker intuition can be found in a well-known monograph on interacting particle systems: "Is it true that every translation invariant strictly positive spin system on $\mathbb{Z}^d$ with finite range has an invariant measure which is a Gibbs state? This is plausible ... [because] the strict positivity of the rates should imply that an invariant measure is somewhat smooth." [251, p. 224] On this same conjecture, another mathematician says: "We couldn't prove in general the existence of a stationary Gibbs measure, although this is very likely to hold." [229, p. 408] As discussed in Section 4.5.4, this conjecture is still an open problem, but there is good reason to suspect that it is false. (These examples, together with those of the preceding footnote, illustrate the difference between physicists and mathematicians: both often have erroneous intuitions, but the mathematicians state them explicitly.)

It is a "big" set in a very weak sense, namely that of being dense in the weak topology. In fact, the Gibbs measures for *finite-range* continuous interactions are already dense:

**Proposition 4.18** *Assume that the single-spin space $\Omega_0$ is a compact metric space, and that the* a priori *single-spin measure $\mu_x^0$ gives nonzero measure to every open set of $\Omega_0$. Then*
$$\mathcal{G}_{finite} \equiv \bigcup_{\Phi \in \mathcal{B}_{finite}} \mathcal{G}_{inv}(\Pi^\Phi)$$
*is dense in $M_{+1,inv}(\Omega)$ in the weak topology.*

PROOF. The proof goes in three steps: First, the ergodic measures of finite entropy density (relative to $\mu^0$) are dense in $M_{+1,inv}(\Omega)$ [Proposition 2.61(e)]. Secondly, Israel [206] has shown, using the Bishop-Phelps theorem, that each ergodic measure of finite entropy density is an (extremal) equilibrium measure for some interaction $\Phi^* \in \mathcal{B}^0$ (see item 2 in Section 2.6.7). Finally, the finite-range interactions form a dense subset $\mathcal{B}_{finite} \subset \mathcal{B}^0$; and it follows from a theorem of Lanford and Robinson [231] (see also Sokal [332]) that every extremal equilibrium measure for $\Phi^* \in \mathcal{B}^0$ can be approximated in the weak topology by equilibrium measures for interactions $\Phi_n \in \mathcal{B}_{finite}$ with $\|\Phi_n - \Phi^*\|_{\mathcal{B}^0} \to 0$. ∎

We emphasize that density in the weak topology is an extremely weak property: it means only that an arbitrary measure $\mu \in M_{+1,inv}(\Omega)$ can be *approximated* arbitrary closely, with regard to any *finite* family of *local* observables, by a measure in $\mathcal{G}_{finite}$. In particular, the long-range-order properties of the approximating measures can be totally different from those of the limiting measure $\mu$. Thus, Proposition 4.18 is very far from saying that "most" measures are Gibbsian.

In a more profound sense we expect that $\mathcal{G}$ is in fact a rather "small" subset of $M_{+1,inv}(\Omega)$. For example, we conjecture:

**Conjecture 4.19** (a) *$\mathcal{G}$ is a set of first Baire category in $M_{+1,inv}(\Omega)$. [That is, $\mathcal{G}$ is a countable union of sets which are nowhere dense in $M_{+1,inv}(\Omega)$.]*

(b) *$\mathcal{G} \cap \mathrm{ex}M_{+1,inv}(\Omega)$ is a set of first Baire category in $\mathrm{ex}M_{+1,inv}(\Omega)$. [Here "ex" denotes the extreme points, i.e. the ergodic measures.]*

First Baire category is a classic notion of "smallness" in topology [285].

We can make some small steps toward proving Conjecture 4.19(a):

**Proposition 4.20** *$\mathcal{G}$ has empty interior.*

**Proposition 4.21** *Assume that the single-spin space $\Omega_0$ is a compact metric space, and that the* a priori *single-spin measure $\mu_x^0$ gives nonzero measure to every open set of $\Omega_0$. Let $S$ be a compact subset of $\mathcal{B}^0$, and let $\mathcal{E}_S$ be the set of equilibrium measures for interactions in $S$. Then $\mathcal{E}_S$ is a compact subset of $M_{+1,inv}(\Omega)$.*

**Corollary 4.22** *Assume that the single-spin space $\Omega_0$ is a compact metric space, and that the* a priori *single-spin measure $\mu_x^0$ gives nonzero measure to every open set of $\Omega_0$. If $S$ is a $\sigma$-compact subset of $\mathcal{B}^0$, with $S \subset \mathcal{B}^1$, then $\mathcal{E}_S = \mathcal{G}_S \equiv \bigcup_{\Phi \in S} \mathcal{G}_{inv}(\Pi^\Phi)$ is $\sigma$-compact and of first Baire category in $M_{+1,inv}(\Omega)$. In particular, this occurs for $S = \mathcal{B}_h$ with $h \gtrapprox 1$.*

PROOF OF PROPOSITION 4.20.   Let $\mu \in \mathcal{G}$ and $\nu \in M_{+1,inv} \setminus \mathcal{G}$. Then, by Proposition 2.48(b), $(1-\lambda)\mu + \lambda\nu \notin \mathcal{G}$ for $0 < \lambda \le 1$. But $(1-\lambda)\mu + \lambda\nu \to \mu$ weakly as $\lambda \downarrow 0$. Hence $\mathcal{G}$ cannot contain any open neighborhood of $\mu$. [In this proof we could equally well have taken $\nu$ to be a Gibbs measure for an interaction not physically equivalent to the one for which $\mu$ is Gibbsian, and then apply Proposition 2.48(b) and the Griffiths-Ruelle theorem.] ■

PROOF OF PROPOSITION 4.21.   $M_{+1,inv}(\Omega)$ is compact, so we need only show that $\mathcal{E}_S$ is closed. Let $\mu_n$ be an equilibrium measure for $\Phi_n \in S$, with $\mu_n \to \mu$ weakly. Then, since $S$ is compact, there exists a subsequence $\Phi_{n_i}$ that converges (in $\mathcal{B}^0$ norm) to some $\Phi \in S$. But then $\mu$ is an equilibrium measure for $\Phi$. ■

PROOF OF PROPOSITION 4.22.   The first statement is an immediate consequence of Propositions 4.20 and 4.21. The second statement follows from Proposition 2.39(b). ■

We thank S.R.S. Varadhan for suggesting these latter results and sketching the proofs.

# 5 Discussion

## 5.1 Numerically Observed Discontinuities of the RG Map

### 5.1.1 Statement of the Problem

In several Monte Carlo renormalization group (MCRG) studies [33, 233, 72, 163], it has been found that the numerically computed renormalization transformation $\mathcal{R}: H \mapsto H'$ is *discontinuous* at a first-order phase-transition surface.[64] However, this behavior is

---

[64]The models in which this behavior has been (at least tentatively) observed include the two-dimensional Ising model at low temperature [72], the 10-state Potts model in two dimensions [72], the 3-state Potts model in three dimensions [33], the $Z_2$ lattice gauge theory in four dimensions [163] and the $U(1)$ lattice gauge theory in four dimensions [233, 72]. However, in a more recent study of the two-dimensional Ising model at low temperature [165], the observed discontinuity was always less than the estimated truncation error, and it decreased as more terms were included in the renormalized Hamiltonian; this was interpreted as evidence against a discontinuity in the exact renormalization map.

*rigorously excluded* by our Second Fundamental Theorem (Theorem 3.6). In this section we would like to offer our interpretation of the numerically observed discontinuities.

A MCRG study [337, 338] proceeds as follows: We choose an original Hamiltonian $H$, and generate a long sequence of random samples $\omega_1, \omega_2, \ldots$ from the Gibbs measure $\mu = \text{const} \times e^{-H}$ using some Monte Carlo procedure. On each of these "original-spin" configurations $\omega_i$ we apply the renormalization map $T$ to generate the corresponding block-spin configuration $\omega_i'$. In this way we have generated a random sample $\omega_1', \omega_2', \ldots$ from the renormalized measure $\mu' = \mu T$. It is now *assumed* that $\mu'$ is the Gibbs measure for some renormalized Hamiltonian $H'$ belonging to a fixed finite-parameter family $H(\lambda_1, \ldots, \lambda_N)$, and some statistical method [339, 164, 9] is employed to estimate the unknown parameters $\lambda_1, \ldots, \lambda_N$.

Such a procedure has three sources of error:

1) Statistical error arising from the finite Monte Carlo sample.

2) Systematic error arising from the finite lattice size. (We take the point of view that our goal is to learn about the behavior of the *infinite-volume* system.)

3) Systematic error arising from truncation of the renormalized Hamiltonian: $\mu'$ may not be (in fact, in almost all cases is not) a Gibbs measure for any Hamiltonian in the assumed $N$-parameter family. We include here the possibility — studied in detail in Section 4 — that $\mu'$ is not the Gibbs measure for *any* reasonable Hamiltonian.

It is useful to study these three sources of error *separately*. In particular, we would like to study the problem of truncation of the renormalized Hamiltonian, independently of the problems of statistical and finite-size errors. Therefore, we begin by formulating an *idealized model* of the parameter-estimation problem in which we assume that the experimenter knows *exactly* the expectation values of an appropriate set of observables (to be specified later) in the *infinite-volume* renormalized measure $\mu'$. This idealized situation can be approximated to arbitrary accuracy with sufficient computer time, by making long Monte Carlo runs on large systems. (In principle we should then discuss the *stability* of our theory relative to small statistical or finite-size errors. But we feel that our considerations are still too preliminary to justify entering into such technicalities.)

### 5.1.2 An Idealized Model of Parameter Estimation

Let us first consider the parameter-estimation problem in a general probabilistic (= statistical-mechanical) context, without regard (for the moment) to the renormalization-group application. Let, therefore, $(\Omega, \mathcal{F})$ be an arbitrary measurable space, let F be some family of probability measures on $(\Omega, \mathcal{F})$, and let $\rho$ be another probability measure on $(\Omega, \mathcal{F})$. We wish to find the measure in F which is in some sense "closest to" (or "best approximates") $\rho$. How should we define "closeness"? Any definition is, of course, somewhat arbitrary, but we claim that the following definition is very natural:

The measure in F closest to $\rho$, denoted $\rho^{\mathsf{F}}$, is the one which minimizes the relative entropy $I(\rho|\cdot)$, assuming that a minimizer with finite relative entropy exists (it may or may not be unique).

Note that the unknown measure is taken here as the reference measure (second argument) in the relative entropy.[65] In support of this definition, we cite the following properties:

1) If $\rho \in \mathsf{F}$, then $\rho^{\mathsf{F}} = \rho$, uniquely. This is a rather trivial property, but it is at least a *necessary* condition for any reasonable definition of "closeness".

2) Suppose that one generates a large random sample from $\rho$, and constructs maximum-likelihood estimates [325] based on the (false) assumption that the sample arose from some measure in F. In the large-sample limit, this maximum-likelihood estimate will converge to $\rho^{\mathsf{F}}$. This can be proven under suitable technical hypotheses [196], but it is easy to see intuitively why it is true: the relative entropy $I(\rho|\nu)$ is, up to an additive constant, precisely minus the mean (under the true measure $\rho$) of the log likelihood function:

$$\begin{aligned} I(\rho|\nu) &= \int d\rho \, \log \frac{d\rho}{d\nu} \\ &= \text{const} - \int d\rho \, \log \text{``}d\nu\text{''} \, , \end{aligned} \quad (5.1)$$

so maximizing the likelihood is equivalent to minimizing the relative entropy. Thus, $\rho^{\mathsf{F}}$ is the estimate that would be generated by an experimenter possessing an *infinite* random sample from $\rho$ and using the *optimal* estimation method (namely, maximum likelihood).[66] (Note also that the maximum-likelihood estimate for a *finite* sample $\omega_1, \ldots, \omega_n$ is the measure in F closest to the *empirical* measure $L_n \equiv n^{-1} \sum_{i=1}^{n} \delta_{\omega_i}$ for the given sample. This close relation between maximum-likelihood estimation and relative entropy has been noticed by previous authors.)

Suppose now that the set F is of the Boltzmann-Gibbs form (= exponential family)

$$\mathsf{F} = \left\{ \nu_\lambda \equiv Z(\lambda)^{-1} \exp\left[-\sum_{i=1}^{N} \lambda_i H_i\right] \mu^0 \colon \lambda \in \mathbb{R}^N \right\} \quad (5.2)$$

for some specified family $H_1, \ldots, H_N$ and *a priori* measure $\mu^0$. We can assume without loss of generality that the functions $1, H_1, \ldots, H_N$ are linearly independent ($\mu^0$-a.e.).

---

[65]This follows Čencov [61]. (Note, however, that Čencov's notation for the arguments of $I(\cdot|\cdot)$ is the reverse of ours.) By contrast, Csiszar [68, 69] considers the quite different problem in which the unknown measure is the first argument in the relative entropy.

[66]This assertion is perhaps somewhat misleading: The maximum-likelihood method is optimal as regards *statistical* errors (in the large-sample limit) [325], while here we are concerned with the *systematic* errors due to truncation. Indeed, the problem here is to *define* what we mean by the "optimal" truncation. In any case, we claim that maximum-likelihood estimation is a sensible idealized model of what a good experimenter would actually do if he/she could.

Then the relative entropy

$$I(\rho|\nu_\lambda) \;=\; \log Z(\lambda) + \sum_{i=1}^{N} \lambda_i \int H_i \, d\rho \;+\; \text{const} \tag{5.3}$$

is a strictly convex function of $\lambda$; in particular, the measure $\rho^{\mathsf{F}}$ is unique if it exists at all[67], and it is defined by the conditions

$$\langle H_i \rangle_{\nu_\lambda} \;=\; \langle H_i \rangle_\rho \quad \text{for all } i = 1, \ldots, N \;. \tag{5.4}$$

The foregoing theory is adequate for parameter estimation in *finite-volume* systems; but for infinite-volume systems it is inapplicable, because the relative entropy of two translation-invariant measures is in nearly all cases $+\infty$. The problem here is that, as discussed in Section 2.6, the relative entropy in volume $\Lambda$ typically grows proportionally to the volume (unless the two measures happen to be Gibbs measures for the same interaction). This volume factor is uninteresting in the present context, because it does not depend on the parameters $\lambda$ over which we want to optimize. Therefore, it makes sense to just divide out this volume factor and minimize the relative entropy *density*. That is, if $\mathsf{F} \subset M_{+1,inv}(\Omega)$ and $\rho \in M_{+1,inv}(\Omega)$, we define:

> The measure in $\mathsf{F}$ closest to $\rho$, denoted $\rho^{\mathsf{F}}$, is the one which minimizes the relative entropy density $i(\rho|\,\cdot\,)$, assuming that these relative entropy densities are well-defined and that a minimizer with finite relative entropy density exists (it may or may not be unique).

Suppose now that $\mathsf{F}$ is the set of translation-invariant Gibbs measures for interactions $\Phi \in V$ (and *a priori* measure $\mu^0$), where $V = \mathrm{span}(\Phi_1, \ldots, \Phi_N)$ is some specified finite-dimensional linear subspace of $\mathcal{B}^1$. Then, from (2.96) we have

$$\begin{aligned} i(\rho|\nu) &= p(-f_\Phi|\mu^0) + \int f_\Phi \, d\rho + i(\rho|\mu^0) \\ &\equiv F_\rho(\Phi) \end{aligned} \tag{5.5}$$

whenever $\nu$ is a Gibbs measure for $\Phi$. Thus, $i(\rho|\nu)$ depends on $\nu$ only via the interaction $\Phi$; in fact, $F_\rho(\Phi)$ is precisely the amount by which the pair $(\rho, \Phi)$ fails to satisfy the variational principle. Therefore, if one Gibbs measure for $\Phi$ happens to minimize $i(\rho|\,\cdot\,)$, then all Gibbs measures for $\Phi$ do so. So the *measure* in $\mathsf{F}$ closest to $\rho$ may not be unique. Nevertheless, the corresponding *interaction* is necessarily unique (modulo physical equivalence) if it exists at all[68], because $F_\rho$ is strictly convex on $V \subset \mathcal{B}^1$

---

[67]The minimizer $\rho^{\mathsf{F}}$ could fail to exist: Consider, for example, $\Omega = \{-1, 1\}$, $\mu^0 = \frac{1}{2}(\delta_{-1} + \delta_{+1})$, $\rho = \delta_{-1}$, $N = 1$ and $H_1(\omega) = \omega$. Then the minimum is "at $\lambda = +\infty$"; there is no minimizer at finite $\lambda$.

[68]Again, a minimizer could fail to exist, if the minimum is "at infinity". This occurs, for example, if $\rho$ is a *ground-state* measure for some interaction $\Phi \in V$: then $F_\rho(\beta \Phi) \to 0$ as $\beta \to +\infty$.

(Proposition 2.59). This is good, because it is after all the interaction that we would like to estimate. Now, an interaction $\Phi^*$ minimizes $F_\rho\!\restriction\! V$ if and only if $\rho\!\restriction\! f[V]$ is a tangent functional at $f_{\Phi^*}$ to the pressure restricted to $f[V]$. [Here $f[V]$ denotes the image of $V$ under the map $\Psi \mapsto f_\Psi$; it is a linear subspace of $C(\Omega)$.] Now, by the Hahn-Banach theorem[69], every tangent functional to $p\!\restriction\! f[V]$ can be extended to a tangent functional to $p$, i.e. to an equilibrium (= Gibbs) measure. It follows that an interaction $\Phi^*$ minimizes $F_\rho\!\restriction\! V$ if and only if there exists a translation-invariant Gibbs measure $\nu$ for $\Phi^*$ such that
$$\langle f_{\Phi_i}\rangle_\nu \;=\; \langle f_{\Phi_i}\rangle_\rho \quad \text{for all } i=1,\ldots,N\;. \tag{5.6}$$

What happens if we consider larger and larger subspaces of interactions? Let $V_1 \subset V_2 \subset \ldots$ be an increasing sequence of finite-dimensional linear subspaces of $\mathcal{B}^1$, whose union is dense in $\mathcal{B}^1$. Let $\Phi_n^*$ be the interaction in $V_n$ that minimizes $F_\rho\!\restriction\! V_n$. Then it is natural to conjecture the following:

**Conjecture 5.1** *(i) If $\rho$ is a Gibbs measure for some interaction $\Phi \in \mathcal{B}^1$, then $\Phi_n^* \to \Phi$ in $\mathcal{B}^1$ norm as $n \to \infty$.*

*(ii) If $\rho$ is not a Gibbs measure for any interaction in $\mathcal{B}^1$, then $\|\Phi_n^*\|_{\mathcal{B}^1} \to \infty$ as $n \to \infty$.*

We are not able to prove this much (and we suspect that it may not be true without additional hypotheses). Regarding conjecture (i), what we can prove is the following:

**Proposition 5.2** *Let $\rho$ be an ergodic translation-invariant measure of finite entropy density (relative to $\mu^0$); and let $V_1 \subset V_2 \subset \ldots$ be an increasing sequence of subsets of $\mathcal{B}^0$, whose union is dense in $\mathcal{B}^0$. Then:*

*(a) There exists an interaction $\widehat{\Phi} \in \mathcal{B}^0$ (not necessarily in $\mathcal{B}^1$!) for which $\rho$ is an equilibrium measure.*

*(b) Let $\widehat{\Phi}$ be any interaction in $\mathcal{B}^0$ for which $\rho$ is an equilibrium measure. Then there exists a sequence $\widehat{\Phi}_n \in V_n$ which converges to $\widehat{\Phi}$ in $\mathcal{B}^0$ norm. If, in addition, $\widehat{\Phi}$ belongs to some space $\mathcal{B}_h \subset \mathcal{B}^0$ and $\cup_{n=1}^\infty V_n$ is dense in $\mathcal{B}_h$ (in $\mathcal{B}_h$ norm), then there exists a sequence $\widehat{\Phi}_n \in V_n$ which converges to $\widehat{\Phi}$ in $\mathcal{B}_h$ norm.*

*(c) Let $\widehat{\Phi}$ be any interaction in $\mathcal{B}^0$ for which $\rho$ is an equilibrium measure, and let $(\widehat{\Phi}_n)$ be any sequence converging to $\widehat{\Phi}$ in $\mathcal{B}^0$ norm. Then $F_\rho(\widehat{\Phi}_n) \to 0$. [In particular, we have $\lim_{n\to\infty} \inf_{\Phi \in V_n} F_\rho(\Phi) = 0$.]*

*(d) Conversely, let $(\widehat{\Phi}_n)$ be any sequence for which $F_\rho(\widehat{\Phi}_n) \to 0$ and which converges in $\mathcal{B}^0$ norm, say to $\widehat{\Phi}_\infty$. Then $\rho$ is an equilibrium measure for $\widehat{\Phi}_\infty$. In particular, if $(\|\widehat{\Phi}_n\|_{\mathcal{B}^1})$ is bounded, then $\widehat{\Phi}_\infty \in \mathcal{B}^1$ and $\rho$ is a Gibbs measure for $\widehat{\Phi}_\infty$.*

---

[69]The required version of the Hahn-Banach theorem [313, p. 157, A.3.2] can be deduced easily from the separating-hyperplane version ([315, p. 46, Theorem II.3.1] or [310, p. 58, Theorem 3.4(a)]) by considering epigraphs.

This shows that if $\rho$ is a Gibbs measure for $\Phi \in \mathcal{B}^1$, then there exists a sequence $\widehat{\Phi}_n \in V_n$ of *approximate* minimizers which converges (in $\mathcal{B}^1$ norm) to $\Phi$. Unfortunately, there is no guarantee that the *exact* minimizers $\Phi_n^*$ (if such exist) converge to $\Phi$; they might fail to converge, or they might converge instead to some interaction $\widehat{\Phi} \in \mathcal{B}^0 \setminus \mathcal{B}^1$ for which $\rho$ is an equilibrium (but not Gibbs!) measure. It is an important *open problem* to find conditions under which conjecture (i), or something like it, can be proven.

**Remark.** It is certainly possible for a sequence $\widehat{\Phi}_n$ of approximate minimizers of $F_\rho$ to fail to converge even when $\rho$ is a Gibbs measure for some interaction in $\mathcal{B}^1$. Consider, for example, an Ising model: take $\rho = \mu^0 =$ product measure, and let $\widehat{\Phi}_n$ be a ferromagnetic two-body interaction $n^{-d} f(n^{-1}(x-y))\sigma_x \sigma_y$, where $f$ is some fixed nonnegative smooth function with $0 < \int f(x)\, d^d x \le 1$. In such a situation, $F_\rho(\widehat{\Phi}_n) = F_{\mu_0}(\widehat{\Phi}_n) = p(-f_{\widehat{\Phi}_n}|\mu^0)$. But then $F_{\mu_0}(\widehat{\Phi}_n) \to 0$ by the Lebowitz-Penrose theorem [241] [345, Appendix C], which tells us that the so-called "Kac limit" $\lim_{n\to\infty} p(-f_{\Phi_n}|\mu^0)$ is the high-temperature mean-field pressure, which is 0 in our normalization. On the other hand it is obvious that $\mu_0$ is the Gibbs measure for the absolutely summable interaction $\Phi = 0$, and that nevertheless the interactions $\widehat{\Phi}_n$ do not converge in $\mathcal{B}^0$ or any of its subspaces $\mathcal{B}_h$. [Moreover, the measures $\mu_n$ defined by the interactions $\widehat{\Phi}_n$ converge to the product measure $\mu_0$.] We fear that something similar could happen also for the exact minimizers $\Phi_n^*$, unless the spaces $V_n$ are very carefully chosen.

On the other hand, we can almost prove conjecture (ii):

**Proposition 5.3** *Let $h\colon \mathcal{S} \to [1, \infty)$ be a translation-invariant weight function, and let $\rho$ be a translation-invariant measure which is* not *an equilibrium measure for any interaction in $\mathcal{B}_h$. Let $(\widehat{\Phi}_n)$ be any sequence in $\mathcal{B}_h$ for which $F_\rho(\widehat{\Phi}_n) \to 0$. Then at least one (and possibly both) of the following two statements is true:*

*(a)* $\lim_{n\to\infty} \|\widehat{\Phi}_n\|_{\mathcal{B}_h} = \infty$.

*(b)* $(\widehat{\Phi}_n)$ *does not converge in $\mathcal{B}^0$ norm.*

*Moreover, if the single-spin space $\Omega_0$ is finite and $h \gtrsim 1$, then statement (a) is always true.*

This comes very close to proving conjecture (ii): the only possible escape clause is that the sequence $(\widehat{\Phi}_n)$ might have no limit at all (even in $\mathcal{B}^0$ norm), even though $(\|\widehat{\Phi}_n\|_{\mathcal{B}^1})$ is bounded. This can happen, for example, if the interactions become longer-and-longer-ranged but with bounded total strength.[70] In such a case one must have $\|\widehat{\Phi}_n\|_{\mathcal{B}_h} \to \infty$ in every space $\mathcal{B}_h$ of "short-range interactions" (Definition 2.38), e.g. $h(X) = \operatorname{diam}(X)^\epsilon$ with $\epsilon > 0$.

---

[70]This situation is reminiscent of "mean-field-like" interactions. However, in such situations one usually expects (and in some cases can prove, as in the previous remark) that the limiting measure $\rho$ *is* Gibbsian for some interaction $\Phi \in \mathcal{B}^1$ which has picked up a magnetic field. We wish to thank Bob Griffiths and Bob Swendsen for a discussion of this point.

PROOF OF PROPOSITION 5.2.   (a) is a special case of a theorem of Israel [206, Theorem V.2.2(a)]. (b) follows from the density of $\cup_{n=1}^{\infty} V_n$ in $\mathcal{B}^0$ (or $\mathcal{B}_h$). (c) follows from the (Lipschitz) continuity of the function $F_\rho$ in $\mathcal{B}^0$ norm. (d) is also a consequence of the continuity of $F_\rho$, since the hypotheses imply that $F_\rho(\widehat{\Phi}_\infty) = 0$. The last statement is a consequence of Proposition 2.39(a) applied to $\mathcal{B}_h = \mathcal{B}^1$.   ∎

PROOF OF PROPOSITION 5.3.   Suppose that $(\widehat{\Phi}_n)$ converges in $\mathcal{B}^0$ norm to $\widehat{\Phi}_\infty$. Then $F_\rho(\widehat{\Phi}_\infty) = 0$, so $\rho$ is an equilibrium measure for $\widehat{\Phi}_\infty$. By hypothesis this means that $\widehat{\Phi}_\infty \notin \mathcal{B}_h$, i.e. $\|\widehat{\Phi}_\infty\|_{\mathcal{B}_h} = \infty$. Now assume that $\|\widehat{\Phi}_n\|_{\mathcal{B}_h} \not\to \infty$; then there is a subsequence of $(\widehat{\Phi}_n)$ on which the $\mathcal{B}_h$ norm is bounded, say by $M$; but by Proposition 2.39(a) this implies that $\|\widehat{\Phi}_\infty\|_{\mathcal{B}_h} \leq M$, a contradiction. This proves that either (a) or (b) [or both] must be true.

Finally, suppose that the single-spin space is finite, that $h \gg 1$, and that there is a subsequence of $(\widehat{\Phi}_n)$ on which the $\mathcal{B}_h$ norm is bounded, say by $M$. Then by Proposition 2.39(b) there exists a sub-subsequence which converges in $\mathcal{B}^0$ norm to some $\widehat{\Phi}_\infty$ with $\|\widehat{\Phi}_\infty\|_{\mathcal{B}_h} \leq M$; and $\rho$ is an equilibrium measure for $\widehat{\Phi}_\infty$; but this contradicts the hypothesis of the proposition.   ∎

**Remarks.** 1. Similar ideas appear in the work of Hugenholtz [199].

2. A partially alternate proof of the second half of Proposition 5.3, when $\rho$ is ergodic (or a *finite* convex combination of ergodic measures), goes as follows: By the Bishop-Phelps theorem [206, Corollary V.2.1] there exists $\widetilde{\Phi}_n \in \mathcal{B}^0$ with $\|\widetilde{\Phi}_n - \widehat{\Phi}_n\|_{\mathcal{B}^0} \leq C_\rho F_\rho(\widehat{\Phi}_n)$ such that $\rho$ is an equilibrium measure for $\widetilde{\Phi}_n$ [if $\rho = \sum_{i=1}^{m} \alpha_i \rho_i$ is the ergodic decomposition of $\rho$, then $C_\rho = (2 \min_{1 \leq i \leq m} \alpha_i)^{-1}$]. Now assume that $\|\widehat{\Phi}_n\|_{\mathcal{B}_h} \not\to \infty$, so that there is a subsequence of $(\widehat{\Phi}_n)$ on which the $\mathcal{B}_h$ norm is bounded, say by $M$. We have then shown that there exist interactions $\widetilde{\Phi}_n$ in $\mathcal{B}^0$, arbitrarily close to $\{\Phi: \|\Phi\|_{\mathcal{B}_h} \leq M\}$, for which $\rho$ is an equilibrium measure. When this $\mathcal{B}_h$-ball is compact in $\mathcal{B}^0$ (i.e. $\Omega_0$ finite and $h \gg 1$), then it is easy to show that there exists an interaction *in* this ball for which $\rho$ is an equilibrium measure [we've done it above, by extracting a sub-subsequence of $(\widehat{\Phi}_n)$ convergent to $\widehat{\Phi}_\infty$; for what it's worth, the corresponding sub-subsequence of $(\widetilde{\Phi}_n)$ also converges to $\widehat{\Phi}_\infty$]. Unfortunately, if the $\mathcal{B}_h$-ball is not compact, we do not see any way to conclude that there exists an interaction *in* the ball (or even in $\mathcal{B}_h$) for which $\rho$ is an equilibrium measure. So this method still does not suffice to prove conjecture (ii).

### 5.1.3 Application to the Renormalization Group

Now let us apply these ideas to the renormalization group, by taking $\rho$ to be the renormalized measure $\mu'$. We assume that the experimenter uses the scheme described in the previous section to construct an estimated renormalized interaction $\Phi'_n \in V_n$. We continue to ignore statistical and finite-size errors.

The expected behavior of the estimates $\Phi'_n$ depends critically on whether $\mu'$ is Gibbsian or non-Gibbsian. Assuming the validity of Conjecture 5.1 (or something like it), we have the following scenario:

*Case (i): $\mu'$ is Gibbsian for $\Phi' \in \mathcal{B}^1$.* Then we expect the estimated renormalized interactions $\Phi'_n$ to converge in $\mathcal{B}^1$ norm to $\Phi'$. Now, by the First Fundamental Theorem (Theorem 3.4), the renormalized measures arising from distinct phases of the original model must be Gibbsian for the *same* interaction $\Phi'$. Therefore, any observed multi-valuedness of the RG map must disappear asymptotically as the assumed interaction space $V_n$ grows.

*Case (ii): $\mu'$ is non-Gibbsian.* In this case we expect the estimated renormalized interactions $\Phi'_n$ to diverge in $\mathcal{B}^1$ norm, i.e. $\|\Phi'_n\|_{\mathcal{B}^1} \to \infty$. This behavior is *almost* rigorously proven (Proposition 5.3).

This dichotomy provides, at least in principle, a clear method for distinguishing experimentally the Gibbsianness or non-Gibbsianness of the renormalized measure $\mu'$. Whether it will work in practice is less clear: the proofs of non-Gibbsianness in Section 4 (and of the Fundamental Theorems in Section 3) involve extremely rare events in large volumes; so the distinction between Gibbsianness and non-Gibbsianness might turn out to be visible only with extremely high statistics and when using an extremely large space of renormalized interactions (that is, including interactions involving many spins simultaneously). On the other hand, it is at least conceivable that this dependence on rare events is an artifact of the *proof* and not of the result. It would be interesting, therefore, to perform a high-precision MCRG test, using a large space of renormalized interactions, to compare a case in which $\mu'$ is expected to be Gibbsian (e.g. the $d = 2$ Ising model at a temperature not too far below critical) with a case in which $\mu'$ is expected or proven to be non-Gibbsian (e.g. the $d = 2$ Ising model at low temperature). We are somewhat pessimistic about whether the asymptotic ($n \to \infty$) behavior can be seen with any currently feasible expenditure of resources, but it cannot hurt to try.

In the existing MCRG studies, the interaction space $V$ is usually taken to be quite small: typically $1 \leq \dim V \lesssim 10$. Can we explain the observed discontinuity of the RG map as an artifact of this truncation to a small space of interactions? In our opinion the answer is *yes*. Note first that the estimated renormalized interaction $\Phi'$ is, according to (5.4)/(5.6), just a proxy for the renormalized expectation values $\langle f_\Psi \rangle_{\mu'}$, $\Psi \in V$. These latter expectation values are, of course, discontinuous at a first-order phase-transition surface (and multi-valued on that surface). That in itself does not imply the discontinuity and multi-valuedness of $\Phi'$, because the map from interactions to expectation values is itself discontinuous and multi-valued. However, for most renormalization transformations we expect the renormalized expectation values to be *more* discontinuous than the original expectation values; and it is far from clear that this larger discontinuity can be realized, simultaneously for all observables $f_\Psi$ ($\Psi \in V$), at any interaction in the given space $V$. If it cannot, then the RG map on the space of interactions will *appear* to be discontinuous.

Consider, for example, an Ising model at $h = 0$ and $\beta > \beta_c$, and use the majority-rule transformation. Then the renormalized magnetization $M'$ will undoubtedly be larger than the original magnetization $M = M(\beta, 0^+)$, since minorities tend to get outvoted. Now suppose that one is using, as in the work of Decker, Hasenfratz and Hasenfratz [72, Section 4], only a single renormalized coupling $h'$, with $\beta'$ fixed to equal $\beta$. (That is, $V$ is a *one-dimensional* affine subspace.) Then one will *inevitably* find a renormalized coupling $h' > 0$ for the image measure $\mu'_+$ (resp. $h' < 0$ for $\mu'_-$), since only in this way can one account for a renormalized magnetization $M' = M(\beta, h') > M$. Decker *et al.* do recognize this objection, and try to argue that allowing $\beta'$ to vary would not produce an effect large enough to account for the observed discontinuity in $h'$, but we do not find their argument convincing.

The situation is more subtle if one considers the *two-dimensional* space of couplings $\beta'$ and $h'$. Then one has to choose the pair $(\beta', h')$ so as to match the observed renormalized magnetization $M'$ *and* the observed renormalized energy $E'$. To do this, let us determine first the unique value $\beta'_*$ such that the renormalized energy can be matched at zero magnetic field, i.e. $E' = E(\beta'_*, 0)$. Then we ask how the renormalized magnetization $M'$ compares to the spontaneous magnetization at $\beta'_*$:

(a) If $M' > M(\beta'_*, 0^+)$, then it is impossible to match both $M'$ and $E'$ at zero magnetic field. Therefore, the renormalized coupling $h'$ will be found to be $> 0$ (resp. $< 0$) for the image measure $\mu'_+$ (resp. $\mu'_-$).

(b) If $M' \leq M(\beta'_*, 0^+)$, then $M'$ and $E'$ can be matched by taking $\beta = \beta'_*$, $h' = 0$. [If $M' < M(\beta'_*, 0^+)$, this entails using a *mixed* phase $\nu$ in (5.6), but that is perfectly legitimate. It corresponds to the minimum of $F_\rho \restriction V$ occurring at a point of non-differentiability.]

We are unable to decide *a priori* between these two possibilities; it seems to be a detailed dynamical question.

One approach is to compute the low-temperature expansion of $M'$ and $E'$, and compare them to the corresponding expansions for $E(\beta, 0)$ and $M(\beta, 0)$. This would answer the question at sufficiently low temperature. We are indebted to Jesús Salas [314] for performing this computation, for the two-dimensional Ising model at $h = 0$ using majority rule on $2 \times 2$ blocks (with a random tie-breaker). Setting $u = e^{-4\beta}$, $M = \langle \sigma_0 \rangle$ and $E = \langle \sigma_0 \sigma_{(1,0)} \rangle$, Salas finds:

$$M'(u, 0^+) = 1 - 4u^3 - 32u^4 + O(u^5) \tag{5.7a}$$
$$E'(u, 0) = 1 - 8u^3 - 63u^4 + O(u^5) \tag{5.7b}$$

These are to be compared with the well-known results

$$M(u, 0^+) = 1 - 2u^2 - 8u^3 - 34u^4 + O(u^5) \tag{5.8a}$$
$$E(u, 0) = 1 - 4u^2 - 12u^3 - 36u^4 + O(u^5) \tag{5.8b}$$

Matching the energies, we find

$$u'_* = \sqrt{2} u^{3/2} + \frac{63\sqrt{2}}{16} u^{5/2} - 3u^3 + O(u^{7/2}) \,. \tag{5.9}$$

Plugging this into $M$, we have

$$M(u'_*, 0^+) = 1 - 4u^3 - \frac{63}{2}u^4 - 4\sqrt{2}u^{9/2} + O(u^5) , \qquad (5.10)$$

which is equal to $M'(u, 0^+)$ at leading order and greater than $M'(u, 0^+)$ at order $u^4$. We conclude that at low temperature $M'$ and $E'$ *can* be matched at $h = 0$ in the two-dimensional Ising model with the $2 \times 2$ majority-rule transformation. However, we do not know what will happen with other transformations.

Similar remarks apply in the case of higher-dimensional interaction spaces $V$. While we are unable to prove that a discontinuity will inevitably be observed, neither do we see any reason to believe that the renormalized expectation values $\langle f_\Psi \rangle_{\mu'}$ can always be matched, simultaneously for all $\Psi \in V$ *and all phases* $\mu'$, by some interaction $\Phi' \in V$. Therefore, we must expect that *typically* the observed RG map will be multi-valued and discontinuous at a first-order phase-transition surface, purely as an artifact of the truncation of the renormalized interaction. Of course, if the image measure $\mu'$ is Gibbsian, then this discontinuity should go to zero asymptotically as the assumed interaction space $V_n$ grows. If $\mu'$ is non-Gibbsian, then we expect the estimated renormalized interactions $\Phi'_n$ to diverge in $\mathcal{B}^1$ norm, and it is perfectly likely that the $\Phi'_n$ corresponding to different phases will diverge *in different ways*.

We think that this explains the numerically observed discontinuities of the RG map, irrespective of whether the renormalized measure $\mu'$ is Gibbsian or not.

## 5.2 A Remark on Dangerous Irrelevant Variables

The renormalization-group description of critical behavior in its simplest form seems to imply hyperscaling relations such as

$$d\nu = \gamma' + 2\beta \qquad (5.11)$$
$$d\nu = 2\Delta_4 - \gamma \qquad (5.12)$$
$$d\nu = 2 - \alpha \qquad (5.13)$$
$$\delta = \frac{d + 2 - \eta}{d - 2 + \eta} \qquad (5.14)$$

where $\nu, \alpha, \beta, \gamma, \gamma', \delta, \Delta_4, \eta$ are critical exponents and $d$ is the spatial dimension. It is a well-known fact, however, that hyperscaling does *not* hold for systems above their upper critical dimension $d_u$: for $d > d_u$ the critical exponents are expected to be those of mean-field theory, and these exponents satisfy the hyperscaling relations only *at* $d = d_u$. Indeed, the hyperscaling relations (5.11)–(5.14) have been *proven rigorously to fail* for Ising-like models in dimension $d > 4$ [2, 130, 7, 13, 6, 115].

The traditional explanation of hyperscaling — and of its failure — is the following [117, 118, 254]: Under an RG transformation $H' = \mathcal{R}(H)$ with linear scale factor $l$, the correlation length $\xi$ and free energy density $f$ transform as

$$\xi(H) = l\xi(H') \qquad (5.15a)$$
$$f(H) = g(H) + l^{-d} f(H') \qquad (5.15b)$$

where $g$ is nonsingular. (In fact, for most RG maps these identities are only approximate.) Near a fixed point $H^*$ we parametrize the Hamiltonian by scaling fields $g_1, g_2, \ldots$ with eigenvalues $l^{y_1}, l^{y_2}, \ldots$; the variable $g_i$ is said to be *relevant* (resp. *irrelevant*) if $y_i > 0$ (resp. $y_i < 0$). The critical surface corresponds to setting all the relevant scaling fields to zero. We can assume without loss of generality that $g_1$ is a relevant variable ($y_1 > 0$). The asymptotic scaling laws then read

$$\xi(g_1, g_2, \ldots) \approx l\xi(l^{y_1}g_1, l^{y_2}g_2, \ldots) \qquad (5.16a)$$

$$f_{sing}(g_1, g_2, \ldots) \approx l^{-d} f_{sing}(l^{y_1}g_1, l^{y_2}g_2, \ldots) \qquad (5.16b)$$

Making the choice $l = g_1^{-1/y_1}$,[71] we obtain

$$\xi(g_1, g_2, g_3, \ldots) \approx |g_1|^{-1/y_1} \xi\left(\pm 1, \frac{g_2}{|g_1|^{y_2/y_1}}, \frac{g_3}{|g_1|^{y_3/y_1}}, \ldots\right) \qquad (5.17a)$$

$$f_{sing}(g_1, g_2, g_3, \ldots) \approx |g_1|^{d/y_1} f_{sing}\left(\pm 1, \frac{g_2}{|g_1|^{y_2/y_1}}, \frac{g_3}{|g_1|^{y_3/y_1}}, \ldots\right) \qquad (5.17b)$$

If now $g_i$ is an *irrelevant* variable ($y_i < 0$), then $g_i/|g_1|^{y_i/y_1} \to 0$ as $g_1 \to 0$. It *appears at first glance*, therefore, that for the purpose of determining the leading scaling behavior, the quantity $g_i/|g_1|^{y_i/y_1}$ on the right-hand sides of (5.17) can be replaced by zero. For example, if only the first two fields are relevant (the case of an ordinary critical point), we would obtain

$$\xi(g_1, g_2, g_3, g_4, \ldots) \approx |g_1|^{-1/y_1} \xi\left(\pm 1, \frac{g_2}{|g_1|^{y_2/y_1}}, 0, 0, \ldots\right) \qquad (5.18a)$$

$$f_{sing}(g_1, g_2, g_3, g_4, \ldots) \approx |g_1|^{d/y_1} f_{sing}\left(\pm 1, \frac{g_2}{|g_1|^{y_2/y_1}}, 0, 0, \ldots\right) \qquad (5.18b)$$

In particular, suppose that we set $g_1 = t$ (the temperature deviation from criticality) and $g_2 = h$ (the magnetic field). Then (5.18a) yields the scaling behavior of the correlation length:

$$\xi(t, h=0, g_3, g_4, \ldots) \sim \begin{cases} t^{-\nu} & \text{as } t \to 0^+ \\ (-t)^{-\nu'} & \text{as } t \to 0^- \end{cases} \qquad \text{with } \nu = \nu' = 1/y_t . \qquad (5.19)$$

Likewise, (5.18b) and its derivatives yield the scaling behavior for the thermodynamic quantities:

(a) Differentiating (5.18b) twice with respect to $t$ and setting $h = 0$, we obtain the critical exponents for the specific heat: $\alpha = \alpha' = 2 - d/y_t$. Combining this with (5.19) yields the hyperscaling law $d\nu = 2 - \alpha$.

---

[71] For a position-space RG map, this can of course be done only approximately, since $l$ must be a power of the basic block size $b$. However, this is good enough for the purpose of obtaining critical exponents: by choosing $l$ within a factor $b$ of the desired value, one obtains the desired equality *within a bounded multiplicative constant*.

(b) Differentiating (5.18b) once or twice with respect to $h$, then setting $h = 0$, we obtain the critical exponents for the spontaneous magnetization and the susceptibility: $\beta = (d - y_h)/y_t$ and $\gamma = \gamma' = (2y_h - d)/y_t$. Combining this with (5.19) yields the hyperscaling law $d\nu = \gamma' + 2\beta$.

(c) Differentiating (5.18b) four times with respect to $h$, then setting $h = 0$ (with $t > 0$), we obtain the critical exponent for the four-point cumulant: $2\Delta_4 + \gamma = (4y_h - d)/y_t$. Combining this with (5.19) and the formula for $\gamma$ yields the hyperscaling law $d\nu = 2\Delta_4 - \gamma$.

(Relations for exponents on the critical isotherm can be obtained in a similar manner by setting $g_1 = h$ and $g_2 = t = 0$.) However, Fisher [117][72] pointed out that *this reasoning is correct only if $f(g_1, g_2, g_3, g_4, \ldots)$ and its low-order derivatives have finite limits as $g_3, g_4, \ldots \to 0$ when $g_1 = \pm 1$*. If $f$ or one of its low-order derivatives diverges as $g_3, g_4, \ldots \to 0$, then the hyperscaling relations can fail. A variable $g_i$ which is irrelevant in the RG sense but which provokes a divergence of the free energy density (or one of its low-order derivatives) is termed a *dangerous irrelevant variable*. We emphasize that the free energy is here being evaluated *well away from the critical point*, namely at $g_1 = \pm 1$.

The standard example of such a behavior is the $\varphi^4$ model in dimension $d > 4$. Here the fixed point is Gaussian, with relevant fields $g_1 = t$ and $g_2 = h$; the $\varphi^4$ coupling constant $g_3 = u$ is irrelevant in the RG sense. However, the Gaussian model is unstable at nonzero magnetic field on the critical isotherm, and also at zero magnetic field below the critical temperature, and the irrelevant $\varphi^4$ term is needed to stabilize it. A mean-field calculation (which is expected to give the correct scaling for $d > 4$) predicts that the free energy diverges as $u \downarrow 0$, as

$$f(t = -1, h, u) \approx u^{-1} W(h u^{1/2}) \qquad (5.20)$$
$$f(t = 0, h, u) \approx u^{-1/3} h^{4/3} \qquad (5.21)$$

where $W$ is a well-behaved function. Inserting this behavior into (5.17b), one finds *modified* hyperscaling laws which differ from (5.11)–(5.14) and which are consistent with the mean-field exponents. This behavior occurs because the fixed point $H^*$ is on the *boundary* of the stability region, and the free energy diverges as this boundary is approached.[73]

Here we would like to make the trivial observation that such a blow-up of the free energy is possible *only* in models with unbounded Hamiltonians (such as the $\varphi^4$ model). Indeed, we know that for absolutely summable interactions ($\Phi \in \mathcal{B}^1$), the free energy density is a *Lipschitz continuous* function of the interaction (Propositions 2.56 and 2.58). This means that *the free energy density and its first derivatives are always bounded*. This situation prevails in all physically sensible models of *bounded* spins.

---

[72]See also Fisher [118, Appendix D] and Ma [254, Section VII.4].

[73]A qualitatively similar behavior is expected to occur also in dimension $d = 4$. Here the dangerous irrelevant variable $g_3 = u$ is only *marginally irrelevant* (i.e. $y_3 = 0$, but second-order effects make $g_3$ irrelevant), so that the violations of hyperscaling are only *logarithmic*.

These considerations do not quite rule out the possibility of dangerous irrelevant variables: in principle it could happen that $f(\pm 1, g_2, g_3, \ldots)$ and its first derivatives are bounded, but that *higher* derivatives blow up. This would cause some or all of the hyperscaling relations to fail.[74] This is indeed what happens in the $XY$ model in dimension $d = 4 - \epsilon$ (and probably also $d = 3$) if we let $g_3$ be the coefficient of a $\cos n\theta$ single-site term, where $n$ is even and $\geq 4$ [271, 10]. Such a term is irrelevant in the RG sense (at least if $\epsilon$ is small enough), but for $T < T_c$ it suppresses the Goldstone modes. In $d < 4$ these modes give rise to a divergent longitudinal as well as transverse susceptibility in the pure $XY$ model [106, 243, 105], so that $(\partial^2 f/\partial h^2)(t = -1, h = 0, g_3)$ is finite for $g_3 \neq 0$ but blows up as $g_3 \to 0$ (presumably at the rate $\sim g_3^{-\epsilon/2}$ [271]). This means that the model with $g_3 \neq 0$ belongs to a $Z_n$-symmetric but $SO(2)$-nonsymmetric universality class — which naively would not exist — and that in this universality class the relation $\gamma' = (2y_h - d)/y_t$ [step (b) above] fails. As a consequence, the scaling law $\gamma' = \gamma$ fails, and is replaced by $\gamma' = \gamma - \frac{1}{2}\epsilon y_3/y_t > \gamma$ [271].

Other cases in which an apparently irrelevant term (in the RG sense) changes the phase diagram have been studied in [50, 10, 350].

However, we have not been able to construct any plausible Ansatz for such a behavior in an Ising-to-Ising RG map for the Ising model in dimension $d > 4$. Nor do we know of any plausible candidate for the dangerous irrelevant variable. (In the Ising language there is no term in the Hamiltonian corresponding to the "$\varphi^4$ coupling"; such a term is built into the *a priori* single-spin measure.)

We conclude that the dangerous-irrelevant-variables scenario is probably *not* the correct description of what is happening in the Ising model in dimension $d > 4$, at least in the context of an Ising-to-Ising RG map. On the other hand, we know that the hyperscaling relations (5.11)–(5.14) do fail for Ising models in dimension $d > 4$. Therefore, one of the *other* assumptions made in the conventional RG theory must fail when applied to Ising-to-Ising RG maps in dimension $d > 4$.

For large-cell RG maps ($b \to \infty$), the results summarized in Section 4.4 show that what fails is the Gibbsianness of the fixed-point measure $\mu^*$. Now there is a very close similarity between the dangerous-irrelevant-variables scenario and the non-Gibbsianness proof for large-cell RG maps: both hinge on the fact that a massless Gaussian field is unstable to magnetic-field perturbations. This reasoning suggests that non-Gibbsianness of the fixed-point measure may occur also for iterated Ising-to-Ising transformations with fixed block size $b$ (e.g. majority rule or the Kadanoff transformation). If this were the case, then the RG map $\mathcal{R}$ from Hamiltonians to Hamiltonians would be ill-defined at the critical Ising model, and the putative fixed-point Hamiltonian $H^*$ would simply not exist.

---

[74]Fisher [118, p. 134] states that the derivation of the hyperscaling relations relies implicitly on the assumption that the free energy $f(\pm 1, g_2, g_3, g_4, \ldots)$ has a well-defined finite limit as $g_3, g_4, \ldots \to 0$. However, this statement is slightly misleading, because it is too weak: in fact, as is clear from (a)–(c) above, to derive a hyperscaling relation one needs to know that at least the *second derivative* of $f$ with respect to $t$ or $h$ has a good limit.

# 6 Conclusions and Open Questions

## 6.1 Conclusions

### 6.1.1 How Much of the Standard Picture of the RG Map is True?

We can classify the evidence regarding the validity or failure of the standard picture of RG transformations in three categories:

1) *Positive results.* Some RG maps are well-defined in parts of the one-phase region. The published proofs refer to the following cases:

   (i) *High-field results.* Decimation and Kadanoff transformations for absolutely summable lattice-gas [173] and Ising-spin [207] interactions.

   (ii) *High-temperature results.* Decimation [207, 212], Kadanoff [207] and averaging [212, 56] transformations for absolutely summable Ising-spin interactions.

   (iii) *Small-field results.* These results refer to decimation transformations of the Ising model in any dimension [258]: For any *fixed* temperature and *nonzero* value of the magnetic field, there exists a minimum block size $b_{min}$ beyond which the renormalization transformation is well-defined (the minimum block size diverges as a power of $1/h$ when $h \to 0$).

   (iv) *Results in one dimension.* The decimation transformation is well-defined in dimension $d = 1$ for lattice-gas interactions with many-body and long-range couplings satisfying the summability condition $\sum_{A \ni 0}(\text{diam}A + 1)|A|^{-1}\|\Phi_A\|_\infty < \infty$ [59]. For instance, this includes all the two-body Ising interactions decreasing strictly faster than $1/r^2$.

These results are, however, of limited interest, as they correspond to well-understood regions of the phase diagram, deep within the regime in which uniqueness of the Gibbs measure, and even analyticity of the free energy and correlation functions, can be proven. If these were the only positive results, then one would conclude, in agreement with Griffiths and Pearce's pessimist [172], that the method only works where one does not really need it (and, we may add, sometimes not even there, given Theorem 4.8).

We can also mention, as positive results (of a sort), our Fundamental Theorems of Section 3 which say that the RG map is single-valued and continuous — in accordance with the standard picture — *if it exists at all*.

2) *Non-negative results.* There is at present no evidence of RG pathologies above or at the critical temperature for models strictly below the upper critical dimension $d_u$ (= 4 for Ising-like models). In fact, there exist models for which the critical point has been rigorously studied using the standard RG prescription: the hierarchical models [218, 32, 219] and the Gross-Neveu model [149]. The hierarchical models present the most faithful transcription of Wilson's prescription, but from our point of view they are somewhat artificial as the possible pathologies are removed "by hand". The Gross-Neveu model is fermionic, and thus has no direct probabilistic interpretation. We also should mention here some very interesting preliminary results [214] indicating that for the two-dimensional Ising model at zero field, the majority-rule transformation might be well-defined at (as well as slightly below) the critical temperature. These results are partially rigorous and partially numerical, and so far they concern only some selected (albeit judiciously selected) block-spin configurations. We feel that this work provides some support for the standard picture, but its results are still inconclusive.

3) *Negative results.* There are pathologies at low temperature (not only at zero magnetic field) in all dimensions, and quite possibly at the critical point in dimension $d \gtrsim d_u$. In the former case these pathologies consist in the non-Gibbsianness of the renormalized measure, that is, in the impossibility of constructing a renormalized Hamiltonian after even a *single* RG transformation. In Sections 4.1–4.3 we have shown examples of such pathologies for all the standard real-space transformations (decimation, Kadanoff, majority-rule, averaging). The range of temperatures where these pathologies are proven to exist does not include the critical temperature, but on the other hand the pathological region extends off the phase-coexistence curve, i.e. to nonzero (and in some cases large) magnetic field (Section 4.3.6). Finally, in Sections 4.4 and 5.2 we have given arguments indicating that for $d \gtrsim 4$ there may be pathologies at the critical point. In these latter cases our arguments suggest that the *fixed-point* Hamiltonian may be ill-defined.

Taken together, these results suggest that non-Gibbsianness may be the normal situation for RG maps at low temperature and/or near a first-order phase-transition surface, or at the critical point in high dimensions. This is in direct conflict with the conventional RG ideology (compare the first paragraph of the Introduction).

### 6.1.2 Responses to Some Objections

Many of our colleagues, upon hearing our results, have initially reacted by saying: "If $\mu'$ is not Gibbsian for some interaction in $\mathcal{B}^1$, then that just means it is Gibbsian for some interaction *not* in $\mathcal{B}^1$. You have to use a larger space of interactions." This view seems *a priori* reasonable — and it is even conceivable that it is correct — but unfortunately things are not quite so simple. Before asserting that $\mu'$ is Gibbsian for some interaction $\Phi' \notin \mathcal{B}^1$, one first has to define what it *means* for a measure to be "Gibbsian" for a non-absolutely-summable interaction. Our notion of Gibbs measure relies on the DLR equations, and if the interaction fails to be absolutely summable (or at least convergent), then these equations simply do not make sense. It is thus incumbent on the advocate of "larger interaction spaces" to make precise what is the correspondence between measures and interactions that is to substitute for the DLR equations (and be equivalent to them when the interaction is absolutely summable).

Now, as is usual when one is looking for the solution of an equation, there are two complementary aspects — existence and uniqueness — and one wants preferably for both properties to hold. The existence is favored by enlarging the space of possible solutions, while the uniqueness is favored by narrowing it; and it is far from clear, *a priori*, whether there exists a space in which the solutions both exist and are unique. These general remarks can be exemplified in our statistical-mechanical problem. Within the class of Feller specifications (and hence *a fortiori* within $\mathcal{B}^1$), the Griffiths-Ruelle theorem (Theorem 2.15, Corollary 2.18 and Proposition 2.59) guarantees the uniqueness of the specification for a given measure $\mu$ (and hence the uniqueness modulo physical equivalence of the interaction). But the existence may fail, as we showed through numerous examples in Section 4. On the other hand, if we enlarge the space of allowed specifications by dropping the Feller ($\approx$ quasilocality) property, then the existence holds but the uniqueness fails spectacularly (see the Remark at the end of Section 2.3.4). Similarly, if we enlarge the space of allowed interactions from $\mathcal{B}^1$ to $\mathcal{B}^0$, and generalize "Gibbs measure" to "equilibrium measure", then every ergodic measure of finite entropy density is the equilibrium measure for some interaction, but the uniqueness again fails spectacularly (see item 2 in Section 2.6.7). One certainly cannot develop a satisfactory RG theory in such pathological spaces.

Furthermore, we have given strong arguments that any physically reasonable specification must be *quasilocal*, at least in systems of bounded spins (see Section 2.3.3). On the other hand, in Sections 4.1–4.3 and 4.5.2 we have proven directly that the renormalized measures are not consistent with *any* quasilocal specification. So even if there were to exist quasilocal specifications corresponding to interactions not in $\mathcal{B}^1$, such specifications could not be of any relevance for our renormalization-group problem.[75]

Our final objection to considering spaces of interactions larger than $\mathcal{B}^1$ is that $\mathcal{B}^1$ is already too big! Indeed, the standard RG ideology [365] is that the RG flow should take place in some space of "short-range" interactions, e.g. interactions which decay exponentially or at least like a sufficiently large power (e.g. $|x|^{-p}$ with $p \geq d+1$). This ideology is not a mere whim, but results from the need to explain universality of critical behavior: one needs to have an interaction space in which the unstable manifold of a given fixed point is *finite-dimensional* (i.e. there are finitely many relevant scaling fields). Now, such a behavior is impossible in a space of long-range interactions (such as $\mathcal{B}^1$), since in general the critical exponents will be altered by any perturbation that decays like $|x|^{-(d+2-\epsilon)}$ with $\epsilon >$ the critical exponent $\eta$ of the original model [292, Section 10.2]. Moreover, even the qualitative phase diagram is unstable to long-range but summable pair interactions [349, 332, 208]: that is, the Gibbs phase rule cannot hold in $\mathcal{B}^1$ or even in any $\mathcal{B}^n$. In order to have any hope of constructing a satisfactory RG theory, it is necessary to work in a space of "short-range" interactions, such as the space $\mathcal{B}_h$ for some $h \gtrsim 1$.

A second comment which is often made is the following: "The RG map is always well-defined as a map from measures to measures; the pathologies come from trying to lift it to a map from Hamiltonians to Hamiltonians. So why not just stick with the RG map (1.1) on the space of measures?"

---

[75]Note also that Sullivan [336] and Kozlov [222] have *almost* proven that every quasilocal specification arises from an interaction in $\mathcal{B}^1$: see Theorem 2.12 and the Remarks following it, plus the Remark at the end of Section 2.4.9.

This is a sensible question, which was already raised by Griffiths and Pearce [173, p. 534–535], and our answer is essentially the same as theirs: Many interesting things *can*, indeed, be learned by studying the action of RG transformations on measures. For linear RG transformations, this is an ancient branch of probability theory that goes back to Gauss' and DeMoivre's investigations of the central limit theorem for independent random variables, and which continues to this day in studies of central and non-central limit theorems for dependent random fields [201, 180, 273, 274, 58]; it is closely related to studies of triviality and non-triviality for scaling limits in statistical mechanics and quantum field theory [323, 71, 115]. For nonlinear RG transformations, this study is only beginning [281, 193], but we expect it to be fruitful as well. Unfortunately, not all of the RG theory can be carried out on the space of measures alone. For example, the critical exponent $\gamma$ measures the rate of divergence of the susceptibility as the temperature approaches the critical temperature. Now, the susceptibility is the integral of the 2-point correlation function, and thus can be read off the measure; while the (inverse) temperature is the coefficient of some term in the Hamiltonian (e.g. the nearest-neighbor term in the case of the Ising model). Therefore, the exponent $\gamma$ can be deduced only from a theory that *relates* the measure to the Hamiltonian (or interaction); it cannot be deduced solely from an RG map acting on the space of measures. The same goes for the exponent $\nu$, which measures the rate of divergence of the correlation length as the temperature approaches criticality. On the other hand, the exponent *ratio* $\gamma/\nu$ measures the *relative* rate of divergence of two different aspects of the 2-point correlation function, and so *can* potentially be deduced from a measures-to-measures RG map. Likewise, the critical exponent $\eta$ measures the rate of decay of the 2-point function at the critical point, making no reference whatsoever to the temperature; therefore, it too can potentially be deduced from a measures-to-measures RG map. It follows that the scaling law $\gamma/\nu = 2 - \eta$ also lies potentially within the purview of a measures-to-measures RG theory.

### 6.1.3 Where Does All This Leave RG Theory?

After the more-or-less cold exposition of facts of Section 6.1.1, and the additional clarification (pre-emptive defense) of the previous subsection, let us present some general remarks about the consequences of the present work for the RG enterprise.

We think that there is already a substantial body of evidence indicating that the *conventional* RG theory, in its narrow sense of Hamiltonian-to-Hamiltonian maps, needs to be reexamined. However, this does not, in itself, detract in any way from the value and significance of the RG ideas that have pervaded much of today's statistical mechanics and quantum field theory. The RG *philosophy* — interpreted broadly to include various kinds of "multi-length-scale" and "coarse-graining" arguments — has been, and will continue to be, our main tool to analyze the otherwise inaccessible "intermediate temperature" regions, which fall beyond the reach of series expansions or perturbative arguments and yet are the regions in which the most interesting phenomena take place.

The main issue in the proper application of RG theory to a particular problem is the choice of variables in which to express the model, along with the choice of the RG map. This was already understood by the founding fathers of the field. It corresponds to what Michael Fisher calls "aptness or focusability" of the transformation, and his own words are especially clear:

> For any given Hamiltonian or class of Hamiltonians there is not just one renormalization group — "*the* renormalization group" as some people say — but rather there are many that might be introduced, and one must question, for example, whether the process is best carried out in real space or momentum space and so on. A "good" renormalization group must be "apt" or appropriate for the problem at hand, and it must, in particular, "focus" properly on the critical phenomena of interest. [118, page 82]

Let us mention an illustrative example: The usual transformations involving averaging (or other kinds of "voting") over square blocks are designed mostly having ferromagnetic systems in mind. They are efficient for selecting the zero-momentum modes, which are indeed the modes that become critical in an ordinary ferromagnetic transition. On the other hand, these transformations are unsuitable for studying antiferromagnets because they do not distinguish the oppositely magnetized sublattices. This was remarked by van Leeuwen [357], who showed how a more careful design of the block shapes could overcome this deficiency (his proposal is depicted in Figure 2(d)). Thus, while most people imagine RG maps as acting in a huge space of Hamiltonians — including regions exhibiting various different types of phase transitions (ferromagnetic, antiferromagnetic and many others) — it is unlikely that any single RG map can exhibit well-behaved fixed points corresponding to all of these transitions. Rather, one must "custom-make" the RG map for each new physical situation.

In this regard, our work — building on that of Griffiths, Pearce and Israel [172, 173, 171, 207] — can be considered an extension of the preceding observations: "aptness" and "focusability" are needed not just to ensure the usefulness of the map, but even its very *existence*. On the other hand, the success stories of rigorous RG studies teach us that the search for this "aptness" may require a very open-minded attitude, in the sense that, in many cases, the appropriate variables are not necessarily spin variables and, in fact, not even local objects. Indeed, with the exception of hierarchical [218, 32, 219] and fermionic [149] models, rigorous RG studies have *not* implemented the strict Wilson prescription involving an RG transformation of Hamiltonians written in terms of spin variables. Rather, they have employed a combination of spin variables and polymer ensembles [147, 148, 150, 151, 181, 188] or a pure polymer ensemble [53] when studying critical phenomena, or an ensemble of Peierls-like contours [145, 146, 44] when studying first-order phase transitions. More generally, "multi-scale" and "coarse-graining" ideas have been used in a wide variety of problems, including:

- ultraviolet stability of the $\varphi_3^4$ [159, 28] and Yang-Mills$_4$ [19] quantum field theories;

- the Kosterlitz-Thouless transition in the two-dimensional $XY$ and related models [134, 135];

- the ferromagnetic transition in the one-dimensional $1/r^2$ Ising model [137];

- confinement in the three-dimensional $U(1)$ lattice gauge theory [166];

- localization for random Schrödinger operators [138];

- the phase transition in plaquette percolation [4];

- the intersection properties of ordinary random walks and of Brownian motion [3, 112, 115]; and

- the critical behavior of self-avoiding walks [52, 185, 187, 186], percolation [183, 182] and branched polymers [184] in high dimensions.

In many of these examples, the "coarse-graining" is applied at the level of objects with some geometric content, such as random walks, clusters, surfaces, contours, etc.

Thus, our work is in no way an attack on the essential physical ideas behind the RG approach. It simply points out the need for a more general definition of their scope.

### 6.1.4 Towards a Non-Gibbsian Point of View

Let us close with some general remarks on the significance of (non-)Gibbsianness and (non-)quasilocality in statistical physics. Our first observation is that Gibbsianness has heretofore been ubiquitous in equilibrium statistical mechanics because it has been put in *by hand*: nearly all the measures that physicists encounter are Gibbsian because physicists have *decided* to study Gibbs measures! However, we now know that natural operations on Gibbs measures can sometimes lead out of this class: among such operations are some renormalization transformations (Sections 4.1–4.3 and 4.5.2), some nonlinear local functions (Section 4.4), convex combinations (Section 4.5.1), and weak limits (Section 4.5.6). It is thus of great interest to study which types of operations preserve, or fail to preserve, the Gibbsianness (or quasilocality) of a measure. This study is currently in its infancy.

More generally, in areas of physics where Gibbsianness is not put in by hand, one should expect non-Gibbsianness to be ubiquitous. This is probably the case in nonequilibrium statistical mechanics (Section 4.5.4).

Since one cannot expect all measures of interest to be Gibbsian, the question then arises whether there are *weaker* conditions that capture some or most of the "good" physical properties characteristic of Gibbs measures. For example, the stationary measure of the voter model appears to have the critical exponents predicted (under the hypothesis of Gibbsianness) by the Monte Carlo renormalization group [362], even though this measure is provably non-Gibbsian [246].

One may also inquire whether there is a classification of non-Gibbsian measures according to their "degree of non-Gibbsianness". Joel Lebowitz has suggested to us the analogy with the rational and real numbers: although the set of rationals is very "small" in many senses (e.g. first Baire category, zero Lebesgue measure), it is "large" in the weak sense that any real number can be approximated by a sequence of rational numbers; and the irrational numbers can be classified according to the rate at which they can be approximated by rationals (Diophantine approximation). In Section 4.5.6 we conjectured a similar scenario for the Gibbsian measures within the space of all measures. It would then be natural to classify the non-Gibbsian measures according to how well (or how rapidly) they can be approximated by Gibbsian ones.

Finally, there is a philosophical question, raised by one of our colleagues in Rome (to whom we apologize because we cannot remember his name): All mathematical modelling, in any branch of science, involves selecting the "important" variables in the description of a system and neglecting the variables judged "unimportant". In a statistical system this means that the "unimportant" variables are integrated out, i.e. one performs a kind of "decimation" transformation. Now, if the decimated variables are only weakly coupled to the others, then one may hope that the decimation will lead to a Gibbs measure (although rigorous theorems guaranteeing this seem to be lacking). However, one could also fear that the result of the decimation might be a *non-Gibbsian* measure, especially if the decimated variables are strongly coupled to the others. (Such variables might still be deemed "unimportant" if they were believed to affect only uninteresting quantitative details of the problem, without changing the features of interest.) In this case, not only would one be making an approximation in describing the system by a particular "model Hamiltonian", but even the description of the decimated system by *any* Hamiltonian would itself be an approximation. And one would have to investigate how good this approximation is.

## 6.2 Some Open Questions

We end with a list of open questions for future research:

1) Clean up the circle of results connected with the Gibbs Representation Theorem (Theorem 2.12), particularly in the translation-invariant case (Sections 2.3.3, 2.4.9 and A.2).

2) Determine whether $\{\Phi\colon \|\Phi\|_{\mathcal{B}_h} \leq M\} + \mathcal{J}$ is a closed subset of $\mathcal{B}^0$, if $h \not\gg 1$, and in particular for $\mathcal{B}_h = \mathcal{B}^1$ (Sections 2.4.4 and 2.4.6). This affects the ways in which the RT map can blow up at the boundary of its domain (Section 3.3), and arises also in our theory of parameter estimation (Section 5.1.2).

3) Devise a clean general theory for systems of unbounded spins, analogous to the spaces $\mathcal{B}^0$ and $\mathcal{B}^1$ for systems of bounded spins (Sections 2.4.4 and 3.1.4).

4) Investigate rigorously the Gibbsianness or non-Gibbsianness of the renormalized measure in the following models:

   (a) Ferromagnetic Ising model, using the decimation transformation with spacing $b$: does the cutoff temperature for non-Gibbsianness tend to $J_c$ as $b \to \infty$? (See Section 4.3.2.)

(b) Ferromagnetic Ising model, using the majority-rule transformation with block sizes $b$ not covered by the construction in Section 4.3.4. (For dimension $d \geq 3$, it appears that *no* block sizes $b$ are covered by this construction: see Appendix C.)

(c) Ferromagnetic Ising model at low temperature and nonzero magnetic field, in dimension $d = 2$, using the decimation, Kadanoff or majority-rule transformation.

(d) Antiferromagnetic nearest-neighbor Ising model in a uniform magnetic field, on the paramagnetic-antiferromagnetic critical surface: compare the majority-rule (or Kadanoff) transformation on square ($b \times b$) blocks to the same transformation on van Leeuwen's 5-spin blocks [357, 55].

(e) $q$-state Potts model with $q$ large, at (or near) the first-order phase transition, using either the ordinary "plurality-rule" (or Kadanoff) transformation [305] or the modified transformation including vacancies [279, 305].

(f) Other models at or near a first-order phase transition.

(g) Ferromagnetic Ising model at the critical point in dimension $d > 4$, using a majority-rule (or Kadanoff) transformation with fixed block size $b$ (Section 4.4).

5) Improve/generalize the theorems on non-Gibbsianness of local nonlinear functions of an anharmonic crystal, and in particular try to prove non-quasilocality (Section 4.4).

6) Try to generalize Schonmann's example (Section 4.5.2) to dimensions $d, d'$ other than $d = 2$, $d' = 1$.

7) Prove (or disprove) the existence of measures consistent with the Fortuin-Kasteleyn random-cluster-model specification (4.91); in particular, prove (or disprove) that the infinite-volume limit measures taken with free or wired boundary conditions are consistent with this specification (Section 4.5.3).

8) Investigate the Gibbsianness or non-Gibbsianness of the stationary measure(s) in various stochastic evolutions not satisfying detailed balance (Section 4.5.4).

9) Investigate the abstract properties of the set $\mathcal{G}$ of Gibbsian measures (Sections 4.5.6 and 6.1.4).

10) Investigate rigorously the model of parameter estimation introduced in Section 5.1.2; in particular, try to prove Conjecture 5.1 or some weakened version of it.

11) Make a high-precision MCRG test, using a large space of renormalized interactions, to compare a case in which the renormalized measure is expected to be Gibbsian (e.g. the $d = 2$ Ising model at a temperature not too far below critical) with a case in which the renormalized measure is expected or proven to be non-Gibbsian (e.g. the $d = 2$ Ising model at low temperature) [Section 5.1.3].

12) Clarify the relationship between RG transformations acting on contours [145, 146] or polymers [147, 148, 150, 151, 181, 188, 53], and the traditional RG transformations acting on spins.

13) Discuss the Gibbsianness or non-Gibbsianness of various states of *quantum* lattice systems [253, 355]. Here one problem is to understand better the relationships between the various alternative notions of "Gibbsianness" in the quantum case.

14) Prove Conjecture C.5.

# A   Proofs of Some Theorems from Section 2

## A.1   Proofs and References for Section 2.1

The remarks made in Section 2.1.2 are all well-known results. Here are some references:

(a) $\Omega_0$ compact $\Longrightarrow$ $\Omega$ compact $\Longrightarrow$ every continuous function on $\Omega$ is bounded [309, Proposition 9.4]. The density of $C_{loc}(\Omega)$ in $C(\Omega)$ is an easy consequence of the Stone-Weierstrass theorem [309, Theorem 9.28].

(b) If $\Omega_0$ is discrete, then every local function is continuous; and continuity is preserved under uniform convergence.

(c) This is an immediate consequence of (a) and (b).

**Further Remark.** If the single-spin space $\Omega_0$ is noncompact, there may exist bounded continuous functions which are *not* quasilocal. Hans-Otto Georgii provided us with the following example: take $\mathcal{L} = \Omega_0 = \mathbb{Z}$ and let $f(\sigma) = \sigma_{\sigma_0}$ (!); then let $g$ be a bounded function of $f$, say $g = |f|/(1+|f|)$.

In fact, this construction can be imitated whenever $\Omega_0$ is a noncompact metric space and $\mathcal{L}$ is infinite: Let $f_1, f_2, \ldots \in C(\Omega_0)$ have disjoint supports $S_1, S_2, \ldots$ with $S_i \cap \overline{\bigcup_{j \neq i} S_j} = \emptyset$ and $\|f_i\|_\infty = 1$ (such functions are easily constructed using Urysohn's lemma); let $x_0, x_1, x_2, \ldots$ be distinct sites in $\mathcal{L}$; let $g \in C(\Omega_0)$ be non-constant; and define $h(\sigma) = \sum_{n=1}^\infty g(\sigma_{x_n}) f_n(\sigma_{x_0})$.

Standard references for the theory of probability measures on metric spaces are the books of Parthasarathy [289] and Billingsley [29]. Probability measures on general (not necessarily metrizable) topological spaces are treated in [358, 73, 320]. The Riesz-Markov theorem is [309, Theorem 14.8] or [289, Theorems II.5.7 and II.5.8]. The theorem on support of a measure is [289, Theorem II.2.1].

The bounded measurable topology on $M(\Omega)$ and $M_{+1}(\Omega)$ is discussed in [143]. The weak topology on $M_{+1}(\Omega)$ is discussed in detail in [289, 29, 358]; in particular, the topological properties of $M_{+1}(\Omega)$ for different classes of spaces $\Omega$ are discussed in [358, Part II], [289, Section II.6] and [73, Theorem III–60].

If $\Omega_0$ is a separable metric space, then every *uniformly* continuous function on $\Omega$ is quasilocal [157, Remark 2.21(2)]. Since the bounded uniformly continuous functions are sufficient to generate the (ordinary) weak topology (this is the famous "portmanteau theorem" [29, Theorem 2.1]), it follows that the bounded quasilocal topology is stronger than the weak topology. On the other hand, if $\Omega_0$ is also discrete (hence countable), then every quasilocal function is continuous, so the two topologies in fact coincide. See [155, Remark 0.3].

## A.2 Proofs and References for Section 2.3

Proposition 2.7 is essentially [157, Remark 1.24]. Examples 1 and 2 in Section 2.3.3 are [157, Proposition 2.24 and Example 2.25]. Theorem 2.10 and related results are discussed in [157, Section 2.2]. Theorem 2.12 is proven by Kozlov [222]; see also Sullivan [336].

**Remarks.** 1. The following conjectured extensions of Theorem 2.12 appear to be open questions:

(a) If $\Pi$ is quasilocal and nonnull (but not *uniformly* nonnull), and $\Omega_0$ is not finite, does there exist a *uniformly convergent* interaction $\Phi$ such that $\Pi = \Pi^\Phi$? [This theorem might be relevant to models of unbounded spins with *finite-range* interactions.]

(b) If $\Pi$ is quasilocal, uniformly nonnull and *strongly Feller* in the sense that $f \in B(\Omega, \mathcal{F}_\Lambda)$ implies $\pi_\Lambda f \in C_{ql}(\Omega)$, and $\Omega_0$ is not finite, does there exist a *continuous* absolutely summable interaction $\Phi$ such that $\Pi = \Pi^\Phi$?

2. Regarding the relation between quasilocality and the Feller property, the following appears to be an open question: If $\Omega_0$ is compact but not finite, can a Feller specification fail to be quasilocal?

PROOF OF THEOREM 2.15.  Let $\mu$ be consistent with Feller specifications $\Pi_1$ and $\Pi_2$. Then

$$E_\mu(f|\mathcal{F}_{\Lambda^c})(\omega) \;=\; (\pi_{1\Lambda}f)(\omega) \;=\; (\pi_{2\Lambda}f)(\omega) \quad \mu\text{-a.e.} \tag{A.1}$$

for each $f \in C(\Omega)$. Now, since $\mu$ gives nonzero measure to every open set, two continuous functions which agree $\mu$-a.e. must in fact agree everywhere. So we must have $(\pi_{1\Lambda}f)(\omega) = (\pi_{2\Lambda}f)(\omega)$ for all $\omega$. But if the two measures $\pi_{1\Lambda}(\omega, \cdot)$ and $\pi_{2\Lambda}(\omega, \cdot)$ give equal expectations to each continuous function $f$, then they must be equal. ∎

Further examples of pathological non-quasilocal specifications, along the lines of the Remark at the end of Section 2.3.4, are given by Georgii [157, pp. 34–35].

Theorem 2.17 and Corollary 2.18 are proven in [157, Theorem 2.34]. Propositions 2.19 and 2.20 are [157, Proposition 7.9 and Theorem 7.7]. Proposition 2.22 is almost immediate from the definition of Feller specification and weak convergence; for related results, see [157, Sections 4.3 and 4.4]. Proposition 2.23 is proven in [157, Theorem 7.12].

PROOF OF PROPOSITION 2.25.  Recall that $\mu^\omega(\cdot)$ is a regular conditional probability for $\mu$ given $\mathcal{F}_\Delta$, i.e. it depends on $\omega$ only through $\omega_\Delta$; and we are interested only in its restriction to $\mathcal{F}_{\Delta^c}$, i.e. we want to study the measure $\mu^{\omega_\Delta}(d\omega'_{\Delta^c})$. The claim is now that for $\mu$-a.e. $\omega_\Delta$, we have

$$\int \mu^{\omega_\Delta}(d\omega'_{\Delta^c}) \, \pi_\Lambda^{\omega_\Delta}(\omega'_{\Delta^c}, A) \;=\; \mu^{\omega_\Delta}(A) \tag{A.2}$$

for all $A \in \mathcal{F}_{\Delta^c}$ and all $\Lambda \subset \Delta^c$. Both sides of this equation are $\mathcal{F}_\Delta$-measurable. So it suffices to prove that for all $f \in B(\Omega, \mathcal{F}_\Delta)$ we have

$$\int d\mu_\Delta(\omega_\Delta) \, f(\Omega_\Delta) \int \mu^{\omega_\Delta}(d\omega'_{\Delta^c}) \, \pi_\Lambda^{\omega_\Delta}(\omega'_{\Delta^c}, A) \;=\; \int d\mu_\Delta(\omega_\Delta) \, f(\Omega_\Delta) \, \mu^{\omega_\Delta}(A) \,. \tag{A.3}$$

Now the right-hand side of (A.3) is $\int f\chi_A \, d\mu$, by definition of regular conditional probability. As for the left-hand side, let us rewrite it as

$$\int [d\mu_\Delta(\omega_\Delta) \, \mu^{\omega_\Delta}(d\omega'_{\Delta^c})] \, f(\Omega_\Delta) \, \pi_\Lambda^{\omega_\Delta}(\omega'_{\Delta^c}, A) \,; \tag{A.4}$$

this passage from an iterated integral to a single integral on the product space is justified by [272, Proposition III–2–1]. But the measure in brackets in A.4 is precisely $d\mu(\omega_\Delta, \omega'_{\Delta^c})$; so the left-hand side of (A.3) equals

$$\int d\mu(\omega_\Delta, \omega'_{\Delta^c}) \, f(\Omega_\Delta) \, \pi_\Lambda(\omega_\Delta \times \omega'_{\Delta^c}, A) \,, \tag{A.5}$$

where we have now inserted the definition (2.37) of $\pi_\Lambda^{\omega_\Delta}$. We now use the fact that $\mu$ is consistent with $\Pi$, and that $\Lambda \subset \Delta^c$ (so $\Delta \subset \Lambda^c$); it follows that (A.5) equals $\int f\chi_A \, d\mu$. ∎

## A.3 Proofs and References for Section 2.4

### A.3.1 Van Hove Convergence

PROOF OF PROPOSITION 2.27.   It is easy to see that

$$x \in \Lambda \;\Longrightarrow\; \text{dist}(x, \Lambda^c) \;=\; \text{dist}(x, \partial_1^+\Lambda) \tag{A.6a}$$
$$x \in \Lambda^c \;\Longrightarrow\; \text{dist}(x, \Lambda) \;=\; \text{dist}(x, \partial_1^-\Lambda) \tag{A.6b}$$

Therefore,

$$|\partial_r^-\Lambda| \;\leq\; (2r+1)^d |\partial_1^+\Lambda| \tag{A.7a}$$
$$|\partial_r^+\Lambda| \;\leq\; (2r+1)^d |\partial_1^-\Lambda| \tag{A.7b}$$

It follows that (a)–(c) are equivalent.

Next notice that

$$x \in \Lambda \setminus (\Lambda + a) \;\Longrightarrow\; x \in \Lambda \text{ and } \text{dist}(x, \Lambda^c) \leq |a| \tag{A.8a}$$
$$x \in (\Lambda + a) \setminus \Lambda \;\Longrightarrow\; x \in \Lambda^c \text{ and } \text{dist}(x, \Lambda) \leq |a| \tag{A.8b}$$

Therefore (c) implies (d) and (e). Conversely,

$$\partial_1^- \Lambda \;=\; \bigcup_{|a|=1} [\Lambda \setminus (\Lambda + a)] \tag{A.9a}$$

$$\partial_1^+ \Lambda \;=\; \bigcup_{|a|=1} [(\Lambda + a) \setminus \Lambda] \tag{A.9b}$$

so (d) $\Longrightarrow$ (a) and (e) $\Longrightarrow$ (b).

Finally,

$$\Lambda \triangle (\Lambda + A) \;\subset\; \bigcup_{a \in A} [\Lambda \triangle (\Lambda + \{a\})] \,, \tag{A.10}$$

so (d) and (e) together imply (f). On the other hand, taking $A = \{a\}$ shows trivially that (f) implies (d) and (e).

This completes the proof of equivalence of (a)–(f).

Next we prove that $\lim_{n\to\infty} |\Lambda_n| = \infty$: this follows immediately from (a) and the fact that $|\partial_1^- \Lambda| \geq 1$ whenever $\Lambda$ and $\Lambda^c$ are both nonempty.

Finally, let us prove statement ($\beta$): For each $n$, choose $a_n \in \mathbb{Z}^d$ and $r_n \in \mathbb{Z}_+$ so that $B_{r_n}(a_n) \equiv \{x \in \mathbb{Z}^d \colon |x - a_n| \leq r_n\}$ is a maximum-sized ball contained in $\Lambda_n$. We claim that $\lim_{n\to\infty} r_n = \infty$. *Proof:* Fix any $r > 0$. Since $\lim_{n\to\infty} |\Lambda_n| = \infty$ and $\lim_{n\to\infty} |\partial_r^- \Lambda_n|/|\Lambda_n| = 0$, we clearly have $\lim_{n\to\infty} |\Lambda_n \setminus \partial_r^- \Lambda_n| = \infty$ and hence in particular $\Lambda_n \setminus \partial_r^- \Lambda_n \neq \emptyset$ for all sufficiently large $n$. But $\Lambda_n \setminus \partial_r^- \Lambda_n \neq \emptyset$ is just another way of saying that $r_n \geq r$. ∎

**Remarks.** 1. Many books [312, 206, 40] use a more complicated definition of van Hove convergence, based on a paving of $\mathbb{Z}^d$ by cubes of side $a$. It is easy to see that this definition is equivalent to conditions (a)–(f).

2. What physicists call van Hove convergence is termed *Følner convergence* by mathematicians. Much of the theory extends, in fact, to locally compact amenable (semi)groups [167]. See [223, Section 6.4] for ergodic theorems in this context.

### A.3.2  Translation-Invariant Measures

Proposition 2.30 is [157, Theorem 14.5 and Proposition 14.7]. Proposition 2.31 is [157, Corollary 14.A5 and Theorem 14.A8]. For more information on ergodic theorems, along with some relevant counterexamples, see [223, pp. 222–226]. Proposition 2.32 is proven in [157, Theorem 14.12] or [206, Lemma IV.3.2]; a stronger form will be proven as Proposition 2.61(e) below. Information on the Poulsen simplex can be found in [252, 284].

### A.3.3  A Digression on Subadditivity

An important role in the theory of translation-invariant lattice systems is played by the concept of a *subadditive set function*. Subadditivity arguments will be used to prove the existence of the infinite-volume limit for the pressure, the entropy density, and quantities connected with the quotient norm. We therefore collect here the needed results.

**Definition A.1** *Let $\mathcal{S}$ be the class of all nonempty finite subsets of $\mathbb{Z}^d$, and let $\mathcal{S}^* = \mathcal{S} \cup \{\emptyset\}$. A function $F \colon \mathcal{S}^* \to [-\infty, \infty)$ is called*

- *subadditive if $F(A_1 \cup A_2) \leq F(A_1) + F(A_2)$ whenever $A_1, A_2 \in \mathcal{S}^*$ with $A_1 \cap A_2 = \emptyset$*

- *completely subadditive if $F(A) \leq \sum_{i=1}^{n} \lambda_i F(A_i)$ whenever $A, A_1, \ldots, A_n \in \mathcal{S}^*$ with $\chi_A = \sum_{i=1}^{n} \lambda_i \chi_{A_i}$ and all $\lambda_i \geq 0$*

- *strongly subadditive if $F(A_1 \cup A_2) + F(A_1 \cap A_2) \leq F(A_1) + F(A_2)$ whenever $A_1, A_2 \in \mathcal{S}^*$*

Clearly, complete subadditivity implies subadditivity. The key nontrivial fact is:

**Lemma A.2 ([268, Théorème 2])** *If $F$ is strongly subadditive and $F(\emptyset) \geq 0$, then $F$ is completely subadditive.*

**Remark.** If $F$ is subadditive, then either $F(\emptyset) \geq 0$ or else $F \equiv -\infty$. In our applications we will always have $F(\emptyset) = 0$.

We can now state the two principal theorems on the existence of the infinite-volume limit:

**Proposition A.3** *Let $F \colon \mathcal{S}^* \to [-\infty, \infty)$ be translation-invariant and completely subadditive. Then $\lim_{\Lambda \nearrow \infty} |\Lambda|^{-1} F(\Lambda)$ exists and equals $\inf_{\Lambda \in \mathcal{S}} |\Lambda|^{-1} F(\Lambda)$.*

**Proposition A.4** *Let $F \colon \mathcal{S}^* \to [-\infty, \infty)$ be translation-invariant and subadditive. Then $\lim_{n \to \infty} |\Lambda_n|^{-1} F(\Lambda_n)$ exists for any van Hove sequence $(\Lambda_n)$ satisfying the additional condition $|\Lambda_n|/\mathrm{diam}(\Lambda_n)^d \geq \delta > 0$ for some $\delta > 0$. Moreover, this limit equals $\inf_{n \geq 1} |C_n|^{-1} F(C_n)$.*

We note that ordinary subadditivity is *not* sufficient for the existence of the van Hove limit; a counterexample has been given in [175].

PROOF OF PROPOSITION A.3. This result is stated in [267, Théorème 0] and proven in [268, Corollaire 10], but the proof is rather difficult to follow. For completeness let us give an elementary proof [333]:

Let $A, B \in \mathcal{S}$; without loss of generality let us suppose that $0 \in B$. Now consider the decomposition

$$\chi_A = \sum_{a: B+a \subset A} \frac{1}{|B|} \chi_{B+a} + \sum_{x \in A} \lambda_x \chi_{\{x\}} \tag{A.11}$$

where
$$\lambda_x = \frac{|\{a\colon B+a \ni x \text{ and } B+a \not\subset A\}|}{|B|}\,. \tag{A.12}$$

Clearly $0 \le \lambda_x \le 1$; and by summing (A.11) over $y \in \mathbb{Z}^d$ we find

$$\begin{aligned}
\sum_{x \in A} \lambda_x &= |A| - \Big|\bigcap_{b \in B}(A-b)\Big| \\
&= \Big|A \setminus \bigcap_{b \in B}(A-b)\Big| \qquad \text{[since } 0 \in B\text{]} \\
&\equiv m_B^-(A)\,.
\end{aligned} \tag{A.13}$$

By complete subadditivity and translation-invariance it follows from (A.11) and (A.13) that

$$\begin{aligned}
F(A) &\le \sum_{a\colon B+a \subset A} \frac{F(B+a)}{|B|} + \sum_{x \in A} \lambda_x F(\{x\}) \\
&= \frac{F(B)}{|B|}\Big|\bigcap_{b \in B}(A-b)\Big| + \Big(\sum_{x \in A} \lambda_x\Big) F(\{0\}) \\
&= \frac{F(B)}{|B|}\big[|A| - m_B^-(A)\big] + m_B^-(A)F(\{0\})\,.
\end{aligned} \tag{A.14}$$

Now divide by $|A|$ and take $A \nearrow \infty$ (van Hove): by Proposition 2.27(d) we have $\lim\limits_{A \nearrow \infty} m_B^-(A)/|A| = 0$. Therefore

$$\limsup_{A \nearrow \infty} \frac{F(A)}{|A|} \le \frac{F(B)}{|B|}\,. \tag{A.15}$$

Since this holds for all $B \in \mathcal{S}$, we have

$$\limsup_{A \nearrow \infty} \frac{F(A)}{|A|} \le \inf_{B \in \mathcal{S}} \frac{F(B)}{|B|} \le \liminf_{B \nearrow \infty} \frac{F(B)}{|B|}\,. \tag{A.16}$$

■

PROOF OF PROPOSITION A.4. This is essentially [198, Proposition 4.10]. See also [175, 333]. ■

**Remarks.** 1. The important concept of complete subadditivity was apparently first introduced by Moulin-Ollagnier and Pinchon [267, 268].

2. The proofs given here actually work (after slight notational changes) in an arbitrary discrete amenable group [333]. A slightly different proof of Proposition A.3, also valid for discrete amenable groups, is implicit in [266, proof of Théorème 2]. For an extension to locally compact amenable groups, see [268].

### A.3.4 A Lemma on Sums of Translates

Next we use subadditivity arguments to prove an important lemma concerning the infinite-volume limit of sums of translates of a function $f$. This lemma will play an important role in our study of the quotient seminorms.

First let us introduce a convenient notation: for any $g \in B(\Omega)$, let us define the *maximum, minimum and midpoint values of $g$* by

$$\sup g \equiv \sup_{\omega \in \Omega} g(\omega) \tag{A.17a}$$

$$\inf g \equiv \inf_{\omega \in \Omega} g(\omega) \tag{A.17b}$$

$$\operatorname{mid} g \equiv \tfrac{1}{2}\left[\sup_{\omega \in \Omega} g(\omega) + \inf_{\omega \in \Omega} g(\omega)\right]$$
$$= \tfrac{1}{2}(\sup g + \inf g) \tag{A.17c}$$

Clearly we have

$$\|g\|_\infty = \max(\sup g, -\inf g) \tag{A.18a}$$

$$\|g\|_{B(\Omega)/const} = \tfrac{1}{2}(\sup g - \inf g) = \|g - \operatorname{mid} g\|_\infty . \tag{A.18b}$$

**Lemma A.5** *Let $f \in B(\Omega)$. Then:*

*(a)* $\displaystyle\lim_{\Lambda \nearrow \infty} |\Lambda|^{-1} \sup \sum_{a \in \Lambda} T_a f$ *exists and equals* $\displaystyle\inf_{\Lambda \in \mathcal{S}} |\Lambda|^{-1} \sup \sum_{a \in \Lambda} T_a f$.

*(b)* $\displaystyle\lim_{\Lambda \nearrow \infty} |\Lambda|^{-1} \inf \sum_{a \in \Lambda} T_a f$ *exists and equals* $\displaystyle\sup_{\Lambda \in \mathcal{S}} |\Lambda|^{-1} \inf \sum_{a \in \Lambda} T_a f$.

*(c)* $\displaystyle\lim_{\Lambda \nearrow \infty} |\Lambda|^{-1} \left\|\sum_{a \in \Lambda} T_a f\right\|_\infty$ *exists and equals* $\displaystyle\inf_{\Lambda \in \mathcal{S}} |\Lambda|^{-1} \left\|\sum_{a \in \Lambda} T_a f\right\|_\infty$.

*(d)* $\displaystyle\lim_{\Lambda \nearrow \infty} |\Lambda|^{-1} \left\|\sum_{a \in \Lambda} T_a f\right\|_{B(\Omega)/const}$ *exists and equals* $\displaystyle\inf_{\Lambda \in \mathcal{S}} |\Lambda|^{-1} \left\|\sum_{a \in \Lambda} T_a f\right\|_{B(\Omega)/const}$.

*(e)* $\displaystyle\lim_{\Lambda \nearrow \infty} |\Lambda|^{-1} \operatorname{mid}\left(\sum_{a \in \Lambda} T_a f\right)$ *exists and lies in the interval* $[\inf f, \sup f]$.

PROOF. (a) Consider the set function $F_+(\Lambda) \equiv \sup \sum_{a \in \Lambda} T_a f$, defined for finite subsets $\Lambda \subset \mathbb{Z}^d$. Clearly $F$ is finite-valued and translation-invariant. Moreover, it is completely subadditive (Definition A.1): if $A, A_1, \ldots, A_n \in \mathcal{S}$ with $\chi_A = \sum_{i=1}^n \lambda_i \chi_{A_i}$ and all $\lambda_i \geq 0$, then

$$F(A) \equiv \sup_{\omega \in \Omega} \sum_{a \in A} (T_a f)(\omega)$$
$$= \sup_{\omega \in \Omega} \sum_{i=1}^n \lambda_i \sum_{a \in A_i} (T_a f)(\omega)$$

$$\leq \sum_{i=1}^{n} \lambda_i \sup_{\omega \in \Omega} \sum_{a \in A_i} (T_a f)(\omega)$$

$$\equiv \sum_{i=1}^{n} \lambda_i F_+(A_i) . \tag{A.19}$$

Proposition A.3 then implies that $\lim_{\Lambda \nearrow \infty} |\Lambda|^{-1} F_+(\Lambda)$ exists and equals $\inf_{\Lambda \in \mathcal{S}} |\Lambda|^{-1} F_+(\Lambda)$.

(b) is simply (a) applied to the function $-f$.

(c) Consider the set function $F(\Lambda) \equiv \|\sum_{a \in \Lambda} T_a f\|_\infty$; the proof is then as in (a).

(d) This is an immediate consequence of (a) and (b) together with (A.18b). [Or it can be proven directly by applying complete subadditivity to $F_c(\Lambda) \equiv \|\sum_{a \in \Lambda} T_a f\|_{B(\Omega)/const}$.]

(e) This is an immediate consequence of (a) and (b). ■

**Remark.** For $f \in B_{ql}(\Omega)$ we can prove this lemma by a slightly different argument based on the fact that the set functions $F_+$, $F$ and $F_c$ are "almost additive" (and not merely *sub*additive). Since the argument is virtually identical to that used by Israel in proving the existence of the pressure [206, Theorems I.2.3 and I.2.4], we give only a brief sketch. For simplicity let us consider part (c); the other parts are similar.

Suppose first that $f$ is a bounded *local* function, i.e. that $f \in B(\Omega, \mathcal{F}_X)$ where $\text{diam}(X) < D$. Then it is easily seen that $F(\Lambda \cup \Lambda') = 2F(\Lambda)$ whenever $\Lambda'$ is a translate of $\Lambda$ with $\text{dist}(\Lambda, \Lambda') \geq D$. [Here it is important that the configuration space is a product space, so that arbitrary pairs of configurations in $\Lambda$ and $\Lambda'$ are compatible (i.e. there are no hard-core exclusions). It also seems to be important that $\Lambda'$ be a translate of $\Lambda$: this guarantees that we can choose configurations in $\Lambda$ and $\Lambda'$ that give $\sum_{a \in \Lambda} T_a f$ and $\sum_{a \in \Lambda'} T_a f$ near-maximum values *of the same sign*.] Moreover, for any two sets $\Lambda_1, \Lambda_2$ we obviously have $|F(\Lambda_1) - F(\Lambda_2)| \leq |\Lambda_1 \triangle \Lambda_2| \, \|f\|_\infty$. From these two facts one can prove the van Hove convergence of $|\Lambda|^{-1} F(\Lambda)$: the idea is to pave a large set $\Lambda$ by medium-sized cubes (of side $a$ which will eventually go to infinity) separated by corridors of width $D$. See [206, pp. 10–13] for details. The extension to general $f \in B_{ql}(\Omega)$ is now a routine approximation argument.

### A.3.5 The Quotient Seminorm

In Section 2.4.3 we stated Proposition 2.34 for the case of a *compact metric* single-spin space $\Omega_0$ and for a *continuous* function $f$. Here we prove a more general result in which these two restrictions are lifted:

**Proposition A.6 (= Proposition 2.34')** *Let $f \in B(\Omega)$. Consider the following properties:*

(a) *$f$ has zero mean with respect to every translation-invariant probability measure, i.e. $\int f \, d\mu = 0$ for all $\mu \in M_{+1,inv}(\Omega)$.*

(b) *$f$ has zero mean with respect to every translation-invariant finite signed measure, i.e. $\int f \, d\mu = 0$ for all $\mu \in M_{inv}(\Omega)$.*

(c') $f$ lies in $\mathcal{I}_f \equiv$ closed linear span of $\{f - T_a f \colon a \in \mathbb{Z}^d\}$.

(c'') $f$ lies in $\mathcal{I}_{B(\Omega)} \equiv$ closed linear span of $\{g - T_a g \colon g \in B(\Omega),\, a \in \mathbb{Z}^d\}$.

(d) $\lim_{n \to \infty} n^{-d} \left\| \sum_{a \in C_n} T_a f \right\|_\infty = 0$.

(e) $\lim_{\Lambda \nearrow \infty} |\Lambda|^{-1} \left\| \sum_{a \in \Lambda} T_a f \right\|_\infty = 0$.

Then (a) $\iff$ (b) $\impliedby$ (c') $\iff$ (c'') $\iff$ (d) $\iff$ (e). Moreover, if $\Omega_0$ is a compact metric space and $f \in C(\Omega)$, then all these properties are equivalent. [In this case property (c) of Proposition 2.34 is intermediate between (c') and (c''), hence also equivalent.]

PROOF. (a) $\implies$ (b): If $\mu \in M_{inv}(\Omega)$, then $\mu_+, \mu_- \in M_{inv}(\Omega)$ [otherwise the Jordan decomposition of $\mu$ into positive and negative parts wouldn't be unique]. So every $\mu \in M_{inv}(\Omega)$ is a linear combination of two measures in $M_{+1,inv}(\Omega)$.

(b) $\implies$ (a): Trivial.

(c') $\implies$ (c''): Trivial.

(c'') $\implies$ (e): Assume that $f = g - T_a g$ with $g \in B(\Omega)$. Then

$$\left\| \sum_{x \in \Lambda} T_x f \right\|_\infty = \left\| \sum_{x \in \Lambda} T_x g - \sum_{x \in \Lambda + a} T_x g \right\|_\infty$$
$$\leq |\Lambda \triangle (\Lambda + a)|\, \|g\|_\infty \tag{A.20}$$

where $\triangle$ denotes symmetric difference. By Proposition 2.27, $|\Lambda \triangle (\Lambda + a)|/|\Lambda| \to 0$ as $\Lambda \nearrow \infty$ (van Hove). This proves the claim for functions $f$ of the given form. The same obviously holds for finite linear combinations. It is then routine to pass to norm limits.

(e) $\implies$ (d): Trivial.

(d) $\implies$ (c'): $h_n \equiv f - n^{-d} \sum_{a \in C_n} T_a f$ lies in the linear span of $\{f - T_a f \colon a \in \mathbb{Z}^d\}$, and $\lim_{n \to \infty} \|h_n - f\|_\infty = 0$.

(d) $\implies$ (b): Since $\mu$ is translation-invariant, $\mu(f) = n^{-d} \sum_{a \in C_n} \mu(T_a f)$ for all $n$. Hence $|\mu(f)| \leq \|\mu\|\, \|n^{-d} \sum_{a \in C_n} T_a f\|_\infty$. Now let $n \to \infty$.

(b) $\implies$ (c) $\implies$ (c''), if $\Omega_0$ is compact and $f \in C(\Omega)$: Suppose that $f \notin \mathcal{I}_{C(\Omega)} \equiv$ closed linear span of $\{g - T_a g \colon g \in C(\Omega),\, a \in \mathbb{Z}^d\}$. Then, by the Hahn-Banach theorem, there exists $l \in C(\Omega)^*$ such that $l \upharpoonright \mathcal{I}_{C(\Omega)} \equiv 0$ and $l(f) = 1$. By the Riesz-Markov theorem, $l$ arises from some $\mu \in M(\Omega)$ and $l \upharpoonright \mathcal{I}_{C(\Omega)} \equiv 0$ means precisely that $\mu \in M_{inv}(\Omega)$. But then (b) implies that $l(f) = 0$, a contradiction. ∎

**Remarks.** 1. Variants of this Proposition seems to be well known [198, p. 454] (see also [66, pp. 39–40] for a similar argument), but we have not been able to find a published proof. See also [313, Exercise 7.2] for a related result.

2. We do not know whether (a)–(b) are equivalent to (c')–(e) in general; or if not, under what minimal extra conditions this equivalence can be proven. For aesthetic reasons, if no other, it would be desirable to resolve this question.

Next we prove an analogue of Proposition A.6 in which we quotient out constant functions:

**Proposition A.7** *Let $f \in B(\Omega)$. Consider the following properties:*

(a) *$f$ has the same mean with respect to every translation-invariant probability measure, i.e. $\int f \, d\mu = \int f \, d\nu$ for all $\mu, \nu \in M_{+1,inv}(\Omega)$.*

(b) *$f$ has zero mean with respect to every translation-invariant finite signed measure of zero total mass, i.e. $\int f \, d\mu = 0$ for all $\mu \in M_{inv}(\Omega)$ satisfying $\mu(\Omega) = 0$.*

(c') *$f$ lies in $\mathcal{I}_f + const \equiv$ closed linear span of $\{f - T_a f\colon a \in \mathbb{Z}^d\}$ and constant functions.*

(c'') *$f$ lies in $\mathcal{I}_{B(\Omega)} + const \equiv$ closed linear span of $\{g - T_a g\colon g \in B(\Omega), a \in \mathbb{Z}^d\}$ and constant functions.*

(d) $\displaystyle\lim_{n\to\infty} n^{-d} \Big\| \sum_{a \in C_n} T_a f \Big\|_{B(\Omega)/const} = 0.$

(e) $\displaystyle\lim_{\Lambda \nearrow \infty} |\Lambda|^{-1} \Big\| \sum_{a \in \Lambda} T_a f \Big\|_{B(\Omega)/const} = 0.$

*Then* (a) $\iff$ (b) $\impliedby$ (c') $\iff$ (c'') $\iff$ (d) $\iff$ (e). *Moreover, if $\Omega_0$ is a compact metric space and $f \in C(\Omega)$, then all these properties are equivalent.*

PROOF. (a) $\implies$ (b): If $\mu \in M_{inv}(\Omega)$ with $\mu(\Omega) = 0$, then $\mu_+, \mu_- \in M_{inv}(\Omega)$ with $\mu_+(\Omega) = \mu_-(\Omega) = \lambda \geq 0$. If $\lambda = 0$ we are done; if $\lambda > 0$, apply (a) to the measures $\lambda^{-1}\mu_+, \lambda^{-1}\mu_- \in M_{+1,inv}(\Omega)$.

(b) $\implies$ (a): Just apply (b) to $\mu - \nu$.

(c') $\implies$ (c''): Trivial.

(c'') $\implies$ (e): Assume that $f = g - T_a g + c$ with $g \in B(\Omega)$ and $c \in \mathbb{R}$. Then

$$\Big\| \sum_{x \in \Lambda} T_x f \Big\|_{B(\Omega)/const} = \Big\| \sum_{x \in \Lambda} T_x g - \sum_{x \in \Lambda + a} T_x g + c|\Lambda| \Big\|_{B(\Omega)/const}$$
$$\leq \Big\| \sum_{x \in \Lambda} T_x g - \sum_{x \in \Lambda + a} T_x g \Big\|_\infty$$
$$\leq |\Lambda \triangle (\Lambda + a)| \, \|g\|_\infty \,. \qquad (A.21)$$

The rest is as in Proposition A.6.

(e) $\implies$ (d): Trivial.

(d) $\implies$ (c'): Let $c \equiv \lim_{n\to\infty} n^{-d} \operatorname{mid}\big(\sum_{a \in C_n} T_a f\big)$ as guaranteed by Lemma A.5(e). Then $h_n \equiv f - n^{-d} \sum_{a \in C_n} T_a f$ lies in the linear span of $\{f - T_a f\colon a \in \mathbb{Z}^d\}$, and

$$\limsup_{n\to\infty} \|(h_n + c) - f\|_\infty = \limsup_{n\to\infty} \Big\| n^{-d} \sum_{a \in C_n} T_a f - c \Big\|_\infty$$
$$\leq \limsup_{n\to\infty} n^{-d} \Big\| \sum_{a \in C_n} T_a f \Big\|_{B(\Omega)/const}$$
$$= 0 \,. \qquad (A.22)$$

(d) $\implies$ (b): A trivial modification of the corresponding proof in Proposition A.6.

(b) $\implies$ (c) $\implies$ (c''), if $\Omega_0$ is compact and $f \in C(\Omega)$: Same as in Proposition A.6, but use the subspace $\mathcal{I}_{C(\Omega)} + const$ in place of $\mathcal{I}_{C(\Omega)}$; the signed measure $\mu$ will then have zero total mass. ∎

Next we prove a strengthened version of Proposition 2.35. Again we can allow an arbitrary (not necessarily compact) single-spin space $\Omega_0$, and an arbitrary (not necessarily continuous or quasilocal) function $f$. The only subtlety is that in this case we must choose the correct definition of $\mathcal{I}$, since (a)–(b) and (c')–(e) are not necessarily equivalent. The right definition turns out to be (c')–(e).

**Proposition A.8 (= Proposition 2.35')** *Let $f \in B(\Omega)$. Then*

$$\lim_{\Lambda \nearrow \infty} |\Lambda|^{-1} \Big\| \sum_{a \in \Lambda} T_a f \Big\|_\infty \;=\; \inf_{\Lambda \in \mathcal{S}} |\Lambda|^{-1} \Big\| \sum_{a \in \Lambda} T_a f \Big\|_\infty \tag{A.23a}$$

$$= \; \|f\|_{B(\Omega)/\widetilde{\mathcal{I}}} \tag{A.23b}$$

*and*

$$\lim_{\Lambda \nearrow \infty} |\Lambda|^{-1} \Big\| \sum_{a \in \Lambda} T_a f \Big\|_{B(\Omega)/const} \;=\; \inf_{\Lambda \in \mathcal{S}} |\Lambda|^{-1} \Big\| \sum_{a \in \Lambda} T_a f \Big\|_{B(\Omega)/const} \tag{A.24a}$$

$$= \; \|f\|_{B(\Omega)/(\widetilde{\mathcal{I}}+const)} \tag{A.24b}$$

*for all closed linear subspaces $\widetilde{\mathcal{I}}$ satisfying $\mathcal{I}_f \subset \widetilde{\mathcal{I}} \subset \mathcal{I}_{B(\Omega)}$.*

PROOF. In Lemma A.5(c,d) we have proven the existence of the limits and their equality to the corresponding infima. Now we want to identify the limits with the quotient seminorms.

Let us denote by $L_f$ the limit (A.23a). Clearly $L_f \leq \|f\|_\infty$. Moreover, by Proposition A.6 (c') $\implies$ (e), $L_f = L_{f'}$ whenever $f - f' \in \mathcal{I}_{B(\Omega)}$; hence $L_f \leq \|f\|_{B(\Omega)/\mathcal{I}_{B(\Omega)}} \leq \|f\|_{B(\Omega)/\widetilde{\mathcal{I}}}$.

To prove the reverse inequality, note that by an easy corollary of the Hahn-Banach theorem [306, Corollary 3 of Section III.3] there exists $l \in B(\Omega)^*$ such that $\|l\| \leq 1$, $l(f) = \|f\|_{B(\Omega)/\widetilde{\mathcal{I}}}$ and $l\!\restriction\!\widetilde{\mathcal{I}} \equiv 0$. On the other hand, for every $l \in B(\Omega)^*$ that annihilates $\widetilde{\mathcal{I}} \supset \mathcal{I}_f$ we have

$$l(f) \;=\; l\Big(n^{-d} \sum_{a \in C_n} T_a f\Big) \;\stackrel{n \to \infty}{\longrightarrow}\; \leq L_f \, \|l\|. \tag{A.25}$$

Hence $\|f\|_{B(\Omega)/\widetilde{\mathcal{I}}} \leq L_f$. This proves (A.23b).

A completely analogous argument handles (A.24): it suffices to replace $\widetilde{\mathcal{I}}$ and $\mathcal{I}_{B(\Omega)}$ everywhere by $\widetilde{\mathcal{I}} + const$ and $\mathcal{I}_{B(\Omega)} + const$, respectively. ∎

**Remark.** The proof given here of (A.23) is a slight elaboration of one sketched by Hugenholtz [198, p. 454]; by using *complete* subadditivity we are able to deduce the full van Hove convergence.

### A.3.6 Closed and Compact Sets in $\mathcal{B}^0$

PROOF OF PROPOSITION 2.39. (a) We shall actually prove something slightly stronger, namely that $\{\Phi\colon \|\Phi\|_{\mathcal{B}_h} \leq M\}$ is closed in the product topology $\prod_{X \in \mathcal{S}} C(\Omega_X)$ (which is weaker than the $\mathcal{B}^0$ norm topology). So let $(\Phi_n)$ be a sequence in $\{\Phi\colon \|\Phi\|_{\mathcal{B}_h} \leq M\}$, and let $\Phi$ be another interaction; and suppose that $\|(\Phi_n)_X - \Phi_X\| \to 0$ for each $X$. (This would occur, in particular, if $\Phi_n \to \Phi$ in $\mathcal{B}^0$ norm.) Then

$$\begin{aligned}
\|\Phi\|_{\mathcal{B}_h} &= \sum_{X \ni 0} \frac{h(X)}{|X|} \|\Phi_X\| = \sum_{X \ni 0} \frac{h(X)}{|X|} \lim_{n \to \infty} \|(\Phi_n)_X\| \\
&\leq \liminf_{n \to \infty} \sum_{X \ni 0} \frac{h(X)}{|X|} \|(\Phi_n)_X\| \\
&= \liminf_{n \to \infty} \|\Phi_n\|_{\mathcal{B}_h} \\
&\leq M \;, \quad\quad\quad\quad\quad\quad\quad\quad\quad (A.26)
\end{aligned}$$

where in the key inequality we have used Fatou's lemma.

(b) Let $(\Phi_n)$ be a sequence in $\{\Phi\colon \|\Phi\|_{\mathcal{B}_h} \leq M\}$. Since the single-spin space is finite, each space $C(\Omega_X)$, $X$ finite, is finite-dimensional. Therefore, by compactness of the ball in $C(\Omega_X)$ together with the usual diagonal argument, we can extract a subsequence $(\Phi_{n'})$ such that $(\Phi_{n'})_X$ converges (in $\|\cdot\|_\infty$ norm) for each $X$, say to $\Phi_X$. Let $\Phi = \{\Phi_X\}$. In part (a) we have shown that $\|\Phi\|_{\mathcal{B}_h} \leq M$. Now we wish to show that $\Phi_{n'} \to \Phi$ in $\mathcal{B}^0$ norm. So fix $K < \infty$; we then have

$$\begin{aligned}
\|\Phi_{n'} - \Phi\|_{\mathcal{B}^0} &= \sum_{\substack{X \ni 0 \\ h(X) < K}} \frac{1}{|X|} \|(\Phi_{n'})_X - \Phi_X\| + \sum_{\substack{X \ni 0 \\ h(X) \geq K}} \frac{1}{|X|} \|(\Phi_{n'})_X - \Phi_X\| \\
&\leq \sum_{\substack{X \ni 0 \\ h(X) < K}} \frac{1}{|X|} \|(\Phi_{n'})_X - \Phi_X\| + 2M/K \;. \quad\quad (A.27)
\end{aligned}$$

Since $h \gtrsim 1$, the first sum is finite; and since $\|(\Phi_{n'})_X - \Phi_X\| \to 0$ for each $X$, we have

$$\limsup_{n'} \|\Phi_{n'} - \Phi\|_{\mathcal{B}^0} \leq 2M/K \;. \quad\quad (A.28)$$

Since $K$ may be taken arbitrarily large, we are done. ■

**Remarks.** 1. For a converse to part (b), see Proposition A.10 below.

2. One might ask whether (a) and (b) can be extended to more general closed bounded sets in $\mathcal{B}_h$ (not just balls). The answer is no, in general: a closed bounded convex set in $\mathcal{B}_h$ need not be closed (much less compact!) in $\mathcal{B}^0$, if $h$ is unbounded. *Example:* Let $\{A_n\}$ be a sequence of finite subsets of $\mathbb{Z}^d$, in which each equivalence class modulo translation occurs at most once, and satisfying $\lim_{n \to \infty} h(A_n) = +\infty$. Let $\Phi_n$ be defined by

$$(\Phi_n)_A = \begin{cases} 1/h(A_n) & \text{if } A \text{ is a translate of } A_n \\ 0 & \text{otherwise} \end{cases} \quad\quad (A.29)$$

Now, for each sequence $\lambda \in \ell^1$, let $\Phi^\lambda = \sum_{n=1}^\infty \lambda_n \Phi_n$ (this sum is absolutely convergent in $\mathcal{B}_h$). Then $\|\Phi^\lambda\|_{\mathcal{B}_h} = \|\lambda\|_{\ell^1}$ and $\|\Phi^\lambda - \Phi^{\lambda'}\|_{\mathcal{B}_h} = \|\lambda - \lambda'\|_{\ell^1}$. (That is, $\lambda \mapsto \Phi^\lambda$ is an isometric isomorphism of $\ell^1$ onto a closed linear subspace of $\mathcal{B}_h$.) Now let $S = \{\Phi_n\}$, and let

$$T = \{\Phi^\lambda \colon 0 \le \lambda_n \le 1 \, \forall n, \, \sum_{n=1}^\infty \lambda_n = 1\} . \tag{A.30}$$

$T$ is the closed convex hull of $S$ in $\mathcal{B}_h$. Now, $\|\Phi\|_{\mathcal{B}_h} = 1$ for all $\Phi \in T$, so $0 \notin T$. On the other hand, $0$ *does* belong to the closure of $T$ in $\mathcal{B}^0$, since $\lim_{n \to \infty} \|\Phi_n\|_{\mathcal{B}^0} = 0$.

The natural setting for discussing the spaces $\mathcal{B}^0$ and $\mathcal{B}_h$ is that of *weighted $\ell^1$ direct sums of Banach spaces*. Let $Y_1, Y_2, \ldots$ be Banach spaces, and let $h \colon \mathbb{N} \to (0, \infty)$. Then we define $\mathbf{Y}_h$ to be the space of sequences $y = (y_1, y_2, \ldots)$, with each $y_i \in Y_i$, for which the norm

$$\|y\|_{\mathbf{Y}_h} \equiv \sum_{i=1}^\infty h(i) \|y_i\|_{Y_i} \tag{A.31}$$

is finite. For $h \equiv 1$ we write $\mathbf{Y}_h = \mathbf{Y}$. It is easy to prove that all the spaces $\mathbf{Y}_h$ are Banach spaces. The canonical projection $p_i \colon \mathbf{Y}_h \to Y_i$ defined by $p_i(y) = y_i$ has norm $1/h(i)$.

We then have the following results:

**Proposition A.9** *The closed ball $\{y \colon \|y\|_{\mathbf{Y}_h} \le M\}$ is closed in the product topology $\prod_i Y_i$.*

**Proposition A.10** *Let $S \subset \mathbf{Y}$. Then the following are equivalent:*

*(a) $S$ has compact closure in $\mathbf{Y}$.*

*(b) $S$ is bounded, $p_i[S]$ has compact closure in $Y_i$ for each $i$, and*

$$\lim_{N \to \infty} \sup_{y \in S} \sum_{i=N}^\infty \|y_i\|_{Y_i} = 0 . \tag{A.32}$$

*(c) $S$ is bounded, $p_i[S]$ has compact closure in $Y_i$ for each $i$, and there exists a function $h \colon \mathbb{N} \to [1, \infty)$ such that $\lim_{i \to \infty} h(i) = +\infty$ and*

$$\sup_{y \in S} \|y\|_{\mathbf{Y}_h} < \infty . \tag{A.33}$$

Note that if $Y_i$ is finite-dimensional, then $S$ bounded $\Longrightarrow p_i[S]$ bounded $\Longrightarrow p_i[S]$ has compact closure in $Y_i$.

The proof of Proposition A.9 is completely analogous to that of Proposition 2.39(a). Let us sketch the proof of Proposition A.10:

(a) $\Longrightarrow$ (b): Let $\bar{S}$ be compact in $\mathbf{Y}$. Then clearly $p_i[\bar{S}] \supset p_i[S]$ is compact in $Y_i$. Moreover, for each $\epsilon > 0$ there exists a finite set $y^{(1)}, \ldots, y^{(n)} \in \mathbf{Y}$ such that $S \subset \bigcup_{k=1}^{n} B(y^{(k)}, \epsilon)$. It follows that

$$\sup_{y \in S} \sum_{i=N}^{\infty} \|y_i\|_{Y_i} \leq \epsilon + \max_{1 \leq k \leq n} \sum_{i=N}^{\infty} \|y_i^{(k)}\|_{Y_i} . \tag{A.34}$$

Taking $N \to \infty$, we get

$$\limsup_{N \to \infty} \sup_{y \in S} \sum_{i=N}^{\infty} \|y_i\|_{Y_i} \leq \epsilon . \tag{A.35}$$

Since $\epsilon$ was arbitrary, the proof is complete.

(b) $\Longrightarrow$ (c): Choose $N_1 < N_2 < \ldots$ such that

$$\sup_{y \in S} \sum_{i=N_m}^{\infty} \|y_i\|_{Y_i} \leq 3^{-m} . \tag{A.36}$$

Now define

$$h(i) = \begin{cases} 1 & \text{for } i < N_1 \\ 2^m & \text{for } N_m \leq i < N_{m+1} \end{cases} \tag{A.37}$$

Then, for all $y \in S$,

$$\begin{aligned}
\|y\|_{\mathbf{Y}_h} \equiv \sum_{i=1}^{\infty} h(i) \|y_i\|_{Y_i} &= \sum_{i=1}^{N_1-1} \|y_i\|_{Y_i} + \sum_{m=1}^{\infty} 2^m \sum_{i=N_m}^{N_{m+1}-1} \|y_i\|_{Y_i} \\
&\leq \sum_{i=1}^{N_1-1} \|y_i\|_{Y_i} + \sum_{m=1}^{\infty} \left(\frac{2}{3}\right)^m \\
&\leq \|y\|_{\mathbf{Y}} + 2 .
\end{aligned} \tag{A.38}$$

Since $S$ is, by hypothesis, bounded in $\mathbf{Y}$, this proves the claim.

(c) $\Longrightarrow$ (a): The proof is essentially identical to that of Proposition 2.39(b). ∎

To apply this to our statistical-mechanical setup, let $(X_i)$ be a sequence of nonempty finite subsets of $\mathbb{Z}^d$ in which each equivalence class modulo translation is represented once and only once. Setting $Y_i = C(\Omega_{X_i})$, it is easy to see that $\mathcal{B}^0$ and $\mathcal{B}_h$ are isometric to the direct-sum spaces $\mathbf{Y}$ and $\mathbf{Y}_h$, respectively. Therefore, Proposition A.10 (a)$\Longrightarrow$(c) tells us that any compact subset of $\mathcal{B}^0$ is contained in the ball $\{\Phi: \|\Phi\|_{\mathcal{B}_h} \leq M\}$ for some $h \gtrsim 1$ and some $M < \infty$. A similar result can be found in [208, Lemmas 1 and 2].

PROOF OF PROPOSITION 2.43.  This is an immediate consequence of Proposition 2.39(b), together with the following well-known fact: if $A$ and $B$ are subsets of a Banach space $X$, with $A$ compact and $B$ closed, then $A + B$ is closed. ∎

### A.3.7 Physical Equivalence

Here we prove Theorem 2.42 on the equivalence of the two notions of physical equivalence (DLR and Ruelle). Since both senses of physical equivalence are statements about the difference $\Phi - \Phi'$, it suffices to consider the case $\Phi' = 0$.

PROOF OF THEOREM 2.42, DLR $\implies$ RUELLE. We wish to measure how strongly $H_\Lambda^\Phi(\omega_\Lambda, \omega_{\Lambda^c})$ depends on $\omega_{\Lambda^c}$. Let us therefore define the *oscillation of $H_\Lambda^\Phi$ with respect to $\omega_{\Lambda^c}$* by

$$\begin{aligned}
\operatorname{osc}_{\Lambda^c}(H_\Lambda^\Phi) &\equiv \sup_{\substack{\omega, \omega' \in \Omega \\ \omega_\Lambda = \omega'_\Lambda}} |H_\Lambda^\Phi(\omega) - H_\Lambda^\Phi(\omega')| \\
&= \sup_{\omega_\Lambda, \omega_{\Lambda^c}, \omega'_{\Lambda^c}} |H_\Lambda^\Phi(\omega_\Lambda, \omega_{\Lambda^c}) - H_\Lambda^\Phi(\omega_\Lambda, \omega'_{\Lambda^c})| .
\end{aligned} \tag{A.39}$$

Considering now the definition $H_\Lambda^\Phi(\omega) = \sum_{A:\, A \cap \Lambda \neq \emptyset} \Phi_A(\omega)$, it is easy to see that $\operatorname{osc}_{\Lambda^c}(H_\Lambda^\Phi)$ gets contributions only from sets $A$ that intersect both $\Lambda$ and $\Lambda^c$, so that

$$\operatorname{osc}_{\Lambda^c}(H_\Lambda^\Phi) \leq 2 \|W_{\Lambda, \Lambda^c}^\Phi\|_\infty . \tag{A.40}$$

In particular, for $\Phi \in \mathcal{B}^1$ we have

$$\operatorname{osc}_{\Lambda^c}(H_\Lambda^\Phi) \leq o(|\Lambda|) \qquad \text{as } \Lambda \nearrow \infty \text{ (van Hove)} , \tag{A.41}$$

by (2.62a).

Suppose now that $\Phi$ is physically equivalent to 0 in the DLR sense, i.e. that $H_\Lambda^\Phi$ is $\mathcal{F}_{\Lambda^c}$-measurable for all $\Lambda$. (Actually, it suffices to assume this for some van Hove sequence of sets $\Lambda$.) Then $H_\Lambda^\Phi(\omega_\Lambda, \omega_{\Lambda^c})$ is independent of $\omega_\Lambda$, so $\operatorname{osc}_{\Lambda^c}(H_\Lambda^\Phi)$ is equal to the *unrestricted* oscillation

$$\begin{aligned}
\operatorname{osc}(H_\Lambda^\Phi) &\equiv \sup H_\Lambda^\Phi - \inf H_\Lambda^\Phi \tag{A.42} \\
&\equiv 2 \|H_\Lambda^\Phi\|_{B(\Omega)/const} . \tag{A.43}
\end{aligned}$$

Combining (A.41) and (A.43), we conclude that

$$\|H_\Lambda^\Phi\|_{B(\Omega)/const} \leq o(|\Lambda|) \qquad \text{as } \Lambda \nearrow \infty \text{ (van Hove)} . \tag{A.44}$$

By Proposition 2.45(c), we conclude that $\|\Phi\|_{\mathcal{B}^0/(\mathcal{J}+Const)} = 0$, i.e. $\Phi \in \mathcal{J} + Const$ — that is, $\Phi$ is physically equivalent to zero in the Ruelle sense. ∎

PROOF OF THEOREM 2.42, RUELLE $\implies$ DLR. Suppose that the single-spin space $\Omega_0$ is a standard Borel space (e.g. a complete separable metric space), and that $\Phi, \Phi' \in \mathcal{B}^1$ are physically equivalent in the Ruelle sense. Then by [157, Theorems 4.22 and 5.19 and the comments after them], there exists a translation-invariant Gibbs measure for $\Phi$, call it $\mu$. By Corollary 2.68, $\mu$ is an equilibrium measure for $\Phi$. By Proposition 2.65, $\mu$ is an equilibrium measure also for $\Phi'$. By Corollary 2.68 again, $\mu$ is a Gibbs measure for $\Phi'$. But then Corollary 2.18 implies that $\Phi$ and $\Phi'$ are physically equivalent in the DLR sense. ∎

**Remark.** The proof given here of Ruelle $\Longrightarrow$ DLR is aesthetically unsatisfying: the two notions of physical equivalence are statements purely about interactions and Hamiltonians, so there ought to be a purely "algebraic" proof of their equivalence involving *only these concepts*, without dragging in the whole theory of equilibrium measures, Gibbs measures and their equivalence. In particular, it is galling to have to assume that $\Omega_0$ is a standard Borel space, for a result that obviously has nothing to do with topology. However, we have been unable to find such an algebraic proof; we hope that some reader will do so.

### A.3.8 Estimates on Hamiltonians and Gibbs Measures

In Section 2.4.5 we stated Proposition 2.40 for the case of a *compact metric* single-spin space. Here we prove a more general theorem in which this restriction is removed. (We still consider only *continuous* interactions and functions, but that restriction too could be removed if we really cared.)

**Proposition A.11 (= Proposition 2.40')** *The map $[\Phi] \mapsto [f_\Phi]$ is an isometry of $\mathcal{B}^0/\mathcal{J}$ onto $C_{ql}(\Omega)/\mathcal{I}_{ql}$, and of $\mathcal{B}^0/(\mathcal{J} + Const)$ onto $C_{ql}(\Omega)/(\mathcal{I}_{ql} + const)$. Here $\mathcal{I}_{ql} \equiv \mathcal{I} \cap C_{ql}(\Omega)$.*

PROOF. It is convenient (following Ruelle [313, Section 3.2]) to introduce the modified observable
$$f''_\Phi \equiv \sum_{X \ni_{mid} 0} \Phi_X \;, \tag{A.45}$$
where $X \ni_{mid} 0$ denotes that 0 is the $\lfloor (|X|+1)/2 \rfloor^{th}$ element ("middle element") of $X$ in lexicographic order. Clearly $f_\Phi - f''_\Phi \in \mathcal{I}_{ql}$. The advantages of $f''_\Phi$ are due to the following easily verified facts [313, p. 37]:

a) $\{f''_\Phi \colon \Phi \in \mathcal{B}_{finite}\} = C_{loc}(\Omega)$.

b) $\{f''_\Phi \colon \Phi \in \mathcal{B}^0\} = C_{ql}(\Omega)$.

c) For all $f \in C_{ql}(\Omega)$,
$$\|f\|_\infty = \inf_{\Phi \in \mathcal{B}^0 \colon f''_\Phi = f} \|\Phi\|_{\mathcal{B}^0} \;. \tag{A.46}$$

Moreover, for $f \in C_{loc}(\Omega)$ there exists a $\Phi \in \mathcal{B}_{finite}$ that attains this minimum.

In particular, the map $\Phi \mapsto [f_\Phi] = [f''_\Phi]$ is *onto* $C_{ql}(\Omega)/\mathcal{I}_{ql}$.

Now, from $\|f_\Phi\|_\infty \leq \|\Phi\|_{\mathcal{B}^0}$ we easily deduce that

$$\|[f_\Phi]\|_{C(\Omega)/\mathcal{I}} \leq \|[\Phi]\|_{\mathcal{B}^0/\mathcal{J}} . \tag{A.47}$$

To prove the reverse inequality, note that by Proposition A.8 we have, for any $f \in C_{ql}(\Omega)$, $\|[f]\|_{C(\Omega)/\mathcal{I}} = \lim_{\Lambda \nearrow \infty} \left\||\Lambda|^{-1} \sum_{a \in \Lambda} T_a f\right\|_\infty$. Now, by property (c) above, for each $\Lambda$ and each $\epsilon > 0$ we can choose $\Psi \in \mathcal{B}^0$ such that $f''_\Psi = |\Lambda|^{-1} \sum_{a \in \Lambda} T_a f$ and $\|\Psi\|_{\mathcal{B}^0} \leq \left\||\Lambda|^{-1} \sum_{a \in \Lambda} T_a f\right\|_\infty + \epsilon$. (In particular, we have $[f] = [f''_\Psi] = [f_\Psi]$.) Then, by taking $\Lambda \nearrow \infty$ and $\epsilon \downarrow 0$ we conclude that for all $f \in C_{ql}(\Omega)$,

$$\|[f]\|_{C(\Omega)/\mathcal{I}} \geq \inf_{\Psi \in \mathcal{B}^0:\, [f''_\Psi]=[f]} \|\Psi\|_{\mathcal{B}^0} . \tag{A.48}$$

In particular, taking $f = f_\Phi$, we get

$$\|[f_\Phi]\|_{C(\Omega)/\mathcal{I}} \geq \|[\Phi]\|_{\mathcal{B}^0/\mathcal{J}} . \tag{A.49}$$

This proves that the map $[\Phi] \mapsto [f_\Phi]$ is an isometry of $\mathcal{B}^0/\mathcal{J}$ into $C(\Omega)/\mathcal{I}$.

Repeating the same argument with $f$ replaced by $f + c$, and then optimizing over $c$, we conclude that $[\Phi] \mapsto [f_\Phi]$ is also an isometry of $\mathcal{B}^0/(\mathcal{J} + Const)$ into $C(\Omega)/(\mathcal{I} + const)$. ∎

PROOF OF PROPOSITION 2.44.

(a) is easy and well known: see [206, p. 9].

(d) By definition we have

$$H^\Phi_{\Lambda, free} = \sum_{X \subset \Lambda} \Phi_X \tag{A.50}$$

and

$$\begin{aligned} \sum_{x \in \Lambda} T_x f_\Phi &= \sum_{x \in \Lambda} \sum_{X \ni 0} |X|^{-1} T_x \Phi_X \\ &= \sum_{x \in \Lambda} \sum_{X \ni 0} |X|^{-1} \Phi_{X+x} \\ &= \sum_{x \in \Lambda} \sum_{Y \ni x} |Y|^{-1} \Phi_Y \\ &= \sum_Y \frac{|Y \cap \Lambda|}{|Y|} \Phi_Y \end{aligned} \tag{A.51}$$

(the double sum is absolutely convergent and hence can be rearranged freely). Thus

$$H^\Phi_{\Lambda, free} - \sum_{x \in \Lambda} T_x f_\Phi = - \sum_{\substack{X \cap \Lambda \neq \emptyset \\ X \cap \Lambda^c \neq \emptyset}} \frac{|X \cap \Lambda|}{|X|} \Phi_X . \tag{A.52}$$

Taking norms, we have

$$\begin{aligned}
\|H^{\Phi}_{\Lambda, free} - \sum_{x \in \Lambda} T_x f_\Phi \|_\infty &\leq \sum_{\substack{X \cap \Lambda \neq \emptyset \\ X \cap \Lambda^c \neq \emptyset}} \frac{|X \cap \Lambda|}{|X|} \|\Phi_X\|_\infty \\
&= \sum_{x \in \Lambda} \sum_{\substack{X \ni x \\ X \cap \Lambda^c \neq \emptyset}} |X|^{-1} \|\Phi_X\|_\infty \\
&= \sum_{x \in \Lambda} \sum_{\substack{Y \ni 0 \\ (Y+x) \cap \Lambda^c \neq \emptyset}} |Y|^{-1} \|\Phi_Y\|_\infty \\
&= \sum_{Y \ni 0} \frac{|(\Lambda^c - Y) \cap \Lambda|}{|Y|} \|\Phi_Y\|_\infty \quad\quad (A.53)
\end{aligned}$$

Now divide by $|\Lambda|$:

$$|\Lambda|^{-1} \|H^{\Phi}_{\Lambda, free} - \sum_{x \in \Lambda} T_x f_\Phi \|_\infty \leq \sum_{Y \ni 0} \frac{|(\Lambda^c - Y) \cap \Lambda|}{|\Lambda|} \frac{\|\Phi_Y\|_\infty}{|Y|} \, . \quad\quad (A.54)$$

This sum is dominated uniformly in $\Lambda$, since $|(\Lambda^c - Y) \cap \Lambda|/|\Lambda| \leq 1$ and $\Phi \in \mathcal{B}^0$. On the other hand, for each fixed finite set $Y$, we have

$$\begin{aligned}
\frac{|(\Lambda^c - Y) \cap \Lambda|}{|\Lambda|} &\leq \sum_{y \in Y} \frac{|\Lambda \cap (\Lambda^c - y)|}{|\Lambda|} \\
&= \sum_{y \in Y} \frac{|\Lambda \setminus (\Lambda - y)|}{|\Lambda|} \, , \quad\quad (A.55)
\end{aligned}$$

which tends to zero as $\Lambda \nearrow \infty$ (van Hove). Hence, by the dominated convergence theorem, (A.54) tends to zero as $\Lambda \nearrow \infty$ (van Hove).

(b) and (c) are immediate consequences of (d) together with Propositions A.8 and A.11. ∎

**Remark.** See [313, p. 41] for an alternate proof of (c), carried out first for $\Phi \in \mathcal{B}_{finite}$ and then extended to $\mathcal{B}^0$ by density.

PROOF OF PROPOSITION 2.45.

(a) is easy and well known: see [206, p. 14] or [157, p. 29].

(d) By definition,

$$H^{\Phi}_\Lambda - H^{\Phi}_{\Lambda, free} = W^{\Phi}_{\Lambda, \Lambda^c} = \sum_{\substack{X \cap \Lambda \neq \emptyset \\ X \cap \Lambda^c \neq \emptyset}} \Phi_X \, . \quad\quad (A.56)$$

Taking norms, we have

$$
\begin{aligned}
\|H_\Lambda^\Phi - H_{\Lambda,free}^\Phi\|_\infty &\leq \sum_{\substack{X \cap \Lambda \neq \emptyset \\ X \cap \Lambda^c \neq \emptyset}} \|\Phi_X\|_\infty \\
&\leq \sum_{x \in \Lambda} \sum_{\substack{X \ni x \\ X \cap \Lambda^c \neq \emptyset}} \|\Phi_X\|_\infty \\
&= \sum_{x \in \Lambda} \sum_{\substack{Y \ni 0 \\ (Y+x) \cap \Lambda^c \neq \emptyset}} \|\Phi_Y\|_\infty \\
&= \sum_{Y \ni 0} |(\Lambda^c - Y) \cap \Lambda|\, \|\Phi_Y\|_\infty\,. \qquad (A.57)
\end{aligned}
$$

The remainder of the argument is completely parallel to the proof of Proposition 2.44(d), but using $\Phi \in \mathcal{B}^1$ rather than $\Phi \in \mathcal{B}^0$. This proves (2.62a).

As for (2.62b), the leftmost term is $o(|\Lambda|)$ as an immediate consequence of (2.62a) and (2.58). The middle term is proven to be $o(|\Lambda|)$ in [157, pp. 320–321]. (That proof is stated only for cubes, but it is valid for arbitrary van Hove sequences.)

(b) is then an immediate consequence of (2.62a) and (2.56). (c) is likewise an immediate consequence of (2.62a) and (2.57). ∎

PROOF OF PROPOSITION 2.46. Let $\mu$ be a Gibbs measure for an interaction $\Phi \in \mathcal{B}^1$ and *a priori* measure $\mu^0$. Then the DLR equation (2.22) states that

$$\frac{d\mu_\Lambda}{d\mu_\Lambda^0}(\omega_\Lambda) = \int d\mu(\tau)\, Z_\Lambda^\Phi(\tau_{\Lambda^c})^{-1}\, \exp[-H_\Lambda^\Phi(\omega_\Lambda \times \tau_{\Lambda^c})]\,, \qquad (A.58)$$

where

$$Z_\Lambda^\Phi(\tau_{\Lambda^c}) = \int \exp[-H_\Lambda^\Phi(\omega_\Lambda \times \tau_{\Lambda^c})] \prod_{x \in \Lambda} d\mu_x^0(\omega_x)\,. \qquad (A.59)$$

Now, by Proposition 2.45(d), we can replace $H_\Lambda^\Phi(\omega_\Lambda \times \tau_{\Lambda^c})$ everywhere by $H_{\Lambda,free}^\Phi(\omega_\Lambda)$, incurring an error which is $o(|\Lambda|)$ uniformly in $\omega$ and $\tau$. Therefore,

$$\left\|\log \frac{d\mu_\Lambda}{d\mu_\Lambda^0} + H_{\Lambda,free}^\Phi + \log Z_{\Lambda,free}^\Phi\right\|_\infty \leq o(|\Lambda|)\,. \qquad (A.60)$$

But $\|H_{\Lambda,free}^\Phi - \sum_{x \in \Lambda} T_x f_\Phi\|_\infty \leq o(|\Lambda|)$ by Proposition 2.44(c), and $|\log Z_{\Lambda,free}^\Phi - |\Lambda| p(\Phi|\mu^0)| \leq o(|\Lambda|)$ by Proposition 2.58(a). Hence

$$\left\|\log \frac{d\mu_\Lambda}{d\mu_\Lambda^0} + \sum_{x \in \Lambda} T_x f_\Phi + |\Lambda| p(\Phi|\mu^0)\right\|_\infty \leq o(|\Lambda|)\,. \qquad (A.61)$$

In particular,

$$\left\|\log \frac{d\mu_\Lambda}{d\mu_\Lambda^0} + \sum_{x \in \Lambda} T_x f_\Phi \right\|_{C(\Omega)/const} \leq o(|\Lambda|) . \tag{A.62}$$

This proves (2.63). This bound is uniform for all $\mu \in \mathcal{G}(\Pi^\Phi)$.

Now let $\mu_1$ (resp. $\mu_2$) be Gibbsian for interactions $\Phi_1$ (resp. $\Phi_2$) in $\mathcal{B}^1$, with the same *a priori* measure $\mu^0$. Combining (A.61) for the two cases, we get

$$\left\|\log \frac{d\mu_{1\Lambda}}{d\mu_{2\Lambda}}\right\|_\infty = \left\|\sum_{x \in \Lambda} T_x f_{\Phi_1 - \Phi_2} + |\Lambda|[p(\Phi_1|\mu^0) - p(\Phi_2|\mu^0)]\right\|_\infty + o(|\Lambda|) . \tag{A.63}$$

But

$$\left\|\sum_{x \in \Lambda} T_x f_{\Phi_1 - \Phi_2}\right\|_\infty = |\Lambda| \, \|\Phi_1 - \Phi_2\|_{\mathcal{B}^0/\mathcal{J}} + o(|\Lambda|) \tag{A.64}$$

by Propositions A.8 and A.11, while

$$\left|p(\Phi_1|\mu^0) - p(\Phi_2|\mu^0)\right| \leq \|\Phi_1 - \Phi_2\|_{\mathcal{B}^0/\mathcal{J}} \tag{A.65}$$

by Propositions 2.56(d,e) and 2.58(a). Hence

$$\left\|\log \frac{d\mu_{1\Lambda}}{d\mu_{2\Lambda}}\right\|_\infty \leq 2|\Lambda| \, \|\Phi_1 - \Phi_2\|_{\mathcal{B}^0/\mathcal{J}} + o(|\Lambda|) . \tag{A.66}$$

But by Propositions 2.56(c) and 2.58(a), the right-hand side of (A.63) is unchanged if we replace $\Phi_1$ by $\Phi_1 + \Psi$ with $\Psi \in Const$ (i.e. if $f_\Psi \in const$). Thus, in (A.66) we can replace the $\mathcal{B}^0/\mathcal{J}$ norm by $\mathcal{B}^0/(\mathcal{J} + Const)$. This proves (2.64).

In a similar way we deduce (2.65) from the two cases of (A.61) together with (A.64). ∎

**Remarks.** 1. We wish to emphasize that (2.65) is an *equality*. This fact plays a crucial role in our proof of the Second Fundamental Theorem (Section 3.3).

2. The proofs of Propositions 2.56 and 2.58 do not use these estimates, so the reasoning is not circular.

## A.4 Proofs and References for Section 2.5

Proposition 2.51 is easy to prove: see e.g. [206, Lemma I.2.2].

PROOF OF PROPOSITION 2.53.

(a), (b), (c) and (g) are proven in [157, Proposition 15.5].

(d) is a trivial generalization of what is proven in [157, Proposition 15.14(1)].

(e) is proven for the bounded measurable topology in [157, Corollary 15.7 and proof of Proposition 15.14(2)]. For the weak topology, see [206, pp. 42–43]; though stated there for compact metric spaces, the proof is in fact valid for arbitrary complete separable metric spaces. See also [157, p. 316].

(f) We know from part (e) that $\{\mu\colon I(\mu|\nu) \leq c\}$ is closed in the bounded measurable topology. In [157, proof of Proposition 15.6] it is shown that the densities $\{d\mu/d\nu\colon I(\mu|\nu) \leq c\}$ are uniformly $\nu$-integrable (see also [73, Theorem II–22]); this implies, by the Dunford-Pettis theorem, that $\{\mu\colon I(\mu|\nu) \leq c\}$ is relatively compact and relatively sequentially compact in the bounded measurable topology ([73, Theorem II–25] or [272, Proposition IV–2–3]). Since the weak topology is weaker than the bounded measurable topology, the last statement is an immediate consequence.

(h) is proven in [102, Theorem 2.1 and Lemma 2.3].

(i) is an abstraction of the usual statement of strong superadditivity [157, Proposition 15.10]. ∎

**Remarks.** 1. Statement (d) is *not* true jointly in $\mu$ and $\nu$. Counterexample: Let $\Omega = \{a, b\}$, $\mu_1 = \nu_2 = \delta_a$, $\mu_2 = \nu_1 = \delta_b$, $\lambda_1 = \lambda_2 = \frac{1}{2}$. Then $I(\mu_1|\nu_1) = I(\mu_2|\nu_2) = +\infty$, while $I(\frac{1}{2}\mu_1 + \frac{1}{2}\mu_2 | \frac{1}{2}\nu_1 + \frac{1}{2}\nu_2) = 0$.

2. For some improvements of (d) if the $\mu_i$ have "almost disjoint" supports, see [26, Proposition 5.1 and Corollary 5.2] and [326, Theorem 2.1].

3. If the $\sigma$-field $\Sigma$ is countably generated, then the set $\{\mu\colon I(\mu|\nu) \leq c\}$ is in fact compact *and metrizable* in the bounded measurable topology: this follows from [73, Theorem II–24].

4. Additional useful properties of the relative entropy are given in [157, Proposition 15.6 and Corollary 15.7].

The finite-volume variational principle (Theorem 2.54) is well known: see e.g. [206, p. 46] or [99, Lemma 2.1].

## A.5 Proofs and References for Section 2.6

### A.5.1 The Infinite-Volume Limit: Proofs

PROOF OF PROPOSITION 2.56. For all but part (e), see [206, Theorems I.2.3 and I.2.4]. Part (e) is an immediate consequence of Proposition 2.34(e). ∎

Proposition 2.57 is [313, Proposition 4.4]. Proposition 2.58 is an immediate consequence of Propositions 2.56 and 2.57 together with the estimates (2.58) and (2.62b).

PROOF OF PROPOSITION 2.59. When $\nu$ is a product measure, this is [157, Corollary 16.15(b)]. When $\nu$ is a Gibbs measure, this follows from the product-measure case together with (2.90). ∎

PROOF OF PROPOSITION 2.61.

The existence of the van Hove limit, and its equality to the supremum, both follow from the strong superadditivity of $I_\Lambda(\mu|\nu)$ as a function of $\Lambda$, when $\nu$ is a product measure [Proposition 2.53(i)]. One way to see this is to note that strong superadditivity

implies complete superadditivity (Lemma A.2); the claim then follows from Proposition A.3. Alternatively, one can make a direct argument using the strong superadditivity [206, Theorem II.2.2].

The affineness is an immediate consequence of Proposition 2.53(c,d), and the lower semicontinuity is an immediate consequence of Proposition 2.53(e) and equation (2.93b); see [206, Theorem II.2.3] or [157, Proposition 15.14].

The proof of (d) employs the following construction: Pave $\mathbb{Z}^d$ by a cube $C_n$ and its disjoint translates. Now, given a translation-invariant measure $\mu$, let $\rho_n$ be a measure which equals $\mu$ when restricted to each of these cubes, and in which the copies of the spins in the various cubes are *rigidly* forced to be equal. Then let $\mu_n = n^{-d} \sum_{a \in C_n} T_a \rho_n$. By construction $\mu_n$ is translation-invariant; and with a little work one can prove that $i(\mu_n|\nu) = i_{max}$. On the other hand, it is easy to see that $\lim_{n \to \infty} \mu_n = \lim_{n \to \infty} \rho_n = \mu$ in the bounded quasilocal topology.

(e) is proven in [206, Lemma IV.3.2].

When $\Omega_0$ is a standard Borel space (e.g. a complete separable metric space, or a Borel subset thereof), the compactness in the bounded quasilocal topology is proven in [157, Proposition 15.14(3)]. (We do not know whether the result is true for more general spaces $\Omega_0$.) Since the weak quasilocal topology is weaker than the bounded quasilocal topology, the last statement is an immediate corollary. ∎

Proposition 2.62 is proven in [157, Theorem 15.30(b)].

**Remark.** Föllmer [122] has given a beautiful formula for $i(\mu|\nu)$ in terms of the relative entropy (*not* relative entropy *density*!) of the conditional distributions of $\mu$ and $\nu$ given the lexicographic past. See also [157, Proposition 15.16 and Theorem 15.20].

Theorem 2.63 is essentially [157, Theorems 15.30(b) and 15.39].

**Remark.** A rather weak converse to Theorem 2.66 is the following: Let $\mu_1, \mu_2 \in M_{+1,inv}(\Omega)$ with $\mu_2$ Gibbsian for $\Phi_2 \in \mathcal{B}^1$, $i(\mu_1|\mu_2) \leq K$ and $\mu_1$ ergodic. Then there exists an interaction $\Phi_1 \in \mathcal{B}^0$ (*not* $\mathcal{B}^1$!) with $\|\Phi_1 - \Phi_2\|_{\mathcal{B}^0} \leq K/2$ such that $\mu_1$ is an equilibrium measure for $\Phi_1$. This can be proven using the Bishop-Phelps theorem [206, Corollary V.2.1]. The same is true if $\mu_1$ is a finite convex combination of ergodic measures, but then the constant $K/2$ is replaced by a worse one.

Theorem 2.67 is proven in [157, Theorem 15.37]. The proof is given there for a sequence of cubes, but the same proof works for an arbitrary van Hove sequence.

PROOF OF COROLLARY 2.68. $\mathcal{G}_{inv}(\Pi^\Phi) \neq \emptyset$, so let $\nu \in \mathcal{G}_{inv}(\Pi^\Phi)$ and use (2.110). Then Gibbs $\Longrightarrow$ equilibrium is Theorem 2.66, and equilibrium $\Longrightarrow$ Gibbs is Theorem 2.67. ∎

### A.5.2 The Infinite-Volume Limit: Counterexamples

As mentioned in Sections 2.6.1 and 2.6.2, the existence of the limits defining the infinite-volume pressure $p(f|\nu)$ and the infinite-volume relative entropy density $i(\mu|\nu)$ is a highly nontrivial problem: contrary to what might be supposed at first glance, these limits do *not* always exist. The first counterexamples bearing on this problem are due to Kieffer [216]. Here we give a simplified version of Kieffer's counterexample, due to Sokal [331]:

Let $\Omega = \{-1, 1\}^{\mathbb{Z}}$. Let $\nu_n$ be the measure which gives weight $1/2n$ to each of the periodic sequences of period $2n$ consisting of $n$ 1's followed by $n$ $-1$'s. Let $\nu$ be the convex combination $\sum_{n=1}^{\infty} a_n \nu_n$. We shall show that for a suitable choice of the coefficients $\{a_n\}$:

(a) For the function $f(\omega) = \omega_0$, the pressure $\lim_{k \to \infty} k^{-1} \log \int \exp\left(\sum_{i=1}^{k} \omega_i\right) d\nu(\omega)$ does not exist.

(b) For the measure $\mu = \delta_+ \equiv$ delta measure concentrated on the sequence of all $+1$'s, the relative entropy density $\lim_{k \to \infty} k^{-1} I_{\{1,\ldots,k\}}(\mu|\nu)$ does not exist.

PROOF OF (A). Let $g_n(k) \equiv \int \exp\left(\sum_{i=1}^{k} \omega_i\right) d\nu_n(\omega)$. It is easy to see that $g_n$ is a periodic function of period $2n$, and satisfies the (crude) bounds

$$\frac{1}{2n} e^{F_n(k)} \leq g_n(k) \leq e^{F_n(k)}, \tag{A.67}$$

where

$$F_n(k) \equiv n - |k \,(\mathrm{mod}\, 2n) - n| \tag{A.68}$$

is the sawtooth function taking the value 0 at $k = 0, 2n, 4n, \ldots$ and the value $n$ at $k = n, 3n, 5n, \ldots$. Hence

$$g(k) \equiv \int \exp\left(\sum_{i=1}^{k} \omega_i\right) d\nu(\omega) \;=\; \sum_{n=1}^{\infty} a_n g_n(k)$$

$$\begin{cases} \geq \sum_{n=1}^{\infty} \frac{a_n}{2n} e^{F_n(k)} \\ \leq \sum_{n=1}^{\infty} a_n e^{F_n(k)} \end{cases} \tag{A.69}$$

Now choose the sequence $\{a_n\}$ to have huge gaps:

$$a_n = \mathrm{const} \times \begin{cases} e^{-\alpha n} & \text{if } n = 2^l \text{ for some integer } l \\ 0 & \text{otherwise} \end{cases} \tag{A.70}$$

where $\alpha > 0$ will be chosen later. Then for $k = 2^l$ we have the lower bound

$$g(k) \geq a_k\, g_k(k) \geq \frac{a_k}{2k} e^{F_k(k)} = \frac{1}{2k} e^{(1-\alpha)k} \tag{A.71}$$

and hence
$$\liminf_{l \to \infty} \frac{1}{2^l} \log g(2^l) \geq 1 - \alpha . \tag{A.72}$$

On the other hand, for $2^l < k < 2^{l+1}$ we have the upper bound

$$\begin{aligned}
g(k) &\leq \sum_{n=1}^{2^l} a_n e^n + \sum_{n=2^{l+1}}^{\infty} a_n e^k \\
&\qquad [\text{using } F_n(k) \leq \min(n,k)] \\
&\leq \left(\sum_{n=1}^{\infty} a_n\right) e^{2^l} + \left(\sum_{n=2^{l+1}}^{\infty} a_n\right) e^k \\
&\leq e^{2^l} + \text{const} \times e^{-\alpha 2^{l+1}} e^k \\
&\qquad [\text{const depends on } \alpha \text{ only}] \\
&\leq \text{const} \times \exp[\max(2^l, k - \alpha 2^{l+1})] . \tag{A.73}
\end{aligned}$$

Defining $\beta = k/2^l$ (so that $1 < \beta < 2$), we find

$$\frac{1}{k} \log g(k) \leq \frac{\text{const}}{k} + \max\left(\frac{1}{\beta}, 1 - \frac{2\alpha}{\beta}\right) . \tag{A.74}$$

Now choose any $0 < \alpha < \frac{1}{2}$. Then $\max(1/\beta, 1 - 2\alpha/\beta)$ is minimized at $\beta = \beta^* \equiv 2\alpha + 1$ (which satisfies $1 < \beta^* < 2$) and takes the value $1/(2\alpha + 1)$ there. By choosing $k = \lfloor 2^l \beta^* \rfloor$ and letting $l \to \infty$, we conclude that

$$\limsup_{l \to \infty} \frac{1}{\lfloor 2^l \beta^* \rfloor} \log g(\lfloor 2^l \beta^* \rfloor) \leq \frac{1}{2\alpha + 1} . \tag{A.75}$$

Since $1/(2\alpha + 1) < 1 - \alpha$ when $0 < \alpha < \frac{1}{2}$, it follows from (A.72) and (A.75) that $\lim_{k \to \infty} k^{-1} \log g(k)$ does not exist. ∎

PROOF OF (B). It is easy to see that

$$\begin{aligned}
h(k) \equiv I_{\{1,\ldots,k\}}(\delta_+|\nu) &= -\log \nu(\omega_1 = \ldots = \omega_k = +1) \\
&= -\log \sum_{n=k}^{\infty} a_n \frac{n-k+1}{2n} . \tag{A.76}
\end{aligned}$$

Let us again take

$$a_n = \text{const} \times \begin{cases} e^{-\alpha n} & \text{if } n = 2^l \text{ for some integer } l \\ 0 & \text{otherwise} \end{cases} \tag{A.77}$$

Then for $k = 2^l$ we have

$$h(k) \leq -\log \frac{a_k}{2k} = \alpha k + \log(2k) . \tag{A.78}$$

On the other hand, for $2^l < k < 2^{l+1}$ we have

$$\begin{aligned}
h(k) &= -\log \sum_{m=l+1}^{\infty} e^{-\alpha 2^m} \frac{2^m - k + 1}{2^{m+1}} \\
&\geq -\log \sum_{m=l+1}^{\infty} e^{-\alpha 2^m} \\
&\geq \text{const} + \alpha 2^{l+1}
\end{aligned} \qquad (A.79)$$

[const depends on $\alpha$ only]

Thus

$$\limsup_{l \to \infty} \frac{1}{2^l} h(2^l) \leq \alpha \qquad (A.80a)$$

$$\liminf_{l \to \infty} \frac{1}{2^l + 1} h(2^l + 1) \geq 2\alpha \qquad (A.80b)$$

So for any $\alpha > 0$ we conclude that $\lim_{k \to \infty} k^{-1} h(k)$ does not exist. ∎

We note also that Varadhan [359] and Newman [275] have given an example of a mixing Gaussian process for which the pressure does not exist.

## B  Low-Temperature Phase Diagrams and Pirogov-Sinai Theory

### B.1  Generalities on Phase Diagrams

The central problem in equilibrium statistical mechanics is the description of the set of Gibbs measures for a given interaction. More generally, *families* of interactions (or of specifications) are considered, with members labelled by certain parameters: inverse temperature[76] $\beta$, magnetic field, chemical potential, etc. The ultimate goal is then to describe the set of Gibbs measures, in particular the number of extremal Gibbs measures, as a function of these parameters. The partition of the parameter space into regions with different numbers of extremal Gibbs measures is called a *phase diagram* of the family of interactions, and the manifolds delimiting such regions are called *phase-transition manifolds*.

A natural approach to the difficult problem of determining the full phase diagram is to fix first some of the parameters so that the resulting "restricted" phase diagram is amenable to a comparatively simple analysis. Then, one studies whether this phase diagram is "stable", that is, whether a small change in the fixed parameters produces only a small deformation of the diagram keeping unaltered the main properties of the extremal Gibbs measures.

---

[76]As we want to explicitly discuss the role of this parameter, throughout this appendix we un-absorb $\beta$ from interactions and Hamiltonians.

The most widely used "restricted" phase diagrams are the high-temperature ($\beta = 0$) and low-temperature ($\beta = \infty$) limits. In the former, the situation is particularly simple: The finite-volume Gibbs distribution (2.20) becomes for $\beta = 0$ just the product measure $\prod_{x \in \Lambda} d\mu_x^0$ independently of the boundary condition. Hence, there is a unique Gibbs measure, namely $\mu^0 = \prod_{x \in \mathcal{L}} d\mu_x^0$, which corresponds to independent spins, the one at site $x$ distributed according to the a priori measure $\mu_x^0$. [Note that for translation-invariant Gibbs measures the same conclusion follows from the variational principle (2.105a), which for $f_\Phi = 0$ requires $i(\mu|\mu^0) = \inf i(\,\cdot\,|\mu^0) = 0$, hence $\mu = \mu^0$.] It is well known that this infinite-temperature phase diagram is stable in a suitable space of interactions: for $\beta$ small the Gibbs measure remains unique and it corresponds to weakly dependent spins. This has been proven for lattice-gas [140] or, more generally, spin-1/2 [205] interactions in $\mathcal{B}^1$, and for general interactions in $\mathcal{B}^2$ [84, 177]. It is not known whether it is true for general interactions in $\mathcal{B}^1$.

The phase diagram for the zero-temperature limit is, in general, more complicated to describe; its stability is the subject of Pirogov-Sinai theory. In this appendix we give a brief overview of the conclusions of this theory with an eye on the applications needed in Section 4. Its understanding requires, of course, a proper grasp of the basic notions involved in the construction of zero-temperature phase diagrams. As remarked already in the seminal work of Ruelle [311], the formalism for zero-temperature statistical mechanics has some important differences with the one for finite temperatures reviewed in Section 2. Moreover, the nomenclature adopted throughout the existing literature is often a source of confusion, with different authors assigning different meanings to the same words. Therefore, for the convenience of the reader and to fix the terminology, we start with a review of the zero-temperature formalism. For this part of the appendix, the reference closest to our needs — and from which we have taken many of the ideas — is the review by Dobrushin and Shlosman [95]. However, for the sake of consistency with the rest of our work, we adopt a nomenclature slightly different from theirs. We shall parenthetically contrast these differences both for the benefit of the reader familiar with [95] and as a token of the confusing state of the nomenclature.

Let us state once and for all that *in this appendix, we consider only the case of periodic interactions and finite single-spin space, i.e. $|\Omega_0|$ finite*. Moreover, except in Sections B.2.9 and B.4.4, *the interactions are assumed to be of finite range*.

## B.2  Zero-Temperature Lattice Systems. General Formalism

Heuristically, as $\beta \to \infty$ only configurations with minimal energy "survive", the others being exponentially damped by the Boltzmann factor. However, in the general theory of zero-temperature statistical mechanics — as in statistical mechanics quite generally — the central objects are not individual configurations but rather probability measures describing a random distribution of configurations [303]. Just as for non-zero temperature, such measures can be defined either via specifications or via a variational principle.

### B.2.1 Zero-Temperature Gibbs Measures

Let us start with the approach based on specifications. We see that the $\beta \to \infty$ limit of the finite-volume Gibbs distribution (2.20) with a fixed boundary condition produces a measure concentrated on the configurations of "minimal energy" for the given boundary condition and giving equal probability to each such configuration. That is, for any interaction $\Phi$, any finite volume $\Lambda$ and any boundary condition $\tau \in \Omega_{\Lambda^c}$, we have

$$\lim_{\beta \to \infty} \pi_{\Lambda,\tau}^{\beta\Phi}(A) = \frac{\mu_\Lambda^0(A \cap \Omega_{\Lambda,\tau}^\Phi)}{\mu_\Lambda^0(\Omega_{\Lambda,\tau}^\Phi)} \equiv \pi_{\Lambda,\tau}^{\Phi,T=0}(A) \tag{B.1}$$

for all sets $A \in \mathcal{F}_\Lambda$, where $\Omega_{\Lambda,\tau}^\Phi$ is the set of configurations $\omega_\Lambda$ in $\Lambda$ minimizing the energy $H_\Lambda^\Phi(\omega_\Lambda \times \tau_{\Lambda^c})$:

$$\Omega_{\Lambda,\tau}^\Phi = \left\{ \omega_\Lambda \colon H_\Lambda^\Phi(\omega_\Lambda \times \tau_{\Lambda^c}) = \inf_{\widetilde{\omega}_\Lambda \in \Omega_\Lambda} H_\Lambda^\Phi(\widetilde{\omega}_\Lambda \times \tau_{\Lambda^c}) \right\} . \tag{B.2}$$

We call $\Pi^{\Phi,T=0} = (\pi_{\Lambda,\tau}^{\Phi,T=0})_{\Lambda \in \mathcal{S}}$ the *zero-temperature specification* (or ground-state specification [95]) for the interaction $\Phi$.

**Definition B.1** *A* zero-temperature Gibbs measure *for $\Phi$ is a measure consistent with the specification (B.1).*

We remark that the specifications (B.1) are quasilocal (since we only consider finite-range interactions), but not uniformly nonnull, hence they are not Gibbsian. Therefore, zero-temperature Gibbs measures happen not to be honest Gibbs measures. In fact, the possibility of including (B.1) in the general framework is one of the advantages of introducing the general notion of specification (Section 2.3.1), rather than just the more restricted (and popular) class of Gibbsian specifications (Section 2.3.2). The zero-temperature Gibbs measures for a given interaction $\Phi$ form a (weakly) closed — hence (weakly) compact — convex subset of the compact metric space $M_{+1}(\Omega) \subset M(\Omega)$. Therefore, by Choquet's theorem [293] any such measure can be written as the barycenter of a probability measure concentrated on the extreme points. In fact, the general theory of specifications guarantees us that this decomposition into extremal measures is *unique* [157, Theorem 7.26], i.e. that the set of zero-temperature Gibbs measures is a simplex.

### B.2.2 Ground-State Configurations. Support Properties of Zero-Temperature Gibbs Measures

The specifications (B.1) satisfy

$$\pi_\Lambda(\omega_\Lambda | \omega_{\Lambda^c}) = 0 \quad \text{unless } \omega_\Lambda \in \Omega_{\Lambda,\omega_{\Lambda^c}}^\Phi . \tag{B.3}$$

This property implies that the zero-temperature Gibbs measures — which satisfy $\mu\pi_\Lambda = \mu$ for every finite set $\Lambda$ — are supported by the set of configurations $\omega$ such that $\omega_\Lambda \in \Omega^\Phi_{\Lambda,\omega_{\Lambda^c}}$ for *every* finite $\Lambda$, i.e. is of configurations that minimize the local energy when they themselves are the boundary condition. Configurations with this property are called *ground-state configurations*. By (B.2) they can be characterized as those configurations whose energy cannot be lowered by any change involving only a finite number of spins. That is, $\omega \in \Omega$ is a ground-state configuration for an interaction $\Phi$ if and only if for every $\Lambda$ and every configuration $\omega'$ such that $\omega_{\Lambda^c} = \omega'_{\Lambda^c}$, we have

$$H^\Phi_\Lambda(\omega') - H^\Phi_\Lambda(\omega) \;\equiv\; \sum_{\substack{A \subset S \\ A \cap \Lambda \neq \emptyset}} [\Phi_A(\omega') - \Phi_A(\omega)] \;\geq\; 0 \;. \tag{B.4}$$

The set of ground-state configurations is closed (hence compact) because the conditions (B.4) involve finite-volume Hamiltonians which are continuous functions of the configurations. This fact of being a closed set justifies the use above of the expression "is supported by" (= "its support is a subset of"). We recall that the support of a measure $\mu$ is the smallest closed set of full measure (Section 2.1.3).

The fact of being supported on configurations satisfying (B.4) is not equivalent to being consistent with the specifications (B.1) — it is *weaker*. The more general measures characterized only by this support property turn out to play an important role in the study of the stability of zero-temperature phase diagrams (Theorem B.12 below). Inspired by [95], we call these measures *w-* (for weak) *zero-temperature measures*. Formally:

**Definition B.2** *A w-zero-temperature measure for an interaction $\Phi$ is a measure $\mu$ satisfying*

$$\mu(\{\text{ground-state configurations for } \Phi\}) \;=\; 1 \;. \tag{B.5}$$

In Section B.2.7 we discuss a natural limit process that produces w-zero-temperature measures, and we present an example (for the Ising antiferromagnet with a magnetic field) in which this limit process produces a translation-invariant w-zero-temperature measure which is *not* a zero-temperature *Gibbs* measure.

Obviously the set of w-zero-temperature measures for a given interaction $\Phi$ is (weakly) closed — hence (weakly) compact and convex. The extreme points are simply the delta measures $\delta_\omega$ concentrated on a single ground-state configuration $\omega$. This set is therefore trivially a simplex.

The previous discussion can be summarized in the following way:

**Theorem B.3** *Every zero-temperature Gibbs measure is a w-zero-temperature measure for the corresponding interaction, i.e. it satisfies (B.5).*

This theorem constitutes the precise version of the idea that only configurations with minimal energy "survive" at zero temperature.

### B.2.3 Rigid Ground-State Configurations

The set of ground-state configurations is in general rather large. Already the ferromagnetic Ising model provides a rich illustration. This model has exactly two periodic (in fact translation-invariant) ground-state configurations: the all-"+" and the all-"−" configurations. But in addition it presents infinitely many non-periodic configurations exhibiting interfaces between "+" and "−" spins. In all dimensions we have the flat-interface configurations:

$$\omega_x = \begin{cases} +1 & \text{for } x_1 \geq 0 \\ -1 & \text{for } x_1 < 0 \end{cases} \tag{B.6}$$

(and translated, 90°-rotated and 180°-rotated versions of this). In higher dimensions we have a growing zoo: For dimensions $d \geq 2$ we have configurations with interfaces in the form of staircases; for $d \geq 3$ there appear configurations resembling "books on a table" or "books on a staircase" [95]. See this last reference for a partial catalogue.

Not all these configurations are equally relevant for zero- and low-temperature phase diagrams. We can distinguish three mutually exclusive categories roughly representing different (for us decreasing) levels of relevance. We shall call them *rigid*, *convivial* and *superfluous*. The rigid configurations are usually the most important ones (albeit not the most numerous); they are associated to deterministic zero-temperature Gibbs measures:

**Definition B.4** *For a given interaction $\Phi$, a ground-state configuration $\omega$ is called rigid [8] if the measure $\delta_\omega$ concentrated on $\omega$ is a zero-temperature Gibbs measure for $\Phi$, i.e. is consistent with the specification (B.1).*

A simple calculation proves the following:

**Proposition B.5** *A ground-state configuration $\omega$ is rigid if and only if*

$$|\Omega^\Phi_{\Lambda,\omega_{\Lambda^c}}| = 1 \tag{B.7}$$

*for all finite $\Lambda$.*

In words, this theorem says that $\omega$ placed as a boundary condition determines uniquely the minimal-energy configuration inside any given volume (thereby justifying the qualifier "rigid"). Equivalently, any local change of $\omega$ produces a strictly positive change of energy. Usual phase-diagram studies — in particular Pirogov-Sinai theory — deal only with these deterministic zero-temperature Gibbs measures and their low-temperature perturbations. (*Warning*: Reference [95] reserves the name "ground-state configurations" only for the rigid ones.)

For the Ising model (ferromagnetic, zero magnetic field), it is clear that the all-"+" and all-"−" configurations satisfy (B.7) and hence they are rigid in any dimension. The case of the non-periodic ground-state configurations (flat-interface, staircase-interface, etc.) is more delicate. There is, however, a simple argument [95] showing that if $\omega$ is a ground-state configuration for the $d$-dimensional Ising model, then its cylindrical extension to an extra dimension — defined by $\widetilde{\omega}_{(x_1,\ldots,x_d,x_{d+1})} \equiv \omega_{(x_1,\ldots,x_d)}$ — is a *rigid* ground-state configuration for the $(d+1)$-dimensional Ising model. Indeed, if we think of the extra dimension as "vertical", any local change of $\widetilde{\omega}$ consists of a *finite* stack of local changes of $\omega$. The bottom and top $d$-dimensional sections of this stack face sections where the configuration is equal to $\omega$ without changes. Thus, some of the corresponding "vertical" bonds join antiparallel spins, which produces a strictly positive contribution to the change in energy. This proves (B.7) and hence the rigidity of $\widetilde{\omega}$.

As a consequence of this argument, we conclude that the flat-interface configurations are rigid for $d \geq 2$, the staircase-interface ones are rigid for $d \geq 3$, and so on. The proof that the rigidity does *not* extend *below* such dimensions requires further arguments. We shall comment on this below.

**Remark.** Rigidity of a ground-state configuration $\omega$ does not exclude its belonging *also* to the support of some zero-temperature Gibbs measure that is not deterministic. For instance, if there is more than one rigid configuration, one can of course take convex combinations of the corresponding delta-measures. The possibility of a less trivial example will be discussed below, after (B.10).

### B.2.4 Convivial Configurations. Zero-Temperature Entropy

However, not all is deterministic in zero-temperature life. Our next type of configurations are those that belong only to the support of a non-deterministic zero-temperature Gibbs measure. We recall that the support of a measure is the complement of the union of all the zero-measure open sets (that is, the smallest closed set with full measure).

**Definition B.6** *For a given interaction, a ground-state configuration $\omega$ is called convivial if $\delta_\omega$ is not a zero-temperature Gibbs measure but there exists a zero-temperature Gibbs measure having $\omega$ in its support.*

These ground-state configurations, which individually have little or no weight but are relevant as an ensemble, and the associated non-deterministic Gibbs measure supported on such an ensemble, are probably not what the physicist-in-the-street has in mind when thinking about zero temperature. One expects them in cases where there is a large degeneracy in the ground state. The precise concept measuring such degeneracy is the *zero-temperature entropy* (also called *residual entropy*). For the sake of completeness, we briefly review the definition and principal properties of this quantity. Our main reference is the classic article by Aizenman and Lieb [8].

There are some subtleties involved in the right notion of zero-temperature entropy. Heuristically, its computation requires a limit process: one must compute (or measure) a sequence of low-temperature entropies and take the limit as the temperature goes to zero. The so-called "third law of thermodynamics" claims that such a limit must be zero; such behavior is indeed seen in simple models, but *not* always. Its violation must be interpreted as signaling a large "degeneracy of the ground state". The formalization of these ideas requires a consideration of the role of the infinite-volume limit. As pointed out by some authors (see references in [8]), the volume must be sent to infinity *before* taking the limit $T \to 0$. But in this case, one must consider with some care the boundary conditions. If, motivated by the "ground-state-energy" (= variational) approach [see eq. (B.14) below], one works with pre-fixed — for instance free — boundary conditions, then there are examples where the contribution of some excited configurations survives the zero-temperature limit, so that the residual entropy seems to be measuring more than just the "degeneracy of the ground state". The correct way to consider the boundary conditions, and hence the right definition of "degeneracy", was pointed out by Aizenman and Lieb [8]. At the same time, they provided a remarkable formula for the zero-temperature entropy purely in terms of zero-temperature concepts, with no reference to limits from finite temperatures. We shall take this formula as the definition. For a finite set $\Lambda$ and an interaction $\Phi$, let us denote $\mathcal{G}_\Lambda^\Phi$ the set of restrictions to $\Lambda$ of the ground-state configurations for $\Phi$.

**Definition B.7** *The* zero-temperature entropy *for an interaction $\Phi$ is the limit*

$$s_\Phi \;=\; \lim_{\Lambda \nearrow \infty} \frac{1}{|\Lambda|} \log |\mathcal{G}_\Lambda^\Phi| \;. \tag{B.8}$$

In words, this formula says that a system has non-zero residual entropy iff the number of distinct ground-state configurations, as viewed within a finite volume, grows exponentially with this volume. Following [244], it is suggestive to call such models *super-degenerate*. Intuitively, this feature requires the presence of competing interactions to produce a sufficient large number of ground-state configurations. Indeed, it can be proven [8] that all ferromagnetic models have zero residual entropy.

The key result establishing the connection between non-zero residual entropy and existence of convivial ground-state configurations is the following. To abbreviate, for a translation-invariant (or periodic) measure $\mu$ we denote $s(\mu) \equiv -i(\mu|\mu^0) + \log|\Omega_0|$, where $i(\mu|\nu)$ is the relative entropy density defined in Section 2.6.2, and $\mu^0$ is the product over all sites of normalized counting measure. (This the the physicists' usual entropy, which is defined relative to *unnormalized* counting measure on the single-spin space $\Omega_0$ — this accounts for the additive constant $\log|\Omega_0|$.)

**Proposition B.8** *Fix an interaction $\Phi$. Then:*

*(a) If $\mu$ is a translation-invariant w-zero-temperature measure for $\Phi$, then $s(\mu) \leq s_\Phi$.*

*(b) There exists for $\Phi$ a translation-invariant zero-temperature Gibbs measure $\mu$ such that $s(\mu) = s_\Phi$.*

We summarize below the results on which this proposition is based (Proposition B.13 and Theorems B.11, B.15 and B.17 part (b)). We note that if the support of $\mu$ is a finite set, then $s(\mu) = 0$ [$\mu$ is of the form $\sum_i \lambda_i \delta_{\omega_i}$, hence $s(\mu) \leq -(1/|\Lambda|) \sum_i \lambda_i \log \lambda_i \to 0$]. Therefore, we conclude the following:

**Proposition B.9** *A super-degenerate system with finitely many rigid ground-state configurations exhibits infinitely many convivial ground-state configurations.*

This proposition covers all the cases we know of in which the existence of convivial ground-state configurations has been proven. Consider, for example, the model with spins $\omega_i = -1, 0, 1$ and Hamiltonian

$$H_\Lambda = \sum_{|i-j|=1} (\omega_i^2 - \omega_j^2)^2 \; . \tag{B.9}$$

The ground-state configurations for this model are all the configurations with no spin equal to zero, and the all-"0" configuration. The zero-temperature entropy for this model is exactly $\log 2$. As the all-"0" configuration is the only rigid one, we conclude, by the previous proposition, that there must be infinitely many convivial ground-state configurations. Another important example is the Ising model with nearest-neighbor antiferromagnetic coupling of strength $J$ and magnetic field $h = 2d|J|$. Its ground-state configurations are those in which no two nearest-neighbor spins are simultaneously "$-$", a fact that produces a non-zero residual entropy. There are no rigid configurations, hence the proposition implies the existence of many convivial ones. For this model, such a fact can be proven also by a different argument which yields some additional insight. Indeed, by identifying a "$-$" spin with the presence of a particle, the ensemble of ground-state configurations — with the associated conditional probabilities giving equal weight to all of them — is seen to correspond to the grand-canonical ensemble for the ideal lattice gas with nearest-neighbor exclusion and chemical potential equal to zero. Using a beautiful computer-assisted proof, Dobrushin, Kolafa and Shlosman [87] proved that such a system has an *unique* Gibbs measure. As none of the ground-state configurations are rigid, this Gibbs measure is non-deterministic and therefore supported on convivial configurations. This example shows a way (in fact, the only one we know of) to interpret and understand the characteristics of non-deterministic zero-temperature Gibbs measures supported on (very many) convivial ground-state configurations: One maps it into a statistical-mechanical problem for another, better understood, equivalent system. As the original ensemble involves configurations satisfying some condition derived from the minimal-energy requirement, this equivalent system will, in general, be a model with exclusions. That is, it will not fit into the general formalism developed in Chapter 2.

Proposition B.9 does not yield any information on models with zero residual entropy, for instance on ferromagnetic systems. In particular, the question remains of whether conviviality *requires* super-degeneracy. A possible counterexample is presented in reference [95]: Consider, for the three-dimensional ferromagnetic Ising model, the ensemble of ground-state configurations that differ only locally (i.e. in finite volumes) from the "zig-zag interface" one:

$$\omega^{\text{zig-zag}}_{(t_1, t_2, t_3)} = \begin{cases} +1 & \text{if } t_1 + t_2 + t_3 > 0 \\ -1 & \text{if } t_1 + t_2 + t_3 \leq 0 \; . \end{cases} \tag{B.10}$$

Such an ensemble can be mapped onto an appropriate solid-on-solid model. If this model can be proven to have at least one Gibbs measure (a problem still open), then it would imply that the above configurations are convivial. We must acknowledge that the standing conjecture [85, 95] is that such Gibbs measures do not exist. Note that if this Gibbs measure exists, then the all-"+" and all-"−" configurations would be at the same time rigid and in the (boundary of the) support of a highly non-deterministic zero-temperature Gibbs measure.

Nevertheless, there is an interesting result (Proposition 4 of [8]) involving models with zero residual entropy:

**Proposition B.10** *If $s_\Phi = 0$, every* translation-invariant *w-zero-temperature measure for $\Phi$ is supported on the set of rigid ground-state configurations.*

That is, if a model with zero residual entropy does in fact possess convivial ground-state configurations, then such configurations can lie in the support only of *non-translation-invariant* zero-temperature Gibbs measures.

**Remark.** On the other hand, Radin [304] has shown examples of super-degenerate systems with a unique translation-invariant w-zero-temperature measure entirely supported on the set of rigid ground-state configurations. In these examples, the set of ground-state configurations does not have any closed translation-invariant proper subset, hence it is formed by all the translates of a *single* (non-periodic) configuration, and limits of such. The non-zero residual entropy implies that any two such translates must differ in infinitely many sites, and hence they all must be rigid ground states. However, this phenomenon can happen only in the presence of *infinite-range* interactions (albeit decreasing arbitrarily fast with the range) [304].

### B.2.5 Superfluous Ground-State Configurations

The last type of ground-state configurations are those that are not in the support of any zero-temperature Gibbs measure. These are obviously the least interesting ones, and we shall call them *superfluous* ground-state configurations. The most immediate example is provided by the one-dimensional ferromagnetic Ising model. Its ground-state configurations are the all-"+", all-"−" and the flat-interface configurations. However, *all* the zero-temperature Gibbs measures are of the form $\lambda \delta_+ + (1-\lambda)\delta_-$; the flat-interface configurations (B.6) are superfluous. Heuristically this is because the interface is free to wander at no energy cost; in the infinite-volume limit it wanders to $\pm\infty$. The proof goes as follows: Denote by $\chi_{+-x_0}$ the indicator function of the configuration which is $+1$ for $x < x_0$ and $-1$ for $x \geq x_0$. Then, for $\Lambda = [-N, N]$, the measures (B.1) yield

$$\pi_\Lambda^{\Phi, T=0}(\chi_{+-x_0}|\tau) \;=\; \begin{cases} 1/(2N+2) & \text{if } \tau_{-(N+1)} = +1, \tau_{N+1} = -1 \\ & \text{and } -N \leq x_0 \leq N+1 \\ 0 & \text{otherwise} \end{cases} \quad (B.11)$$

[The first line is due to the $2N+2$ possible positions $x_0$ for the "kink" (lack of rigidity).] Therefore, if $\mu$ is a zero-temperature Gibbs measure, then for every $N \geq |x_0|$ we have

$$\mu(\chi_{+-x_0}) \;=\; \mu\left(\pi_\Lambda^{\Phi, T=0}(\chi_{+-x_0}|\cdot)\right)$$

$$= \frac{1}{2N+2}\mu(\omega_{-(N+1)} = +1 \text{ and } \omega_{N+1} = -1)$$
$$\leq \frac{1}{2N+2}. \tag{B.12}$$

Letting $N \to \infty$ we conclude that $\mu(\chi_{+-x_0}) = 0$. The same holds, of course, for the flat-interface configurations which go from $-$ to $+$. As the ground-state configurations here form a *countable* set (labelled by $x_0 \in [-\infty, \infty]$ and the polarity of the kink), its measure is the sum of the measure of each of its points. Therefore, (B.12) implies that $\mu$ gives full measure to the set formed only by the all-"+" and all-"−" configurations; the flat-interface configurations are superfluous. Combining this with the results stated above, we conclude that the flat-interface configurations (B.6) are superfluous in $d = 1$, and rigid in $d \geq 2$.

The preceding argument requires not only that there be a growing degeneracy in the position of the interface, but also that the number of ground-state configurations be not too large, i.e. at most countable. This second fact is not true for higher dimensions. In dimension two, for instance, the "staircase-interface" configurations form an uncountable set. To be sure, the set of staircases with *finitely many* stairs is countable, and the above argument can be used to prove that this set has measure zero for all zero-temperature Gibbs measures. This would prove that the support of such measures is always contained in the set formed by the all-"+", all-"−", flat-interface and infinite-staircase configurations (this being a *closed* set whose complement has measure zero). But it does not rule out the occurrence of non-deterministic zero-temperature Gibbs measures supported on infinite-staircase configurations, similarly to what may happen in the three-dimensional Ising model for configurations close to $\omega^{\text{zig-zag}}$. Nevertheless, we must keep in mind that we are primarily interested in those features of zero-temperature phase diagrams that survive at (can tell us something about) *low but nonzero* temperature. Therefore, for the two-dimensional Ising model the possible existence of such a zero-temperature Gibbs measure with support on infinite-staircase configurations is a rather irrelevant issue, since it has been proven [1, 192] that only the Gibbs measures of the form $\lambda \delta_+ + (1-\lambda)\delta_-$ "survive" at non-zero temperatures. The question of non-deterministic Gibbs measures becomes really important only for dimension $d \geq 3$.

### B.2.6 Nonuniqueness of Specifications and Interactions

We shall now comment on one important difference between the zero-temperature and nonzero-temperature formalisms: At zero temperature the "inverse problem" — given a measure, determine the specification and/or the interaction — is no longer well-posed: the map from interactions (or specifications) to zero-temperature Gibbs measures is, in general, many-to-one. This lack of uniqueness appears at three different levels:

A) There are measures consistent with several different zero-temperature specifications simultaneously. Theorem 2.15 does not apply at zero temperature because there are (large) open sets having zero measure for all zero-temperature Gibbs measures. Therefore, by redefining the specification more or less arbitrarily on such open sets we can obtain several different specifications for the same zero-temperature Gibbs measure. Let us present an explicit example. Consider the nearest-neighbor Ising model with formal Hamiltonian $-J \sum_{\langle xy \rangle} \omega_x \omega_y - h \sum_x \omega_x$. Then the measure $\delta_+$ is a zero-temperature Gibbs measure for the following zero-temperature specifications:

1. The specification $\pi^{\Phi_1, T=0}$ where $\Phi_1$ is defined by $J = 0$ and some $h > 0$.

2. The specification $\pi^{\Phi_2, T=0}$ where $\Phi_2$ is defined by some $J > 0$ and $h = 0$.

Nevertheless, $\Pi^{\Phi_1, T=0} \neq \Pi^{\Phi_2, T=0}$ because the former has $\delta_+$ as its only zero-temperature Gibbs measure, while the latter has both $\delta_+$ and $\delta_-$ as zero-temperature Gibbs measures.

B) There are zero-temperature specifications which arise from several non-physically-equivalent interactions (in other words, the notion of physical equivalence becomes meaningless at $T = 0$). We give two examples:

1. Trivial example: Consider any interaction $\Phi$ and any number $\lambda > 0$. Then $\Phi$ and $\lambda \Phi$ are not (usually) physically equivalent, but they have the same zero-temperature specifications.

2. Less trivial example: All Ising-type pair interactions (not necessarily ferromagnetic) such that $h_x > \sum_y |J_{xy}|$ for all $x$ give rise to the same zero-temperature specification, namely the one that for each finite set $\Lambda$ and every boundary condition gives, inside $\Lambda$, the measure concentrated in the all-"+" configuration.

C) The variational principle (Section B.2.8 below) reduces to the minimization of the specific energy, which is not a strictly convex functional on $\mathcal{B}^1$ or any of its subspaces $\mathcal{B}_h$.

### B.2.7 Stability and w-Stability

Zero temperature is in itself unattainable. So one really is interested in those zero-temperature features that "survive" at low but nonzero temperatures. For instance, one is interested in determining which are the measures that can be obtained as a $\beta \to \infty$ limit of positive-temperature Gibbs measures for a *fixed* interaction $\Phi$. We shall refer to these measures as *stable measures* for the interaction $\Phi$. It is simple to check that all these stable measures must be zero-temperature Gibbs measures for $\Phi$:

**Theorem B.11** *Let $\mu_n$ be Gibbs measures for a fixed interaction $\Phi$ and a sequence of inverse temperatures $\beta_n$ with $\beta_n \to +\infty$. If $\mu_n \to \mu$, then the measure $\mu$ is a zero-temperature Gibbs measure for $\Phi$.*

However, not every zero-temperature Gibbs measure for a given potential is necessarily stable. We can illustrate this concept with the case of the Ising model. For $d = 1$, none of the deterministic zero-temperature Gibbs measures ($\delta_+$ and $\delta_-$) are stable. In fact, the only stable measure is $(\delta_+ + \delta_-)/2$. For the Ising model in dimension 2, only the measures of the form $\lambda\delta_+ + (1-\lambda)\delta_-$ are stable. The deterministic Gibbs measures associated with the rigid flat-interface configurations are unstable: at any nonzero temperature, the interface "wanders" to $\pm\infty$ and we are left with a convex combination of the "+" and "−" phases [1, 192]. For dimension 3, the flat-interface measures were proven to be stable by Dobrushin [85] (see also [348]). **Remark:** The low-temperature Gibbs measure for the flat-interface phase seems, at least in numerical experiments, to disappear at a temperature strictly below the critical temperature, giving rise to a roughening transition. If the above-mentioned zero-temperature measure supported near the configuration $\omega^{\mathrm{zig-zag}}$ happens to exist, we could ask two questions: (i) Does it survive for $T > 0$ (stability)?; and, if so, (ii) Does it fail to survive to $T = T_c$?. If the answer to both questions were yes, then the Ising model would exhibit a second roughening transition.

However, as our eventual goal is the study of how the full phase diagram deforms as the temperature is raised, we must consider a more general situation in which the interaction is also varied as the temperature goes to zero. That is, we must consider the more general class of measures that can be obtained as a $\beta \to \infty$ limit of positive-temperature Gibbs measures for interactions $\Phi_n \to \Phi$. We shall refer to such measures as *w-stable measures* for the interaction $\Phi$ [95]. (*Warning:* Reference [328] calls these measures *stable*.) In general, such measures *need not be* zero-temperature Gibbs measures for $\Phi$. For example, if to the antiferromagnetic Ising model with field $h = 2d|J|$ considered above we add an additional field $h_n = \alpha/\beta_n$, we obtain, in the limit $\beta_n \to \infty$, a Gibbs measure corresponding to a lattice gas with chemical potential $\alpha$. All these measures are different among themselves, and different from the unique zero-temperature Gibbs measure for the Ising antiferromagnet in a field $h = 2d|J|$, which corresponds to $\alpha = 0$. A more dramatic example would be to add, to the same model, a field $h = 1/\sqrt{\beta_n}$. The measure obtained in the limit $\beta_n \to \infty$ would then be the measure $\delta_+$ (all the conditional probabilities $\pi_\Lambda^{T=0}$ are equal to $\delta_+$), which is not a zero-temperature Gibbs measure for $\Phi$ because there is no rigid ground-state configuration for this model. This example shows that the notion of w-stability is perhaps a little too general; for interesting applications one usually constrains oneself to *w-stability with respect to a pre-fixed set of perturbed interactions*. In Section B.3.2 we shall make precise the desirable properties of such perturbations.

At any rate, it is immediate that all w-stable measures have the weaker property of being supported on the ground-state configurations for the given interaction, i.e. they are *weak* zero-temperature measures:

**Theorem B.12** *Let $\mu_n$ be Gibbs measures for a sequence of interactions $\Phi_n$ and a sequence of inverse temperatures $\beta_n$ such that $\Phi_n \to \Phi$ and $\beta_n \to +\infty$. If $\mu_n \to \mu$, then*

$$\mu(\{\text{ground-state configurations for }\Phi\}) = 1, \qquad (B.13)$$

*i.e. $\mu$ is a weak zero-temperature measure for $\Phi$.*

The notion of zero-temperature entropy involves a zero-temperature limit, hence it must have something to say about stability. Indeed, Aizenman and Lieb [8] have proven the following:

**Proposition B.13** *If $\mu$ is a stable translation-invariant zero-temperature Gibbs measure for $\Phi$, then*

$$s(\mu) = s_\Phi \ .$$

This result, together with Theorem B.11 and the fact that the set of translation-invariant Gibbs measures is non-empty at all temperatures, proves Proposition B.8 (b) above. In the case of super-degenerate systems (i.e. systems for which $s_\Phi > 0$), Proposition B.13 can be used to rule out the stability of some measures:

**Corollary B.14** *For a super-degenerate system, every translation-invariant zero-temperature Gibbs measure supported on a finite set is unstable.*

For instance, for the system (B.9), the all-"0" Gibbs state is unstable.

### B.2.8 Variational-Principle Approach

The variational-principle approach for zero-temperature measures was historically the first one to be considered [311]. At zero temperature it provides an even simpler criterion than at non-zero temperatures, because it reduces to a minimal-energy condition ($F = E - TS$ reduces to $F = E$ if $T = 0$). For a translation-invariant interaction $\Phi$, it is not hard to show that the limit

$$e_\Phi \equiv \lim_{\Lambda \nearrow \infty} \frac{1}{|\Lambda|} \inf_{\omega \in \Omega_\Lambda} \sum_{A \subset \Lambda} \Phi_A(\omega) \tag{B.14}$$

exists; we call it the *minimal specific energy* (or *ground-state energy*). A translation-invariant measure $\mu$ satisfying

$$\mu(f_\Phi) = e_\Phi \tag{B.15}$$

is called a *zero-temperature equilibrium measure* for the interaction $\Phi$. The definition can be extended to periodic measures if $f_\Phi$ includes an average over all the sites of a basic period: If $\Phi$ is invariant under a subgroup $S$ of $\mathbb{Z}^d$, with $\mathbb{Z}^d/S$ isomorphic to a finite set $P \subset \mathbb{Z}^d$, then one must define $f_\Phi \equiv |P|^{-1} \sum_{x \in P} \sum_{X \ni x} |X|^{-1} \Phi_X$. We shall assume this extension in the sequel. Schrader has proven [319]:

**Theorem B.15**

(a) *Every translation-invariant w-zero-temperature measure for $\Phi$ is a zero-temperature equilibrium measure for $\Phi$, i.e. satisfies (B.15).*

(b) *Conversely, every zero-temperature equilibrium measure for $\Phi$ is a w-zero-temperature measure for $\Phi$.*

That is, for translation-invariant measures to be supported on (local) ground-state configurations is equivalent to having minimal average energy density. We notice that, unlike the finite-temperature case, we do *not* have an equivalence between the variational and the Gibbsian-specifications approaches; only the more general w-measures appear in the previous theorem. The relationship between zero-temperature Gibbs measures and equilibrium measures is much more problematic.

The variational approach yields also a characterization of *periodic* ground-state configurations:

**Theorem B.16**

1. For any periodic configuration $\omega \in \Omega$, the specific energy (energy per site)

$$e_\Phi(\omega) = \lim_{\Lambda \nearrow \infty} \frac{1}{|\Lambda|} \sum_{A \subset \Lambda} \Phi_A(\omega) \qquad (B.16)$$

   exists.

2. The infimum of $e_\Phi(\omega)$ over all periodic configurations $\omega$ is finite and equals the value $e_\Phi$ defined in (B.14).

3. $\omega$ is a periodic ground-state configuration if and only if $e_\Phi(\omega) = e_\Phi$ [328, 95].

For completeness, we mention also two variational principles involving the minimal energy density and the residual entropy:

**Theorem B.17**

(a)
$$e_\Phi = \inf_{\mu \in M_{+1,\text{per}}(\Omega, \mathcal{F})} \mu(f_\Phi) \qquad (B.17)$$

(b) [8]
$$s_\Phi = \sup\{s(\mu) \,|\, \mu \in M_{+1,\text{per}}(\Omega, \mathcal{F}) \text{ and } \mu(f_\Phi) = e_\Phi\} \qquad (B.18a)$$
$$= \sup\{s(\mu) \,|\, \mu \text{ is a w-zero temperature measure for } \Phi\} \,. \qquad (B.18b)$$

In particular, (B.18b) proves Proposition B.8(a). The "inf" in part (a) and the "sup" in part (b) are in fact "min" and "max", respectively. They are realized by the zero-temperature equilibrium measures.

### B.2.9 Infinite Range and Lack of Quasilocality

The variational-principle approach to zero-temperature classical lattice systems can be extended without difficulty to interactions in $\mathcal{B}^0$ [319, 8]. The extension of the DLR approach to infinite-range interactions (e.g. in $\mathcal{B}^1$) is, however, more problematic. In particular, the validity of the important Theorem B.11 is an open question: A sequence of positive-temperature Gibbs measures for $\Phi$ could conceivably converge to a limiting measure that is not consistent with the zero-temperature specification (B.1). If this latter specification were quasilocal, such a phenomenon could not occur [157, Theorem 4.17]; however, for long-range interactions the specification (B.1) is in general not quasilocal. Let us conclude this section with an example showing this lack of quasilocality.

Consider any long-range one-dimensional Ising model with pair interactions $J_{xy} = J_{|x-y|}$ satisfying $\sum_n |J_n| < \infty$. The model has to be truly long-range in the sense that there must be infinitely many nonzero couplings $J_n$; for simplicity of notation we assume that $J_n \neq 0$ for all $n$. We claim that the zero-temperature specification of such a model is non-quasilocal. Indeed, the zero-temperature conditional probability for the spin at the origin satisfies:

$$\pi_{\{0\}}^{\Phi,T=0}(\omega_0 = +1 | \tau) \;=\; \begin{cases} 1 & \text{if } \sum_{x \neq 0} J_x \tau_x > 0 \\ 1/2 & \text{if } \sum_{x \neq 0} J_x \tau_x = 0 \\ 0 & \text{if } \sum_{x \neq 0} J_x \tau_x < 0 \, . \end{cases} \qquad (B.19)$$

To prove that this is not a quasilocal function of the boundary condition $\tau$, we need to show that there exists some $\varepsilon > 0$ for which the following is true: For an infinite sequence of nested finite sets $\Lambda$ there exist two open sets of configurations, $\mathcal{N}_\Lambda$ and $\mathcal{N}'_\Lambda$, formed by configurations which are all identical inside $\Lambda$, but such that

$$\left| \pi_{\{0\}}^{\Phi,T=0}(\omega_0 = +1 | \tau) - \pi_{\{0\}}^{\Phi,T=0}(\omega_0 = +1 | \tau') \right| \;\geq\; \varepsilon \qquad (B.20)$$

if $\tau \in \mathcal{N}_\Lambda$ and $\tau' \in \mathcal{N}'_\Lambda$. Such sets $\mathcal{N}_\Lambda, \mathcal{N}'_\Lambda$ are constructed as follows: Take $\Lambda_N = [-N, N]$ and fix $N_0 > N$ such that

$$\sum_{x > N_0} |J_x| \;<\; |J_{N+1}| \, , \qquad (B.21)$$

and let $\mathcal{N}_\Lambda$ be the set of configurations $\tau$ such that

$$\tau_x \;=\; \begin{cases} +1 & \text{if } 1 \leq x \leq N \\ -1 & \text{if } -N \leq x \leq -1 \\ \operatorname{sgn} J_x & \text{if } N+1 \leq |x| \leq N_0 \\ \text{anything} & \text{if } |x| > N_0 \, . \end{cases} \qquad (B.22)$$

The set set $\mathcal{N}'_\Lambda$ is defined analogously but replacing $\operatorname{sgn} J_x$ by $-\operatorname{sgn} J_x$. We then have:

$$\sum_{x \neq 0} J_x \tau_x = \sum_{|x| > N} J_x \tau_x = \begin{cases} 2 \displaystyle\sum_{|x|=N+1}^{N_0} |J_x| + 2 \sum_{|x|>N_0} J_x \tau_x > 0 & \text{for } \tau \in \mathcal{N}_\Lambda \\ -2 \displaystyle\sum_{|x|=N+1}^{N_0} |J_x| + 2 \sum_{|x|>N_0} J_x \tau_x < 0 & \text{for } \tau \in \mathcal{N}'_\Lambda, \end{cases} \tag{B.23}$$

where the last inequalities follow from (B.21). Therefore, by (B.19),

$$\left| \pi_{\{0\}}^{\Phi,T=0}(\omega_0 = +1|\tau) - \pi_{\{0\}}^{\Phi,T=0}(\omega_0 = +1|\tau') \right| = 1 \tag{B.24}$$

if $\tau \in \mathcal{N}_\Lambda$ and $\tau' \in \mathcal{N}'_\Lambda$, and the specification is not quasilocal.

## B.3 Phase Diagrams

### B.3.1 Regular Phase Diagrams

The words "phase diagram" are usually associated with nice pictures in which two conditions are satisfied:

1) Only *periodic* extremal Gibbs measures are considered. We emphasize that the order of the qualifiers has been carefully chosen: the measures relevant here are those *extremal Gibbs* measures that *happen to be periodic*; we are *not* referring to the measures that are extremal among the periodic ones (this latter is a larger and less-well-behaved class). For short, we shall call these measures *pure phases*, but we emphasize that this embodies a double change with respect to the terminology adopted in the rest of this paper: First, we consider all periodic Gibbs measures on the same footing, whether they are invariant under the whole translation group $\mathbb{Z}^d$ or merely a nontrivial $d$-dimensional subgroup of it. Second, we invert the order of the qualifiers, that is, we call *pure phase* an extremal measure in the sense of (ii) in Section 2.4.9, rather than in the more customary sense (iii).

We shall fulfill this condition throughout the rest of this appendix: by "phase diagram" we will mean the partition of a certain parameter space into regions with a given number and type of pure phases.

2) The Gibbs phase rule [363] is obeyed. Let us explain in a little more detail what this means. An example of a phase diagram satisfying the Gibbs phase rule is presented in Figure 13 below: There is a point where three pure phases coexist (*point of maximal coexistence*), from which there emanate three lines where two pure phases coexist, which in turn bound three open regions in which there is only one periodic extremal Gibbs measure. Such a phase diagram will be called *regular*. More generally, an *r-regular phase diagram* consists of [170, Appendix A]:

(1) a point of maximal coexistence where $r$ pure phases coexist;

(2) $r$ one-dimensional open manifolds, each bounded by this maximal-coexistence point, where exactly $r - 1$ phases coexist;

(3) $r(r-1)/2$ two-dimensional open manifolds, each bounded by pairs of the previous one-dimensional manifolds, where exactly $r - 2$ pure phases coexist;

⋮

($r$) $r$ open $(r - 1)$-dimensional manifolds, each bounded by the $(r - 2)$-dimensional 2-phase-coexistence manifolds, and such that the closure of their union is the whole parameter space, where there is only one pure phase.

Usually, the pure phases are defined by fixing the boundary conditions according to some *parameter-independent* set $\mathcal{K}$ of reference configurations (or, more generally, measures). Typically, $\mathcal{K}$ is the set of ground-state configurations at the point of maximal coexistence at $T = 0$. One can then label each pure phase according to the boundary condition employed in its definition. One calls the *K-stratum* ($K \subset \mathcal{K}$) the manifold in parameter space where the coexisting phases are precisely those labelled by elements of $K$. For instance, in Figure 13, the different strata are labelled by the boundary conditions "+", "0" and "−". There are, therefore, seven strata: $\{+\}, \{0\}, \{-\}, \{+, 0\}, \{+, -\}, \{0, -\}, \{+, 0, -\}$.

A more abstract (topological) way of visualizing such a phase diagram is provided by the following equivalent characterization: a $r$-regular phase diagram is a diagram that can be homeomorphically mapped onto the boundary of the positive octant in $r$ dimensions,

$$\partial Q_r = \left\{ (t_1, \ldots, t_r) \in \mathbb{R}^d_{\geq 0} : \min_{1 \leq i \leq r} t_i = 0 \right\}, \tag{B.25}$$

in such a way that the point of maximal coexistence corresponds to the origin, the curves of $(r-1)$-phase coexistence correspond to the positive coordinate axes excluding the origin, ... , the open sets with only one pure phase correspond to the $(r - 1)$-dimensional coordinate hyperplanes excluding their $(r - 2)$-dimensional boundaries. In brief, the different strata are mapped into the different submanifolds of the boundary of the $r$-octant.

General phase diagrams need not obey the Gibbs phase rule. A typical situation is for some of the pure phases to always appear together throughout the diagram. Such a situation is called a *degeneracy*, and it is usually associated to some symmetry of the system (if no symmetry can explain it, the degeneracy is called *fortuitous*). The addition of further interactions (not respecting the symmetry) can produce a phase diagram without degeneracy. These extra interactions are said to *break the degeneracy* of the pure phases in question. An interaction is said to *completely break the degeneracy* of the pure phases if its addition yields a regular phase diagram.

### B.3.2   Zero-Temperature Regular Phase Diagrams

For zero-temperature phase diagrams, it is relatively simple to give conditions on the extra interactions needed to ensure a regular phase diagram. Indeed, at zero tem-perature degeneracy means equal specific energy for all values of the parameters, and its breaking involves adding interactions producing a different set of specific energies for each of the initially degenerate pure phases. This is usually done perturbatively, that is, each additional interaction is multiplied by an overall "turn-on" parameter. Suppose we start with an interaction $\Phi_0$ having $r$ degenerate zero-temperature pure phases $\mu_1, \ldots \mu_r$. Then, to completely break the degeneracy one usually considers $r-1$ additional interactions $\Phi_1, \ldots, \Phi_{r-1}$ and constructs the "perturbed" interactions

$$\Phi_{\overline{\lambda}} = \Phi_0 + \sum_{i=1}^{r-1} \lambda_i \Phi_i . \tag{B.26}$$

[Examples: (i) For the Ising model at zero field, $\lambda_1 = h$; (ii) for the Blume-Capel interaction defined by (B.34) below, $\lambda_1 = g$ and $\lambda_2 = h$ in the "perturbation" (B.35).] The parameters $\overline{\lambda} = (\lambda_1, \ldots, \lambda_{r-1})$ usually take values in a certain neighborhood of the origin. The degree of degeneracy for the perturbed interaction $\Phi_{\overline{\lambda}}$ depends on the $r$-tuple of specific energies

$$\overline{e}(\overline{\lambda}) = (e_{\Phi_{\overline{\lambda}}}(\mu_1), \ldots, e_{\Phi_{\overline{\lambda}}}(\mu_r)) . \tag{B.27}$$

In fact, if we denote
$$Q(\overline{\lambda}) = \{i : \mu_i \text{ minimizes } e_{\Phi_{\overline{\lambda}}}(\mu)\} . \tag{B.28}$$

then the strata of the zero-temperature phase diagram are the sets

$$S_K = \{\overline{\lambda} : Q(\overline{\lambda}) = K\} \tag{B.29}$$

for each subset of labels $K \subset \{1, \ldots, r\}$. The perturbed interaction completely breaks the degeneracy if the phase diagram formed by the strata (B.28) is $r$-regular.

It is of interest to translate the requirement of regularity into conditions on the perturbations $\Phi_i$. One way to do it is to notice that, as the specific energy depends linearly on the parameters $\lambda_i$, it can be written in the form

$$\overline{e}(\overline{\lambda}) = \sum_{i=1}^{r-1} \lambda_i \overline{e}(\Phi_i) , \tag{B.30}$$

with

$$\overline{e}(\Phi_i) = (e_{\Phi_i}(\mu_1), \ldots, e_{\Phi_i}(\mu_r)) . \tag{B.31}$$

One of the conditions for the phase diagram to be $r$-regular is that the origin $\overline{\lambda} = 0$ be the only point of maximal coexistence. This implies that no nonzero vector of the form (B.30) can have all its coordinates equal. A little bit of linear algebra shows that all the other conditions for regularity are satisfied if the vectors $\{\overline{e}(\Phi_i)\}_{1 \leq i \leq r-1}$ are, in addition, linearly independent. Therefore, *the perturbations $\Phi_1, \ldots \Phi_{r-1}$ completely break the degeneracy of $\Phi_0$ if and only if the vectors $\overline{e}(\Phi_i)$ are linearly independent and they do not span the vector $(1, \ldots, 1) \in \mathbb{R}^r$*.

Alternatively, if we resort to the previous geometrical description of regularity, we conclude that it is equivalent to require that the vector $\overline{e}(\overline{\lambda})$ — shifted so it always has at least one coordinate equal to zero — sweeps over the boundary $\partial Q_r$ of the positive octant. Precisely stated, if we denote

$$\widetilde{e}_{\overline{\lambda}}(\mu_i) \;=\; e_{\Phi_{\overline{\lambda}}}(\mu_i) - \min_{1 \leq j \leq r} e_{\Phi_{\overline{\lambda}}}(\mu_j) \,, \tag{B.32}$$

*the perturbation $\Phi_{\overline{\lambda}}$ completely breaks the degeneracy of $\Phi_0$ if and only if the map*

$$\overline{\lambda} \mapsto (\widetilde{e}_{\overline{\lambda}}(\mu_1), \ldots \widetilde{e}_{\overline{\lambda}}(\mu_r)) \tag{B.33}$$

*is one-to-one. In other words, if such a map is a bijection from a neighborhood of $0 \in \mathbb{R}^{d-1}$ to a neighborhood of $0 \in \partial Q_r$. For each particular value of $\overline{\lambda}$, the coexisting pure phases are those $\mu_i$ with $\widetilde{e}_{\overline{\lambda}}(\mu_i) = 0$.*

### B.3.3 Low-Temperature Phase Diagrams. Scope of Pirogov-Sinai Theory

If nature is fair, one expects that low-temperature phase diagrams look very similar to the corresponding zero-temperature ones. This is not always so, however, and the question of stability or w-stability of Gibbs measures is an important issue. Pirogov-Sinai theory has been precisely designed to single out some important cases in which indeed the low-temperature diagrams are only a small deformation of the ones at zero temperature. When the theory applies, one is guaranteed that the regularity of the diagram is preserved at least for small temperatures; and, furthermore, that the low-temperature pure phases look very "similar" to the zero-temperature ones.

As an input to the Pirogov-Sinai theory one must determine the zero-temperature phase diagram and show that two key hypotheses are satisfied. The first hypothesis refers to the number of zero-temperature deterministic pure phases: In its original version [296, 297], Pirogov-Sinai (PS) theory applies to a system with a finite-range periodic interaction, exhibiting a *finite number of periodic rigid ground-state configurations*. (This has subsequently been generalized to some extent: see Section B.4.4.) The second hypothesis is that the interaction satisfy the so-called "Peierls condition", to be stated more precisely below, which roughly requires that for each rigid periodic ground-state configuration the energy cost of introducing a droplet of spins aligned as in a *different* ground state must grow typically as the area of the boundary of the droplet. This condition allows the energy cost of creating excitations to beat the entropy gain, preserving the long-range order observed at zero temperature. However, the Peierls condition has this desired effect only for $d \geq 2$. The trouble is that for $d = 1$ the size of the boundary of a set does not grow with its volume. Therefore, *Pirogov-Sinai theory is not applicable to one-dimensional models*. On the other hand, for $d \geq 2$ the Peierls condition is certainly stronger than necessary[77]: there exist models with a finite number of rigid periodic ground-state configurations which have a non-trivial phase diagram and which do not satisfy the Peierls condition [290, 264]. Nevertheless, the Peierls condition applies in a large number of interesting models, and allows a very precise description of the low-temperature behavior.

---

[77]In some sense it is the *strongest possible* condition: see the comments after Definition B.19 below.

The output of the theory is a family of results involving extensions to non-zero temperatures. The main result of the theory is that for a system satisfying the Peierls condition the phase diagram involving these periodic deterministic measures is stable: As the temperature increases, the coexistence manifolds deform continuously (in fact analytically). Moreover, the theory makes rigorous the intuitive picture of what each low-temperature pure phase looks like: its typical configurations consist of a "sea" of spins aligned as in the ground-state configuration with small and sparse "islands" of overturned spins.

We remark that the theory does not have anything to say about the stability of the (possibly infinitely many) *non-periodic* ground-state configurations and the zero-temperature Gibbs measures they support (but see Section B.4.4). Other techniques are needed to show, for example, that the flat-interface ground-state configurations (B.6) — which are rigid for $d \geq 2$ — are unstable for $d = 2$ [141, 1, 192] and stable for $d \geq 3$ [85, 348].

Moreover, the original version of PS theory gives only very limited information as to the specifics of the deformation of the phase diagram; in particular it does not produce a useful criterion to determine which pure phases are stable for the different regions of the zero-temperature phase diagram. That is, it does not tell us *in which direction* the phase boundaries move when the temperature is raised from zero. Therefore, for interactions $\Phi$ lying *on* a phase-transition manifold of the zero-temperature phase diagram, the original PS theory does not tell us in which phase(s) $\Phi$ ends up at $T > 0$; that is, it does not tell us which one(s) of the coexisting zero-temperature pure phases is/are stable, and which are only w-stable for the family of interactions adopted. However, Slawny's extension of PS theory [329] provides this additional information.

To clarify this point, let us borrow a very instructive example from the review by Slawny [329]. Consider the spin-1 Blume-Capel model defined by the formal Hamiltonian

$$H_0 = \tfrac{1}{2} \sum_{\langle xy \rangle} (\omega_x - \omega_y)^2 , \qquad (B.34)$$

where $\omega_x = -1, 0, 1$ and the sum is over pairs of nearest-neighbor sites in $\mathbb{Z}^d$, $d > 1$. Such a model has three periodic (in fact translation-invariant) ground-state configurations: all-"+", all-"0" and all-"−". They are all rigid. To obtain a 3-regular phase diagram one can consider, for instance, the family of interactions defined by the formal Hamiltonians

$$H(g, h) = H_0 - g \sum_x \omega_x^2 - h \sum_x \omega_x . \qquad (B.35)$$

The corresponding zero-temperature phase diagram is presented in Figure 13(a). Pirogov-Sinai theory tells us that for $T > 0$ low enough the phase diagram is just a continuous deformation of the one depicted, but to conclude that such deformations look as in Figure 13(b) we need some extra information which is not directly obtainable from PS theory, although it is probably contained in it. This extra information is presented ex-plicitly, for instance, in Slawny's theory of asymptotics of phase diagrams [329]. From the latter diagram we see, for instance, that of the three deterministic pure phases of $H(g = 0, h = 0)$ only the all-"0" is stable, while the other two pure phases are w-stable. The well-studied ferromagnetic Ising model provides an example of an exceptional nature: its phase diagram remains undeformed at low temperatures; for all values of the magnetic field the periodic zero-temperature Gibbs measures are stable.

## B.4 Pirogov-Sinai Theory

We summarize now the main aspects of PS theory. In the first two subsections we carefully discuss the basic hypotheses required by the theory; in the third subsection we present a somewhat detailed account of the results (for finite-range interactions). Of course, we omit all proofs; these can be found in the references cited. As already pointed out, the theory does not apply for $d = 1$, therefore *in the rest of this appendix we restrict ourselves to $d \geq 2$*.

### B.4.1 Boundary of a Configuration. The Peierls Condition

Typical configurations of a low-temperature pure phase are expected to be small fluctuations around those of a corresponding zero-temperature pure phase. These fluctuations result in the appearance of "droplets" ("bubbles", "islands") of spins aligned according to a different zero-temperature pure phase — or, more generally, a "metastable phase" [369] as we discuss below. These droplets are surrounded by a transitional region or "boundary" of sets of spins not aligned according to any zero-temperature pure phase, which therefore raises the energy of the configuration. The probability of such fluctuations is determined by the competition between two factors: the energy cost of introducing a boundary and the entropy gain due to the different possible shapes and locations of the droplets. If the energy cost is large enough to overcome, at low temperatures, the entropy gain, then each zero-temperature pure phase gives rise to a low-temperature one which differs only in the presence of few and small droplets of overturned spins. In particular, this would prove that there are precisely as many coexisting pure phases at low temperatures as there are at zero temperature, and hence that there is a phase transition. This type of argument was first introduced by Peierls [291, 82, 168] to prove the existence of a phase transition in the $d$-dimensional Ising model for $d \geq 2$, and hence it is often referred to as the "Peierls argument". Pirogov-Sinai theory is a generalization (and, thus, a more abstract version) of such an argument.

To formulate the Peierls argument in a rigorous form we need a criterion to determine when the energy cost of a boundary is "large enough" to defeat the entropy gain. The Peierls condition is precisely one such (sufficient) criterion. It relies, however, on a suitable definition of the "boundary" of a configuration, which is not a uniquely defined concept. In fact, two complementary notions are introduced at this stage: the *boundary*, which roughly corresponds to the collection of sites where the spins are misaligned, and the *contours*, which are the different components of this boundary *together* with the corresponding spin configurations. The latter allow a complete determination of the energy of a given configuration.

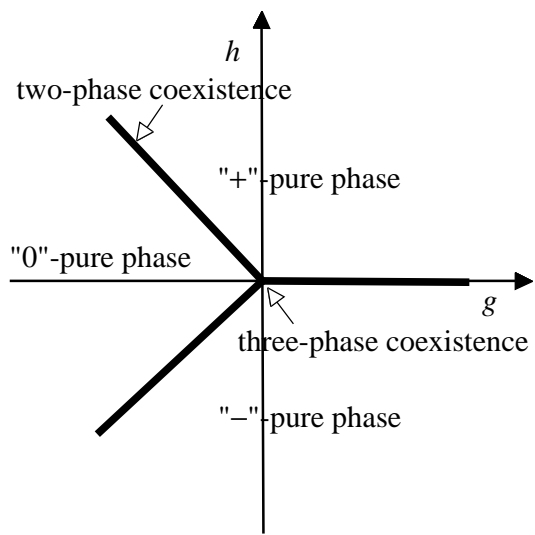

(a)

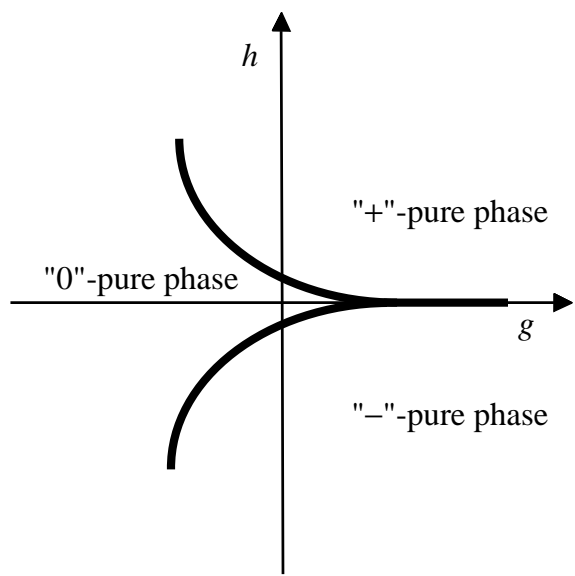

(b)

Figure 13: Phase diagrams of the model with interaction (B.35). (a) Zero temperature. (b) Low temperature.

Let us motivate the general definitions via examples. In the original case of the ferromagnetic Ising model, the boundary of a configuration can, for instance, be defined as all the pairs of nearest-neighbor sites with opposite spins. As each such pair contributes equally to the energy of the configuration, regardless of which spin of the pair is up and which is down, one does not need to specify the actual configuration on the boundary to compute the energy. Therefore contours are defined with no reference to configurations, by considering the polyhedral surface formed by plaquettes perpendicular to the bonds joining misaligned spins, and taking its connected components [82, 168]. The same definition of boundary works for the Blume-Capel model (B.34), but to compute the energy we now must specify the configuration of each pair of misaligned spins, as different combinations have different energies. The definition of contours requires hence to consider polyhedra labeled by the configuration on the immediately adjacent (internal and external) shells of spins. The next complication appears for models with interactions extending beyond nearest neighbors and/or involving more than two spins at a time. An example of practical interest is the antiferromagnet on a face-centered cubic lattice [329, p. 145 and references therein]. Such models require "thicker" boundaries and contours defined specifying the configurations of larger groups of spins.

Therefore, to define boundary and contours in a general fashion, we must check whether sets of spins are aligned or misaligned, but this checking has to be done on sufficiently large collections of spins at a time. Following closely [328, Chapter II], we consider a set $\mathcal{K} = \{\underline{\omega}^{(1)}, \ldots, \underline{\omega}^{(r)}\}$ of *periodic* configurations ($r \geq 1$). For the standard statement of the Peierls condition $\mathcal{K}$ will be the set of periodic (deterministic) ground-state configurations of some interaction, but the definition can be done (and *must* be done, as we shall discuss in next section), for general sets of periodic configurations. Let us call $\mathcal{K}$ the set of *reference configurations* [369]. We also consider for some fixed $a \geq 0$ the cubes $W_a(x) = \{y \in \mathbb{Z}^d: , |y_i - x_i| \leq a \text{ for } 1 \leq i \leq d\}$ — the *sampling cubes*.

**Definition B.18** *The boundary of a configuration $\omega$ — with respect to the set of reference configurations $\mathcal{K}$ and sampling cubes $W_a(x)$ — is the set of sites*

$$\partial \omega = \bigcup_{x \in \mathbb{Z}^d} \left\{ W_a(x) \colon \omega|_{W_a(x)} \neq \underline{\omega}|_{W_a(x)} \ \forall \underline{\omega} \in \mathcal{K} \right\}. \tag{B.36}$$

Typically we will consider configurations $\omega$ equal to some $\underline{\omega} \in \mathcal{K}$ except for a finite set of spins. In this situation the boundary is a finite set.

Let us now state the simplest and most popular version of Peierls condition; in the following section we discuss a more general definition. We consider an interaction $\Phi_0$ and, for each fixed $\underline{\omega} \in \mathcal{K}$ construct the relative Hamiltonian

$$H^{\Phi_0}(\omega|\underline{\omega}) = \sum_{A: A \subset \mathbb{Z}^d \text{ finite}} [\Phi_{0A}(\omega) - \Phi_{0A}(\underline{\omega})], \tag{B.37}$$

defined only for configurations $\omega$ coinciding with $\underline{\omega}$ except on a finite set. Let us denote $\mathcal{G}^{\text{per}}_{T=0}(\Phi_0)$ the set of periodic ground-state configurations of $\Phi_0$.

**Definition B.19** *The interaction $\Phi_0$ satisfies the (original) Peierls condition if there exists a constant $\rho_0 > 0$ such that for each $\underline{\omega} \in \mathcal{G}_{T=0}^{\mathrm{per}}(\Phi_0)$*

$$H^{\Phi_0}(\omega|\underline{\omega}) \geq \rho_0 |\partial\omega| \tag{B.38}$$

*for every configuration $\omega$ coinciding with $\underline{\omega}$ except possibly on a finite set of sites. Here $\partial\omega$ is the boundary of $\omega$ with respect to $\mathcal{K} = \mathcal{G}_{T=0}^{\mathrm{per}}(\Phi_0)$ and sampling cubes defined by some fixed choice of $a \geq 0$.*

We shall call a constant $\rho_0$ satisfying (B.38) a *Peierls constant* for the interaction $\Phi_0$ (and the chosen $\mathcal{K}$ and $a$). The Peierls condition immediately implies that each periodic ground-state configuration is rigid, and hence defines a *deterministic* zero-temperature pure phase (Section B.2.3). The converse is not true [290, 264]. We also notice that an upper bound of the form $H^{\Phi_0}(\omega|\underline{\omega}) \leq \tilde{\rho}_0 |\partial\omega|$ is always true, hence the Peierls condition is basically a requirement for the energy cost to grow as fast as possible with the size of the boundary of the configuration. There are important models where this is not true, i.e. in which the energy cost grows more slowly than the area of the boundary: for example, the balanced model [132, 48] and the ANNNI model [48, and references therein]).

We notice that the *validity* of the Peierls condition does not depend on the particular choice of the parameter $a \geq 0$ adopted for the definition of the boundary, but the actual value of the Peierls constant does. Indeed, following [328] we notice that if $\partial'\omega$ indicates the boundary defined via sampling cubes $W_{a'}$ with $a' \geq a$ (other cases left to the reader), then

$$\partial\omega \subset \partial'\omega \subset \cup_{x \in \partial\omega} W_{a'}(x) \tag{B.39}$$

thus,

$$|\partial\omega| \leq |\partial'\omega| \leq (2a'+1)^d |\partial\omega|. \tag{B.40}$$

Therefore, different choices of $a$ change the actual value of $\rho_0$:

$$\frac{\rho_0}{(2a'+1)^d} \leq \rho_0' = \rho_0 \sup_\omega \frac{|\partial\omega|}{|\partial'\omega|} \leq \rho_0, \tag{B.41}$$

but not its nonzero character. One has the freedom of adjusting $a$ according to future convenience. However, the actual value of $\rho_0$ is related to the range of temperatures where PS theory is valid (this range is proportional to $\rho_0$). Hence, for the sake of quantitative predictions one should employ a value of $\rho_0$ as large as possible, which means $a$ as small as possible. An extremely favorable case is exemplified by the ferromagnetic Ising model, and its generalizations to higher spins, for which the boundary of configurations can be defined via polyhedra of "zero width" [and we may even have equality in (B.38)]. Strictly speaking, the corresponding "zero-width" (or *thin*) contours are not included in the formalism to be introduced below, but we shall keep them within our discussion through appropriate comments.

The actual verification of the Peierls condition is a model-dependent, generally nontrivial, procedure. The starting point is, in principle, the determination of all periodic ground-state configurations — an often tedious process. A slight simplification follows from the observation that *if we find that (B.38) is satisfied for some finite set $\mathcal{K}$ of periodic configurations, automatically these must be all the periodic ground states.* Indeed, (B.38) implies that such configurations $\underline{\omega}$ are ground states, and if there were others (B.38) would not be satisfied because arbitrarily large boundaries could be constructed without extra energy cost, simply by interposing regions occupied by the ground states not accounted for. In practice, this observation is of little help, as the determination of ground states is made using some sort of contour ideas, so checking the Peierls condition and finding the ground-state configurations are almost simultaneous processes (however, see [215]). The only real shortcut available is a sufficient condition due to Holsztynski and Slawny [195] which we will use for almost all the applications in this paper.

**Definition B.20** *A potential $\Phi$ is an m-potential if there exists a configuration $\omega$ simultaneously minimizing each $|A|$-body function:*

$$\Phi_A(\omega) = \min_{\widetilde{\omega} \in \Omega} \Phi_A(\widetilde{\omega}) \qquad \forall A \in \mathcal{S} . \tag{B.42}$$

For such an interaction let us denote by $\mathcal{G}_{T=0}(\Phi)$ the (nonempty) set of configurations minimizing all $\Phi_A$. The sufficient condition is:

**Theorem B.21 (Holsztynski-Slawny)** *A finite-range m-potential $\Phi$ with $\mathcal{G}_{T=0}(\Phi)$ finite satisfies the Peierls condition.*

Resorting to an alternative — and suggestive — terminology, we can say that an $m$-potential is one for which there are ground states "satisfying" all bonds. An immediate example is any Ising model with ferromagnetic interactions ($\Phi_A = -J_A \sigma^A$ with $J_A \geq 0$ for all $A$): clearly the all-"+" configuration simultaneously minimizes all $\Phi_A$. The opposite case is that of the potentials with "frustration", i.e. for which every configuration has bonds that give an energy contribution larger than the minimum possible ("frustrated bonds"). However, these notions of $m$-potentials and "frustration" must be taken modulo physical equivalence, because equivalent potentials have the same statistical-mechanical properties. This adds an extra twist to the matter. A popular example is the antiferromagnetic Ising model in a triangular lattice. It is easy to see that when the model is given its usual formulation in terms of two-spin interactions, no configuration can "satisfy" simultaneously the three bonds of a triangular plaquette. But this seemingly frustrated potential can equivalently be written by considering the triangular plaquettes themselves as the bonds, with an energy contribution obtained by a suitable combination of the contributions of the original two-spin bonds around the plaquette. In this formulation the model is now an $m$-potential (although one cannot use Theorem B.21 because there are infinitely many periodic ground-state configurations). In this regard, probably the most difficult aspect of the application of this very convenient theorem is the verification of whether the potential of interest can be rewritten as (i.e. is physically equivalent to) an $m$-potential. It would be very nice to complement Theorem B.21 with some simple sufficient criterion for an interaction to be physically equivalent to an $m$-potential, but this may not be an easy task. For instance, the natural conjecture that every finite-range translation-invariant interaction is equivalent to a (translation-invariant finite-range) $m$-potential is false [263].

At any rate, once the $m$-potential character has been verified, Theorem B.21 is an extremely convenient tool. It has, however, an important drawback: its proof is not constructive, so it does not provide any explicit expression for the Peierls constant. Therefore, arguments based on this theorem do not allow any determination of the range of temperatures where the PS theory remains valid.

### B.4.2 Contours. The Generalized Peierls Condition

In the presence of the Peierls condition for an interaction $\Phi_0$, the usual Peierls argument can be repeated for those zero-temperature pure phases of $\Phi_0$ for which the entropy factor can be shown to grow at most exponentially with the size of the boundary. Indeed, the Peierls condition ensures that the energy cost grows as least as fast but with an exponent including a factor $\beta$, hence the energy cost beats the entropy gain for large enough $\beta$, and only small boundaries are present. However, this energy-beats-entropy phenomenon is in general *not* true for all the pure phases, only for the stable ones. It turns out that to obtain a situation in which the entropy is beaten by the energy for *all* the rigid periodic ground-state configurations of $\Phi_0$ — and hence all of them coexist — one must consider a perturbed interaction $\Phi = \Phi_0 + \widetilde{\Phi}$ for a suitably adjusted $\widetilde{\Phi}$ (shift in the point of maximal coexistence). In general, not all the ground-state configurations for $\Phi_0$ are ground-state configurations for $\Phi$, hence this process of "tuning" $\Phi$ requires us to consider a set $\mathcal{K}$ not reduced just to configurations with minimal $\Phi$-energy.

Another reason to generalize the Peierls condition appears when studying whole regions of the phase diagram. In such a situation one is interested in estimates valid *uniformly* throughout the region; but a uniform Peierls condition, as stated in Definition B.19, is not in general possible. For example, consider the Ising model in the presence of a strictly positive magnetic field. The only ground-state configuration is the all-"+" — to be denoted $\omega^{(+)}$ — and hence $|\partial\omega|$ is proportional to the number of "−" present. For instance, for the configurations $\omega^W$ equal to $+1$ everywhere except inside a cube $W$, the relative energy is $H(\omega^W|\omega^{(+)}) = 2J|\partial W| + 2h\,\mathrm{vol}(W)$, while $|\partial\omega^W| \sim \mathrm{vol}(W)$. A simple calculation shows that for the Peierls condition to be valid for all these $\omega^W$ we need $\rho \lesssim h$. Thus the (original) Peierls condition is not satisfied uniformly in a neighborhood of the point $h = 0$, which is precisely the most interesting region.

A generalized Peierls condition must, therefore, allow configurations that are not necessarily ground states and also must satisfy some "uniformity" requirement. Such a condition is already contained in the work by Pirogov and Sinai, where the main results are shown to be consequence of a further generalized condition for the perturbed $\Phi$ that follows from the the Peierls condition satisfied by $\Phi_0$. Zahradník [366] was, however, the first to point out that this generalized condition is a more natural starting point from the conceptual point of view. We initially had a more concrete motivation: the uniformity requirement is important for our example of the Kadanoff transformation (Section 4.3.3). In fact, this application demands only a particular case of uniformity (Corollary B.25 below), and, moreover, the result we need is exactly given by a theorem due to Zahradník (Theorem B.30 below). However, we shall take here the time to discuss the uniformity issue in some generality, because we feel that it has not been sufficiently emphasized in the literature.

Let us first introduce the notion of contour. We fix a set $\mathcal{K}$ of reference configurations and a choice of sampling cubes (value of $a$). The idea is to decompose the boundary in components: two sets $A$ and $B$ of sites are called *connected* if $\text{dist}(A, B) \leq 1$ in lattice units. A *contour* of a configuration $\omega$ is a pair $\Gamma = (M, \omega_M)$ where $M$ is a maximally connected component of the boundary of $\omega$. The set $M$ is often called the *support* of the contour $\Gamma$. At this point we start introducing constraints on the size of the sampling cubes. We require:

(C1) The value $2a + 1$ must be strictly larger than all the periods of the reference configurations $\underline{\omega} \in \mathcal{K}$.

Such a requirement implies the following *extension property* (nomenclature taken from [329]): if a configuration $\omega$ coincides with a reference configuration $\underline{\omega} \in \mathcal{K}$ on the sampling cube $W_a(x)$ and with $\underline{\omega}' \in \mathcal{K}$ on the cube $W_a(y)$ with $\text{dist}(x, y) \leq 1$, then $\underline{\omega} = \underline{\omega}'$. This has the key consequence that we can reconstruct uniquely a configuration $\omega$ starting from its family of contours.

Each contour with a finite support divides $\mathbb{Z}^d \setminus M$ into several disconnected components: One of them is unbounded, and is called the *exterior* of the contour; the others are bounded and are collectively called the *interior* of the contour. Each of these components has a reference configuration associated to it, namely that of the sampling cubes centered on sites adjacent to the support of the contour. The contour is a $\underline{\omega}$-*contour* if its exterior corresponds to the reference configuration $\underline{\omega}$. On the other hand, the $\underline{\omega}^{(i)}$-interior — denoted $\text{Int}_{\underline{\omega}^{(i)}}(\Gamma)$ — is the union of the components of the interior of $\Gamma$ associated to a reference configuration $\underline{\omega}^{(i)}$. In general, $\omega$ will have other contours besides $\Gamma$, some of which may be in the interior of $\Gamma$. Hence $\omega$ may not coincide with $\underline{\omega}^{(i)}$ on the whole $\text{Int}_{\underline{\omega}^{(i)}}$. The generalized Peierls condition is a requirement on the minimum energy cost of introducing a contour. This can be estimated by considering the configuration $\omega^\Gamma$ that has $\Gamma$ as its *only* contour. If $\Gamma = (M, \omega_M)$ is a $\underline{\omega}$-contour, $\omega^\Gamma$ coincides with $\underline{\omega}$ on the exterior of $\Gamma$, with $\underline{\omega}^{(i)}$ on the whole $\text{Int}_{\underline{\omega}^{(i)}}$, and with $\omega_M$ on the support $M$.

Let us introduce now a periodic interaction $\Phi$. The energy cost of the $\underline{\omega}$-contour $\Gamma$ is given by the relative energy $H^\Phi(\omega^\Gamma | \underline{\omega})$, which can be decomposed in the form:

$$H^\Phi(\omega^\Gamma | \underline{\omega}) \;=\; \Psi(\Gamma) + \sum_{i=1}^{r} [e_\Phi(\underline{\omega}^{(i)}) - e_\Phi(\underline{\omega})] |\text{Int}_{\underline{\omega}^{(i)}}| \;. \qquad (B.43)$$

The second term in the RHS is, up to terms proportional to $|\partial M|$, the energy contribution due to the configurations in the interior of $\Gamma$. This term is absent if all the $\underline{\omega}^{(i)}$ are ground-state configurations of $\Phi$. The *contour functional* $\Psi(\Gamma)$ is *defined* by the identity (B.43); it is roughly equal to $\sum_{A\subset M}[\Phi_A(\omega^\Gamma) - \Phi_A(\underline{\omega})]$, but it also includes the just mentioned terms proportional to $|\partial M|$.

**Definition B.22** *An interaction $\Phi$ satisfies the generalized Peierls condition — with respect to a set $\mathcal{K}$ of reference configurations — if there exists a constant $\rho > 0$ such that for each contour $\Gamma = (M, \omega_M)$.*

$$\Psi(\Gamma) \geq \rho|M| \qquad (B.44)$$

The (original) Peierls condition (B.38) corresponds to the particular case in which $\mathcal{K} = \mathcal{G}_{T=0}^{\text{per}}(\Phi)$. We remark that this generalized condition is sometimes called just "Peierls condition", or "Gerzik-Pirogov-Sinai" condition. We shall also call a *Peierls constant* — for the interaction $\Phi$ — a constant $\rho$ satisfying (B.44).

To understand why Definition B.22 has the desired uniformity, let us return to the example of the Ising model with magnetic field $h > 0$. We must now consider $\mathcal{K} = \{\omega^{(+)}, \omega^{(-)}\}$, where $\omega^{(+)}$ and $\omega^{(-)}$ are the all-"+" and all-"−" configurations respectively. We notice, however, that $\omega^{(-)}$ is not a ground state. With this choice of $\mathcal{K}$, the contours can be taken to be "thin" as in the zero-field case, and we have that for any $\omega^{(+)}$-contour $\Gamma$

$$H^\Phi(\omega^\Gamma | \omega^{(+)}) = 2J|\partial W| + 2h|\text{Int}_{\omega^{(-)}}(\Gamma)|, \qquad (B.45)$$

while for an $\omega^{(-)}$-contour

$$H^\Phi(\omega^\Gamma | \omega^{(-)}) = 2J|\partial W| - 2h|\text{Int}_{\omega^{(+)}}(\Gamma)|. \qquad (B.46)$$

So, comparing with (B.43) we see that the generalized Peierls condition is satisfied with $\rho = 2J$, uniformly in $h$. We see that this uniformity is gained by including the extra configuration $\omega^{(-)}$ which is not a ground state, but rather could be interpreted as a "metastable state".

In general, the uniformity property of the generalized Peierls condition is a consequence of an estimate valid for sampling cubes larger than the period and range of the interaction; that is, we impose the following extra condition on the sampling cubes:

(C2) The value $2a + 1$ must be strictly larger than the period and the range of the interaction $\Phi$.

We emphasize that due to requirements (C1) and (C2), the value chosen for $a$ — that is, the definition of the contours — depends on the set $\mathcal{K}$ and on the interaction(s) present. The reader should keep this in mind especially because, to keep formulas simple to read, the notation will not make this dependence explicit. In particular, a change in the interaction — for instance the addition of an arbitrarily small perturbation — will require the redefinition of the contours.

Under condition (C2),

$$\left| \sum_{A: A \cap \left[\text{Int}_{\underline{\omega}^{(i)}} \cup \partial_{\text{ext}}\text{Int}_{\underline{\omega}^{(i)}}\right] \neq \emptyset} \Phi_A(\omega^\Gamma) - |\text{Int}_{\underline{\omega}^{(i)}}| \, e_\Phi(\underline{\omega}^{(i)}) \right|$$
$$\leq 2(2a+1)^d \|\Phi\|_{\mathcal{B}^0} |\partial_{\text{ext}}\text{Int}_{\underline{\omega}^{(i)}}| \qquad (B.47)$$

which implies the following key estimate. If $\Gamma = (M, \omega_M)$ is a a $\underline{\omega}$-contour, then for any periodic interaction $\widetilde{\Phi}$

$$\left| H^{\widetilde{\Phi}}(\omega^\Gamma | \underline{\omega}) - \sum_{i=1}^{r} |\text{Int}_{\underline{\omega}^{(i)}}| [e_{\widetilde{\Phi}}(\underline{\omega}^{(i)}) - e_{\widetilde{\Phi}}(\underline{\omega})] \right| \leq 2(2a+1)^d \|\widetilde{\Phi}\|_{\mathcal{B}^\circ} |M| . \qquad (B.48)$$

Therefore:

**Theorem B.23 (Uniformity property)** *If a periodic interaction $\Phi_0$ satisfies the generalized Peierls condition with constant $\rho$, then for any interaction $\widetilde{\Phi}$ with $\|\widetilde{\Phi}\|_{\mathcal{B}^\circ} \leq c\rho/(2a+1)^d$, the sum $\Phi_0 + \widetilde{\Phi}$ satisfies the generalized Peierls condition with constant $\rho(1-2c)$.*

Another useful result, which basically follows from (B.48), is the following [328, Lemma 2.2]:

**Proposition B.24** *Consider a periodic interaction $\Phi_0$ satisfying the original Peierls condition (B.38) with constant $\rho_0$. Then, for any other periodic interaction $\widetilde{\Phi}$*

$$\|\widetilde{\Phi}\|_{\mathcal{B}^\circ} < \rho/(2a+1)^d \implies \mathcal{G}^{\text{per}}_{T=0}(\Phi_0 + \widetilde{\Phi}) \subset \mathcal{G}^{\text{per}}_{T=0}(\Phi_0) . \qquad (B.49)$$

We present two corollaries of Theorem B.23. For our study of the Kadanoff transformation we need the following trivial consequence:

**Corollary B.25** *Consider a periodic interaction $\Phi_0$ satisfying the original Peierls condition (B.38) with constant $\rho_0$, and another interaction $\widetilde{\Phi}$ such that $\mathcal{G}^{\text{per}}_{T=0}(\Phi_0) = \mathcal{G}^{\text{per}}_{T=0}(\Phi_0 + \widetilde{\Phi})$. Then if $\|\widetilde{\Phi}\|_{\mathcal{B}^\circ} \leq c\rho_0/(2a+1)^d$, the sum $\Phi_0 + \widetilde{\Phi}$ satisfies the original Peierls condition with constant $\rho_0(1-2c)$.*

However, the corollary more often used is:

**Corollary B.26** *If $\Phi_0$ satisfies the original Peierls condition with constant $\rho_0$ [and $\mathcal{K} = \mathcal{G}^{\text{per}}_{T=0}(\Phi_0)$], then a "perturbed" interaction $\Phi = \Phi_0 + \widetilde{\Phi}$ satisfies the generalized Peierls condition with constant $\rho_0(1-2c)$ [and the same $\mathcal{K}$] if $\|\widetilde{\Phi}\|_{\mathcal{B}^\circ} \leq c\rho_0/(2a+1)^d$.*

This corollary generalizes what was observed regarding the Ising model in non-zero field.

As the inclusion in (B.49) is in general strict, the last corollary implies that, from the point of view of $\Phi = \Phi_0 + \widetilde{\Phi}$, the uniformity is gained at the cost of including some extra reference configurations that are not ground states (e.g. $\omega^{(-)}$ in the above example). These extra configurations can be interpreted as "metastable states" or "local ground states" for $\Phi$ [369]. On the other hand, any system with a finite number of periodic ground-state configurations ought to satisfy the generalized Peierls condition if one adds *all* the local ground states of the model [369] (or allow more complicated types of reference states).

At the risk of being considered almost patronizing, we emphasize again that the size $a$ in the previous results is chosen so as to satisfy (C1) and (C2) for the *total* interaction $\Phi_0 + \widetilde{\Phi}$. Often, $\Phi_0$ is a simpler or more standard interaction that one studies independently or for which one can borrow results from the literature. The Peierls constant determined in this manner corresponds, hence, to a values of $a$ chosen without reference to anything but $\Phi_0$. When considering in addition perturbations $\widetilde{\Phi}$, no matter how small, this size may need to be redefined to a new value $a'$ suitable for the total interaction. If so, the Peierls constant $\rho_0$ appearing in the previous results is *smaller* than the one initially determined. The simplest procedure at this point, if one does not want to completely redo the analysis with the new definition of contours, is to adopt for $\rho_0$ the initial value divided by $(2a' + 1)^d$ [leftmost inequality in (B.41)]. Note that the Peierls constant chosen in this way goes to zero with increasing range of the perturbations. In fact, in general one can not do much better than this. In particular, it is known (cf. Remark 4 in Section 2.6.7) that arbitrarily weak perturbations with long-range interactions can destroy the phase diagram.

To conclude this section, we observe that the notion of contour can be presented in a slightly more general (and abstract) fashion. Indeed, the key properties supporting the rest of the theory are the unique reconstruction of a configuration from a set of contours [here a consequence of the extension property, requirement (C1)], estimates (B.44) and (B.48), and that the entropy gain be beaten by the energy cost at low temperatures. As long as these properties are satisfied, contours need not be defined via sampling cubes. An illustration of this observation is the use of "thin" contours in ferromagnetic nearest-neighbor Ising models or, more generally, models whose ground-state configurations are constant. The boundary in such a model can be defined as a set of polyhedra, and the contours are non-self-intersecting closed (hyper)surfaces (uniquely defined via suitable fixed prescriptions to handle intersections), labelled by the configurations of the adjacent spins. The labelling allows for a unique reconstruction of the configuration, and the thin contours satisfy estimate (B.44) with $|M|$ replaced by $|\Gamma|$ = area of the polyhedra = number of plaquettes forming its faces, and estimate (B.48) with $a$ determined on the basis of $\widetilde{\Phi}$. Moreover, they have smaller entropy than the "thick" contours. Note, however, that the remark discussed in the previous paragraph is especially relevant in connection with thin contours: in general, if the interaction is perturbed, one *cannot* use the value of $\rho_0$ determined via thin contours; one must, for instance, divide it by a factor $(2a + 1)^d$, where $a$ depends on the perturbation $\widetilde{\Phi}$ considered.

### B.4.3 Results of the Theory

We present here the main results of PS theory. We include some general comments on the underlying ideas, but we do not discuss the details of the proofs. These can be consulted in the bibliography. We mention that there are two approaches to PS theory: the original one, based on "contour models with parameters", and the more recent one, due to Zahradník, based instead on a classification of contours into "stable" and "unstable" ones. References for the first approach are the seminal papers [296, 297], Sinai's book [328], and Slawny's review article [329]. The second approach was introduced in [366]; a concise presentation is given in [35] and a pedagogical one in [369]. This second approach is intuitively more appealing, provides some more information — as for instance the completeness [366] and analyticity [368] of the phase diagram — and has served as a basis for further extensions and applications of the theory [367, 287, 288, 194, 35, 36]. In the comments below we mostly have in mind such an approach.

The essence of Pirogov-Sinai theory — inherited from the Peierls argument — is the definition of maps from the original spin ensemble into ensembles of contours that *interact only by volume-exclusion*, that is, into *gases of contours*. The families of contours in the latter do not necessarily correspond to an actual collection of contours of a spin configuration, because they are not required to "match" exteriors with interiors. For instance, a set of two "−"-contours, one inside the other, is an allowed element of one of the contour ensembles, even when there is no spin configuration having it as its family of contours (in a spin configuration there would be an intermediate "+"-contour). This lack of "matching" requirement makes the contour ensembles much simpler systems to work with. The maps are defined so that each stable pure phase is equivalent to a contour ensemble in the sense that both have the *same distribution of* external *contours*. The low-temperature picture of only small "islands" of overturned spins can then be precisely proven by estimating the probabilities of (external) boundaries using the contour ensembles.

One considers $r$ different contour ensembles, one for each reference configuration $\underline{\omega}^{(i)} \in \mathcal{K}$. The $i$-th ensemble is formed by all the $\underline{\omega}^{(i)}$-contours interacting only via the restriction of being separated by 2 or more lattice units. The statistical weight of each contour is given by an activity $\exp[-F_\beta^{(i)}(\Gamma)]$ with a functional $F_\beta^{(i)}(\Gamma)$ determined via a relation (formulas (1.14) or (1.19) in [366]) that roughly compares the "work needed to install a contour" [369] in the spin and contour ensembles. [In the original PS theory, some extra weights $\exp[b^{(i)}|\text{Int}(\Gamma)|]$ are assigned to the external contours [328], and both the "parameters" $b^{(i)}$ and the functional $F^{(i)}$ are also determined by comparing "works" (formula (2.43) in [328]). We prefer to follow here Zahradník's approach in which the "parameter degree of freedom" is absorbed into the functional $F_\beta^{(i)}$.]

Each contour ensemble is a statistical-mechanical system of its own, which can be studied without any reference to the original spin system. Properties of these contour models can then be transcribed into results for the spin system via the identification between the ensembles. This is the usual policy in the standard expositions of the theory, all of which include an "interlude" in which abstract contour ensembles are analyzed *per se* (Sections 7 to 9 in Chapter 2 of [328], Section 2 in [366], etc). Basically, contour models are studied via cluster-expansion techniques: this is the method of choice for systems at "high temperature" or "low density". All the contour ensembles satisfy one of the key ingredients of the Peierls argument: *the entropy factor grows at most exponentially with the size of the contours* [328, Lemma 2.7] (this fact is false for $d = 1$!). Therefore, there is a marked difference according to whether the functional $F_\beta^{(i)}$ defining the contour activity satisfies a bound of the form

$$F_\beta^{(i)}(\Gamma) \geq \tau_\beta^{(i)} |M| \tag{B.50}$$

with $\tau_\beta^{(i)} > 0$. If this is the case, it is customary to say that $F_\beta^{(i)}$ is a $\tau_\beta^{(i)}$-*functional.* [For the original PS approach, the big difference is whether the corresponding parameter $b^{(i)}$ is zero; all the functionals $F^{(i)}$ in the PS approach are $\tau$-functionals.]

The contour models defined by $\tau$-functionals enjoy several remarkable properties *if $\tau$ is large enough to overcome the entropy growth*. This growth is characterized by an exponential factor bounded by [328, Lemma 2.7]

$$\alpha = \max\{d\log(2a+1),\ \log|\Omega_0| + 3^d\}. \tag{B.51}$$

If the contour model has a convergent cluster expansion, which occurs at least if [328, Lemma 2.8 and Propositions 2.1 and 2.2]

$$\tau_\beta^{(i)} \geq 4\alpha, \tag{B.52}$$

then it has a well-defined thermodynamic limit, with a well-defined pressure and infinite-volume probability measure. For this measure, infinite contours have zero probability of occurrence, more generally, the probability for a given contour to be present decreases exponentially with the size of its support. Moreover, the measure satisfies exponential mixing conditions for disjoint families of external contours. (See, for instance, Sections 7-9 of [328].) Furthermore, each of such contour measures is equivalent to a Gibbs measure in the spin system: *if $F_\beta^{(i)}$ is a $\tau_\beta^{(i)}$-functional, with $\tau_\beta^{(i)} \geq 4\alpha$, then the (infinite-volume) probability density of external contours of the contour ensemble is equal to that of the Gibbs measure — at inverse temperature $\beta$ — of the spin model defined by the $\underline{\omega}^{(i)}$ boundary condition.* Thus, this Gibbs measure inherits the sparsity of (external) contours characterizing the contour ensemble and its mixing properties. It is, therefore, an extremal periodic Gibbs measure (pure phase) which is only a small perturbation of the reference (in fact ground-state) configuration $\underline{\omega}^{(i)}$. The precise result of this argument is:

**Theorem B.27 (Pirogov-Sinai-Zahradník)** *Assume $d \geq 2$. If a finite-range periodic interaction $\Phi$ satisfies the generalized Peierls condition (B.44) with respect ot a finite set of periodic reference configurations $\mathcal{K} = \{\underline{\omega}^{(1)}, \ldots, \underline{\omega}^{(r)}\}$, then there exist $\beta_0 < \infty$ such that for each $\beta \geq \beta_0$*

*(a) All the pure phases are Gibbs measures $\mu_\beta^{(i)}$ defined by the boundary conditions $\underline{\omega}^{(i)}$ with $F_\beta^{(i)}$ being a $\tau_\beta^{(i)}$-functional. In this case, $\tau_\beta^{(i)} \to \infty$ as $\beta \to \infty$.*

*(b) Each pure phase $\mu_\beta^{(i)}$ is concentrated on configurations with finite boundaries and moreover, the probability that a given boundary be present tends to zero as $\beta \to \infty$. More precisely, if $\Gamma = (M, \omega_M)$*

$$\mu_\beta^{(i)}\{\Gamma\ external\ contour\} \leq e^{-\tau_\beta^{(i)}|M|}. \tag{B.53}$$

This theorem was proved by Pirogov and Sinai (see for instance [328, Propositions 2.6 and 2.2]), except for the word "all" in Part (a), which was incorporated by Zahradník [366] ("completeness"). One of the consequences of this completeness is that if the interaction $\Phi$ has a *unique* periodic ground-state configuration, and it satisfies the Peierls condition, then there is also a unique pure phase at low temperature. In fact, Martirosyan [260] has proven that, in this situation, in $d \geq 2$ there are no other extremal Gibbs measures, periodic or not.

The parameters $\tau_\beta^{(i)}$ characterizing the pure phases are of the form (see the proof of Proposition 2.3 in [328])

$$\tau_\beta^{(i)} \geq \beta\rho - \epsilon^{(i)} \qquad \text{with } \epsilon^{(i)} \leq 2e^{-\tau_\beta^{(i)}} 3^d \,, \tag{B.54}$$

where $\rho$ is the PS-constant of the interaction $\Phi$. From this expression one can obtain some (far from optimal) estimates of the parameters involved. Indeed, for example we can choose

$$\tau_\beta^{(i)} = \tau_\beta \tag{B.55}$$

with $\tau_\beta$ satisfying

$$\tau_\beta = \beta\rho - 2e^{-\tau_\beta} 3^d \,. \tag{B.56}$$

Then, by the requirement (B.52) we obtain the bound

$$\tau_\beta \geq \beta\rho - \frac{1}{2e} \geq 4\alpha \tag{B.57}$$

and hence,

$$\beta_0 = \frac{4\alpha + 1/(2e)}{\rho} \,. \tag{B.58}$$

Note that as $\beta \to \infty$ one also obtains, from (B.56)

$$\tau_\beta = \beta\rho + O(e^{-\beta\rho}) \,. \tag{B.59}$$

In principle, Theorem B.27(a) provides a criterion for the stability of a ground-state configuration, but it is quite useless for practical applications. The work of Zahradník [366, 369] provides a different criterion which could be employed for a computer-based procedure. It is based on the computation of the pressure $\widetilde{p}(F_\beta^{(i)})$ for the $\underline{\omega}^{(i)}$-contour ensemble but including only *small* (or stable) contours. These are contours whose interior volume is at most proportional to the size of the support. At low temperature, the coexisting pure phases are those minimizing

$$h_\beta^{(i)} \equiv e_\Phi(\underline{\omega}^{(i)}) - \widetilde{p}(F_\beta^{(i)}) \,. \tag{B.60}$$

It can be proven [366] that $\widetilde{p}(F_\beta^{(i)}) \to 0$ as $\tau_\beta^{(i)} \to \infty$ (i.e. $\beta \to \infty$), thus the minimizing configurations are ground states. Hence, (B.60) means that the stable ground-state configurations are those maximizing the contour-ensemble pressure; that is, those admitting the larger number of low-energy small contours. A related stability criterion was developed by Slawny [329], employing the *pressure of a gas of elementary excitations* instead of the contour-ensemble pressure. The stable phases are therefore determined as those with the larger number of low-energy excited configurations (*dominant* *ground-state configurations*). This criterion is simpler to apply for paper-and-pencil calculations.

The preceding theorem is the main tool used to prove the stability of the whole phase diagram. Let us denote $U_\varepsilon(\overline{\lambda}') = \{\overline{\lambda} \in \mathbb{R}^{d-1} : \sum_{i=1}^{r-1} |\lambda_i - \lambda_i'| < \varepsilon\}$.

**Theorem B.28 (Pirogov-Sinai-Zahradník)** *Consider a finite-range periodic interaction $\Phi_0$ in dimension $d \geq 2$ such that: (i) it has $r < \infty$ periodic ground-state configurations and (ii) it satisfies the original Peierls condition (B.38). Consider a perturbation $\Phi_{\overline{\lambda}} = \Phi_0 + \sum_{i=1}^{r-1} \lambda_i \Phi_i$, with each $\Phi_i$ periodic and of finite range, that completely breaks the degeneracy of $\Phi_0$. Then there exist positive constants $\beta_0$, $\varepsilon_0$ such that*

*(a) For each $\beta \geq \beta_0$ there exists a nonempty open set $V_\beta \subset \mathbb{R}^{d-1}$ such that for parameters $\overline{\lambda} \in V_\beta$ the phase diagram at inverse temperature $\beta$ is $r$-regular. For each $\overline{\lambda} \in V_\beta$ results (a) and (b) of Theorem B.27 hold for the interaction $\Phi_{\overline{\lambda}}$ taking as reference configurations the ground-state configurations of $\Phi_0$.*

*(b) Moreover, there exists an invertible map*

$$I_\beta : U_{\varepsilon_0}(\overline{0}) \longrightarrow V_\beta \qquad (B.61)$$

*(the "underlying deformation of the parameter space" [369]), which maps each zero-temperature coexistence manifold onto the corresponding coexistence manifold at inverse temperature $\beta$ (more generally, stratum onto stratum). The map $I_\beta$ converges to the identity as $\beta \to \infty$. In fact, it is an homeomorphism, and even $C^\infty$.*

*(c) The phase diagram deforms analytically with temperature in the following* local *sense: Consider a point $(\overline{\lambda}', \beta')$ of the phase diagram, with $|\overline{\lambda}'| < \varepsilon_0$ and $\beta'$ large enough [its minimum value depends on the distance from $\overline{\lambda}'$ to the complement of $U_{\varepsilon_0}(\overline{0})$]. Let $K$ be the set of reference configurations giving rise to the pure phases for the interaction $\Phi_{\overline{\lambda}'}$ at inverse temperature $\beta'$. Then there exists an analytic function*

$$\beta \mapsto \overline{\lambda}(\beta) \qquad (B.62)$$

*such that for $\beta$ close to $\beta'$ and $\overline{\lambda}(\beta)$ close to $\overline{\lambda}'$, the pure phases for the interactions $\Phi_{\overline{\lambda}(\beta)}$ at inverse temperature $\beta$ correspond to the same set $K$ of reference configurations.*

Pirogov and Sinai proved Parts (a) and (b) of the theorem except for the completeness of the phase diagram and the homeomorphic and $C^\infty$ character of $I_\beta$. These additional results are due to Zahradník [366, 369], who also proved Part (c) [368].

As remarked in [369], maps other than (B.62) need not be analytic. In particular, the ($\beta$-dependent) maps $V_\beta \to \partial Q_r$ establishing the regularity of the phase diagram at inverse temperature $\beta$ are, in general, *not* analytic. Such a map can be defined, for example [366, 369], analogously to the zero-temperature map (B.32)–(B.33) but replacing $e_{\Phi_{\overline{\lambda}}}(\mu_i)$ by $h_\beta^{(i)}$ [see (B.60)]:

$$\overline{\lambda} \longmapsto \left( h_\beta^{(1)}(\overline{\lambda}) - \min_{1 \leq i \leq r} h_\beta^{(i)}(\overline{\lambda}), \ldots, h_\beta^{(r)}(\overline{\lambda}) - \min_{1 \leq i \leq r} h_\beta^{(i)}(\overline{\lambda}) \right). \qquad (B.63)$$

[In the original PS formulation, the parameters $b^{(i)}$ played the role of the $h^{(i)}$ here.] Such a map is in general not analytic because if it were it would imply that the free energy of a pure phase could be analytically continued in $\overline{\lambda}$ into the metastability region; and already for the Ising model it has been shown [204] that such an analytic metastable extension does not exist. On the other hand, the map (B.63) is of limited physical significance, because for $\underline{\omega}^{(i)}$ not defining a pure phase, the corresponding quantity $h^{(i)}(\overline{\lambda})$ is only an auxiliary concept, not even uniquely defined [369]. The physically interesting objects are the strata

$$S_K(\beta) = \{\overline{\lambda} : Q_\beta(\overline{\lambda}) = K\} \tag{B.64}$$

for each $K \subset \mathcal{K}$, where

$$Q_\beta(\overline{\lambda}) = \left\{ i : h^{(i)}_\beta(\overline{\lambda}) = \min_{1 \le j \le r} h^{(j)}_\beta(\overline{\lambda}) \right\}. \tag{B.65}$$

These strata deform (locally) analytically with the temperature, by Part (c) of Theorem B.28.

Non-optimal estimations of the limit values $\beta_0$ and $\varepsilon_0$ of Theorem B.28 can be obtained combining Corollary B.26 with (B.58): If $\Phi_0$ satisfies the Peierls condition with constant $\rho_0$, then at most

$$\beta_0 = \frac{4\alpha + 1/(2e)}{\rho_0(1-2c)} \tag{B.66}$$

if at least

$$\varepsilon_0 = \frac{c\rho_0}{(2a+1)^d}. \tag{B.67}$$

These bounds are an explicit example of a general fact about the proof of Theorem B.28. As all the results follow from studying the equivalent contour ensembles, the relevant magnitudes are those actually used to construct these ensembles: the Peierls constant $\rho_0$ and the exponential entropy factor $\alpha$. In addition, the dimension $d$ and the size $a$ of the sampling cubes appear via the uniformity property. As a consequence, we have:

**Corollary B.29** *Consider a family of original interactions $\{\Phi_0(\overline{p})\}_{\overline{p} \in P \subset \mathbb{R}^m}$ satisfying the Peierls condition uniformly that is, with the same constant $\rho_0$ and the same family of periodic ground-state configurations $\mathcal{K}$ for all $\overline{p} \in P$ (e.g. in the conditions of Corollary B.25). Then Theorem B.28 holds also uniformly for all the interactions $\Phi_0(\overline{p})$ [i.e., one can chose the same $\beta_0$ and $\varepsilon_0$ in Parts (a) and (b), and the same $\beta$- and $\overline{\lambda}$- intervals in Part (c)].*

More generally, Theorem B.28 can be extended to situations in which there is a further smooth dependence of the interactions on the extra parameters $\overline{p}$.

**Theorem B.30** *Assume $d \ge 2$ and consider interactions $\Phi_0(\overline{p}), \Phi_1(\overline{p}), \ldots, \Phi_{r-1}(\overline{p})$ depending analytically on parameters $\overline{p}$ taking values on an open set $P \subset \mathbb{R}^m$, and with* *bounded period and range. Assume that there exists a $\overline{p}_0 \in P$ such that (i) $\Phi_0(\overline{p}_0)$ has $r < \infty$ periodic ground-state configurations and (ii) $\Phi_0(\overline{p}_0)$ satisfies the original Peierls condition (B.38). Assume also that for each $\overline{p} \in P$ the perturbation $\Phi_{\overline{\lambda}}(\overline{p}) = \Phi_0(\overline{p}) + \sum_{i=1}^{r-1} \lambda_i \Phi_i(\overline{p})$ completely breaks the degeneracy of $\Phi_0(\overline{p})$. Denote $\overline{\beta} \equiv (\beta, \overline{p})$. Then there exist positive constants $\beta_0$, $\varepsilon_0$ and $\varepsilon_1$ such that the results (a), (b) and (c) of Theorem B.28 hold replacing $\beta$ by $\overline{\beta}$ and the condition "$\beta \geq \beta_0$" by "$\beta \geq \beta_0, |\overline{p} - \overline{p}_0| \leq \varepsilon_1$".*

There are no simple explicit formulas for the values $\beta_0$, $\varepsilon_0$ and $\varepsilon_1$.

As mentioned at the end of the preceding Section B.4.2, the theory can be adapted to slightly more general notions of contours. In particular it applies for the "thin" (polyhedral-like) contours of models with constant ground-state configurations (e.g. the ferromagnetic Ising model, Blume-Capel models, etc). Such contours provide the best (largest) estimate of $\rho_0$ (e.g. for Ising models $\rho_0 = 2J$), and have and entropy growth with an exponential factor bounded by (see, for instance, [169])

$$\alpha_{\text{thin}} = \log(2d - 1) . \tag{B.68}$$

This bound is smaller than (B.51), and therefore yields another source of improvement on the estimates of $\beta_0$ in (B.58) or (B.66). Moreover, in the bound (B.67) for $\varepsilon_0$, we can use $2a$ = range of the perturbation $\sum \lambda_i \Phi_i$; which is the minimal possible choice.

### B.4.4 Extensions of the Theory. The Random Case

Pirogov-Sinai theory has been extended in several directions. For example, we mention the extensions to systems with long-range [287, 288], quasiperiodic [221] and complex [21, 295, 368, 35] interactions; systems with continuous spins [97, 367]; systems on a continuous space [45, 46]; field-theoretical systems [202, 203, 35]; and systems with infinitely many periodic ground-state configurations [46, 47, 80, 79, 77, 179, 78, 48]. Among these are models with energy cost growing slower than the area of the boundary, such as the balanced model [132, 48] and the ANNNI model [48, and references therein]. Also worth mentioning are the applications to the study of interfaces [367, 369, 194], random surfaces [261] and finite-size scaling [36].

These extensions generalize the theory chiefly in two directions. First, ensembles of *interacting* contours are introduced [202, 203, 21, 46, 79, 77, 287, 288, 48]. The interaction among contours (on top of volume exclusion) must be weak, to allow the convergence of the cluster expansions. Second, the set of reference configurations is replaced by a set of *reference measures*. These can be supported on whole classes of ground-state configurations [179] or, more generally, on families of configurations — *restricted ensembles* — suitably chosen so as to include entropy contributions. In many cases, these restricted ensembles are formed by low-energy excitations of ground states [80, 79, 77, 78, 48, 261], but other definitions are in principle possible. For instance, the ensemble for the disordered phase of the large-$q$ Potts model — and for other examples pertaining to the study of liquid-gas phase transitions [46] — is supported on "maximally disordered" configurations [46]. The latter corresponds to a "pure-entropy" restricted ensemble [46], as opposed to the "pure-energy" (just one ground state configuration) or "almost-pure-energy" (ground state plus excitations) used in most of the applications. In general, the restricted ensembles are chosen so to have minimal (restricted) *free* energy at the temperature of interest. Often, interacting contours and reference measures are alternative procedures; it is a matter of taste to choose one or the other.

A third direction in which PS theory has been extended — and one which is crucial for our application to the proof of RG pathologies in nonzero magnetic field (Section 4.3.6) — is towards the incorporation of random interactions. In work unpublished so far, Zahradník [370, 371] — generalizing the work of Bricmont and Kupiainen [43, 44] — has proven that for $d \geq 3$ the addition of a small enough random interaction only produces small deformations in the phase diagram. An important issue is the meaning of "small enough". In the original work [370], the random interaction was required to be uniformly small respect to the nonrandom part. Later we were informed [371] that the proof also applies to random contributions that are small *in probability*. We state the later version, which is the one suited to our applications.

**Theorem B.31** *Consider the lattice $\mathbb{Z}^d$, $d \geq 3$, and an interaction $\Phi_{\overline{\lambda}} = \Phi_0 + \sum_{i=1}^{r-1} \lambda_i \Phi_i$ satisfying the hypothesis of Theorem B.28. Add a finite-range random interaction $\Phi^{\mathrm{rdom}} = \left\{ \Phi_A^{\mathrm{rdom}}(\,\cdot\,, \kappa) \right\}_{A \in \mathcal{S}}$, where $\kappa$ is a random variable with probability distribution $P$, such that the random variables $\Phi_A^{\mathrm{rdom}}(\,\cdot\,, \kappa)$ and $\Phi_{A'}^{\mathrm{rdom}}(\,\cdot\,, \kappa)$ have the same distribution if $A'$ is a translate of $A$ and are independent if $A \cap A' = \emptyset$. Assume, in addition, the following smallness condition: For each $\delta > 0$ there exists $\epsilon(\delta)$ small enough such that*

$$P(|\Phi_A^{\mathrm{rdom}}(\omega, \kappa)| > \delta) \leq \epsilon \qquad (\text{B.69})$$

*for all $A \in \mathcal{S}$, $\omega \in \Omega$. Then, for $\beta$ large enough the phase diagrams for $\Phi_{\overline{\lambda}}$ and $\Phi_{\overline{\lambda}} + \Phi^{\mathrm{rdom}}$ are homeomorphic. More precisely, for $\beta$ large and $\epsilon$ uniformly small, there exists an homeomorphism*

$$L_{\beta, \epsilon} : V'_{\beta, \epsilon} \longrightarrow V''_{\beta, \epsilon} \qquad (\text{B.70})$$

*between two open sets $V'_{\beta, \epsilon}, V''_{\beta, \epsilon} \subset \mathbb{R}^{r-1}$ mapping an $r$-regular $\Phi_{\overline{\lambda}}$ phase diagram onto an $r$-regular $\Phi_{\overline{\lambda}} + \Phi^{\mathrm{rdom}}$ phase diagram. The homeomorphism $L_{\beta, \epsilon}$ tends to the identity as $\epsilon \to 0$ uniformly.*

## B.5 Application to the Examples of Section 4

### B.5.1 General Strategies

In principle, the verification of the Peierls condition for a certain interaction $\Phi_0$ is a two-stage process:

(A) Find all the periodic ground-state configurations of $\Phi_0$. This stage usually involves counting how many "frustrated bonds" each candidate configuration has. This should be followed by a proof showing that indeed no other periodic configuration has the same or less energy density, but this proof is usually omitted because it is either considered to be obvious or too messy to write down. Furthermore, as remarked before Definition B.20, this proof is not really necessary if the Peierls condition (next stage) can be successfully verified. In this regard, the tedious process of finding these reference configurations is a natural candidate for a computer-assisted procedure. However, this may not be possible, in general: the problem of checking whether a given periodic configuration is a ground state may not be algorithmically decidable (see [313, Section 4.15] and [316] and references therein for some related undecidability results, and see also [263]). In any case, this problem may be alleviated in practice if one works in the framework of PS theory. Indeed, the further steps of the theory work as a correcting mechanism: if too few configurations have been found, the Peierls condition will fail and the the configurations of large contours with very low energy density will give a hint of how additional ground-state configurations look like. On the other hand, if too many configurations have been selected, the spurious ones will be eventually ruled out in the sense that they will not give rise to pure phases; they lead to contour ensembles with high free-energy cost, which are not associated to $\tau$-functionals. [We owe this insight to conversations with Miloš Zahradník.]

(B) Devise a suitable notion of contour and show that the energy grows proportionally to its support. This is an extremely model-dependent process.

Often, the determination of ground-state configurations and of contour energies are done simultaneously: The comparison between the energies of different candidate configurations is done already with the help of contours. This is a manifestation of what we have repeatedly commented upon: the contour energy is the essential quantity; the crucial stage is (B). If we manage to show that contour energies are large enough then we do not need to care about the nature of the reference configurations, they are ground states by force. Yet again, the same argument used to prove the appropriate growth of contour energies, usually shows the ground-state character of the configurations. Proving (B) directly is not a substantial saving of misery.

Below, we shall use this canonical approach for decimation and Kadanoff-$p$ prescriptions, where we can resort to the Ising type of contours: polyhedra drawn on the dual lattice with plaquettes perpendicular to each frustrated bond. For the studies on other RG transformations we shall use instead the Holsztynski-Slawny criterion (Theorem B.21) based on the notion of $m$-potentials (Definition B.20). The corresponding procedure usually starts by rewriting the interaction into an equivalent form which is indeed an $m$-potential. This is the hard part of the game; it involves some knowledge of what the ground-state configurations look like. Once the rewriting is done, the actual verification of which configurations have minimal energy is in general simple; one studies one bond at a time. As already remarked, the disadvantage of this approach is that it does not supply a value for the Peierls constant, a fact that, in our case, prevents us from producing any estimate of the range of temperatures for which the pathologies of the RG transformations occur.

We observe that for the Steps 1–3 in the proofs of existence of pathologies, we do not need the full information on phase diagrams provided by PS theory. Rather, we are only interested in the point of maximum coexistence (for Step 1), and regions of uniqueness of Gibbs measure (for Step 2). Moreover, both types of questions refer to *different* systems: The maximum coexistence point must be determined for internal-spin systems obtained when the image spins do not favor any phase of the original system (this amounts, in general, to block spins chosen in an alternating or random way). In these cases, the symmetry-breaking perturbation is chosen simply as a uniform magnetic field and symmetry considerations imply that maximum coexistence is achieved only when this field is zero. Therefore, this symmetry-breaking interaction plays an almost invisible role, and it is only briefly mentioned. The only case in which the symmetry-breaking deserves careful consideration is that of systems which already initially have a magnetic field (Sections 4.3.6 and B.5.7). On the other hand, the uniqueness of Gibbs measures is of interest for internal-spin systems corresponding to block spins chosen so as to definitely favor one of the phases. It is not wise to think of these two internal-spin systems (determined by block spins either favoring or not favoring one pure phase) as living in the same phase diagram with block-spin flipping as symmetry-breaking interaction. The problem is that such an interaction can not be considered a perturbation: it goes by finite steps and hence may throw us out of the small-parameter region $V_\beta$ (see Theorem B.28) of the phase diagram where PS theory holds.

Let us now discuss the different applications starting from the simplest ones.

### B.5.2  Internal-Spin Systems with Unique Ground-State Configurations

In all the applications of Section 4 there is a step (Step 2.2) which involves showing that at low temperature the ensemble of internal (or original) spins has a unique Gibbs measure for some particular choice of image spins. These are all cases in which there is only one (periodic) ground-state configuration, due to the presence of a periodic single-sign magnetic field. The uniqueness of the Gibbs state is, basically, a consequence of PS theory plus Zahradník's completeness result (see the comment immediately following Theorem B.27). However, this would only prove uniqueness *among the periodic Gibbs measures*; we need Martirosyan's extension of this result [260] to prove uniqueness among the set of *all* Gibbs measures.

### B.5.3  Internal Spins under Decimation

In this case we can apply the canonical two-stage process described above to verify the Peierls condition. As discussed in detail in Sections 4.1.2 and 4.2 (Step 0), the ensemble of internal spins for a given image-spin configuration is just the ensemble of configurations of original spins with some of the spins constrained to be fixed. To the latter we can apply the usual Peierls construction of polyhedral contours. We shall therefore use the "thin" notion of contours discussed at the end of the two previous Sections B.4.2 and B.4.3. The *internal-spin* contours are obtained from the original-spin contours by removing the plaquettes adjacent to an image spin [thick lines in Figure 14(a)]. The argument that follows is thus based on comparing internal-spin with original-spin (= internal + image) quantities.

We fix the image spin in the fully alternating "+/−" configuration, so the two obvious candidates to ground-state configurations in the resulting system of *internal spins* are the same as for the original system: $\underline{\omega}^{(+)}$ equal to $+1$ everywhere, and $\underline{\omega}^{(-)}$ equal to $-1$ everywhere. That is, in terms of original spins, $\underline{\omega}^{(+)}$ (resp. $\underline{\omega}^{(-)}$) corresponds to all spins "+" ("−") except for a sublattice of period $2b$ — with $b$ being the decimation spacing — where the spins are flipped. The symmetry-breaking perturbation $\lambda_1 \Phi_1$ can be taken to be a magnetic field at each (internal) site. However, the symmetry of the problem (i.e. of the choice of block spins), implies that the coexistence of the "+" and "−" measures occurs at zero values of this field. We therefore forget about this extra field, and concentrate on proving that the zero-temperature phase diagram deforms little for low temperatures.

If we wish only to prove that $\underline{\omega}^{(+)}$ and $\underline{\omega}^{(-)}$ are all the internal-spin ground-state configurations, we can for instance consider the corresponding set of *original-spin* contours, which is just an array of $2b$-spaced unit cubes surrounding each flipped spin, and show that all other original-spin configurations lead to a system of (original-spin) contours with a larger area. This is not hard to do, for instance we can argue as follows: Observe that every interval parallel to the axis between two nearest-neighbor image spins necessarily contains at least one broken bond, as two nearest-neighbor image spins always have opposite signs. The choice of all internal spins either all $+$ or all $-$ has precisely this minimal choice of exactly one broken bond in each interval, and no other broken bonds.

However, such an argument is not really needed. As commented above, a more convenient approach for our purposes is to show directly that the insertion of a contour in $\underline{\omega}^{(+)}$ has an energy cost proportional to its area. Consider then the (internal-spin) configuration $\omega^\Gamma$ obtained by inserting a "+" (internal-spin) contour $\Gamma$ inside the configuration $\underline{\omega}^{(+)}$, that is, a region of "−" bounded by $\Gamma$ [Figure 14(a)]. Its relative energy can be written in the form:

$$H(\omega^\Gamma | \underline{\omega}^{(+)}) \;=\; 2J|\Gamma| - \Delta E_- + \Delta E_+ \;, \tag{B.71}$$

where $|\Gamma|$ is the area of the contour [number of plaquettes, or length of the thick lines for the two-dimensional example of Figure 14(a)], $\Delta E_-$ is the energy gain due to the fact that the "−" image spins inside of, or visited by, $\Gamma$ acquire "−" neighbors [e.g. the sites $a_i$ in Figure 14(a)], and $\Delta E_+$ is the additional energy due to "+" image spins in $\Gamma$.

To prove the Peierls condition we have to check that the extra contribution $\Delta E_- - \Delta E_+$ is proportional to the area of the contour with a not too large proportionality constant. The intuition is clear: The contributions corresponding to "+" and "−" image spins inside the volume cancel each other, except possibly for a layer of image spins placed close to the contour. This correction is hence of the order of the area of the contour, with proportionality constant given roughly by the inverse of the separation between these image spins. However, this last bound is not applicable if the contour involves few (e.g. one) image spins. In this sense, it is natural to distinguish between contours surrounding and contours avoiding the image spins. While the former "feel"

(a)

(b)

Figure 14: System of internal spins (small circles) under decimation, when the image spins (squares) are fixed in the fully alternating "+/−" configuration. (a) A "−" contour $\Gamma$ (thick lines). (b) The fundamental bonds (bounded by thick lines) of an equivalent $m$-potential.

the sublattice of image spins, the energy contribution of the latter is almost the same as for the usual Ising model. This produces an estimation of the Peierls constant (and hence of the critical temperature), with two competing terms: one depending on the decimation period $b$ (and tending to 0 as $b$ tends to infinity), and another independent of $b$ and close to the Peierls constant for the Ising model (the closer the higher the dimension).

To formalize these ideas we choose one coordinate axis, say the one labelled 1, and perform the cancellations by sweeping in order along it. To abbreviate, let us call "left" the direction in $\mathbb{Z}^d$ towards smaller 1-components, and "right" the opposite direction. Also, we shall say that an image spin is "in" the contour if it is visited by it or it is contained in its volume. More specifically, an "internal spin with $l$ plaquettes in the contour" is an internal spin such that the contour surrounds $l$ of the plaquettes of the unit cube centered on it (e.g. in Fig. 14(a), the internal spin at $a_1$ has 1 plaquette in the contour, the one at $a_4$ has 3, etc.). Note that, in such a case, the energy contribution of the image spin is

$$\Delta E_- = 2|J|l \qquad (B.72a)$$
$$\Delta E_+ = 0 \qquad (B.72b)$$

if it is a "$-$" spin, and the converse for a "$+$" image spin. The cancellation can be done, for instance, as follows: For each line in the left-right direction intersecting the contour, we choose the image spin in the contour further to the left and look for the next image spin in the contour, of opposite sign and located to the right and along the same line. If the "$+$" image spin has the same number of plaquettes or more in the contour than the "$-$", we cancel both contributions (obtaining a lower bound for the energy if the number of plaquettes is not the same). Otherwise we do nothing. We then proceed to the next uncancelled image spin along the same line, always travelling towards the right. Once all the left-right lines have been scanned, we obtain a lower bound for the energy of the form (B.71) but where $\Delta E_-$ and $\Delta E_+$ refer only to a layer of image spins in the contour at a distance not exceeding $b+1$ from it (remaining spins). The energy gain due to these remaining spins can not exceed that of the case in which there are no "$+$" image spins left and all the "$-$" spins have their $2d$ plaquettes in the contour. We therefore bound

$$\Delta E_- - \Delta E_+ \leq 2dJN_- , \qquad (B.73)$$

where $N_-$ is the number of "$-$" image spins inside the above-mentioned layer.

To complete the Peierls bound, we have to relate $N_-$ to the contour area $|\Gamma|$. At this point we must distinguish between contours with $N_- \geq 2$ ("wide contours") and contours with $N_- \leq 1$ ("narrow contours"). For the wide contours, the key observation is that each two "$-$" image spins must be at least a distance $2b$ apart in each coordinate direction and the contour must pass at a distance $b+1$ or less of each of them. Therefore, the number of remaining "$-$" image spins for a given value of $\Gamma$ can not exceed that of the case in which all of the spins are located so as to form a "tube", separated $2b$ units from each other, and $\Gamma$ being the wall of such a "tube":

$$N_- \leq \frac{|\Gamma|}{2(d-1)(b-1)} \,. \tag{B.74}$$

(Note that the bound contains $b-1$, rather than $b$, because the plaquettes corresponding to "+" image spins are not part of the contour.) From (B.71)–(B.74), we conclude that

$$H(\omega^\Gamma|\underline{\omega}^{(+)}) \geq 2J\left[1 - \frac{d}{2(d-1)(b-1)}\right]|\Gamma| \,; \quad N_- \geq 2 \,. \tag{B.75}$$

This bound is not useful for the limit case $d = 2$, $b = 2$; but for it we have already good bounds for the critical temperature (Section 4.1.2).

Let us now consider the narrow contours ($N_- \leq 1$). It is not hard to convince oneself that the worst case is when the contour visits only one "−" image spin, which has $l$ plaquettes in the contour. The contour must then include the opposite $l$ plaquettes plus the plaquettes needed to join these among themselves or/and to the cube so to form a closed surface. One can check that the worst situation (smallest ratio $l/|\Gamma|$) is when the $l$ plaquettes of the cube are not consecutive, in which case the contour includes the $l$ opposite plaquettes and the $2(d-1)l$ plaquettes needed to join them to the cube. Hence,

$$|\Gamma| \geq l[2(d-1) + 1] \,.$$

Therefore, using (B.72),

$$\Delta E_- - \Delta E_+ \leq \frac{2J}{2d-1}|\Gamma|$$

and

$$H(\omega^\Gamma|\underline{\omega}^{(+)}) \geq 2J\left[1 - \frac{1}{2d-1}\right]|\Gamma| \,; \quad N_- \leq 1 \,. \tag{B.76}$$

(Another way to interpret this Peierls bound is by noting that the narrow contour with the least energy cost is the one produced by a single flipped internal spin adjacent to a "−" image spin.)

Formulas (B.75) and (B.76) show both that the configurations $\underline{\omega}^{(+)}$ and $\underline{\omega}^{(-)}$ are indeed the only periodic ground-state configurations and that the Peierls condition is satisfied for the internal-spin system with Peierls constant

$$\rho_0 \geq 2J(1 - M_{d,b}) \,, \tag{B.77}$$

where

$$M_{d,b} = \max\left\{\frac{1}{2d-1}, \frac{d}{2(d-1)(b-1)}\right\} \,. \tag{B.78}$$

This value of $\rho_0$, together with the estimates (B.58) for $\beta_0$ and (B.68) for the entropy factor $\alpha$ of thin contours, implies there is a phase transition at least for

$$\beta \geq \frac{4\log(2d-1) + 1/(2e)}{2J(1 - M_{d,b})} \,. \tag{B.79}$$

This estimate is probably very weak, but at least it increases with $d$ and with $b$ as it should. In fact, if $b$ is large, the alternating fields are very far apart, and the system becomes almost indistinguishable from a zero-field Ising model. Therefore we expect, but can not prove, that this limit temperature approaches the Ising-model critical temperature when $b$ tends to infinity.

Inequality (B.79) determines the range of temperatures for which we can prove that the decimation transformation has pathologies (Sections 4.2 and 4.3.2).

As a warm-up for the following sections, let us sketch how the Peierls condition can be verified in the present example using the Holsztynski-Slawny criterion (Theorem B.21). The argument depends slightly on the decimation spacing $b$. For $b = 2$ the internal-spin interaction is already an $m$-potential because it is just a ferromagnetic two-body interaction in a "diluted" lattice (Sections 4.2 and 4.3.1). For $b \geq 3$ a periodic $m$-potential is obtained by grouping all the bonds inside cubes containing at least one period of the image-spin configuration [Figure 14(b)]. Explicitly, if $\Phi^{\text{internal}}$ is the interaction for the internal-spin system, the equivalent $2b$-periodic $m$-potential $\Phi^{m\text{-pot}}$ has $\Phi_A^{m\text{-pot}} = 0$ unless $A$ is a periodic translate, with period $2b$, of the cube

$$V = \left[ -\left\lfloor \frac{b-1}{2} \right\rfloor, \left\lfloor \frac{3b-1}{2} \right\rfloor \right]^d \tag{B.80}$$

(here $\lfloor \rfloor$ denotes integer part), or of the inter-cube bonds formed by nearest-neighbor pairs $\langle x, y \rangle$ with, say, $x$ in the (internal) boundary of $V$. For these pairs $\Phi_{\{xy\}}^{m\text{-pot}} = \Phi_{\{xy\}} = -J$, and for the cube $V$

$$\Phi_V^{m\text{-pot}} = \sum_{A \subset V} \Phi_A^{\text{internal}}. \tag{B.81}$$

It is not hard to verify that the configurations $\underline{\omega}^{(+)}$ and $\underline{\omega}^{(-)}$ are the only minimizers of the $\Phi_V^{m\text{-pot}}$, and they obviously also minimize the energy of the inter-cube bonds. Therefore, $\Phi^{m\text{-pot}}$ is an $m$-potential with a finite number of ground-state configurations. By Theorem B.21, the Peierls condition is satisfied. No estimation of $\beta_0$ follows from this approach.

### B.5.4 Internal Spins under the Kadanoff Transformation. Uniformity

For this case we have to apply the uniformity results that we so carefully stated above. The Hamiltonian (4.33) can be decomposed in the form

$$H_{\text{eff}} = H_0 + \widetilde{H}_p \tag{B.82}$$

where $H_0$ is the usual Ising Hamiltonian and $\widetilde{H}_p$ corresponds to the interaction $\widetilde{\Phi}(p, \beta)$ defined by

$$(\widetilde{\Phi}_A(p, \beta))(\sigma) = \begin{cases} -(p/\beta)\sigma'_x \sigma_i & \text{if } A = \{i\} \text{ and } i \in B_x \\ (1/\beta) \log 2 \cosh\left(p \sum_{i \in B_x} \sigma_i\right) & \text{if } A = B_x \\ 0 & \text{otherwise}. \end{cases} \tag{B.83}$$

(We recall that in this appendix we are "unabsorbing" $\beta$ which in (4.33) is absorbed *only* in $J$.) We choose $\omega'_{\text{special}}$ as some alternating configuration—with as many pluses as minuses—so that, by symmetry, the coexistence of the "+" and "−" internal-spin Gibbs measures does not require any additional field. That is, as before we forget about symmetry-breaking interactions.

The interaction $\Phi_0$ satisfies the original Peierls condition with $\overline{\rho}_0 = 2J$ (thin contours, $\mathcal{K} = \{\omega^{(+)}, \omega^{(-)}\}$). We can then resort to Corollary B.25 to conclude that the whole interaction $\Phi_0 + \tilde{\Phi}(p, \beta)$ satisfies the original Peierls condition. However, to estimate the Peierls constant we must correct $\overline{\rho}_0$ so as to satisfy (C1) and (C2) for the total interaction. For instance (see remarks at the end of Section B.4.2), we can replace it by

$$\rho_0 = \frac{2\beta J}{(2a+1)^d},$$

with $a$ equal to half the length of the largest side of the block. We then conclude that for each $p$ there exists a value $\beta_1(p)$ defined by

$$\frac{p}{\beta_1} + \frac{1}{\beta_1}\log(2\cosh p|B|) = \frac{2cJ}{(2a+1)^{2d}}, \quad \text{(B.84)}$$

such that for $\beta \geq \beta_1(p)$ the effective internal-spin interaction for the $p$-Kadanoff transformation satisfies the Peierls condition with constant

$$\rho = \frac{2J(1-2c)}{(2a+1)^d}. \quad \text{(B.85)}$$

In the last two formulas, the constant $c$ is arbitrary as long as $0 < c < 1/2$. We shall find an optimal choice below.

At this point we can apply, for instance, Corollary B.29 to obtain that *for each finite $p$ there exists a value $\beta_0(p)$ such that for $\beta \geq \beta_0(p)$ the system of internal spins corresponding to a $p$-Kadanoff transformation has a nontrivial phase diagram with a first-order phase transition between a "+" and a "−" Gibbs measure.* The formulas (B.66) and (B.68) imply the estimate

$$\beta_0 \geq \max\left\{\beta_1(p), [4\log(2d-1) + (2e)^{-1}]\frac{(2a+1)^d}{2J(1-2c)^2}\right\}. \quad \text{(B.86)}$$

From the point of view of this bound, the optimal choice of $c$ is when $\beta_1(p)$ equals the competing term in the RHS of (B.86). This produces the rather ugly-looking bound

$$\beta_0 \geq \frac{(2a+1)^{2d}}{2J} \frac{16L_{p,|B|}^2}{M_d[4L_{p,|B|} + M_d - \sqrt{M_d(M_d + 8L_{p,|B|})}]}, \quad \text{(B.87)}$$

with

$$M_d = 4\log(2d-1) + (2e)^{-1}$$
$$L_{p,|B|} = p + \log 2\cosh p|B|.$$

Formula (B.87) gives a lower bound for the temperature up to which a Kadanoff $p$-transformation exhibits pathologies (Section 4.3.3). This bound goes to zero if $p$ tends to infinity, hence it is useless for majority-rule transformations.

### B.5.5  Internal Spins under Majority Rule

For this case, we use the Holsztynski-Slawny criterion (Theorem B.21). The system of internal spins subjected to the constraint of a doubly-alternating $7 \times 7$ block-spin configuration can be written as a periodic $m$-potential with period 28. The fundamental bonds of this potential are the squares of size $28 \times 28$ depicted in Figure 8(a), and all the bonds connecting neighboring squares. It is straightforward to check that the configurations of Figure 8(b) are the only ones satisfying all the bonds of this $m$-potential. Hence, the system has a finite number (two) of ground states which, by Theorem B.21 and PS theory, give rise to different and coexisting Gibbs measures at low-enough temperature. By symmetry considerations this coexistence takes place at zero values of the symmetry-breaking field. An analogous argument can be used for all the other block-sizes $b_k$ given in (4.35).

### B.5.6  Internal Spins under Block-Averaging Transformations

Again, we resort to the Holsztynski-Slawny criterion (Theorem B.21). The system of internal spins subjected to the constraint of zero average spin in every $2 \times 2$ block can again be written as an m-potential which is periodic (with period 2). The fundamental bonds are the $2 \times 2$ squares and the bonds connecting them, and the only four periodic ground states satisfying every bond are easily seen to be the ones depicted in Figure 9. Thus at sufficiently low temperature PS-theory provides the phase transition needed in Step 1. Again symmetry allows us to dispose of any symmetry-breaking field to follow the coexistence point.

### B.5.7  Internal Spins when $h \neq 0$. Random Field

The result needed to prove the presence of pathologies for non-zero field in the decimation and Kadanoff examples of Section 4.3, is a direct consequence of Zahradník's Theorem B.31. We apply this theorem with $\Phi_0$ equal to the interaction for the system of internal spins with fully alternating image spins, and the symmetry-breaking perturbation $\Phi_1$ taken to be a uniform magnetic field. The random interaction is the random magnetic field induced by those block spins that were flipped from "+" to "−" with probability $\epsilon/(2J)$, as discussed in Section 4.3.6. By Theorem B.31, the resulting low-temperature and low-$\epsilon$ phase diagram is only a small perturbation of the phase diagram for the non-random part, which is itself a small perturbation of its zero-temperature phase diagram. In particular, the ground-state energy for almost all realizations of the Hamiltonian is an almost linear function of the parameters $h$ and $\epsilon$. For $\epsilon$ sufficiently small, the linearity is only weakly violated, and the compensating uniform field (that is, the value of $h$ as a function of $\epsilon$ needed to keep the system on a coexistence surface) is an almost linear — hence strictly increasing — function.

## C  Solution of the Diophantine Equation (4.34)

Consider the pair of Diophantine equations $b^2 = 2c^2 \pm 1$ ($b, c$ integers $\geq 1$). The following intuition was suggested to us by Vincent Rivasseau: If $(b, c)$ satisfies either of these equations, then $b/c$ must be an excellent rational approximation to $\sqrt{2}$, in the sense that

$$\left| b/c - \sqrt{2} \right| = \frac{|b^2/c^2 - 2|}{b/c + \sqrt{2}} \leq \frac{1}{\sqrt{2}\, c^2} \,. \tag{C.1}$$

(Note that, by contrast, for "typical" integer denominators $c$, one has $\inf_{b \in \mathbb{Z}} |b/c - \sqrt{2}| \sim 1/c \gg 1/c^2$.) Now, the best rational approximations to $\sqrt{2}$ can be obtained from the continued fraction [234, 317]

$$\sqrt{2} - 1 = \cfrac{1}{2 + \cfrac{1}{2 + \cfrac{1}{2 + \cdots}}} \,. \tag{C.2}$$

This suggests to consider the recursion

$$x_{n+1} = \frac{1}{2 + x_n} \,, \tag{C.3}$$

which converges to $\sqrt{2} - 1$ as $n \to \infty$ (for any $x_0 > -2$); equivalently, defining $y_n = x_n + 1$, we find the recursion

$$y_{n+1} = \frac{2 + y_n}{1 + y_n} \,, \tag{C.4}$$

which converges to $\sqrt{2}$ (for any $y_0 > -1$). In particular, setting $y_n = b_n/c_n$ with $b_n, c_n$ positive integers, we find the linear recursion

$$b_{n+1} = b_n + 2c_n \tag{C.5a}$$
$$c_{n+1} = b_n + c_n \tag{C.5b}$$

Now this recursion has the remarkable property that

$$b_{n+1}^2 - 2c_{n+1}^2 = -(b_n^2 - 2c_n^2) \,; \tag{C.6}$$

in particular, if $b_n^2 = 2c_n^2 \pm 1$, then $b_{n+1}^2 = 2c_{n+1}^2 \mp 1$. Therefore, if we start from $(b_0, c_0) = (1, 1)$, we generate pairs $(b_n, c_n)$ which satisfy $b_n^2 = 2c_n^2 - 1$ (resp. $b_n^2 = 2c_n^2 + 1$) for even (resp. odd) values of $n$. The explicit formula is

$$b_n = \frac{1}{2}\left[ (1 + \sqrt{2})^{n+1} + (1 - \sqrt{2})^{n+1} \right] \tag{C.7a}$$

$$c_n = \frac{1}{2\sqrt{2}}\left[ (1 + \sqrt{2})^{n+1} - (1 - \sqrt{2})^{n+1} \right] \tag{C.7b}$$

To prove that this sequence constitutes the *complete* set of integer solutions to $b^2 = 2c^2 \pm 1$, we run the iteration (C.5) backwards: given $(b,c)$, we define

$$b' = -b + 2c \qquad (C.8a)$$
$$c' = b - c \qquad (C.8b)$$

and show that repeated application of this map must eventually lead to the pair $(1,1)$.

**Lemma C.1** *Let $b, c \geq 1$ be integers satisfying $b^2 = 2c^2 \pm 1$. Then $b', c'$ are integers satisfying $0 < b' \leq b$, $0 \leq c' < c$ and $b'^2 = 2c'^2 \mp 1$.*

PROOF. (a) $c \geq 1$ implies $b^2 = 2c^2 \pm 1 \geq 2c^2 - 1 \geq 2c^2 - c^2 = c^2$. Hence $b \geq c$ and $c' \geq 0$. Also $b - b' = 2(b - c) \geq 0$, so $b' \leq b$.
(b) $c \geq 1$ implies $b^2 = 2c^2 \pm 1 \leq 2c^2 + 1 \leq 2c^2 + c^2 = 3c^2 < 4c^2$. Hence $b < 2c$ and $b' > 0$. Also $c - c' = 2c - b > 0$, so $c' < c$.
(c) $b'^2 - 2c'^2 = (2c - b)^2 - 2(b - c)^2 = -(b^2 - 2c^2) = \mp 1$. ∎

Since $c$ strictly decreases at each iteration of (C.8), we must eventually reach $c = 1$, hence $b = 1$. Since (C.8) is the inverse of (C.5), the original pair $(b,c)$ must be $(b_n, c_n)$ for some $n$. We have therefore proven:

**Theorem C.2** *A pair of integers $b, c \geq 1$ satisfies the Diophantine equation $b^2 = 2c^2 - 1$ (resp. $b^2 = 2c^2 + 1$) if and only if $(b,c) = (b_n, c_n)$ for some even (resp. odd) integer $n \geq 0$.*

In particular, the block sizes $b$ for which the majority-rule construction in Section 4.3.4 works are $b_2, b_4, b_6, \ldots = 7, 41, 239, 1393, 8119, \ldots$.

**Remarks.** 1. After completing this proof, we learned that it was previously published by Theon of Smyrna [344] circa 130 A.D., and probably goes back to the Pythagorean school [301]; the identity (C.6) is proven geometrically in Euclid's *Elements* (Book II, Proposition 10). The special case $b = 7$, $c = 5$ is mentioned in Plato's *Republic* (546 C), though without the renormalization-group application. For a history, see Heath [191], Dickson [76, Chapter XII], Tannery [343] and Mugler [270].

2. One might wonder about the rational approximants to $\sqrt{2}$ obtained by using Newton's method $y \mapsto \frac{1}{2}(y + 2/y)$. Setting $y = b/c$, we obtain the *nonlinear* recursion $(b,c) \mapsto (b^2 + 2c^2, 2bc) \equiv (\hat{b}, \hat{c})$. It is straightforward to prove by induction that if $(b,c) = (b_n, c_n)$, then $(\hat{b}, \hat{c}) = (b_{2n+1}, c_{2n+1})$. So Newton's method generates a *subsequence* of the continued-fraction sequence.

It is natural to ask whether our construction in Section 4.3.4 can be extended to Ising models in dimension $d \geq 3$. Clearly this works if and only if the block size $b$ and island size $c$ satisfy the Diophantine equation

$$1 + b^d = 2c^d. \qquad (C.9)$$

Unfortunately, we suspect that for $d \geq 3$ there are *no* positive-integer solutions to (C.9) other than $b = c = 1$. The best results we have been able to glean from the mathematical literature are summarized in Theorems C.3 and C.4:

**Theorem C.3** *Let $l$ be a positive integer satisfying any one of the following three conditions:*

(a) *$l = 3$; or*

(b) *$l = 4$; or*

(c) *$l$ is a regular prime[78] such that the exponent of 2 mod $l$ is[79] either $(l-1)/2$ or even, and such that $2^{l-1} \not\equiv 1 \pmod{l^2}$.*

*Let $d$ be any multiple of $l$ (including $l$ itself). Then the only positive-integer solutions to $x^d + y^d = 2z^d$ are $x = y = z$. In particular, the only positive-integer solution to $1 + b^d = 2c^d$ is $b = c = 1$.*

Theorem C.3 has a long history. Obviously, if the theorem holds for any given power $l$, it trivially holds also for multiples of that power. The case $l = 3$ was proven by Euler sometime before 1770 [76, p. 572]; much more general results are now known [265, pp. 126, 203, 220]. The case $l = 4$ is a specialization of a theorem proven by Schopis in 1825 [76, p. 618] [265, p. 18]; again, more general results are now known [265, pp. 271, 274, 276]. The case $l = 5$ was proven in the mid-nineteenth century, but the proper attribution is unclear. Dénes [74] credits Dirichlet [81], but our reading of Dirichlet's paper indicates that he treated numerous cases of $x^5 + y^5 = Az^5$ but *not* $A = 2$ (see also [76, p. 735]). The correct attribution seems to be V.A. Lebesgue in 1843 [235] [76, p. 738]. See also [76, pp. 755–756] and [265, p. 276] for generalizations.

The case (c) was proven by Dénes [74] in 1952. To interpret it, note that the first irregular primes are $37, 59, 67, 101, \ldots$. The first regular primes for which the exponent condition fails are $31, 73, 89, 127, \ldots$. Finally, the only primes $l < 6 \times 10^9$ (and indeed the only ones currently known) for which $2^{l-1} \equiv 1 \pmod{l^2}$ are 1093 and 3511 [308, pp. 263–264]. Thus, the first primes for which condition (c) fails are $31, 37, 59, 67, 73, 89, 101, \ldots$. In particular, Theorem C.3 holds for all exponents $d \leq 100$ except possibly $31, 37, 59, 62, 67, 73, 74, 89$.[80]

**Theorem C.4** *For arbitrary $d \geq 3$, there is at most one positive-integer solution to $1 + b^d = 2c^d$ other than $b = c = 1$.*

---

[78]A prime $l > 3$ is called *regular* if it does not divide any of the numerators of the Bernoulli numbers $B_2, B_4, \ldots, B_{l-3}$ expressed in lowest terms.

[79]The exponent of 2 mod $l$ is the smallest integer $n \geq 1$ such that $2^n \equiv 1 \pmod{l}$.

[80]Dénes' paper [74] contains a list of all primes $l < 619$ for which condition (c) fails. This list has, however, a few mistakes: The primes 389 and 613 should be added to the list of irregular primes [247]; the exponent of 2 mod 281 (resp. mod 563) is 70 (resp. 562); and the final list in his article should read $31, 73, 89, 127, 151, 223, 337, 431, 439, 601$.

Theorem C.4 is a special case of a result of Domar [98], who proves that for arbitrary integers $A, B \geq 1$ and $d \geq 5$, the equation $|Ax^d - By^d| = 1$ has at most two solutions in positive integers $x, y$. See also [111].

The conjectured unsolvability of $x^d + y^d = 2z^d$ for $d \geq 3$ is a special case of an "extended Fermat's last theorem" which might conceivably be true [142, 62]:

**Conjecture C.5** *Let $d$ and $a$ be integers, with $d \geq 3$ and $1 \leq a \leq d$. Then there are no solutions of $x^d + y^d = az^d$ in nonzero integers, except for $x = y = z$ in the case $a = 2$.*

We find it amusing that a real problem in physics should be connected with Fermat's last theorem, but we think that this is an artifact of our method of proof and *not* an intrinsic fact about majority-rule transformations. Indeed, we suspect that Theorem 4.5 holds for all block sizes $b \geq 2$ and all dimensions $d \geq 3$, without regard for subtle number-theoretic properties. But we could be wrong.

# Acknowledgments

We are very grateful to Jean Bricmont, Anna Hasenfratz, Michael Kiessling, Jesús Salas, Dan Stein and Doug Toussaint for many helpful comments on a first draft of this paper. In addition, we wish to thank Pavel Bleher, Carlo di Castro, Joel Cohen, Roland Dobrushin, Michael Fisher, Jürg Fröhlich, Stuart Geman, Hans-Otto Georgii, Antonio González-Arroyo, Bob Griffiths, Robert Israel, Tom Kennedy, Roman Kotecký, Antti Kupiainen, Joel Lebowitz, Christian Maes, Fabio Martinelli, Jaček Miękisz, Chuck Newman, Enzo Olivieri, Charles Radin, Vincent Rivasseau, Roberto Schonmann, Senya Shlosman, Bob Swendsen, Srinivasa Varadhan, Marinus Winnink and Miloš Zahradník for helpful conversations and correspondence.

Two of the authors (R.F. and A.D.S.) thank the Università di Roma "Tor Vergata" and the Rijksuniversiteit Groningen for hospitality while this paper was being written. The other author (A.C.D.v.E.) thanks the ETH–Hönggerberg and the EPFL–Lausanne for hospitality during this same period.

Last but not least, we wish to thank Joel Lebowitz for being willing even to *consider* a paper of this length for publication in the *Journal of Statistical Physics*.

The research of the first author (A.C.D.v.E.) has been made possible by a fellowship of the Royal Netherlands Academy of Arts and Sciences (KNAW). The research of the second author (R.F.) was supported in part by the Schweizer Nationalfonds and by the Fonds National Suisse. The research of the third author (A.D.S.) was supported in part by U.S. National Science Foundation grants DMS–8705599, DMS–8911273 and DMS–9200719 and by U.S. Department of Energy contract DE-FG02-90ER40581.